\documentclass[longauth]{aa}  

\usepackage{graphicx}
\usepackage{txfonts}
\usepackage{subcaption}         % necessary for continued figures, example in section 3
\usepackage{lscape}             % to rotate a single page table, example in appendix.
\usepackage{placeins}           % useful with \FloatBarrier, to keep 
\usepackage{afterpage}
                                
\usepackage[dvipsnames]{xcolor}
\graphicspath{ {./figures/} }
\usepackage[colorlinks=true, citecolor=blue, linkcolor=blue]{hyperref}

\def\e#1{\times 10^{#1}}

\def\msol{\mathrm{M}_\odot}
\def\lsol{\mathrm{L}_\odot}

\def\h2{$\mathrm{H}_2$}
\def\so2{$\mathrm{SO}_2$}
\def\spy{\;\msol~\mathrm{ yr}^{-1}}

\def\citeapos#1{\citeauthor{#1}'s (\citeyear{#1})}  %Defines a possessive citation
\def\hc3n{HC$_3$N}
\def\deg{^\circ}

\newcommand*\edits[1]{{#1}}
\newcommand*\editss[1]{{#1}}

\newcommand*\pigru{{$\pi^1$~Gru}}
\newcommand*\ltpup{{L$_2$~Pup}}
\newcommand\kms{km\,s$^{-1}$}
\newcommand\micron{$\mu$m}

\begin{document}

   \title{ATOMIUM: Continuum emission and evidence of dust enhancement from binary motion}

%   \subtitle{Subtitle}

\author{
T.~Danilovich\inst{\ref{kul},\ref{monash}},
N. Samaratunge\inst{\ref{monash}},
Y.~L.~Mori\inst{\ref{monash}},
A.~M.~S.~Richards\inst{\ref{man}},
A.~Baudry\inst{\ref{bordeaux}},
S.~Etoka\inst{\ref{man}},
M.~Montarg\`es\inst{\ref{Inst:LESIA}},
P.~Kervella\inst{\ref{Inst:LESIA}, \ref{inst:frchile}},
I.~McDonald\inst{\ref{man},\ref{Inst:OpenU}},
C.~A.~Gottlieb\inst{\ref{harvard}},
A.~Wallace\inst{\ref{monash}},
D.~J.~Price\inst{\ref{monash}},
L.~Decin\inst{\ref{kul},\ref{leeds}},
J.~Bolte\inst{\ref{kiel}},
T.~Ceulemans\inst{\ref{kul}},
F.~De~Ceuster\inst{\ref{kul}},
A.~de~Koter\inst{\ref{Inst:Amsterdam},\ref{kul}},
D.~Dionese\inst{\ref{Inst:ULB},\ref{warsaw}},
I.~El~Mellah\inst{\ref{Inst:Univ_Santiago}, \ref{Inst:USACH}},
M.~Esseldeurs\inst{\ref{kul}},
M.~Gray\inst{\ref{man},\ref{thai}},
F.~Herpin\inst{\ref{bordeaux}},
%M.~Jeste\inst{\ref{mpibonn}},
T.~Khouri\inst{\ref{Inst:Chalmers}},
E.~Lagadec\inst{\ref{nice}},
C.~Landri\inst{\ref{kul}},
L.~Marinho\inst{\ref{bordeaux}, \ref{csic}},
K.~M.~Menten\inst{\ref{mpibonn}}$^,$\thanks{Passed away on 30/12/2024},
T.~J.~Millar\inst{\ref{belfast}},
H.~S.~P.~M\"uller\inst{\ref{koln}},
B.~Pimpanuwat\inst{\ref{thai}},
J.~M.~C.~Plane\inst{\ref{leeds}},
R.~Sahai\inst{\ref{Inst:CalTech}},
L.~Siess\inst{\ref{Inst:ULB}},
M.~Van~de~Sande\inst{\ref{Inst:Leiden}},
O.~Vermeulen\inst{\ref{kul}},
K.~T.~Wong\inst{\ref{Inst:Uppsala}},
J.~Yates\inst{\ref{Inst:UCL}},
A.~Zijlstra\inst{\ref{man},\ref{Inst:Macquarie}}
%and the ATOMIUM collaboration$^{x}$
}

% List of institutions
\institute{
Institute of Astronomy, KU Leuven, Celestijnenlaan 200D,  3001 Leuven, Belgium \label{kul}
\and 
School of Physics \& Astronomy, Monash University, Wellington Road, Clayton 3800, Victoria, Australia\\
\email{taissa.danilovich@monash.edu}\label{monash}
%\and ARC Centre of Excellence for All Sky Astrophysics in 3 Dimensions (ASTRO 3D), Clayton 3800, Australia
\and 
JBCA, Department Physics and Astronomy, University of Manchester, Manchester M13 9PL, UK \label{man}
\and 
{Universit\'e de Bordeaux, Laboratoire d'Astrophysique de Bordeaux, 33615 Pessac, France}\label{bordeaux}
\and 
{LIRA, Observatoire de Paris, Universit\'e PSL, Sorbonne Universit\'e, Universit\'e Paris Cit\'e, CY Cergy Paris Universit\'e, CNRS, 92195 Meudon CEDEX, France \label{Inst:LESIA}}
\and
{French-Chilean Laboratory for Astronomy, IRL 3386, CNRS and Universidad de Chile, Casilla 36-D, Santiago, Chile \label{inst:frchile}}
\and 
               Open University, Walton Hall, Milton Keynes, MK7 6AA, UK\label{Inst:OpenU}
            \and
{Harvard-Smithsonian Center for Astrophysics, 60 Garden Street, Cambridge, MA 02138, USA\label{harvard}}
\and
            {Department of Mathematics, Kiel University, Heinrich-Hecht-Platz 6, 24118 Kiel, Germany}\label{kiel}
            \and
            {School of Chemistry, University of Leeds, Leeds LS2 9JT, UK\label{leeds}}
\and 
University of Amsterdam, Anton Pannekoek Institute for Astronomy, 1090 GE, Amsterdam, The Netherlands\label{Inst:Amsterdam}
              \and        
              Institut d'Astronomie et d'Astrophysique, Universit\'e Libre de Bruxelles (ULB), CP 226, 1060, Brussels, Belgium\label{Inst:ULB}
              \and
              Astronomical Observatory, University of Warsaw, Al. Ujazdowskie 4, 00-478 Warsaw, Poland\label{warsaw}
              \and
              Departamento de F\'isica, Universidad de Santiago de Chile, Av. Victor Jara 3659, Santiago, Chile\label{Inst:Univ_Santiago}
              \and
              Center for Interdisciplinary Research in Astrophysics and Space Exploration (CIRAS), USACH, Chile\label{Inst:USACH}
            \and
            {National Astronomical Research Institute of Thailand, Chiangmai 50180, Thailand}\label{thai}
            \and
            Instituto de F\'isica Fundamental, CSIC, C/ Serrano123, 28006 Madrid, Spain \label{csic}
            \and
            {Max-Planck-Institut f{\" ur} Radioastronomie, Auf dem H{\" u}gel 69, 53121 Bonn, Germany}\label{mpibonn}
              \and
              Chalmers University of Technology, Onsala Space Observatory, 43992, Onsala, Sweden\label{Inst:Chalmers}
            \and
            {Universit\'e C\^ote d'Azur, Laboratoire Lagrange, Observatoire de la C\^ote d'Azur, F-06304 Nice Cedex 4, France}\label{nice}
            \and
            {Astrophysics Research Centre, School of Mathematics and Physics, Queen's University Belfast, University Road, Belfast BT7 1NN, UK}\label{belfast}
            \and
            {Universit\"at zu K\"oln, Astrophysik/I. Physikalisches Institut, 50937 K\"oln, Germany}\label{koln}
\and 
California Institute of Technology, Jet Propulsion Laboratory, Pasadena, CA, 91109, USA\label{Inst:CalTech}
              \and
              Leiden Observatory, Leiden University, P.O. Box 9513, 2300 RA Leiden, The Netherlands\label{Inst:Leiden}
\and Theoretical Astrophysics, Department of Physics and Astronomy, Uppsala University, Box 516, SE-751 20 Uppsala, Sweden\label{Inst:Uppsala}
            \and
            University College London, Department of Physics and Astronomy, London, WC1E 6BT, UK\label{Inst:UCL}
              \and
              School of Mathematical and Physical Sciences, Macquarie University, Sydney, New South Wales, Australia\label{Inst:Macquarie}
}

\titlerunning{ATOMIUM: Continuua}
\authorrunning{Danilovich, et al}

   \date{Received 31 March 2025 / Accepted ...}

% \abstract{}{}{}{}{}
% 5 {} token are mandatory
 
  \abstract
   {Low- and intermediate-mass stars on the asymptotic giant branch (AGB) account for a significant portion of the dust and chemical enrichment in their host galaxy. Understanding the dust formation process of these stars and their more massive counterparts, the red supergiants, is essential for quantifying galactic chemical evolution.}
   {To improve our understanding of the dust nucleation and growth process, we aim to better constrain stellar properties at millimetre wavelengths. To characterise how this process varies with mass-loss rate and pulsation period, we study a sample of oxygen-rich and S-type evolved stars.} 
   {Here we present ALMA observations of the continuum emission around a sample of 17 stars from the ATOMIUM survey. We analyse the stellar parameters at 1.24~mm and the dust distributions at high angular resolutions.}
   {From our analysis of the stellar contributions to the continuum flux, we find that the semi-regular variables all have smaller physical radii and fainter monochromatic luminosities than the Mira variables. Comparing these properties with pulsation periods, we find a positive trend between stellar radius and period only for the Mira variables with periods above 300 days and a positive trend between the period and the monochromatic luminosity only for the red supergiants and the most extreme AGB stars with periods above 500 days.
   We find that the continuum emission at 1.24~mm can be classified into four groups. ``Featureless'' continuum emission is confined to the (unresolved) regions close to the star for five stars in our sample, relatively uniform extended flux is seen for four stars, tentative \editss{elongated} features are seen for three stars, and the remaining five stars have unique or unusual morphological features in their continuum maps. These features can be explained by binary companions to 10 out of the 14 AGB stars in our sample.}
   {Based on our results we conclude that there are two modes of dust formation: well established pulsation-enhanced dust formation, and our newly proposed companion-enhanced dust formation. If the companion is located close to the AGB star, in the wind acceleration region, then additional dust formed in the wake of the companion can increase the mass lost through the dust driven wind. This explains the different dust morphologies seen around our stars and partly accounts for a large scatter in literature mass-loss rates, especially among semiregular stars with small pulsation periods.}


   \maketitle

\section{Introduction}

Cool evolved stars, in particular low and intermediate mass (\mbox{$\sim0.8$--$8~\msol$}) asymptotic giant branch (AGB) stars, are responsible for a significant amount of dust production in the nearby universe \citep{Hofner2018,Decin2021}. During the AGB phase, stars lose material through a pulsation-enhanced, dust-driven wind, with mass-loss rates in the range of $\sim 10^{-8}$--$10^{-4}\spy$ \citep{Hofner2018}. The winds of these stars are also responsible for supplying heavy elements and hence causing the chemical enrichment of their host galaxies \citep{Kobayashi2020}. Among other elements, significant quantities of carbon are produced during the AGB phase, and AGB stars with initial masses in the range $ 1.5\lesssim M_i \lesssim 4~\msol$ \citep{Karakas2016} are thought to progress from oxygen-rich M-type stars (C/O < 1), to S-type stars (C/O $\sim1$), to carbon-rich stars (C/O > 1).

Red supergiant stars (RSGs) are the more massive counterparts of AGB stars, with progenitors $\gtrsim 9~\msol$. They also produce significant amounts of dust \citep{Levesque2017}, possibly through more episodic ejections \citep[e.g.  the Great Dimming of Betelgeuse,][and the ejecta of VY~CMa, \citealt{Humphreys2024}]{Montarges2021}. 
To fully understand these processes and the amount and rate at which material is returned to the interstellar medium (ISM), we first need to understand how dust is formed, its chemical composition, and what affects the rates of dust production.

Dust shells around evolved stars have been imaged at both large and smaller scales, for example using space telescopes such as IRAS \citep{Young1993a}, ISO \citep{Izumiura1996,Hashimoto1998}, \textsl{Spitzer} \citep{Ueta2006} and \textsl{Herschel}/PACS \citep{Cox2012}. In particular, \cite{Cox2012} resolved dust emission around a substantial sample of evolved stars, including bow shock emission where circumstellar dust is colliding with the ISM.
Spectroscopy in the mid infrared has enabled us to gain some understanding of the composition of dust around evolved stars \citep[e.g.][]{Waters1999,Hony2009,Justtanont2013,Sloan2016}. These observations all focus on relatively large scales and are best suited to characterising historical dust formation, over the past hundreds or thousands of years, around evolved stars.

At smaller angular scales, recent developments in adaptive optics have enabled dust imaging in the optical and near infrared at very high resolution. For example, \cite{Montarges2023} imaged a sample of 14 evolved stars (a subset of the sample we present in this work) at resolutions of $\sim 30$~mas. They presented maps of scattered polarised light which showed complex interactions in the inner circumstellar environments (within 10 to a few 100s of au). Even higher resolution imaging has been carried out for a few individual stars, spatially resolving their surfaces and showing non-uniform photospheres \citep{Ohnaka2016,Paladini2018,Khouri2020}, and convective time scales on the order of a month for at least one star \citep[R~Dor,][]{Vlemmings2024a}. These findings are in agreement with high resolution 3D models of stellar convective envelopes, which show patchy dust formation \citep{Hofner2019} and dust forming in the wake of atmospheric shocks \citep{Freytag2023}.
High resolution ALMA observations of individual stars have also revealed inhomogeneities in the inner circumstellar envelope in the continuum and in molecular emission \citep{Velilla-Prieto2023,Baudry2023,Gottlieb2022}, further emphasising the non-uniform formation of dust at the surface of an AGB star.

%\cite{Dharmawardena2018,Dharmawardena2019}?
%
%ATOMIUM
%
%\blue{finish intro}
%
%Stellar continuum but also mm continuum has a dust contribution

ATOMIUM (ALMA Tracing the Origins of Molecules In dUst-forming oxygen-rich M-type stars) is an ALMA Large Programme with the aim of understanding the chemistry and dust formation of evolved stars. \cite{Decin2020} and \cite{Gottlieb2022} elaborate on these goals and present some of the first results of the project, and \cite{Wallstrom2024} present an overview of the molecular inventory of the line data. In this work we present an overview and analysis of the continuum data for the whole sample.

%Here we present a comprehensive analysis of the continuum emission of 14 AGB stars and 3 RSGs, observed with ALMA at high resolution. 
This paper is arranged as follows.
In Sect.~\ref{sec:obs} we present the sample, describe the data reduction and initial calculations performed on the data set. In Sect.~\ref{sec:contim} we present the continuum images. Further analysis is done in Sect.~\ref{sec:analysis} and our results are discussed in Sect.~\ref{sec:discussion}. We summarise our conclusions in Sect.~\ref{sec:concl}.

\section{Observations}\label{sec:obs}

\begin{table*}
	\centering\small
	\caption{Key properties of stars in our sample.}
	%\red{Should RA, Dec be to more decimal points? OH30 was to 5 for individual arrays, mean to 4}
	\label{tab:stars}
	\begin{tabular}{ccrccccc}
		\hline
Star	&	Right Ascension	& 	Declination	&	Distance	&	Variability	&	Period	& Observed	& Fig.	\\
	&	(ICRS)	& (ICRS)\phantom{00} &	[pc]	&	type	&	[days]	& configurations	& 	\\
\hline
GY Aql	&	19:50:06.3148	&	 $-$7:36:52.189	&	410	(40, 40)	$^{a}$	&	Mira	&	464	&	Extended, mid, compact, ACA	&	\ref{fig:gyaql}	\\
R Aql	&	19:06:22.2567	&	  8:13:46.678	&	266	(85,52)	$^{a}$	&	Mira	&	268.8	&	Extended, mid, compact, ACA	&	\ref{fig:raql}	\\
IRC$-$10529	&	20:10:27.8713	&	 $-$6:16:13.740	&	930	(70, 60)	$^{a}$	&	Mira	&	670	&	Extended, mid, compact, ACA	&	\ref{fig:irc-10529}	\\
W Aql	&	19:15:23.3781	&	 $-$7:02:50.330	&	380	(49, 68)	$^{a}$	&	Mira	&	488	&	Extended, mid, compact	&	\ref{fig:waql}	\\
SV Aqr	&	23:22:45.4002	&	$-$10:49:00.188	&	445	(65, 90)	$^{a}$	&	\phantom{*} SRb *	&	93, 231.8\edits{*}	&	Extended, mid, compact	&	\ref{fig:svaqr}	\\
U Del	&	20:45:28.2500	&	 18:05:23.976	&	333	(10, 11)	$^{b}$	&	SRb	&	120, 1163	&	Extended, mid, compact	&	\ref{fig:udel}	\\
$\pi^1$ Gru	&	22:22:44.2696	&	$-$45:56:53.006	&	164	(12, 10)	$^{b}$	&	SRb	&	195.5, \edits{5750}*	&	Extended, mid, compact, ACA	&	\ref{fig:pi1gru}	\\
U Her	&	16:25:47.4514	&	 18:53:32.666	&	271	(19, 21)	$^{a}$	&	Mira	&	405.9	&	Extended, mid, compact, ACA	&	\ref{fig:uher}	\\
R Hya	&	13:29:42.7021	&	$-$23:16:52.515	&	126	(2, 3)	$^{a}$	&	Mira	&	359	&	Extended, mid, compact	&	\ref{fig:rhya}	\\
T Mic	&	20:27:55.1797	&	$-$28:15:39.553	&	175	(15, 19)	$^{a}$	&	SRb	&	352, 178	&	Extended, mid, compact, ACA	&	\ref{fig:tmic}	\\
S Pav	&	19:55:14.0055	&	$-$59:11:45.194	&	184	(16, 17)	$^{a}$	&	SRa	&	390	&	Extended, mid, compact	&	\ref{fig:spav}	\\
IRC+10011	&	01:06:25.9883	&	 12:35:52.849	&	720	(30, 30)	$^{a}$	&	Mira	&	651	&	Extended, mid, compact	&	\ref{fig:irc+10011}	\\
V PsA	&	22:55:19.7228	&	$-$29:36:45.038	&	299	(11, 12)	$^{b}$	&	SRb	&	148\edits{*}	&	Extended, mid, compact	&	\ref{fig:vpsa}	\\
RW Sco	&	17:14:51.6867	&	$-$33:25:54.544	&	560	(25, 30)	$^{b}$	&	Mira	&	388.45	&	Extended, mid, compact, ACA	&	\ref{fig:rwsco}	\\
\hline
KW Sgr	&	17:52:00.7282	&	$-$28:01:20.572	&	2183	(304, 406)	$^{b}$	&	SRc	&	695	&	Extended, mid	&	\ref{fig:kwsgr}	\\
VX Sgr	&	18:08:04.0460	&	$-$22:13:26.621	&	1570	(270, 270)	$^{c}$	&	SRc	&	755	&	Extended, mid, compact	&	\ref{fig:vxsgr}	\\
AH Sco	&	17:11:17.0159	&	$-$32:19:30.764	&	1735	(200, 286)	$^{b}$	&	SRc	&	735	&	Extended, mid	&	\ref{fig:ahsco}	\\
		\hline
	\end{tabular}
\tablefoot{AGB stars are listed in the top part of the table and RSG stars are listed in the bottom part of the table. 
The positions are for the epoch 2019 June 23 -- 2019
July 12.
The numbers in parentheses in the distance column are the lower and upper uncertainties on the distances. In the period column, the primary period is listed first, followed by the secondary period if there is one (Appendix \ref{app:periods}).
(*) indicates an updated designation discussed in Appendix \ref{app:periods}. The ``Fig.'' column gives the figure number of the plots showing the continuum images from the extended, combined and mid array configurations. 
%Plots for the continuum observed with the compact array and the ACA are given in Figs.~\ref{fig:compact} and \ref{fig:aca}, respectively. 
\textbf{References.} ($^{a}$) \cite{Andriantsaralaza2022}; ($^{b}$) \cite{Bailer-Jones2021}; ($^{c}$) \cite{Chen2007}.}
\end{table*}

\subsection{ATOMIUM sample}\label{sec:sample}

The ATOMIUM sample consists of 17 evolved stars, of which 14 are AGB stars and three are RSGs (VX Sgr, AH Sco and KW Sgr). The sample was constructed so as to cover a range of mass-loss rates and variability types (e.g. Mira variables and semiregular variables). Of the AGB stars, 12 are oxygen-rich (M-type, C/O~$<1$) and two are S-type stars with C/O~$\sim 1$ (W~Aql and \pigru). Coincidentally, the two S-type stars are also known to have binary companions \citep{Ramstedt2011,Mayer2014}, though \pigru\ has been proven to be a triple system \citep{Homan2020,Montarges2025,Esseldeurs2025}. The AGB star R~Hya also has a companion star, but at a projected separation of $\sim 21\arcsec \approx 2700$~au  \citep[R Hya B,][aka Gaia DR3 6195030801634430336]{Smak1964,Kervella2022}, it is too distant to have a significant impact on the circumstellar envelope.

Table~\ref{tab:stars} gives an overview of the stellar properties. We describe our choice of distances and periods in the following subsections.

\subsubsection{Distances}

The distances given in Table~\ref{tab:stars} were collected from the literature and represent the best current estimates and uncertainties. Where available, we used the estimates and uncertainties provided by \cite{Andriantsaralaza2022}, who used several methods tailored to AGB stars, otherwise, we use the results of \cite{Bailer-Jones2021} based on Gaia observations (with geometric priors).
%, who calculate two distances from Gaia data; one based on a geometric (direction-dependent) prior and one on a photo-geometric prior, the latter also considering the colour of the star. 
%Where both distances were given, we took their mean (they generally differed by $<10$~pc) and used the most conservative estimates of their uncertainty. 
The exception is VX Sgr, for which both the geometric and photo-geometric priors gave distances of 4.6 and 4.1 kpc, with uncertainties in excess of 1 kpc \citep{Bailer-Jones2021}. This is significantly larger than the value of $1.57\pm0.27$ kpc found by \cite{Chen2007} using maser proper motions. The discrepancy most likely arises from the surface variability of the star and its large angular size, which combine to make it difficult to determine the photocentre. Our adopted distance of 1570~pc is consistent with water maser proper motions \citep{Murakawa2003} and with membership of the Sgr OB1 association \citep{Humphreys1972}.
AH~Sco is another RSG for which a similar maser proper motion method gives a distance of $2.26\pm0.19$ kpc \citep{Chen2008}, in better agreement with the \cite{Bailer-Jones2021} distance of $1739^{+282}_{-241} $~pc. For the third RSG in our sample, KW~Sgr, the Gaia distance of $2171_{-314}^{+450}$~pc \citep{Bailer-Jones2021} is in agreement with the previously estimated distance of $2.4\pm0.3$ kpc based on its membership of the Sgr OB5 association \citep{MelNik2009,Arroyo-Torres2013}.
%\blue{R Aql distance is very uncertain from Miora's paper. She's checking why it was excluded from the PL calculation in her paper.}

\subsubsection{Variability}

We list periods and variability types in Table~\ref{tab:stars} for all the stars in our sample. The variability types and periods were taken from the International Variable Star Index (VSX\footnote{VSX: \url{www.aavso.org/vsx/index.php}}). These periods were also further verified by cross-checking the VSX period with light curves from the American Association of Variable Star Observers (AAVSO\footnote{AAVSO: \url{www.aavso.org}}), All Sky Automated Survey \citep[ASAS,][]{Pojmanski2002} and the ASAS-SN Variable Stars Database \citep{Shappee2014,Jayasinghe2019} using the Lomb-Scargle periodogram \citep{VanderPlas2018}, as the time series are unevenly sampled. Among the AGB stars we have a mix of Mira variables and semi-regular (SR) variables of types SRa and SRb. The RSGs are all SRc, as expected. The SRb stars aside from V~PsA have a secondary period as well as a primary period listed in Table \ref{tab:stars}.

In Appendix~\ref{app:periods} we discuss individual stellar periods in more detail. To briefly summarise the key points, 
SV~Aqr was reported in earlier ATOMIUM papers as a long period variable \citep[e.g.][]{Gottlieb2022}. For the present work, we looked further into the pulsations of SV~Aqr and were able to define a primary period of 93 days, as described in Appendix~\ref{app:periods}.
For \pigru, we identified a long secondary period (LSP, of $\sim15$ years), also described in Appendix~\ref{app:periods}.
R~Aql and R~Hya have been reported to have shortening periods \citep{Wood1981,Greaves2000,Zijlstra2002,Joyce2024}, possibly owing to undergoing thermal pulses in the recent past (a few hundred or so years ago). This is also discussed further in Sect.~\ref{sec:tps}.

\subsection{ALMA 12m Array}\label{sec:12mobs}

Observations were taken with three configurations of the ALMA 12m Array in Band 6 over the frequency range 214--270 GHz as part of the ATOMIUM Large Programme (2018.1.00659.L, PI: L. Decin). Typical resolutions for the extended, mid and compact array configurations are 25 mas, 200 mas, and 1\arcsec. \edits{In Table~\ref{tab:stars} we note which stars were observed with which configurations, and give a reference to which figure their continuum images are shown in (Fig.s~\ref{fig:gyaql}--\ref{fig:ahsco}). All the compact configuration images are shown in Fig.~\ref{fig:compact}.} 
A detailed discussion of the data reduction, \edits{including self-calibration,} is given in \cite{Gottlieb2022} and a complete list of the spectral line IDs and their properties is given in \cite{Wallstrom2024}. The maximum recoverable scales (MRS, \edits{the scale at which smooth regions of flux can be recovered with confidence}) are $\sim0.4$--$0.6\arcsec$ for the extended array data, $1.5\arcsec$ for the mid array data, and 8--$10\arcsec$ for the compact array data. This is discussed further in Sect.~\ref{sec:resolved-out} and more precise MRS for each star and configuration are given in Table E.2 of \citet{Gottlieb2022}.
In this paper we focus on the continuum data, which we now analyse in detail. However, the basic parameters of the individual array continuum observations, including observation dates, are given in Table E.2 of \cite{Gottlieb2022}. \edits{In Tables \ref{tab:extended}, \ref{tab:mid}, and \ref{tab:compact} we give an overview of the continua observed with the extended, mid and compact array configurations.}

The ALMA 12m array data taken in the compact configuration (lowest resolution) were observed for the shortest time spans and so have lower signal to noise (S/N) and  positional precision.  For each star and array configuration, we identified and excluded spectral lines and self-calibrated the phase-referenced continuum data, as described in \cite{Gottlieb2022}. The flux scale of individual observations should be accurate to 5--7\% at Band 6 \citep{Cortes2023} and our observations met this.
Where there were discrepancies \edits{due to calibration uncertainty, we rescaled an individual configuration if it was inconsistent or, if all configurations disagreed, we took the mid configuration flux densities as the most reliable standard to rescale the other configurations. This flux comparison was done for baseline lengths common between configurations, hence was not affected by missing spacings.} The resulting flux scale is nominally accurate to $\sim10\%$ although detailed examination \citep[e.g. of R~Hya by][]{Homan2021} suggests that it may be better. 

We fitted 2-D Gaussian components to  the continuum images after applying phase-reference solutions only.
The accuracy of Gaussian fitting is limited by (beam size)/(S/N), where S/N is $>$50 in all cases; the main astrometric uncertainty comes from transfer of calibration \citep{Gottlieb2022}.
The most accurate astrometric positions ($\sim$2 mas) were obtained for extended configuration images. The mid and compact array data were adjusted to the position of the extended array data before combination. Thus the stellar reference positions are for the epoch 2019 June 23 -- 2019 July 12 \citep[observing dates for each star are given individually in Table E1 of][]{Gottlieb2022}. 
\edits{Overall, data for each star was taken over a period of 6--12 months and the proper motions for each star range from 1.3--69~mas~yr$^{-1}$ \citep{Gaia-DR3-2023}, likely accounting for the most significant positional shifts. The only exception is KW~Sgr, for which one set of mid observations were taken in May 2021, giving a total observation range of 30 months. The proper motion shift was minimal, however, since this distant star has the smallest proper motion of our sample.}

%In combining data from the three configurations, the peak positions (measured initially by fitting 2D Gaussian components) for the mid and compact data were aligned with the position of the extended array data, and the phase centres were centred on the stellar centre.

When we discuss combined data, we are referring to the ALMA 12m array data from the extended, mid and (for most stars) compact configurations that have been combined to improve overall sensitivity while maintaining a high angular resolution. \edits{An overview of the combined data is given in Table \ref{tab:combined}.}
The observed data for each separate array observation were weighted during pipeline calibration according to the integration times, channel widths, and amplitude calibration variance, but no further weighting was done in self-calibration (and flagging was done after channel averaging). The configurations were combined with equal weight. \edits{The combined MRS is generally as good as for the compact configuration data, except for AH Sco and KW Sgr, which were not observed with the compact configuration, and for which the MRS is as good as for the mid configuration data.} 

For any \edits{combined images} with signs of artefacts such as negatives exceeding $4\sigma$, we re-examined the data carefully. Calibration and flux scale errors are symmetric (or anti-symmetric for phase) in the stellar image, whilst angular misalignments tend to show up as anti-symmetric central adjacent errors after \edits{subtracting a uniform disc (see Sect.~\ref{sec:ud})}.  In a few cases we repeated the self-calibration of individual configurations with different solution intervals and other constraints to optimise the continuum.  However, the main cause of artefacts was usually misalignment of the flux or position between configurations, since even within the ALMA tolerances an e.g. 5\% flux scale error exceeds the dynamic range for the brighter stars.  We therefore aligned the configurations to minimise the relative errors.

\subsection{ALMA Compact Array}

A subset of the ATOMIUM sample was also observed in Band 3 with the Morita Array (ALMA Compact Array; project 2019.1.00187.S, PI: T. Danilovich). 
The observations were taken over discrete spectral windows in the frequency range 97--113~GHz. The central frequency for the continuum images is 104.96~GHz. The MRS for these observations fall in the range 64--77\arcsec. The flux scale accuracy of this data is around 5\% \citep{Cortes2023}.
Note that we refer to observations with the Morita Array as ``ACA data'' and to observations with the compact configuration of the ALMA 12m Array as  ``compact data''. The details of the ACA data are given in Table~\ref{tab:aca}. In Fig.~\ref{fig:aca} we show all the continuum images obtained using the ACA.

All the ACA data used comes from the standard ALMA pipeline \edits{(and has not been self-calibrated)} aside from RW~Sco, for which we created a custom dataset that was not primary beam corrected. The reason for this is a confounding nearby source, approximately $27\arcsec$ south of RW~Sco, which is intrinsically brighter than the AGB star once the primary beam correction is applied. This is discussed further in Appendix~\ref{res:rwsco}.
%\blue{Insert result of cross check with Sydney radio catalogue}
We present some line observations obtained concurrently with the ACA continuum in Appendix \ref{sec:acalines}.

%\section{Calculations}\label{sec:calc}

\subsection{Fitting uniform discs}\label{sec:ud}

\begin{table}
	\centering
	\small
	\caption{Results of uniform disc fits.}
	\label{tab:ud}
	\begin{tabular}{cccccc}
		\hline
Star	&	Diameter	&	$F_\mathrm{UD}$ &  rms	& $T_b$\\
	&	[mas]	&	[mJy]&	[$\mu$Jy]	& [K]\\
\hline
GY Aql	&	15.29	(	0.17	)	&	9.5	&	23	&	1227	(	28	)	\\
R Aql	&	16.07	(	0.04	)	&	19.9	&	8	&	2326	(	12	)	\\
IRC$-$10529	&	17.68	(	3.25	)	&	7.9	&	28	&	763	(	281	)	\\
W Aql	&	16.65	(	0.06	)	&	8.0	&	5	&	871	(	7	)	\\
SV Aqr	&\phantom{$\dagger$}	6.65	(	1.54	)	$\dagger$ &	2.5	&	33	& \it	1672	(	775	)	\\
U Del	&	11.25	(	0.18	)	&	7.1	&	10	&	1693	(	53	)	\\
$\pi^1$ Gru	& \phantom{*}	23.97	(	3.39	)	* &	26.5	&	15	&	1392	(	394	)	\\
U Her	&	18.98	(	2.95	)	&	15.0	&	13	&	1257	(	391	)	\\
R Hya	&	31.87	(	3.16	)	&	64.8	&	57	&	1926	(	381	)	\\
T Mic	&	21.41	(	0.04	)	&	29.7	&	13	&	1956	(	8	)	\\
S Pav	&	20.42	(	0.03	)	&	27.9	&	10	&	2020	(	6	)	\\
IRC+10011	&	17.93	(	0.29	)	&	12.7	&	20	&	1192	(	39	)	\\
V PsA	&	11.86	(	0.09	)	&	9.3	&	9	&	1996	(	29	)	\\
RW Sco	&	1.06	(	3.52	)	$^\ddagger$ &	5.0	&	20	&	...		                   		\\
\hline															
KW Sgr	&	6.87	(	6.61	)	$^\ddagger$ &	2.8	&	8	& \it	1791	(	3447	)	\\
VX Sgr	&	16.81	(	0.08	)	&	16.9	&	19	&	1805	(	17	)	\\
AH Sco	&	15.80	(	0.09	)	&	7.9	&	11	&	953	(	11	)	\\
\hline
	\end{tabular}
%\begin{flushleft}
\tablefoot{$F_\mathrm{UD}$ is the flux density. The central frequency used for the brightness temperature, $T_b$, calculation is 241.75 GHz (1.24 mm). The values in parentheses are the uncertainty on the UD diameter or $T_b$. The uncertainties on the flux density and $T_b$ do not contain the flux scale uncertainties which could be up to 10\%, see Sect.~\ref{sec:12mobs}.
%The uncertainty on the UD flux density includes the rms but is dominated by the calibration uncertainty which, conservatively, is about 10\% for the combined data. 
The rms are measured close to the star where there may be dynamic range limitations, see Sect.~\ref{sec:contim}. (*) A two-component fit was used for \pigru; see text for details and flux density of secondary component. ($\dagger$) SV Aqr parameters are \edits{from an image-plane fit to} the extended array data only; see text for details. ($^\ddagger$) RW Sco and KW Sgr have low S/N  and small angular sizes, which is why the UD diameter uncertainties are so large; see text for details. Brightness temperatures given in italic text are very uncertain but are included for completion. The diameter uncertainty for RW~Sco is too large to estimate the brightness temperature.}
%\end{flushleft}
\end{table}

%\begin{figure}
%\centering
%	\includegraphics[width=0.4\textwidth]{Diameter_vs_flux.pdf}
%    \caption{INSERT HERE The grey dashed line is a fit to the data excluding the two points with diameters $<10$~mas (RW Sco and KW Sgr). See text for details.}
%    \label{fig:udfluxdiam}
%\end{figure}

\edits{For the Band 6 observations of each star, we show the amplitudes of the visibilities as a function of $uv$ distance in Fig.~\ref{fig:udfits}. These plots generally indicate a compact component and, in many cases, some extended features, the latter discussed in more detail in Sect.~\ref{sec:contim}.}
\edits{At radio frequencies, e.g. wavelengths of cm to sub-mm, the main compact contribution to the continuum emission is expected to come from electron-neutral free-free emission in an extended radio photosphere, above the stellar photosphere \citep[][]{Reid1997,Matthews2015,Vlemmings2019}. Emission offset from the stellar position, assumed to be the continuum peak, most likely originates from dust surrounding the stellar photosphere and radiosphere (see Sect.~\ref{sec:contim}).}

To estimate the stellar sizes and flux densities, we fit a Uniform Disc (UD) model to the calibrated, line-free combined Band 6 visibilities for each source. We tested Gaussian fits but found higher residuals indicating worse fits. Trying to fit a UD + a point source at the stellar position also gave worse results, including relatively bright residuals within the stellar disc diameter. This suggests that any stellar-related flux not fit by a UD is not concentrated in a single point and is not symmetric, i.e. has a random or more complex and blotchy distribution, 
%whether close to/against the star, or more extended, 
which cannot be fit in the uv plane. We also found that the irregularities in the sources are greater than any differences between a uniform disc or a limb-darkened disc (where a limb-darkened disc is similar to what \cite{Bojnordi-Arbab2024} found from their models). 

For the UD fits we used the CASA task \texttt{uvmodelfit} \edits{to fit a single UD to the data.}
In the case of AH Sco, RW Sco  and \pigru, we used the \texttt{uvmultifit} package \citep{Marti-Vidal2014}, \edits{which has many capabilities, including the ability to fit multiple components and more detailed constraints at low S/N, which allowed us to ensure that the diameter and flux did not simply match pre-set limits.
For our sample of evolved stars, we report only single component fits as adding more components, whether  points or extended emission, did not converge to a stable solution, suggesting that any star-spots and/or extended dust cannot be described by simple, symmetric components.} 
For AH Sco and IRC+10011, we used the extended configuration data only as there are irregularities on short spacings.
\edits{These are probably due to large-scale flux which is irregularly or asymmetrically distributed, possibly simply contributing a small excess on short baselines but smooth enough on large scales to be resolved out for long baselines. For most stars such irregularities did not affect the fit, but for AH Sco (because of low S/N) and IRC+10011 (significant resolved-out flux, see Sect.~\ref{sec:resolved-out}), the errors in the fit were much greater if short baselines were included.}
For RW Sco we forced the position to be at the phase centre, as it has very weak continuum. In each case, the starting model was the \edits{literature} optical diameter and peak flux density from Tables 1 and E.2 in \cite{Gottlieb2022}. All parameters (peak, position, diameter, ellipticity, position angle) were allowed to vary except for the special case of \pigru, for which a close companion (\pigru~C, separation $\sim0.038\arcsec$) was reported by \cite{Homan2021}. Hence, for \pigru\ we first subtracted the companion and then fixed the position and ellipticity (none) to avoid contamination by the companion residuals. This two-component fit was done using only the extended array data, but we find the UD model agrees with the full dataset, except at the shortest baselines (see Fig.~\ref{fig:udfits}). \pigru~C is too faint and compact to be resolved even by component fitting so we do not give its size (see Fig.~\ref{fig:pi1gru}, in which it is comparable to the extended data beam size). We determine an integrated flux density for \pigru~C of 2.78 mJy.

For SV Aqr, attempts to fit to the visibilities for all configurations failed to resolve the star and could only measure the flux density within the central 15~mas with any confidence. The result of this fit is shown in Fig.~\ref{fig:udfits}. However, to retrieve more reliable stellar parameters, we performed a fit \edits{in the image plane} using the extended configuration data only, which retrieved the parameters with S/N~$>5$. It is these parameters that are included in Table \ref{tab:ud}.

The fitting routines report the uncertainties in the fits. The disc fits were almost circular ($r=$~minor/major axis ratio >0.9) for most sources, except RW Sco and KW Sgr (which have low S/N) and IRC$-$10529 ($r=0.87$), and U~Her ($r=0.89$), for which the diameter listed in Table \ref{tab:ud} is the longer axis.
We assume that the stars are unlikely to be elliptical to this degree, but instead that deviations are either due to elliptical beams (Table~\ref{tab:combined}) or stellar surface irregularities probably distributed randomly.  We thus take the ellipticity as a contribution to the uncertainty in the fit.  The other contributions are the observational errors. Phase errors contribute a size uncertainty proportional to the (beam size)/(S/N), and the flux density errors are mostly dominated by the flux scale uncertainty except for the weakest sources where the noise is significant \citep[see][]{Gottlieb2022}.

The resulting UD parameters are given in Table \ref{tab:ud} and the visibility amplitudes with the fitted UD models are shown in Fig.~\ref{fig:udfits}. The error bars on the visibilities represent the observational scatter of the visibility amplitudes in each bin. The shaded regions represent the fitting errors with the edges of the shaded region representing the upper and lower extremes of these uncertainties. The rms values given in Table \ref{tab:ud} do not represent the total uncertainty on the flux and should be added in quadrature with a 10\% calibration uncertainty on the flux, though in some cases this uncertainty appears to be much better (based on consistency between observations with different calibrators). 
%\red{Can we be more specific here? Otherwise maybe remove the last part of the sentence.} %Note that the shading is used to make these limits clearer and does not directly represent the uncertainties, which are rather represented as the error bars on the visibilities.

We were unable to find good fits for KW Sgr or RW Sco because these two sources have low S/N and small angular sizes, consistent with being unresolved. (Note that the other stars, aside from SV~Aqr, have S/N > 50.) We still list the best fitting parameters for these stars in Table \ref{tab:ud} and show the fits in Fig.~\ref{fig:udfits} but note the very large uncertainties on the UD diameters. 
%Aside from these two sources, the disc fits were almost all circular, with the ratios of the minor and major axes $r>0.9$, except for IRC$-$10529 ($r=0.87$), SV~Aqr ($r=0.79$), U~Her ($r=0.89$), for which the diameter listed in Table \ref{tab:ud} is the longer axis. Non-circular beams are likely contributors to these asymmetries (see Table~\ref{tab:combined}).
In Table \ref{tab:stars}, we give the right ascension and declination (ICRS) determined from the UD fits, which in all cases are within 1 mas of the astrometric positions (see Sect.~\ref{sec:12mobs} and \citealt{Gottlieb2022}).
%The mid and compact array data were adjusted to the position of the extended array data before combination. Thus the stellar reference positions are for the epoch 2019 June 23 -- 2019 July 12, ICRS. In all cases the fitted centre offsets are $<1$ mas.
Preliminary UD sizes were reported for 5 of the ATOMIUM sources previously \citep[R~Aql, U~Her, R~Hya, T~Mic, and S~Pav,][see also \citealt{Homan2021} for R~Hya]{Baudry2023}. Those sizes were within 1\% of what we find in the present work, except for R~Hya where the size difference is $\sim 10\%$. \edits{The improvement in the R~Hya fit comes from the adjustments to the relative positions and fluxes described in Sect.~\ref{sec:12mobs}, which enabled us to improve the dynamic range from $\sim1500$ to 6500.}

Based on our UD fits we also calculated brightness temperatures for our sample, using 241.75~GHz as the central frequency of our combined data. An explanation of this calculation is given in Appendix \ref{sec:Tbright}.
%This was done using the formula \citep{Vlemmings2017}
%\begin{equation}
%T_b = \frac{1.96\e{4} S_\mathrm{mJy} c^2}{\nu^2 \theta^2}
%\end{equation}
%where $S_\mathrm{mJy}$ is the UD flux density (in mJy), $\theta$ is the UD diameter (in arcsec), $\nu=241.75$~GHz is the central frequency of our combined data, and $c$ is the speed of light. 
The resultant brightness temperatures are given in Table~\ref{tab:ud}, excluding RW~Sco, for which the UD fit values were too uncertain to implement. The uncertainties we find for KW~Sgr and SV~Aqr are very large, but we list the obtained values in the table for completion. The flux densities of these sources are the lowest in our sample. Note that KW~Sgr is our most distant target. For the sample overall, the derived brightness temperatures are in the range 760--2300~K, all systematically cooler than the effective stellar temperatures collected from the literature (see Table \ref{tab:radii}), but \edits{many are} largely in line with the earlier results of \cite{Vlemmings2019}, 
\edits{for four nearby AGB stars. The theoretical predictions of \cite{Bojnordi-Arbab2024} give phase-dependent brightness temperatures $\gtrsim1900$~K, depending on the model. Where our brightness temperatures are notably lower than this, we assume it is because newly-formed dust, located close to the surface of the star, has been captured by our best-fitting UD. }
%This indicates that some dust is included in our UD fits.
% This is discussed further in \red{Sect.~\ref{sec:discussion} - should refer to specific subsection}.

%\red{Anita to check this paragraph. Do we need more details on the subtraction procedure?}
We subtracted the UD models from the combined visibility data and imaged the residuals, which we show alongside the continuum images in Figures~\ref{fig:gyaql}--\ref{fig:ahsco}.
We weighted the baselines during imaging (see Sect.~\ref{sec:contim}) to provide a larger synthesised beam than was used for the un-subtracted combined images, to give better sensitivity to features with low surface brightness. The exceptions are GY Aql and $\pi^1$ Gru where we use a synthesised beam of comparable size to the combined image in order to resolve the faint, but compact and distinct, offset emission.
The details of the UD subtracted images are given in Table \ref{tab:udsub}. For SV Aqr we show the residual image after subtracting the original UD fit to the combined data \edits{(Fig.~\ref{fig:svaqr})}, which represents contributions from the star and possibly some inner dust. We caution that the resulting features are not as reliable
as for the other stars and should not be over-interpreted.

\section{Continuum imaging}\label{sec:contim}

\begin{figure*}[t!]
	\includegraphics[height=4.9cm]{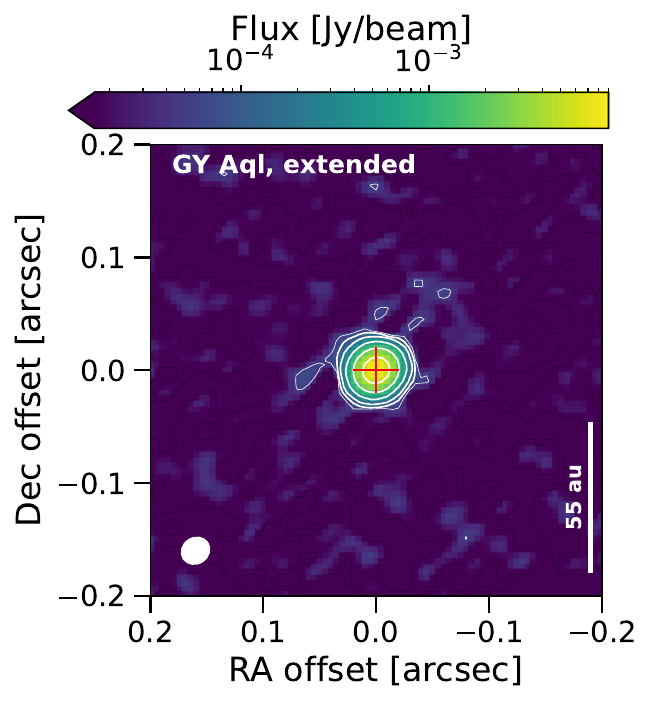}
  	\includegraphics[height=4.9cm]{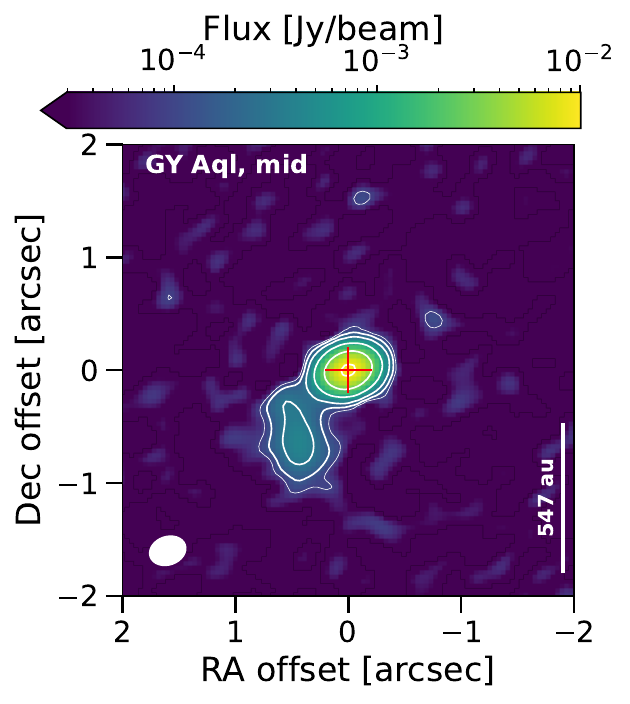}
 	\includegraphics[height=4.9cm]{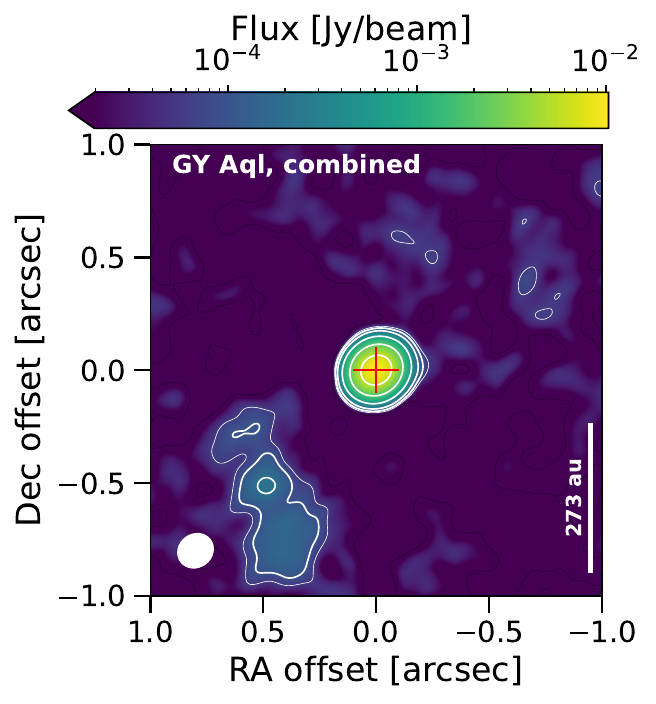}
	\includegraphics[height=4.9cm]{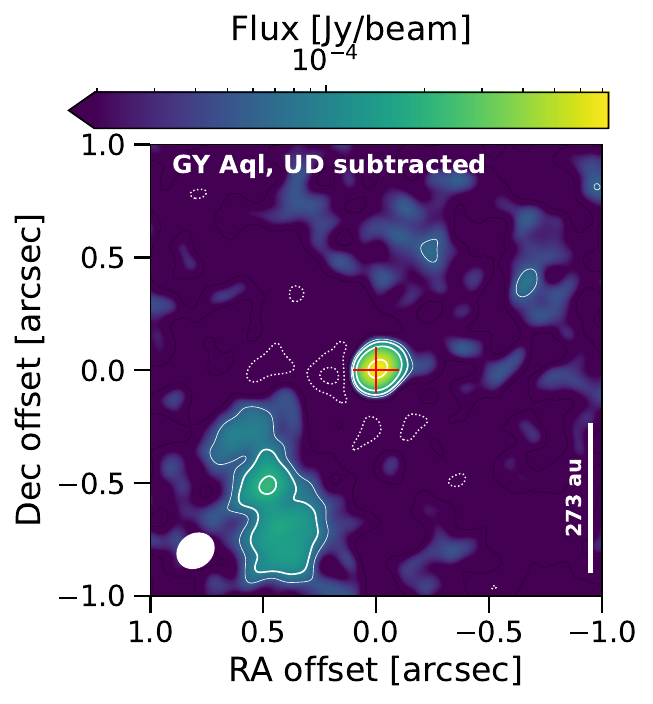}
    \caption{Continuum maps of GY Aql taken with the extended (left), mid (centre left) and combined (centre right) array configurations of ALMA. The rightmost plot shows the residuals after subtracting a uniform disc representing the AGB star (see text for details), highlighting extended dust features. The thin solid contours indicate levels of $3\sigma$, the thick solid contours indicate levels of 5, 10, 30, 100, and 300$\sigma$, and the dotted contours indicate levels of $-3$ and $-5\sigma$. The continuum peak is indicated by the red cross and the synthetic beam is given in the bottom left corner of each image. \edits{The white bars on the lower right give indicative sizes in physical units for a third of the box length, based on the distance in Table~\ref{tab:stars}.} North is up and east is left.}
    \label{fig:gyaql}   
    \vspace{0.2cm}
%\end{figure*}
%
%\begin{figure*}
	\includegraphics[height=5.15cm]{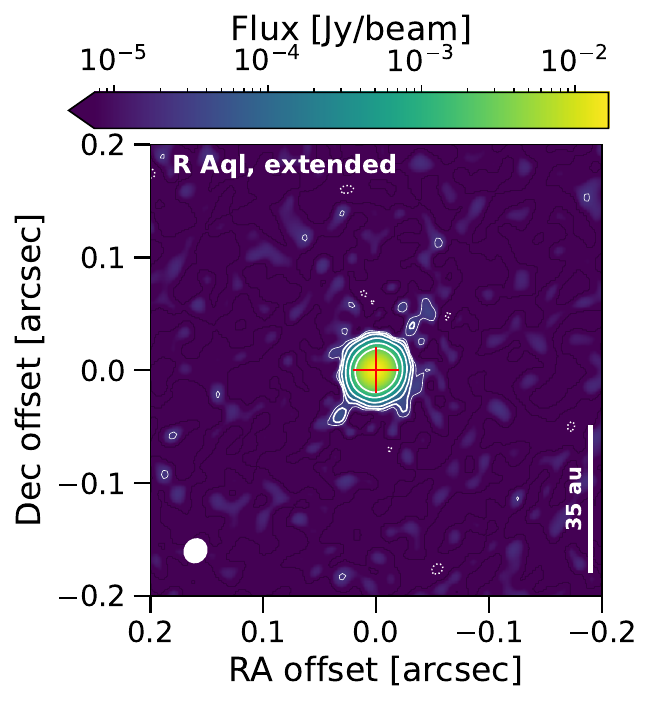}
  	\includegraphics[height=5.15cm]{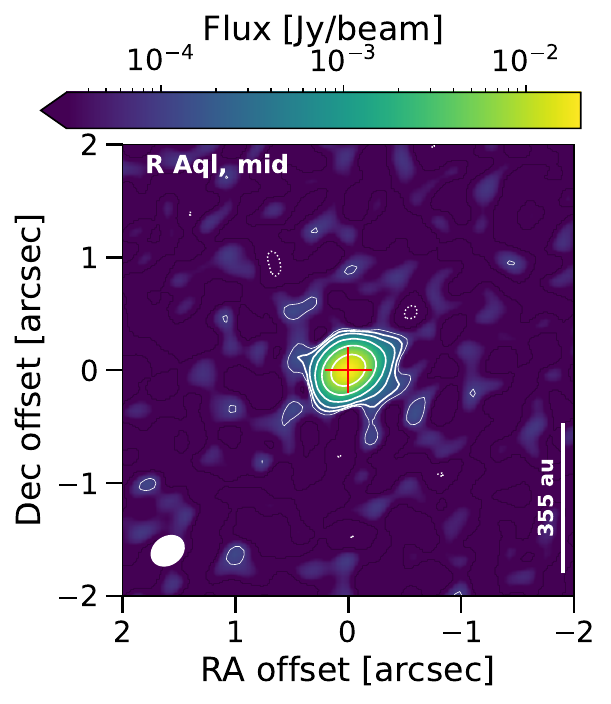}
 	\includegraphics[height=5.15cm]{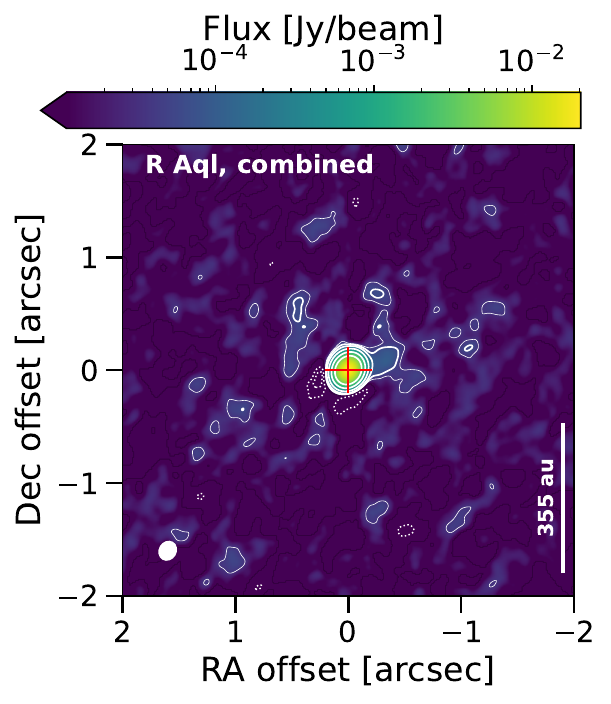}
	\includegraphics[height=5.15cm]{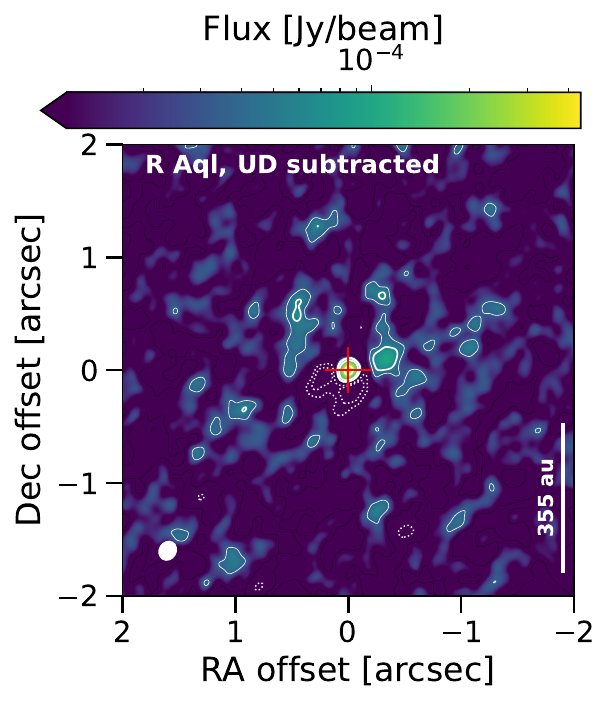}
    \caption{Continuum maps of R~Aql. See caption of Fig.~\ref{fig:gyaql}.}
    %\caption{Continuum maps of R Aql taken with the extended (left), mid (centre left) and combined (centre right) array configurations of ALMA. The rightmost plot shows the residuals after subtracting a uniform disc representing the AGB star. The thin solid contours indicate levels of $3\sigma$, the thick solid contours indicate levels of 5, 10, 30, 100, and 300$\sigma$, and the dotted contours indicate levels of $-3\sigma$. The continuum peak is indicated by the red cross.}
    \label{fig:raql}
\end{figure*}

\afterpage{
\begin{figure*}
	\includegraphics[height=5cm]{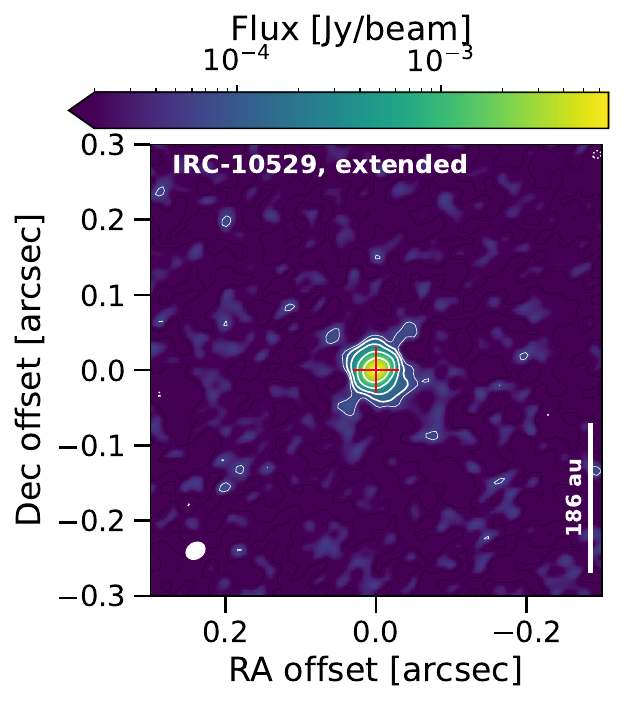}
  	\includegraphics[height=5cm]{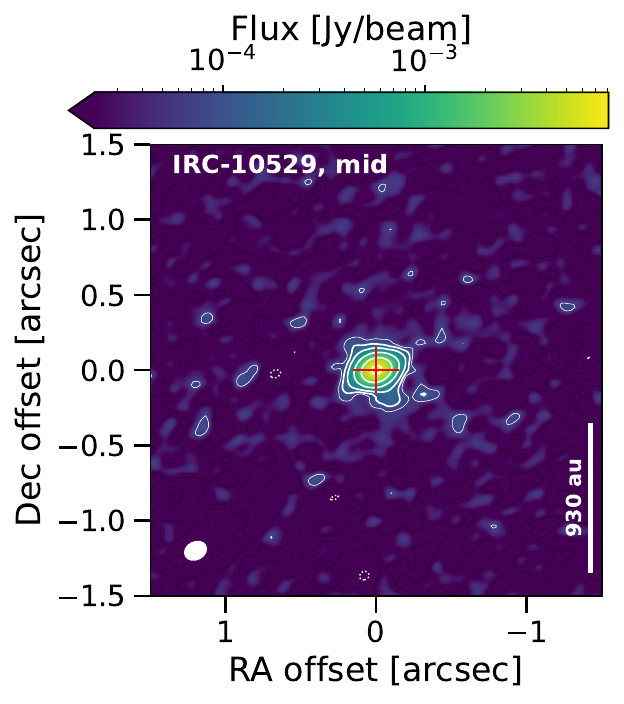}
 	\includegraphics[height=5cm]{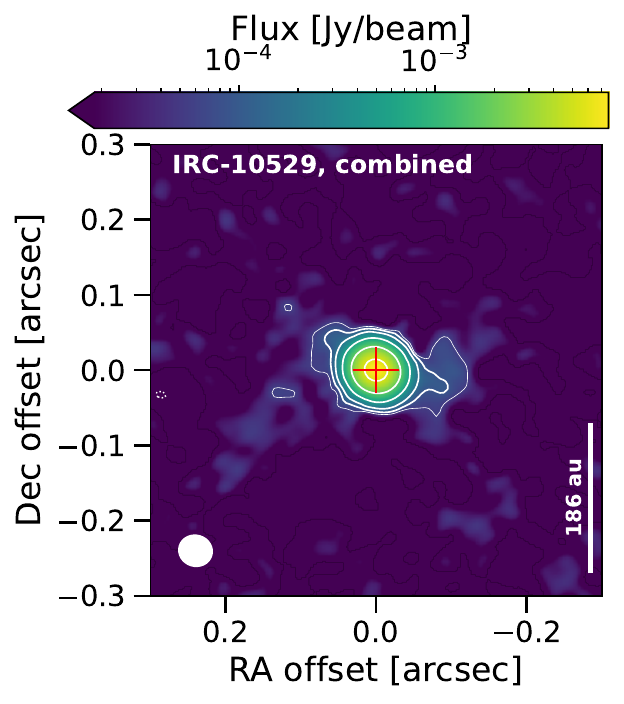}
	\includegraphics[height=5cm]{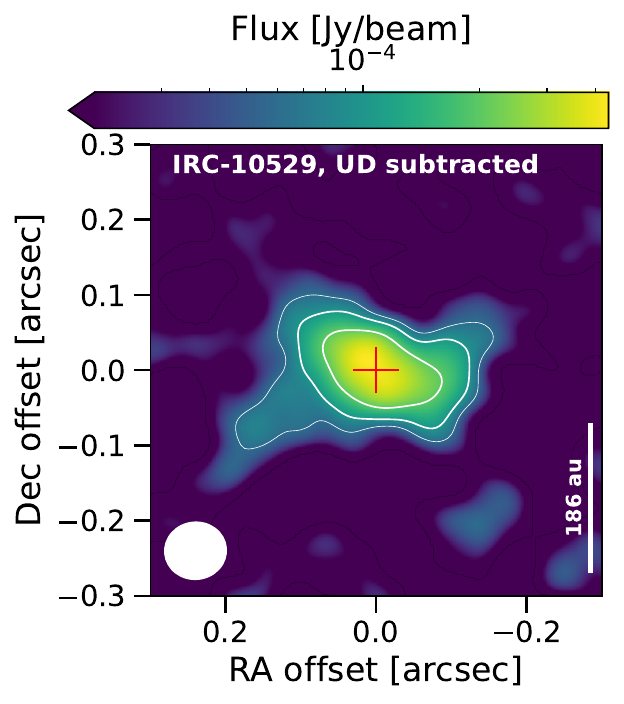}
\caption{Continuum maps of IRC $-$10529. See caption of Fig.~\ref{fig:gyaql}.}
%    \caption{Continuum maps of IRC $-$10529 taken with the extended (left), mid (centre left) and combined (centre right) array configurations of ALMA. The rightmost plot shows the residuals after subtracting a uniform disc representing the AGB star. The thin solid contours indicate levels of $3\sigma$, the thick solid contours indicate levels of 5, 10, 30, 100, and 300$\sigma$, and the dotted contours indicate levels of $-3\sigma$. The continuum peak is indicated by the red cross.}
    \label{fig:irc-10529}
    \vspace{0.5cm}
%\end{figure*}
%
%\begin{figure*}
	\includegraphics[height=5cm]{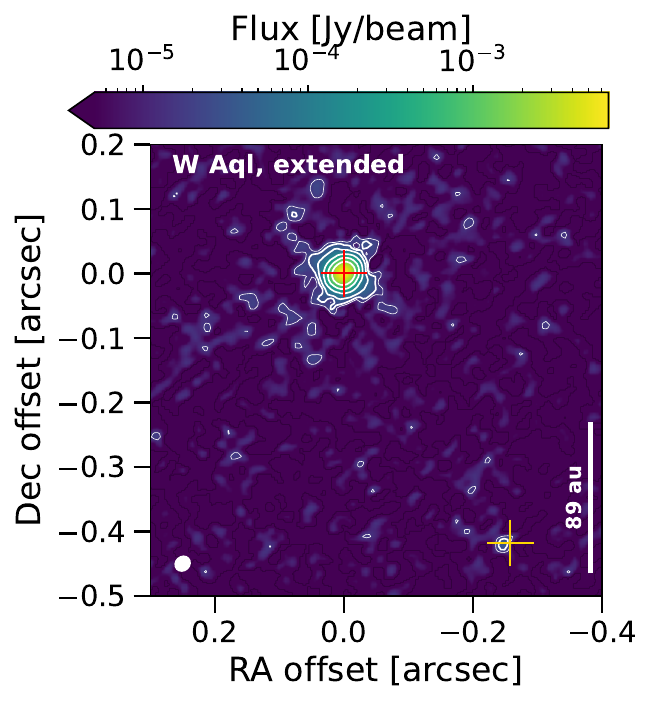}
  	\includegraphics[height=5cm]{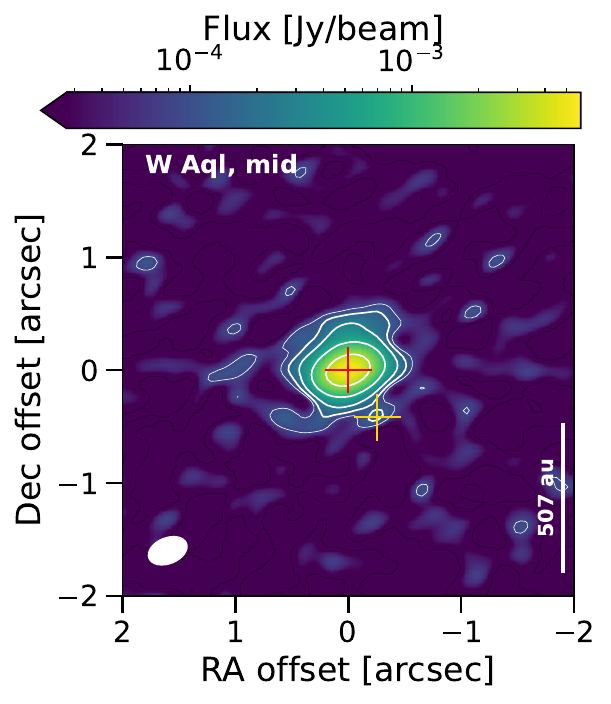}
 	\includegraphics[height=5cm]{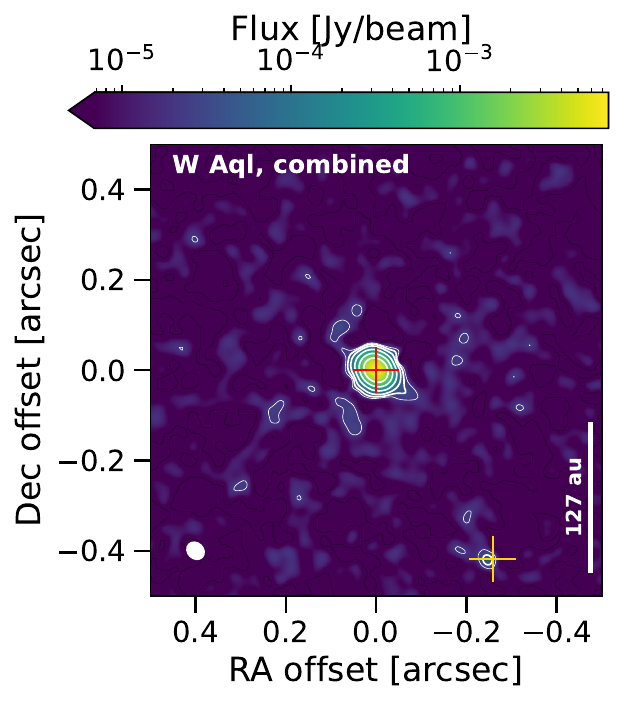}
	\includegraphics[height=5cm]{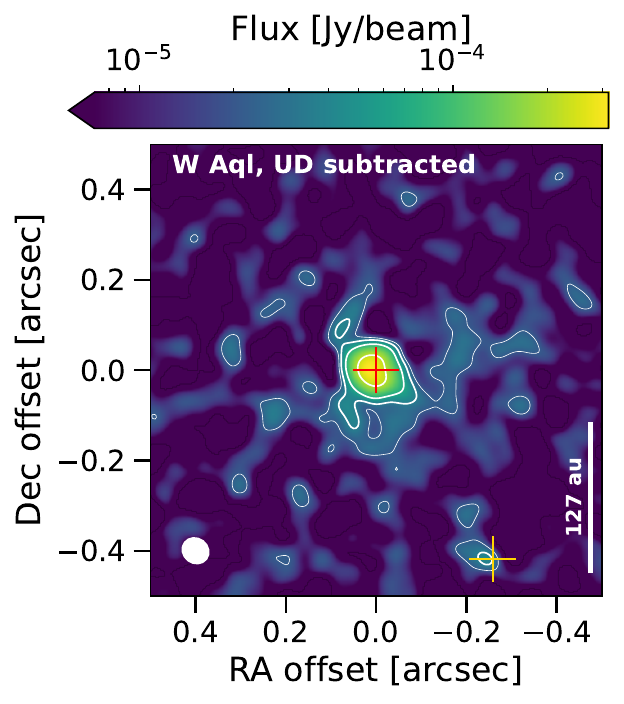}
\caption{Continuum maps of W~Aql. See caption of Fig.~\ref{fig:gyaql}. The position of the F9 main sequence companion is indicated by the yellow cross.}
%    \caption{Continuum maps of W Aql taken with the extended extended (left), mid (centre left) and combined (centre right) array configurations of ALMA. The rightmost plot shows the residuals after subtracting a uniform disc representing the AGB star. The thin solid contours indicate levels of $3\sigma$, the thick solid contours indicate levels of 5, 10, 30, 100, and 300$\sigma$. The continuum peak is indicated by the red cross. The position of the F9 main sequence companion is indicated by the yellow cross.}
    \label{fig:waql}
\end{figure*}
}

\afterpage{
\begin{figure*}
	\includegraphics[height=5cm]{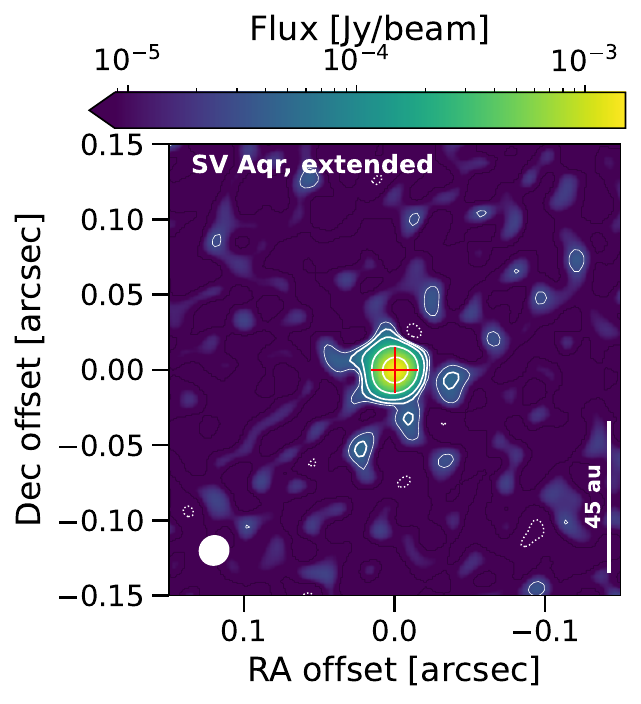}
  	\includegraphics[height=5cm]{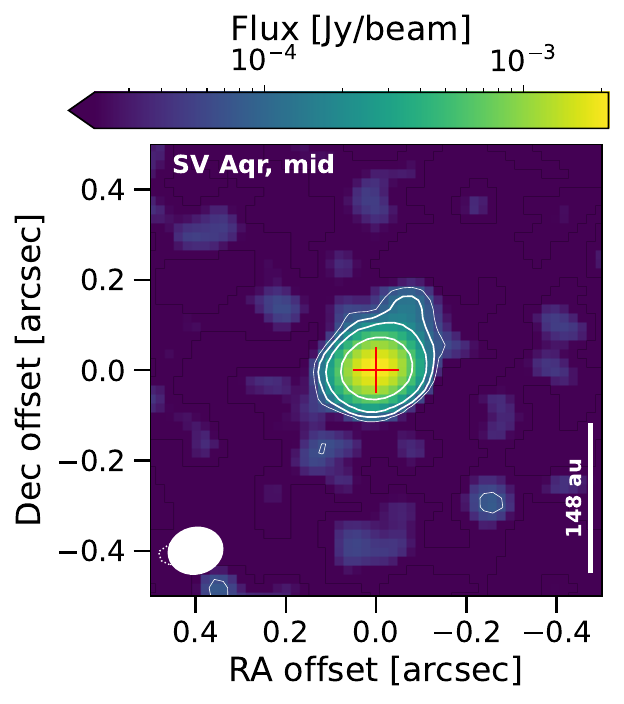}
 	\includegraphics[height=5cm]{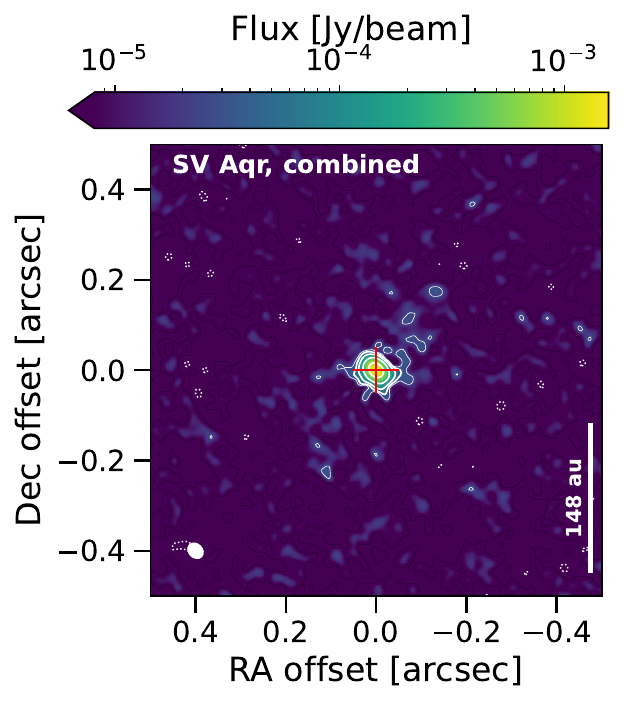}
	\includegraphics[height=5cm]{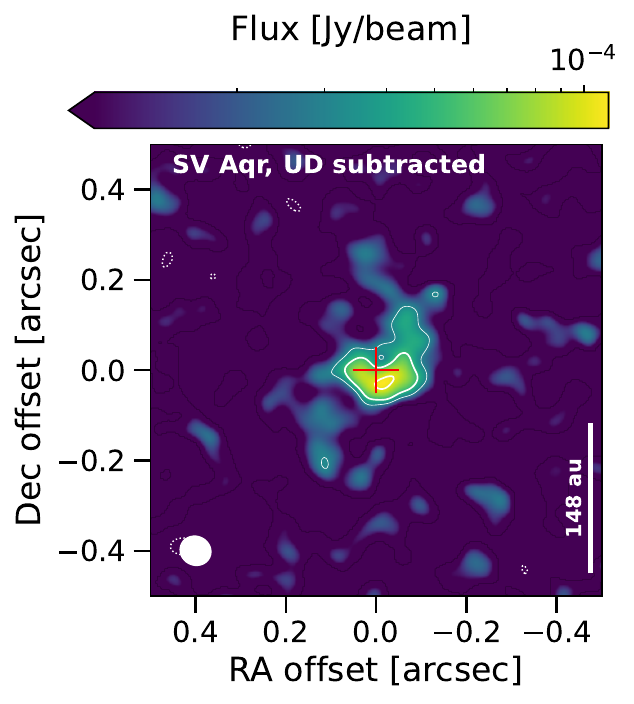}
\caption{Continuum maps of SV Aqr. See caption of Fig.~\ref{fig:gyaql}.}
%    \caption{Continuum maps of SV Aqr taken with the extended (left), mid (centre left) and combined (centre right) array configurations of ALMA. The rightmost plot shows the residuals after subtracting a uniform disc representing the AGB star. The thin solid contours indicate levels of $3\sigma$, the thick solid contours indicate levels of 5, 10, 30, 100, and 300$\sigma$, and the dotted contours indicate levels of $-3\sigma$. The continuum peak is indicated by the red cross.}
    \label{fig:svaqr}
    \vspace{0.5cm}
%\end{figure*}
%
%
%\begin{figure*}
	\includegraphics[height=5cm]{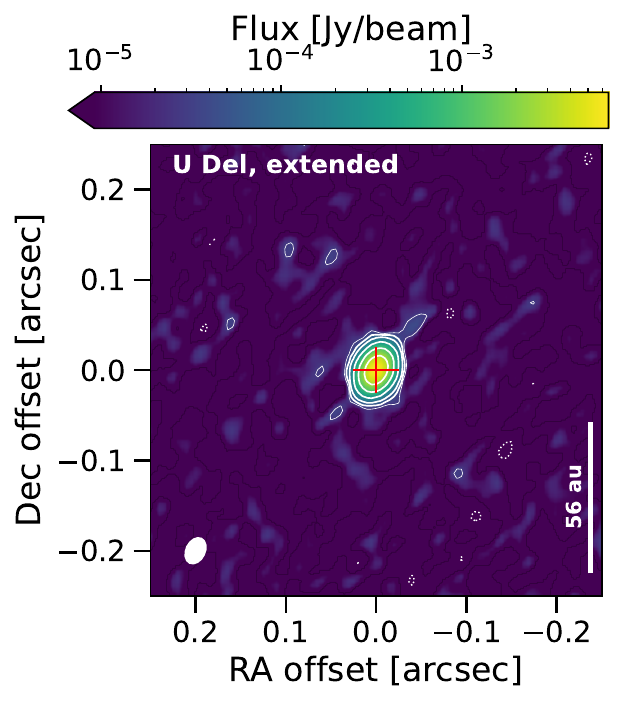}
  	\includegraphics[height=5cm]{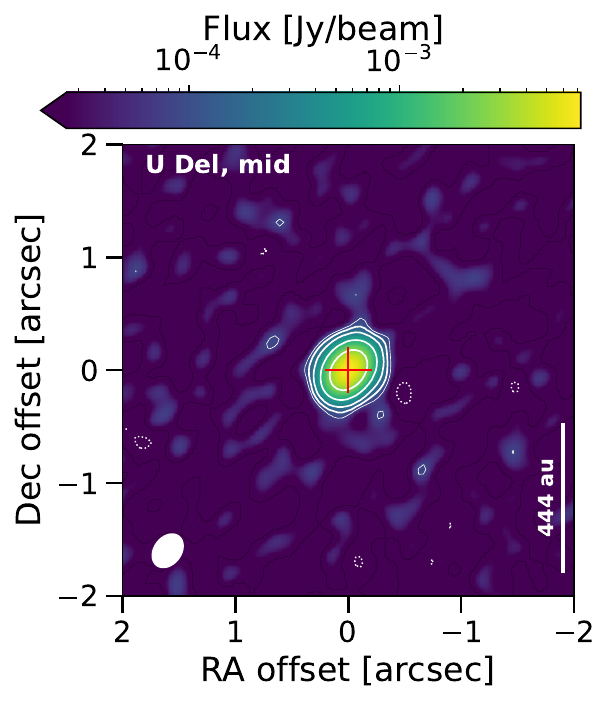}
 	\includegraphics[height=5cm]{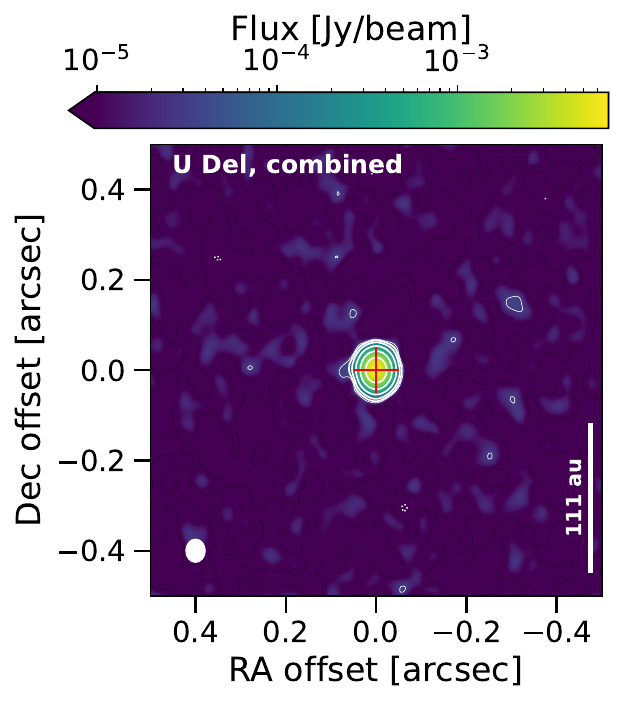}
	\includegraphics[height=5cm]{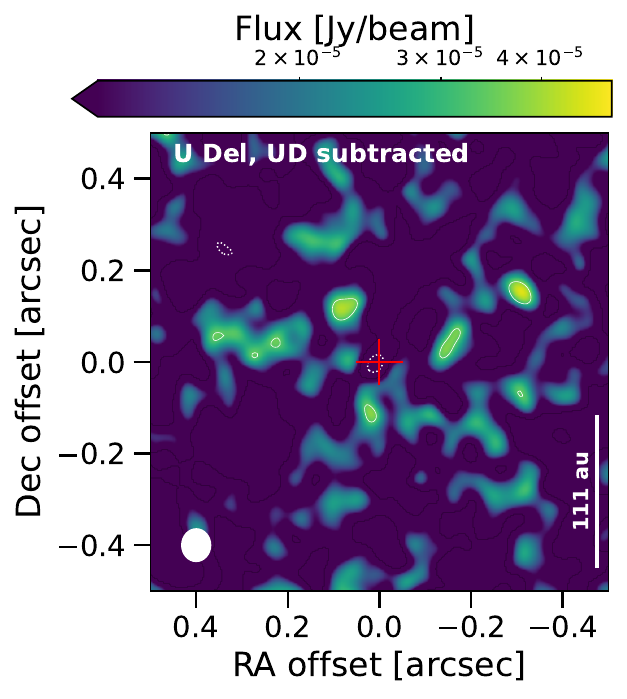}
\caption{Continuum maps of U Del. See caption of Fig.~\ref{fig:gyaql}.}
%    \caption{Continuum maps of U Del taken with the extended (left), mid (centre left) and combined (centre right) array configurations of ALMA. The rightmost plot shows the residuals after subtracting a uniform disc representing the AGB star. The thin solid contours indicate levels of $3\sigma$, the thick solid contours indicate levels of 5, 10, 30, 100, and 300$\sigma$, and the dotted contours indicate levels of $-3\sigma$. The continuum peak is indicated by the red cross.}
    \label{fig:udel}
\end{figure*}
}

\afterpage{
\begin{figure*}
	\includegraphics[height=5cm]{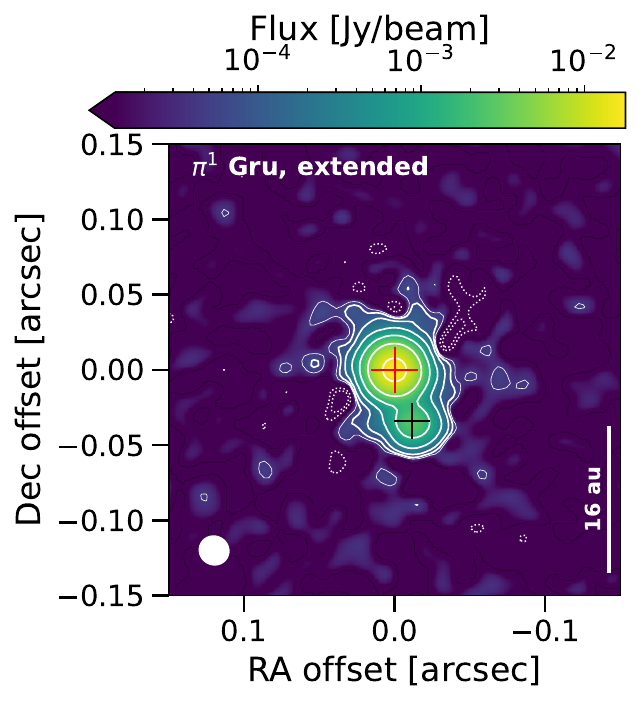}
  	\includegraphics[height=5cm]{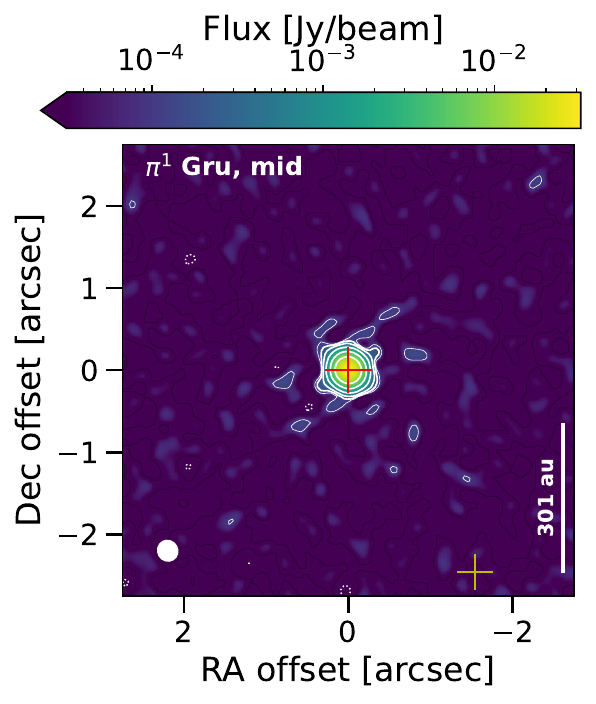}
 	\includegraphics[height=5cm]{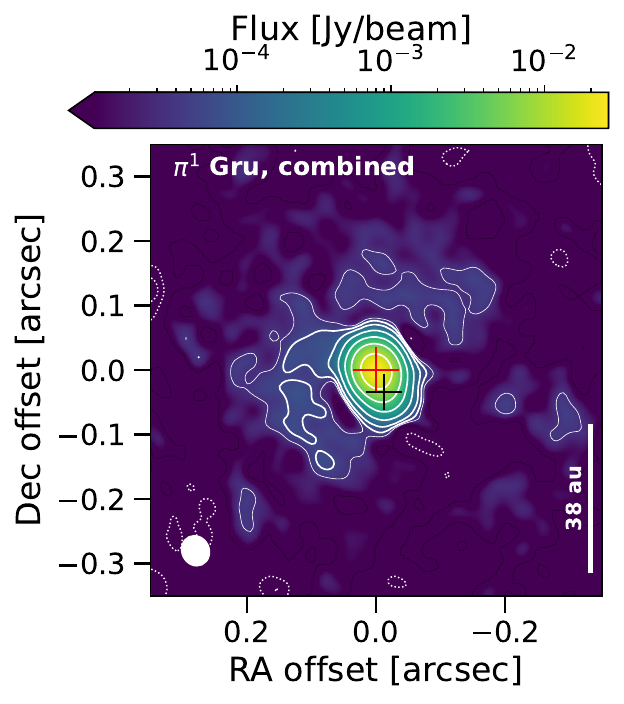}
	\includegraphics[height=5cm]{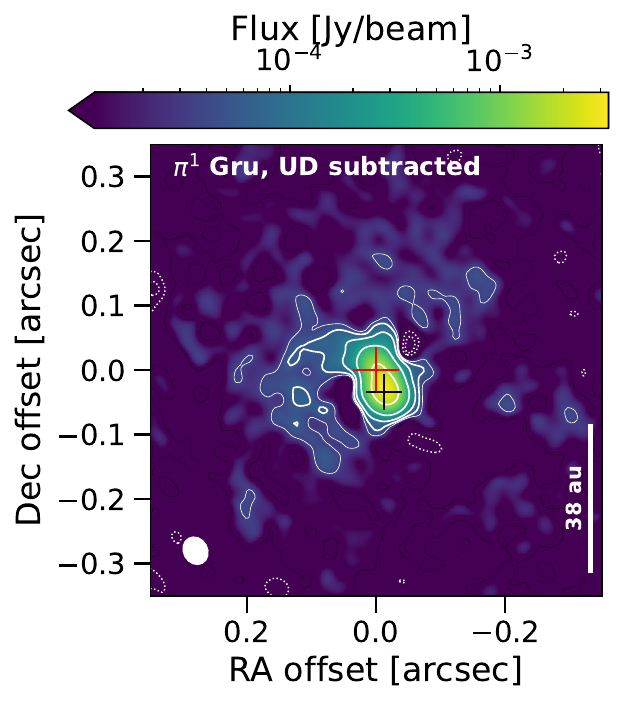}
\caption{Continuum maps of \pigru. See caption of Fig.~\ref{fig:gyaql}. The centre of the secondary peak, corresponding to the C companion, is marked by the black cross on the extended, combined and UD-subtracted plots. The yellow cross on the middle left plot indicates the position of the B companion based on Gaia observations.}
%    \caption{Continuum maps of \pigru\ taken with the extended (left), mid (centre left) and combined (centre right) array configurations of ALMA. The rightmost plot shows the residuals after subtracting a uniform disc representing the AGB star. The thin solid contours indicate levels of $3\sigma$, the thick solid contours indicate levels of 5, 10, 30, 100, 300, and 1000$\sigma$, and the dotted contours indicate levels of $-3$ and $-5\sigma$. The continuum peak is indicated by the red cross. The centre of the secondary peak, corresponding to the C companion, is marked by the black cross on the extended and combined plots. The yellow cross on the mid plot indicates the position of the B companion based on Gaia observations.}
    \label{fig:pi1gru}
    \vspace{0.5cm}
%\end{figure*}
%
%\begin{figure*}
	\includegraphics[height=4.87cm]{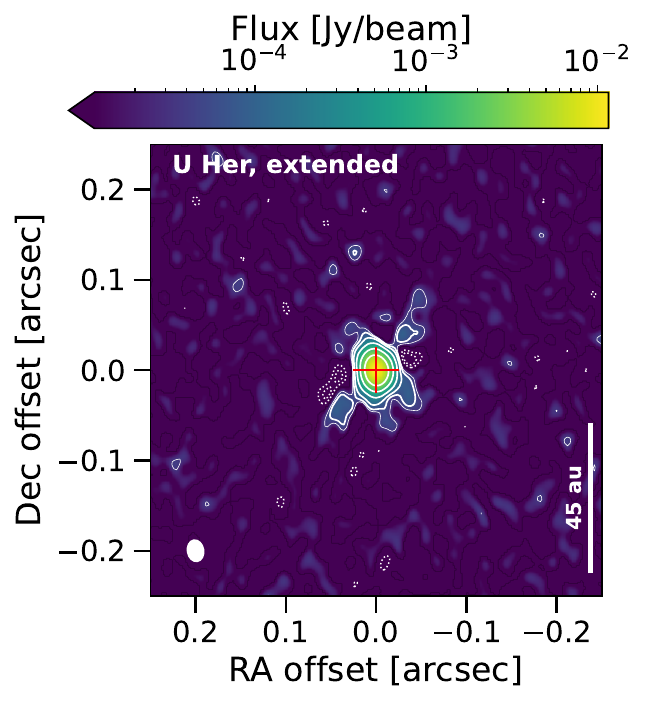}
  	\includegraphics[height=4.87cm]{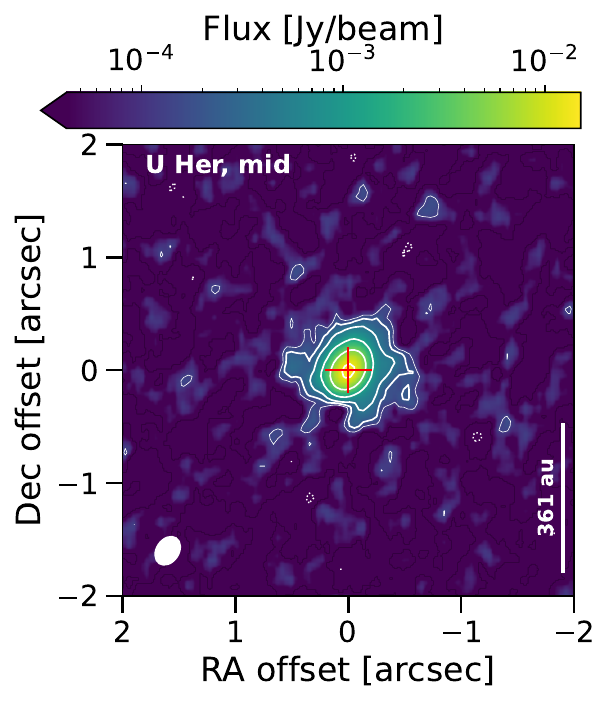}
 	\includegraphics[height=4.87cm]{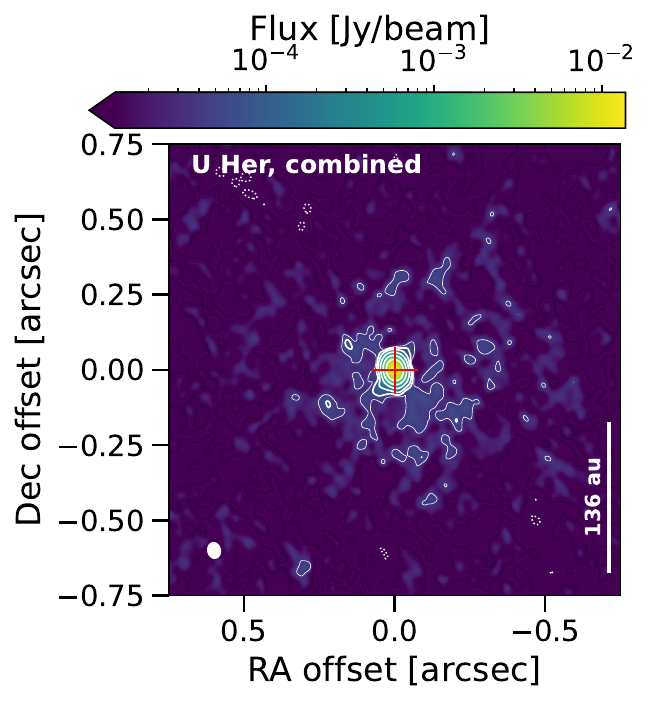}
	\includegraphics[height=4.87cm]{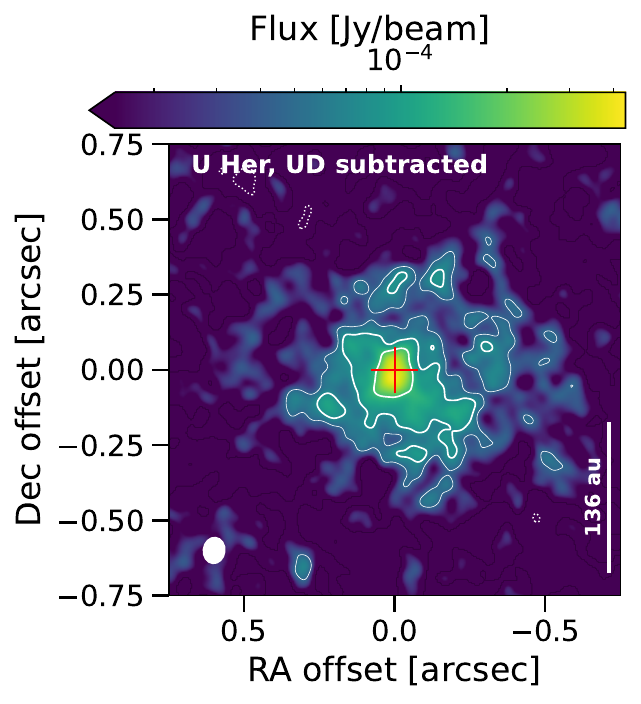}
\caption{Continuum maps of U Her. See caption of Fig.~\ref{fig:gyaql}.}
%    \caption{Continuum maps of U Her taken with the extended (left), mid (centre left) and combined (centre right) array configurations of ALMA. The rightmost plot shows the residuals after subtracting a uniform disc representing the AGB star. The thin solid contours indicate levels of $3\sigma$, the thick solid contours indicate levels of 5, 10, 30, 100, and 300$\sigma$, and the dotted contours indicate levels of $-3$ and $-5\sigma$. The continuum peak is indicated by the red cross.}
    \label{fig:uher}
\end{figure*}
}

\afterpage{
\begin{figure*}
	\includegraphics[height=5cm]{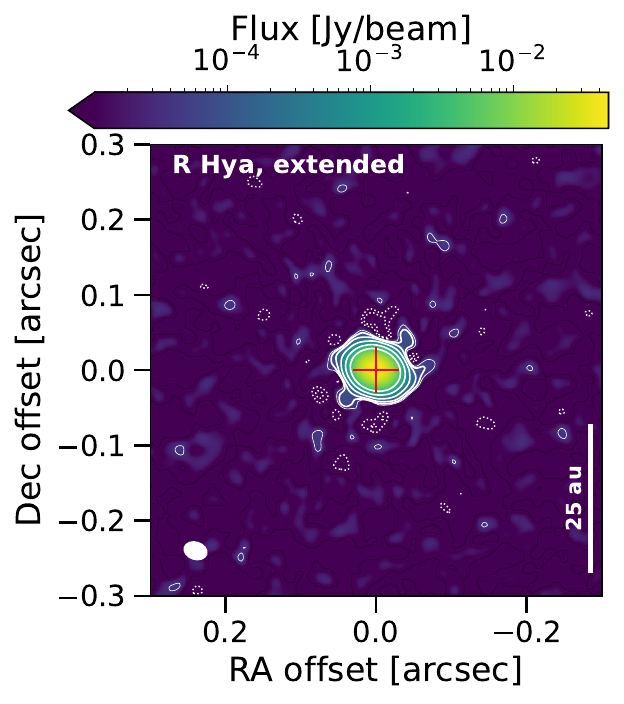}
  	\includegraphics[height=5cm]{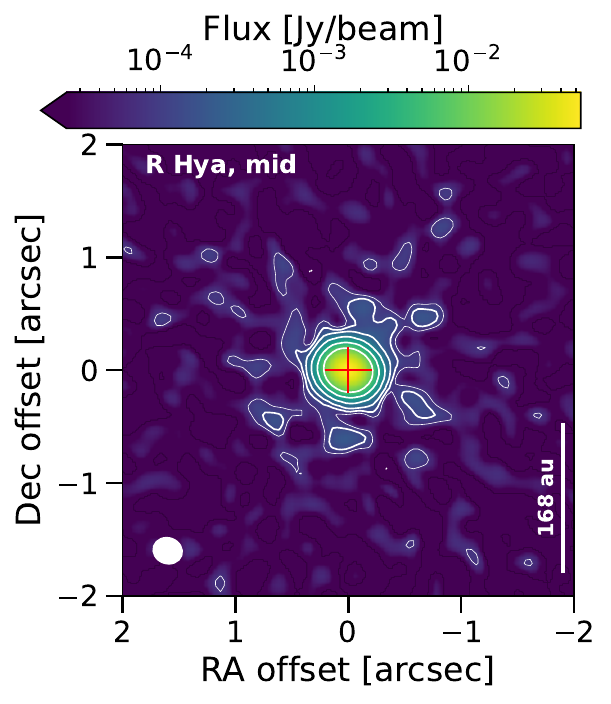}
 	\includegraphics[height=5cm]{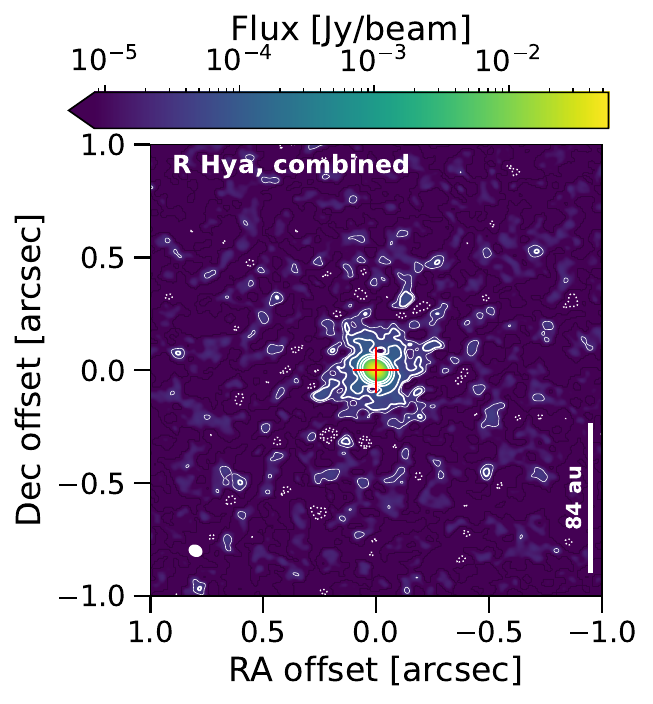}
	\includegraphics[height=5cm]{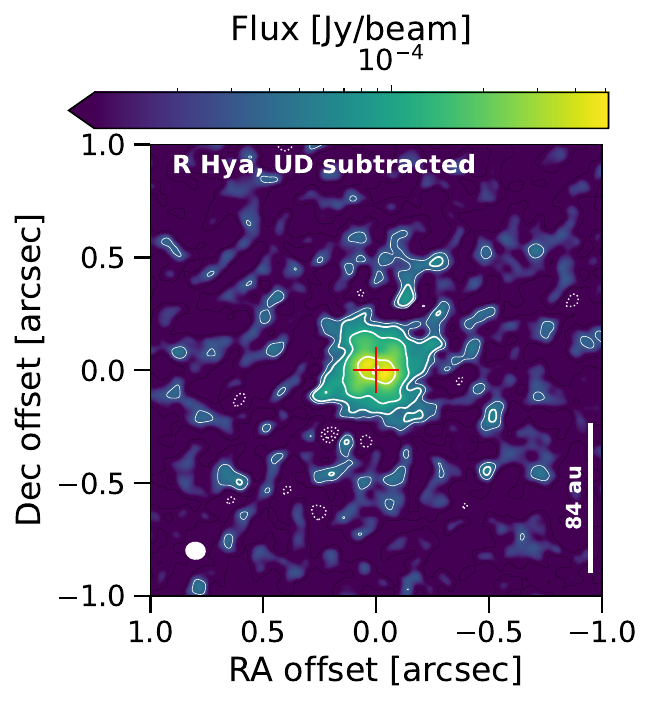}
\caption{Continuum maps of R Hya. See caption of Fig.~\ref{fig:gyaql}.}
%    \caption{Continuum maps of R Hya taken with the extended (left), mid (centre left) and combined (centre right) array configurations of ALMA. The rightmost plot shows the residuals after subtracting a uniform disc representing the AGB star. The thin solid contours indicate levels of $3\sigma$, the thick solid contours indicate levels of 5, 10, 30, 100, and 300$\sigma$, and the dotted contours indicate levels of $-3$ and $-5\sigma$. The continuum peak is indicated by the red cross.}
    \label{fig:rhya}
    \vspace{0.6cm}
%\end{figure*}
%
%\begin{figure*}
	\includegraphics[height=5.1cm]{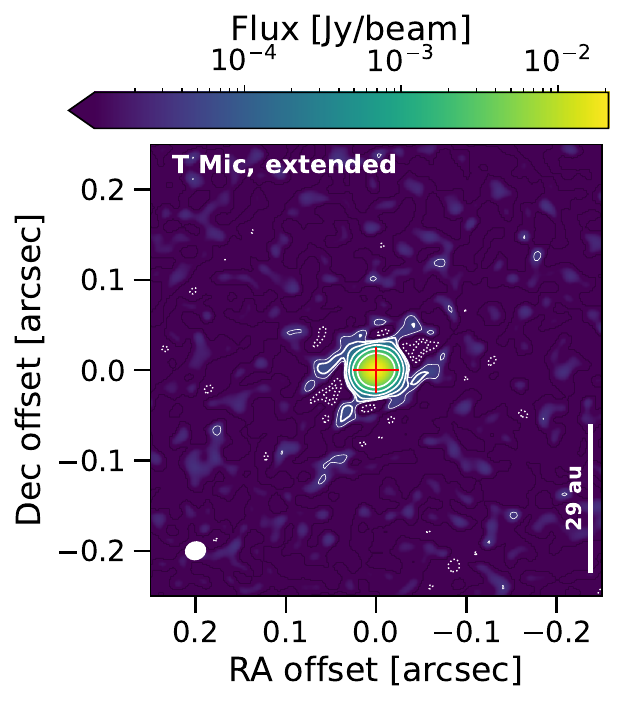}
  	\includegraphics[height=5.1cm]{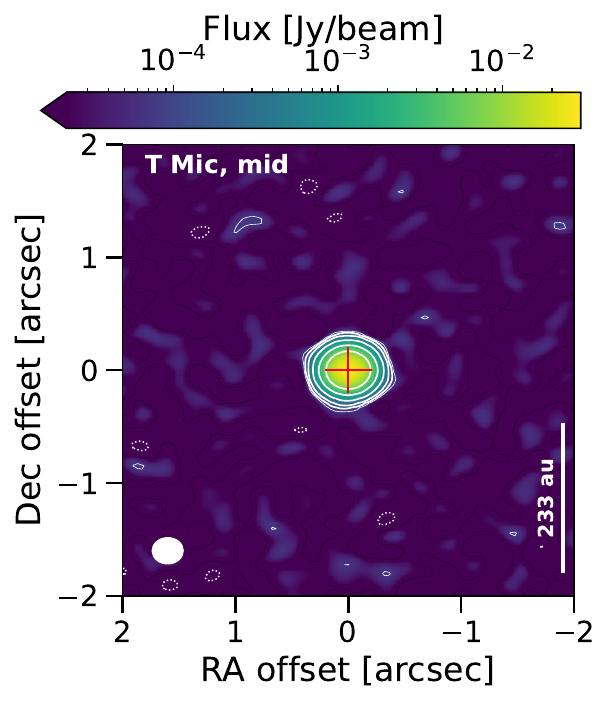}
 	\includegraphics[height=5.1cm]{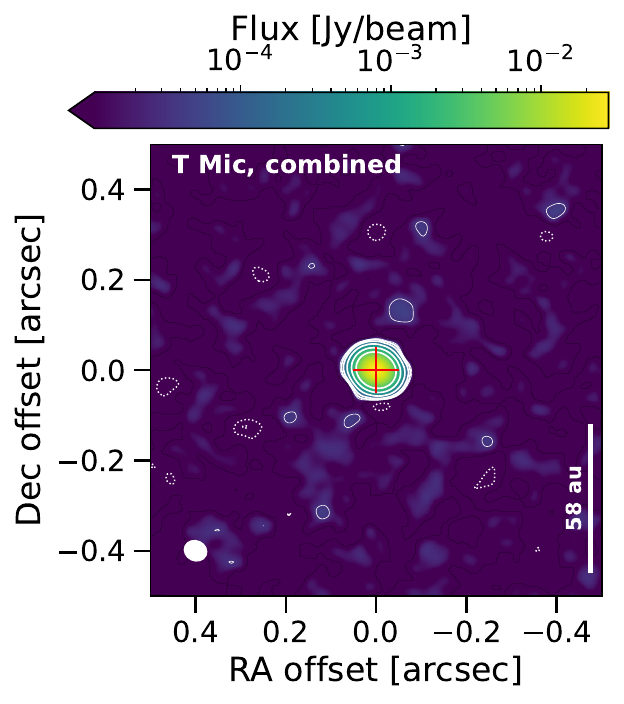}
	\includegraphics[height=5.1cm]{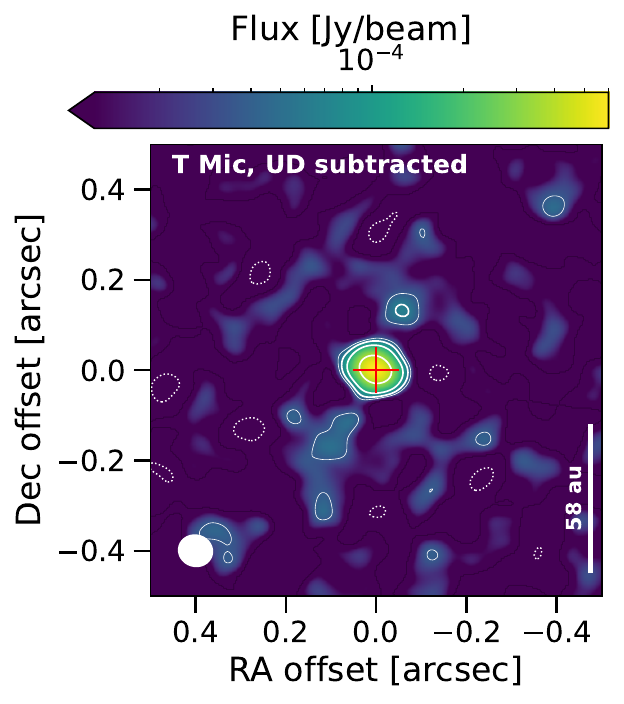}
\caption{Continuum maps of T Mic. See caption of Fig.~\ref{fig:gyaql}.}
%    \caption{Continuum maps of T Mic taken with the extended (left), mid (centre left) and combined (centre right) array configurations of ALMA. The rightmost plot shows the residuals after subtracting a uniform disc representing the AGB star. The thin solid contours indicate levels of $3\sigma$, the thick solid contours indicate levels of 5, 10, 30, 100, and 300$\sigma$, and the dotted contours indicate levels of $-3\sigma$ and $-5\sigma$. The continuum peak is indicated by the red cross.}
    \label{fig:tmic}
\end{figure*}
}

\afterpage{
\begin{figure*}
	\includegraphics[height=5.1cm]{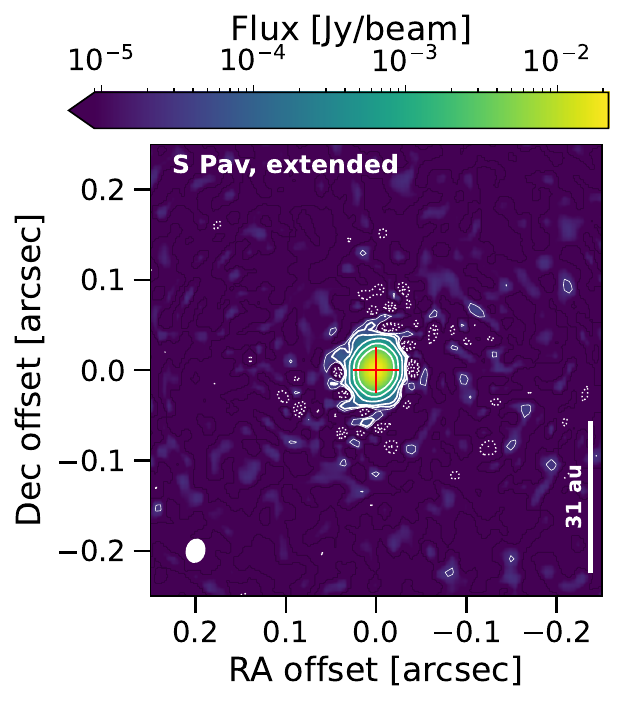}
  	\includegraphics[height=5.1cm]{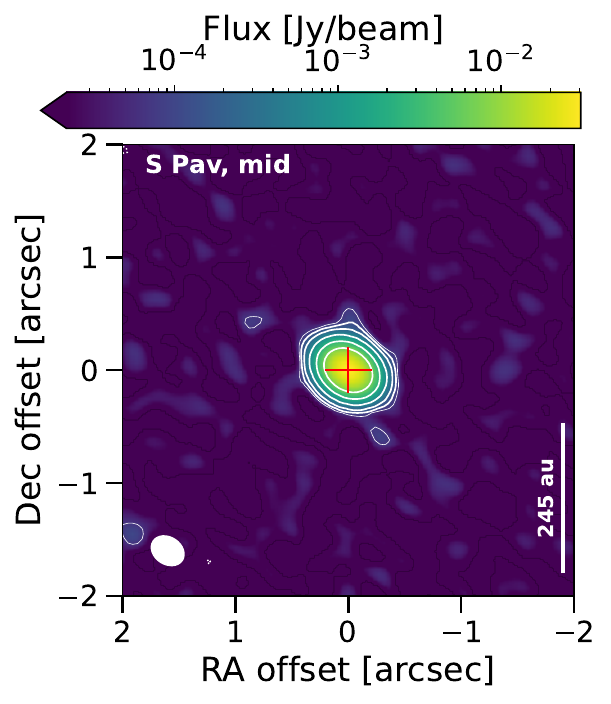}
 	\includegraphics[height=5.1cm]{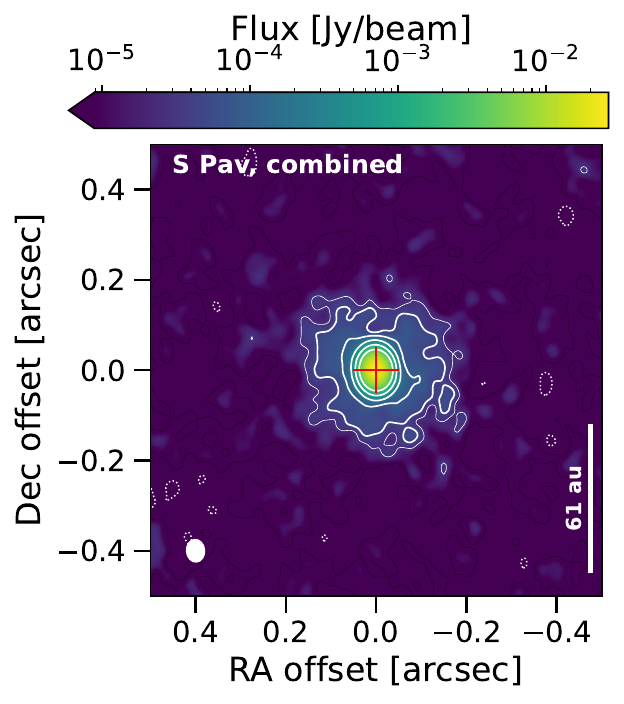}
	\includegraphics[height=5.1cm]{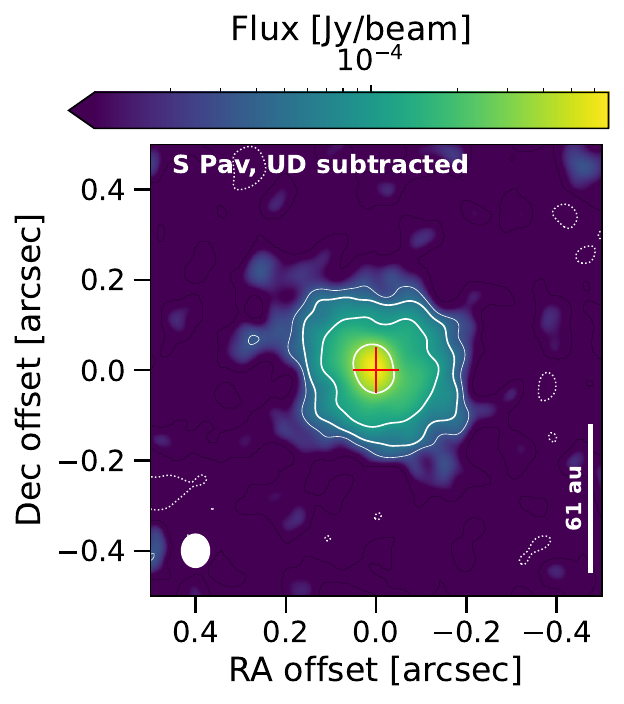}
\caption{Continuum maps of S Pav. See caption of Fig.~\ref{fig:gyaql}.}
%    \caption{Continuum maps of S Pav taken with the extended (left), mid (centre left) and combined (centre right) array configurations of ALMA. The rightmost plot shows the residuals after subtracting a uniform disc representing the AGB star. The thin solid contours indicate levels of $3\sigma$, the thick solid contours indicate levels of 5, 10, 30, 100, and 300$\sigma$, and the dotted contours indicate levels of $-3$ and $-5\sigma$. The continuum peak is indicated by the red cross.}
    \label{fig:spav}
    \vspace{0.7cm}
%\end{figure*}
%
%\begin{figure*}
    \includegraphics[height=4.9cm]{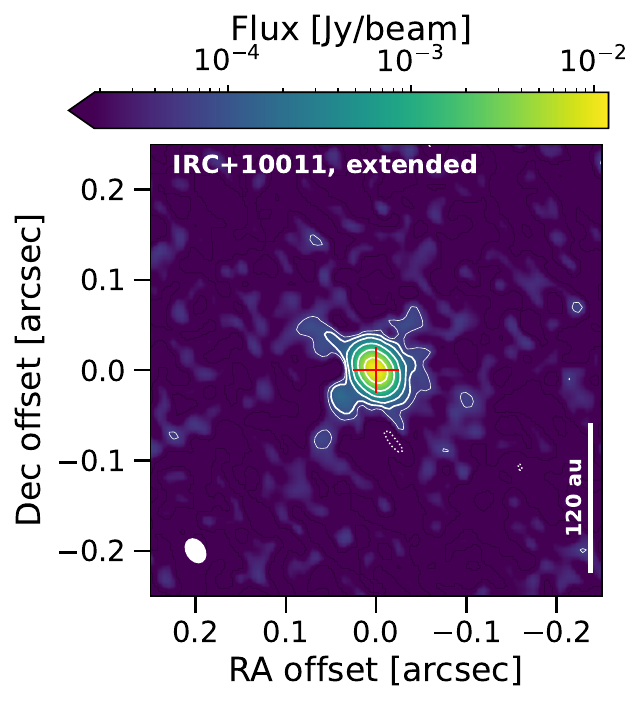}
    \includegraphics[height=4.9cm]{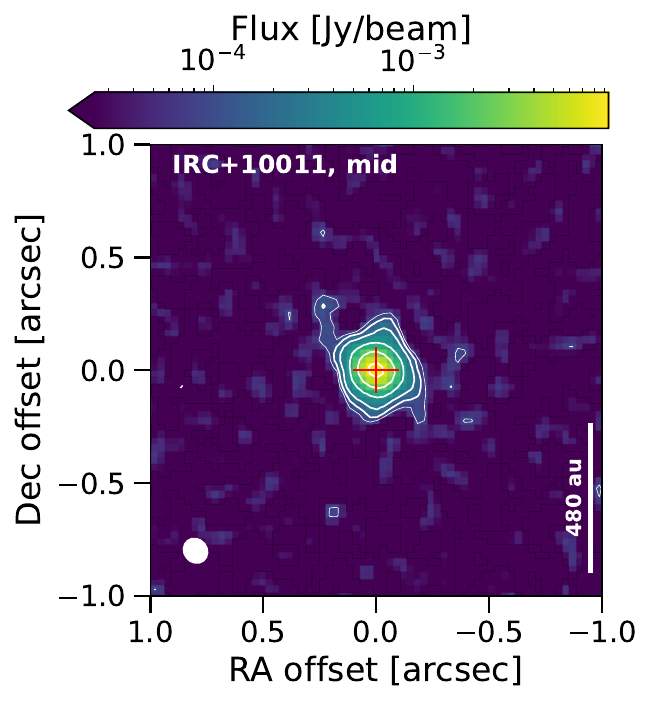}
    \includegraphics[height=4.9cm]{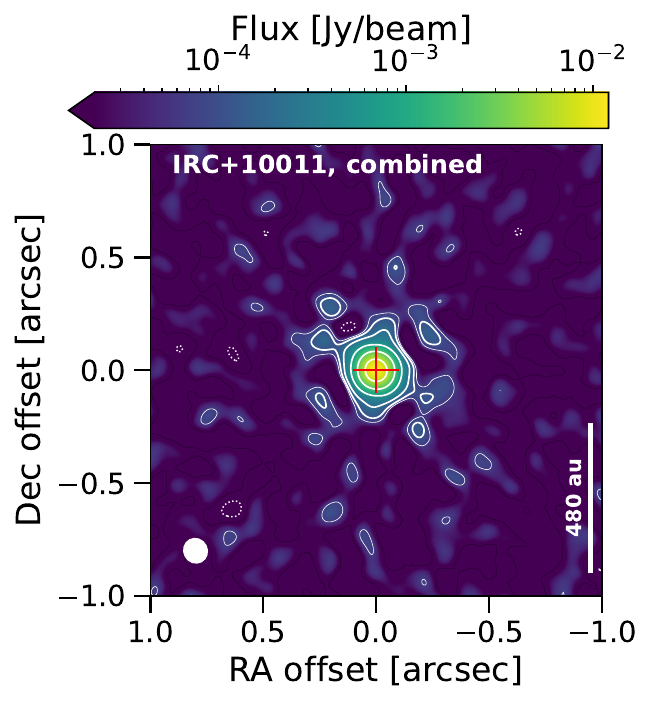}
    \includegraphics[height=4.9cm]{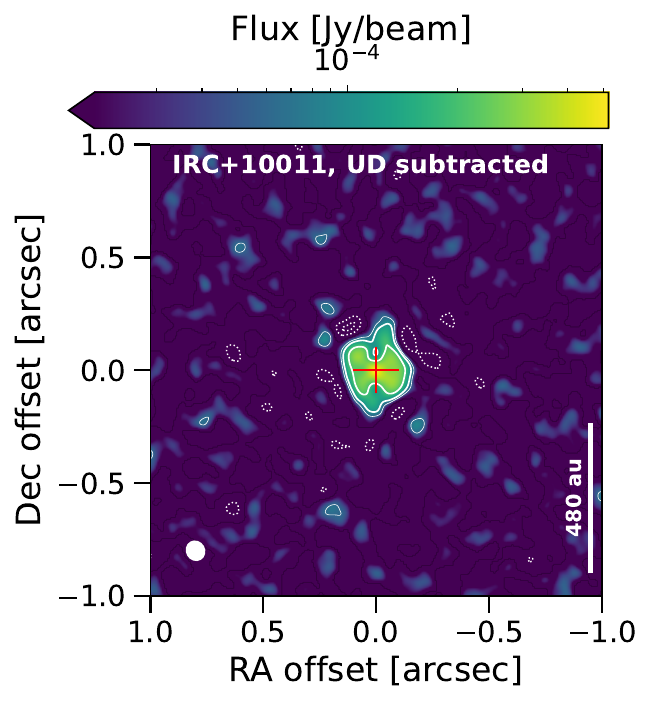}
\caption{Continuum maps of IRC+10011. See caption of Fig.~\ref{fig:gyaql}.}
%    \caption{Continuum maps of IRC +10011 taken with the extended (left), mid (centre left) and combined (centre right) array configurations of ALMA. The rightmost plot shows the residuals after subtracting a uniform disc representing the AGB star. The thin solid contours indicate levels of $3\sigma$, the thick solid contours indicate levels of 5, 10, 30, 100, and 300$\sigma$, and the dotted contours indicate levels of $-3\sigma$. The continuum peak is indicated by the red cross.}
    \label{fig:irc+10011}
\end{figure*}
}

\afterpage{
\begin{figure*}
	\includegraphics[height=5.1cm]{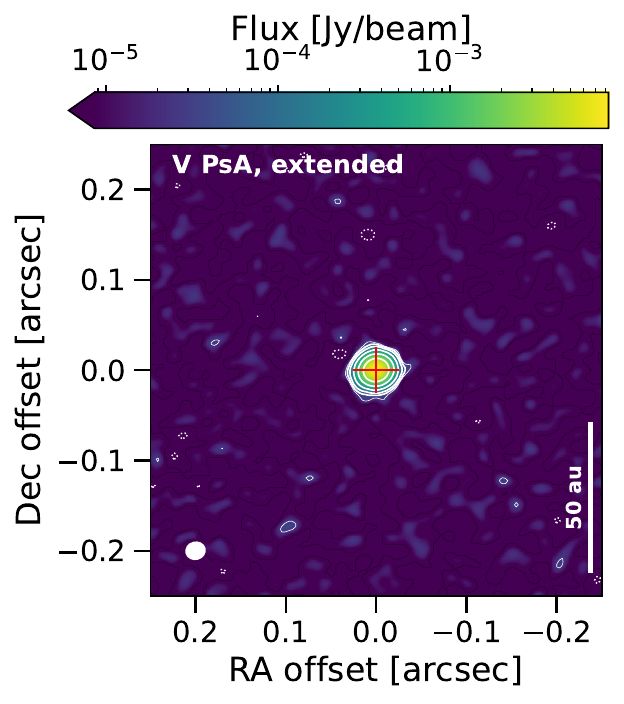}
  	\includegraphics[height=5.1cm]{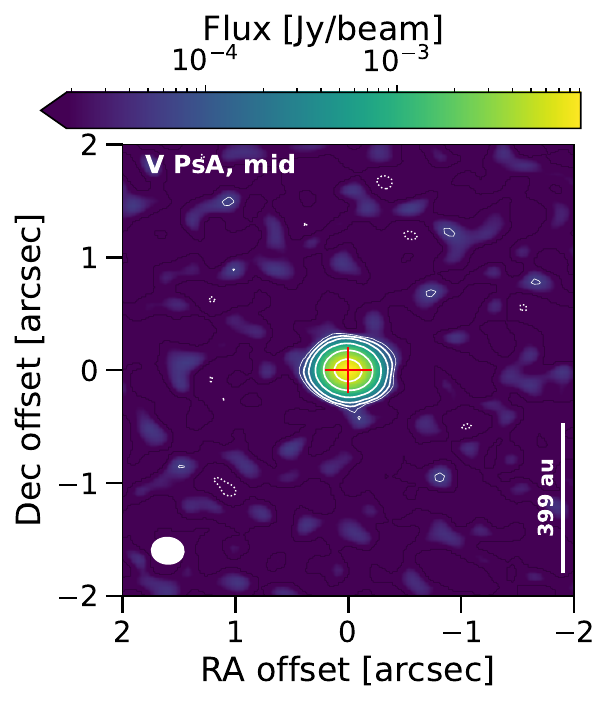}
 	\includegraphics[height=5.1cm]{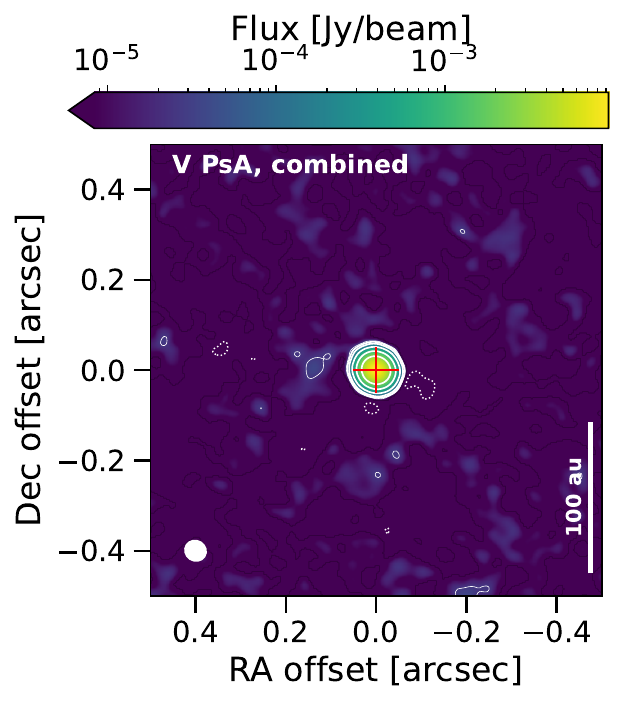}
	\includegraphics[height=5.1cm]{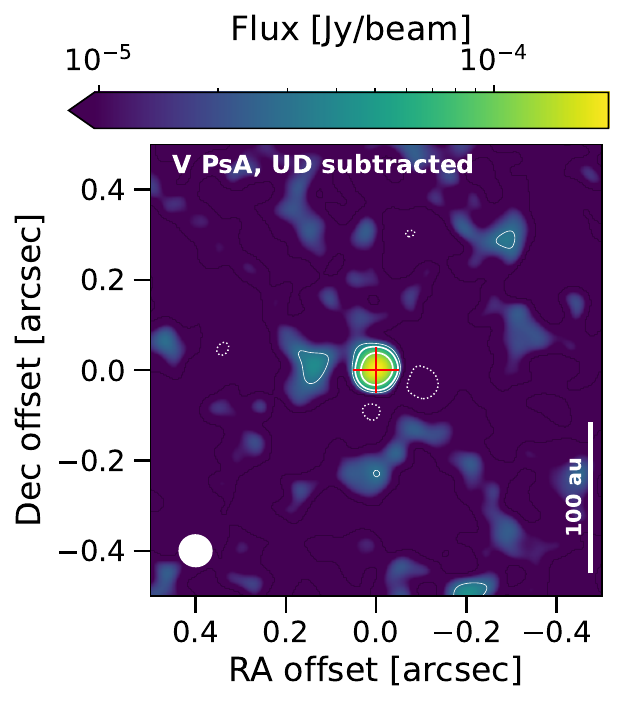}
\caption{Continuum maps of V PsA. See caption of Fig.~\ref{fig:gyaql}.}
%    \caption{Continuum maps of V PsA taken with the extended (left), mid (centre left) and combined (centre right) array configurations of ALMA. The rightmost plot shows the residuals after subtracting a uniform disc representing the AGB star. The thin solid contours indicate levels of $3\sigma$, the thick solid contours indicate levels of 5, 10, 30, 100, and 300$\sigma$, and the dotted contours indicate levels of $-3\sigma$. The continuum peak is indicated by the red cross.}
    \label{fig:vpsa}
    \vspace{0.6cm}
%\end{figure*}
%
%\begin{figure*}
    \includegraphics[height=5cm]{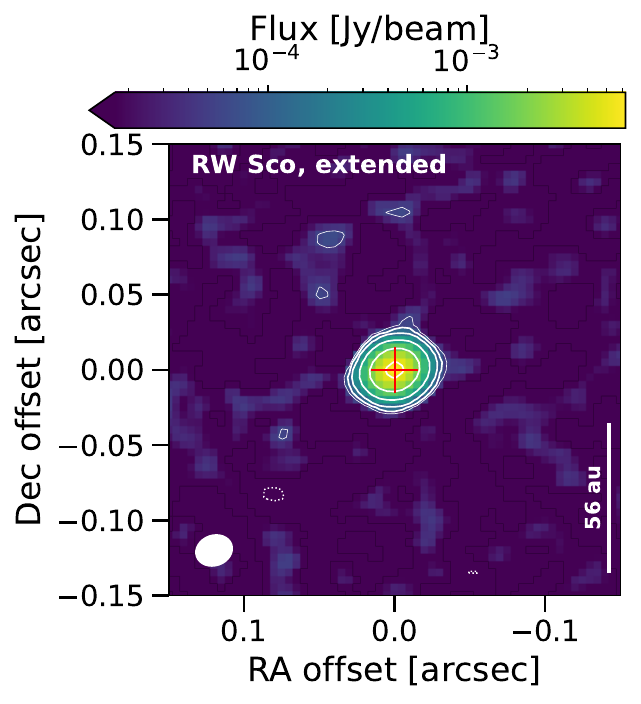}
    \includegraphics[height=5cm]{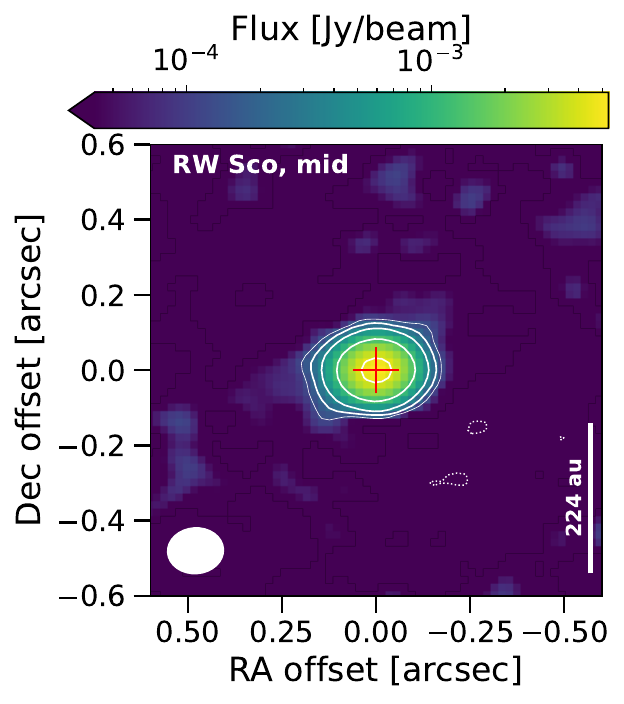}
    \includegraphics[height=5cm]{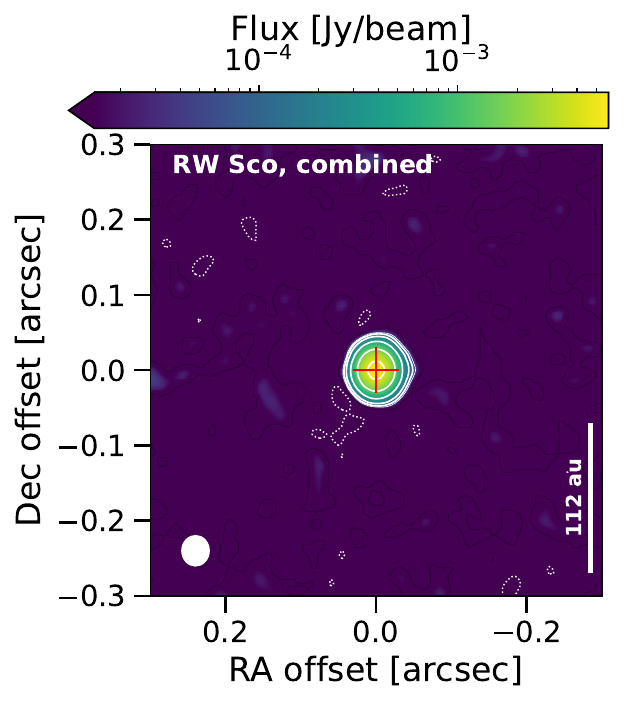}
    \includegraphics[height=5cm]{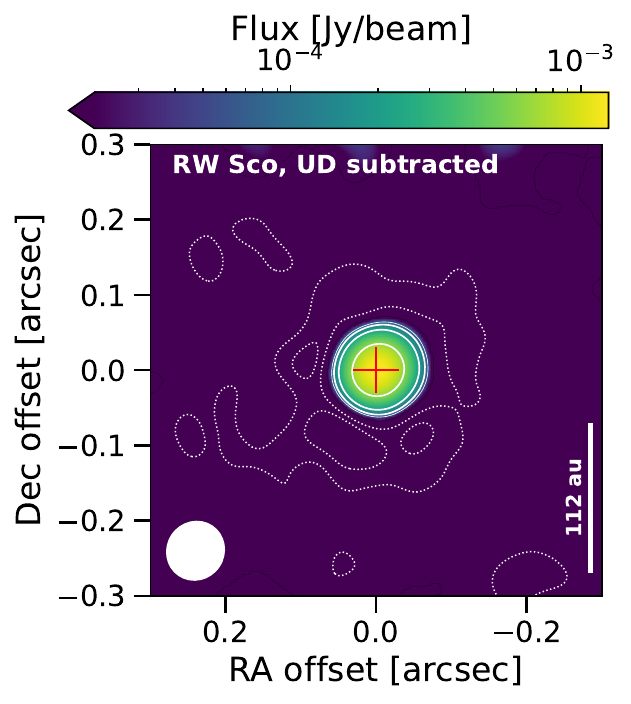}
\caption{Continuum maps of RW Sco. See caption of Fig.~\ref{fig:gyaql}.}
%    \caption{Continuum maps of RW Sco taken with the extended (left), mid (centre left) and combined (centre right) array configurations of ALMA. The rightmost plot shows the residuals after subtracting a uniform disc representing the AGB star. The thin solid contours indicate levels of $3\sigma$, the thick solid contours indicate levels of 5, 10, 30, 100, and 300$\sigma$, and the dotted contours indicate levels of $-3\sigma$. The continuum peak is indicated by the red cross. \blue{Note negative regions in combined.}}
    \label{fig:rwsco}
\end{figure*}
}

\afterpage{
\begin{figure*}
	\includegraphics[height=4.9cm]{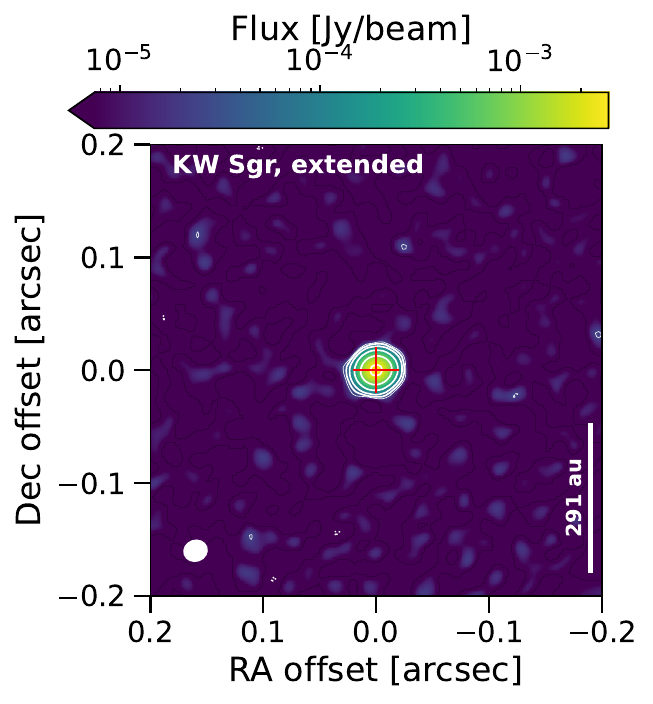}
  	\includegraphics[height=4.9cm]{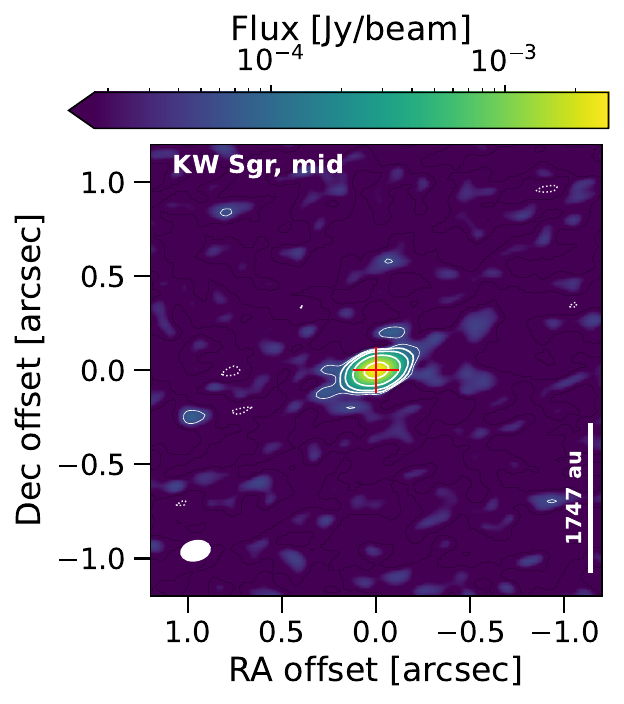}
 	\includegraphics[height=4.9cm]{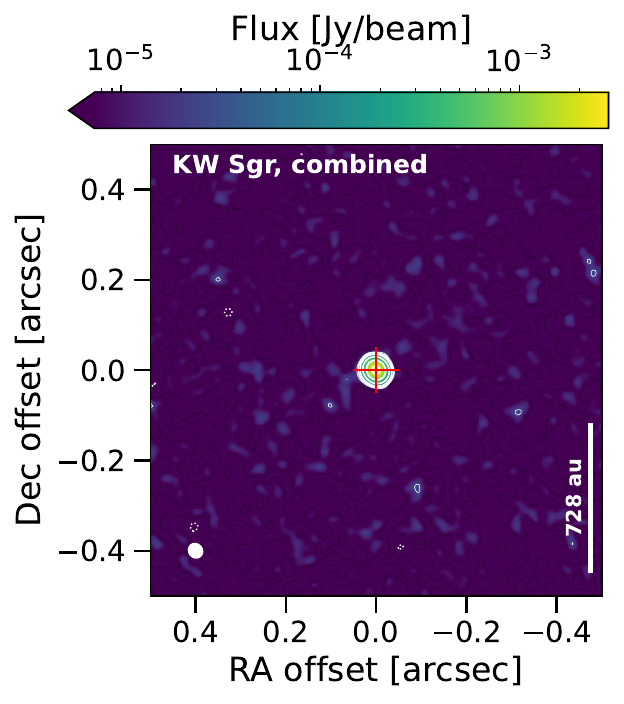}
	\includegraphics[height=4.9cm]{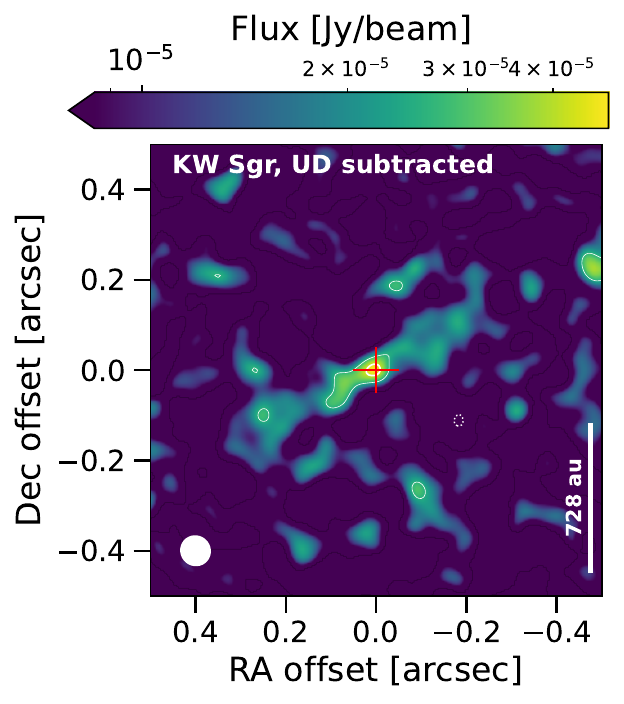}
\caption{Continuum maps of KW Sgr. See caption of Fig.~\ref{fig:gyaql}.}
%    \caption{Continuum maps of KW Sgr taken with the extended (left), mid (centre left) and combined (centre right) array configurations of ALMA. The rightmost plot shows the residuals after subtracting a uniform disc representing the AGB star. The thin solid contours indicate levels of $3\sigma$, the thick solid contours indicate levels of 5, 10, 30, 100, and 300$\sigma$, and the dotted contours indicate levels of $-3\sigma$. The continuum peak is indicated by the red cross.}
    \label{fig:kwsgr}
%\end{figure*}
%
%\begin{figure*}
	\includegraphics[height=4.9cm]{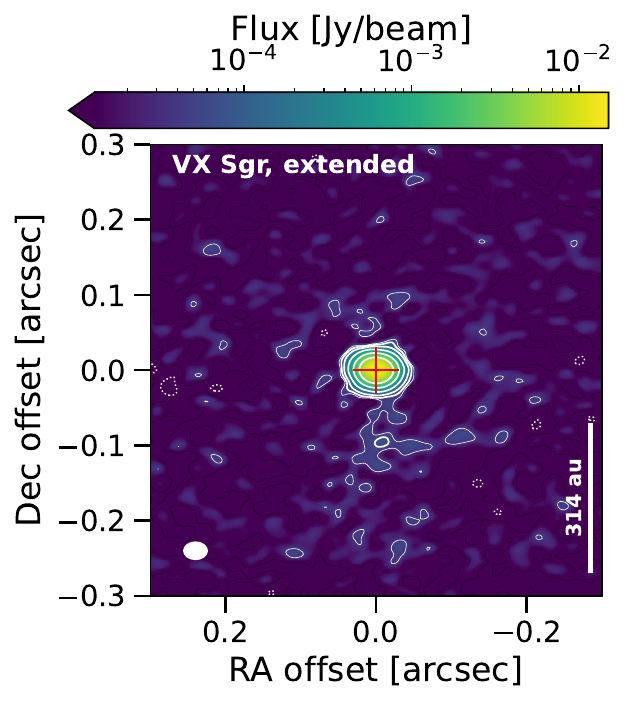}
  	\includegraphics[height=4.9cm]{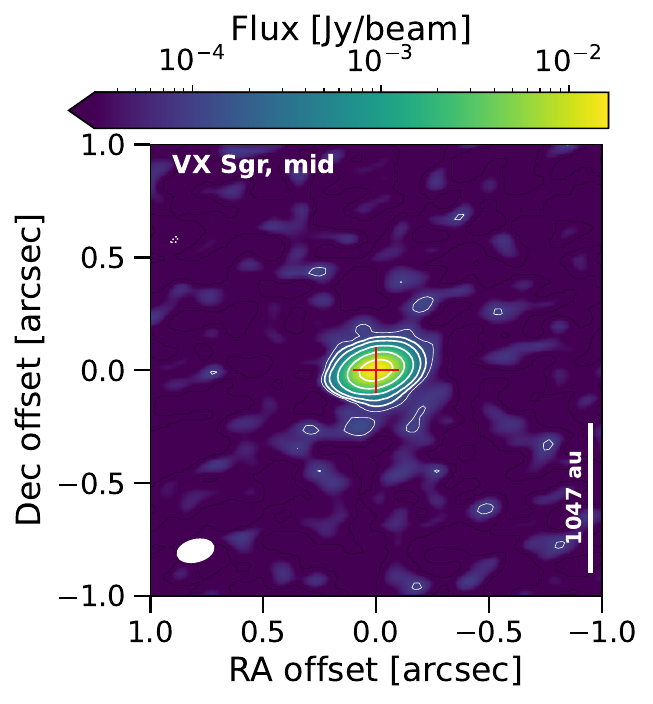}
 	\includegraphics[height=4.9cm]{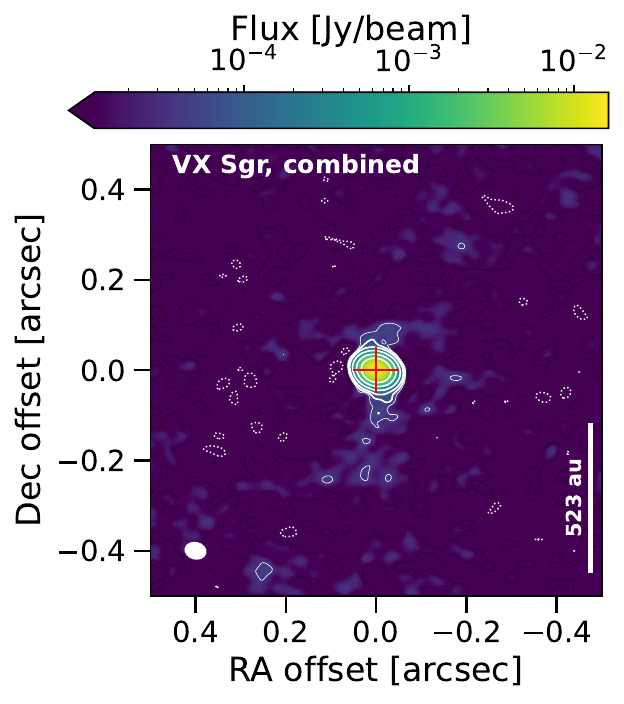}
	\includegraphics[height=4.9cm]{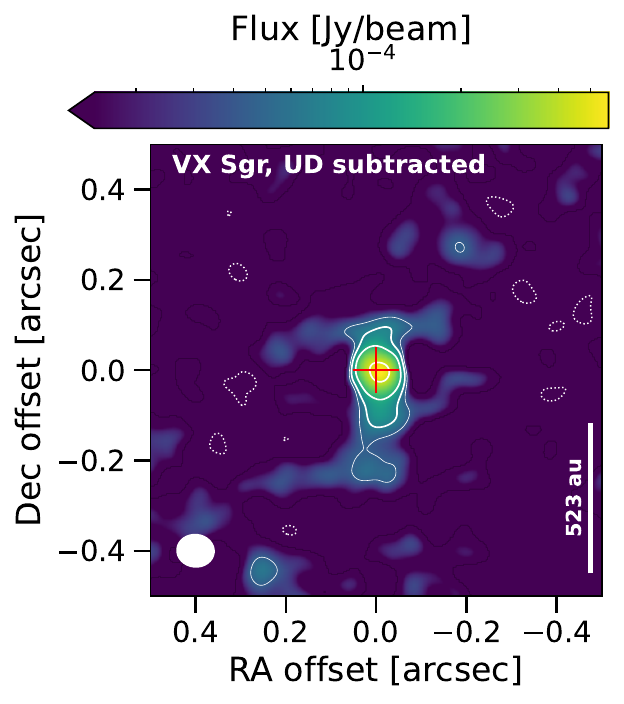}
\caption{Continuum maps of VX Sgr. See caption of Fig.~\ref{fig:gyaql}.}
%    \caption{Continuum maps of VX Sgr taken with the extended (left), mid (centre left) and combined (centre right) array configurations of ALMA. The rightmost plot shows the residuals after subtracting a uniform disc representing the AGB star. The thin solid contours indicate levels of $3\sigma$, the thick solid contours indicate levels of 5, 10, 30, 100, and 300$\sigma$, and the dotted contours indicate levels of $-3\sigma$. The continuum peak is indicated by the red cross.}
    \label{fig:vxsgr}
%\end{figure*}
%
%
%\begin{figure*}
	\includegraphics[height=4.9cm]{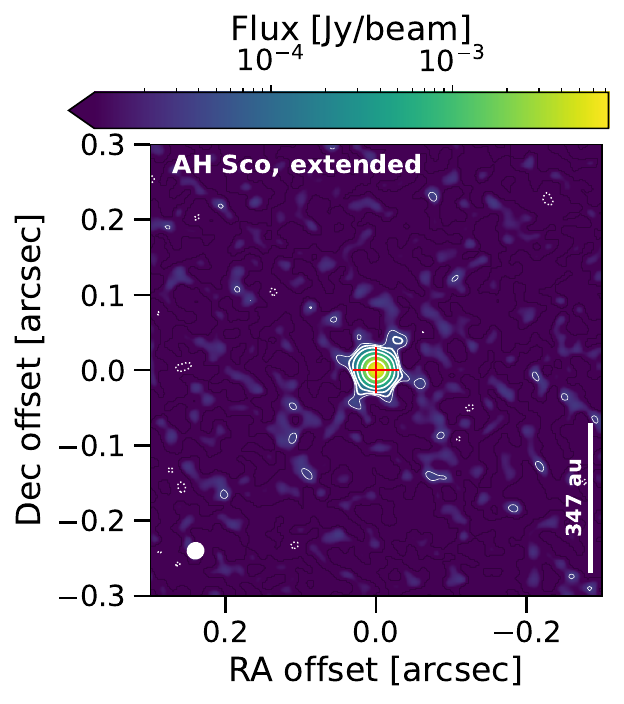}
  	\includegraphics[height=4.9cm]{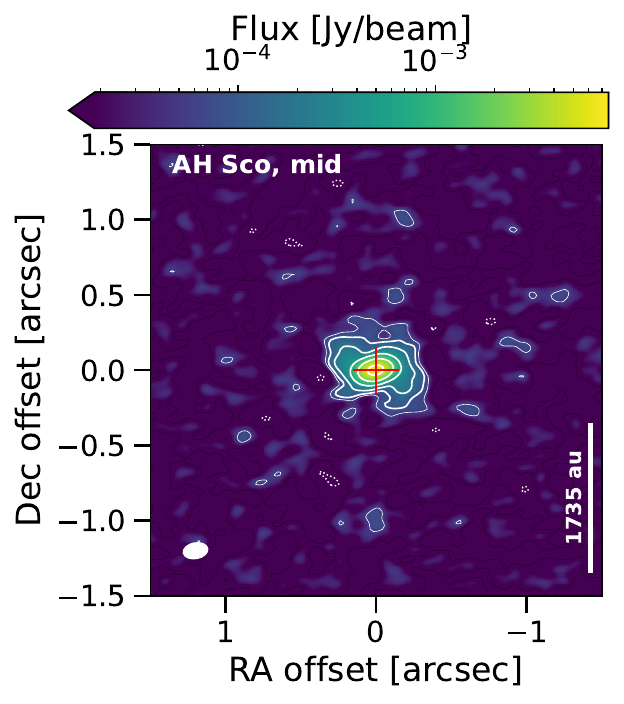}
 	\includegraphics[height=4.9cm]{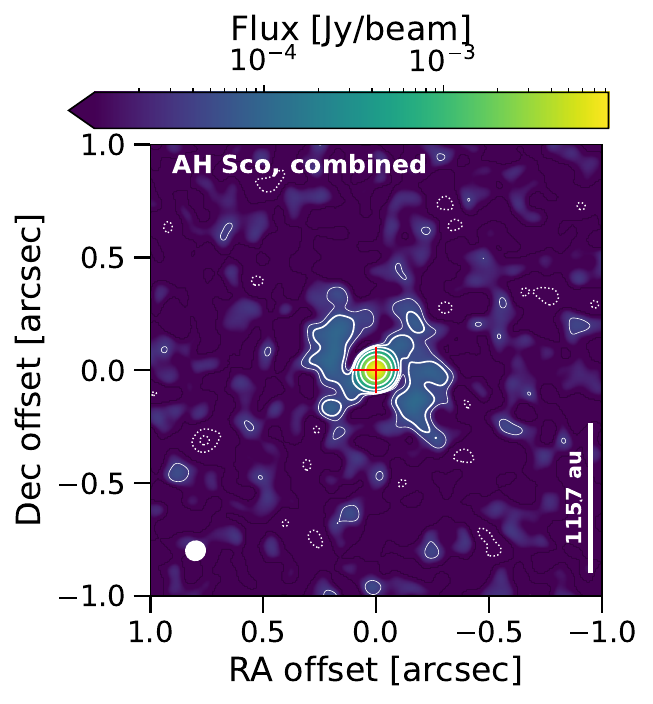}
	\includegraphics[height=4.9cm]{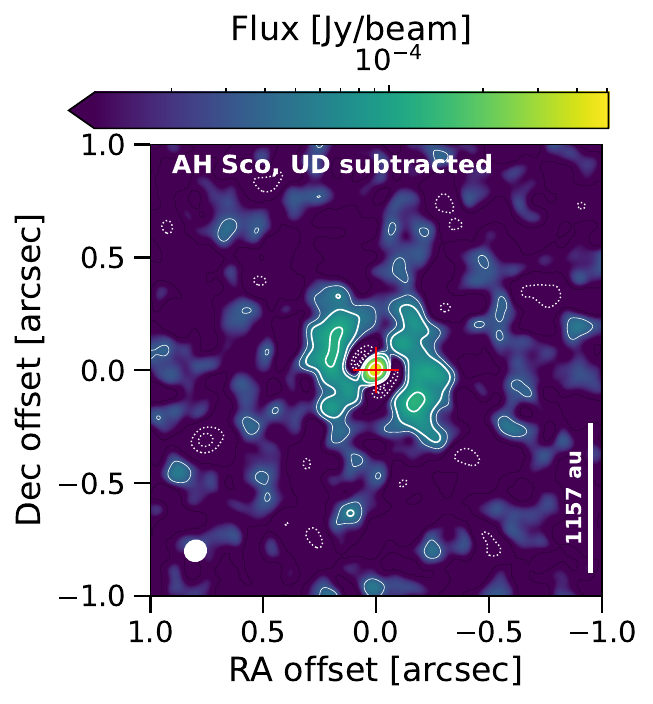}
\caption{Continuum maps of AH Sco. See caption of Fig.~\ref{fig:gyaql}.}
%    \caption{Continuum maps of AH Sco taken with the extended (left), mid (centre left) and combined (centre right) array configurations of ALMA. The rightmost plot shows the residuals after subtracting a uniform disc representing the AGB star. The thin solid contours indicate levels of $3\sigma$, the thick solid contours indicate levels of 5, 10, 30, 100, and 300$\sigma$, and the dotted contours indicate levels of $-3$ and $-5\sigma$. The continuum peak is indicated by the red cross.}
    \label{fig:ahsco}
\end{figure*}
}

%\blue{Anita: Also is it worth commenting under each star where the uv plots show signs of excess on short baselines - e.g. for U Her, R Hya, S Pav, IRC+10011, AH Sco.}

The combined visibility data provide sensitivity to emission on scales from  a few tens of mas up to 8--10\arcsec. However, the visibility plane coverage is not complete nor evenly distributed. To avoid artefacts due to missing spacings between the longest baselines, we applied a slight taper in imaging to weight down their contribution, providing resolutions $\sim$50 mas. Residual calibration errors and antenna position errors limit the dynamic range\footnote{Dynamic range is (peak flux) / (rms).} to about 1000, so for the brightest stars, e.g. R~Hya \edits{(Fig.~\ref{fig:rhya})}, symmetric negative and  positive artefacts may be present at $<0.1$\% of the peak.
Tables \ref{tab:extended}, \ref{tab:mid}, \ref{tab:compact}, and \ref{tab:combined} include the dynamic ranges for for the extended, mid, compact and combined data, respectively.
The ACA continua have relatively low dynamic ranges, from 4 for RW~Sco to 50 for T Mic (Table \ref{tab:aca} \edits{and Fig.~\ref{fig:aca}}). 
%For some stars, the individual and/or combined main array continua have a dynamic range over 1000 .
% and this leads to dynamic range limitations producing convolution errors near the star. Thus, weak, symmetric offset features in the images with high dynamic ranges are unreliable.

Smooth emission on scales much greater than 10\arcsec\ will not be detected, but any extended dust on around this scale may cause bowl-like artefacts due to being detected by just a few baselines. Artefacts may also arise from intermediate under-sampled scales. 
In general, for the brightest stars, if there are regions of negative flux (indicated by dotted contours at the 3 and $5\sigma$ levels), corresponding positive regions at those levels or weaker should be treated with caution. 
The noise rms given in Table \ref{tab:ud}, used to estimate the accuracy of quantities derived from UD fits, are measured close to the stellar position. The background noise far from the star is given in Tables \ref{tab:extended}--\ref{tab:combined} for the extended, mid, compact and combined array data respectively. The combined noise at >~1\arcsec\ is $\lesssim0.01$ mJy.

The continuum images for the extended, mid and combined data are plotted, per star, in Figures \ref{fig:gyaql}--\ref{fig:ahsco}, 
%\ref{fig:gyaql}--\ref{fig:ahsco} and \ref{fig:udel}--\ref{fig:kwsgr}
as indicated for each star in Table~\ref{tab:stars}. We also include in these figures combined images of the residuals after subtracting uniform discs representing the AGB stars and sometimes including dust close to the stars (as described in Sect.~\ref{sec:ud}).
The compact array data are plotted for all observed stars in Fig.~\ref{fig:compact} and the Band 3 ACA data are shown in Fig.~\ref{fig:aca}. In general, the compact configuration images are unresolved or marginally resolved, with some stars exhibiting small ($\lesssim$ beam) irregular extended regions of flux detected at the $3\sigma$ level. We can determine that the bulk of the flux is unresolved because it mimics the elliptical shape of the beam down to the 5 or $10\sigma$ level. All of the ACA images are unresolved, owing to the large beams (see Table~\ref{tab:aca}). 
%The only unusual ACA result is for RW~Sco and this is discussed further in Appendix~\ref{res:rwsco}. 

%In Fig.~\ref{fig:udfits}, we plot the visibility amplitudes of the continuum data for each star against the uv distance and baseline length (the bottom and top axes). These plots also show the UD fits (black curves, see Sect.~\ref{sec:ud}). 
%\edits{Our UD fits (Sect.~\ref{sec:ud} and Fig.~\ref{fig:udfits}) indicated excesses at short baselines} for W~Aql, U~Her, R~Hya, S~Pav, IRC+10011, and AH~Sco, indicating the presence of large scale structures. This is also reflected in the imaging, as discussed in Sect.~\ref{sec:extflux} for U~Her, R~Hya, S~Pav, and IRC+10011 and in Sections \ref{sec:waqlfeatures} and \ref{sec:ahscofeatures} for W~Aql and AH~Sco, respectively.

In the following subsection we discuss \edits{the issue of resolved out flux. Subsequently we discuss the unresolved continuum emission, then} some stars grouped by similar continuum features and some stars individually when they exhibit relatively unique features. These groupings are summarised in Table \ref{tab:class}.

\afterpage{
\begin{table*}
	\centering
	\caption{An overview of the morphological classification of the continuum images.}\label{tab:class}
	\label{tab:mid}
	\begin{tabular}{ccccc}
		\hline
Star	&	Classification & Notes & Sect. & \edits{Fig.}\\
\hline
GY Aql	&	Unique	&	Bar to the southeast	&	\ref{sec:gyaqldisc}	&	\ref{fig:gyaql}	\\
R Aql	&	\edits{Extended, unique}	&	\edits{Some extended emission around the star}	&	\ref{sec:roundconts}	&	\ref{fig:raql}	\\
IRC$-$10529	&	\editss{Elongated}, extended	&	Rectangular or bipolar extensions	&	\ref{sec:bipolar}	&	\ref{fig:irc-10529}	\\
W Aql	&	Extended, unique	&	Continuum flux seen near companion	&	\ref{sec:waqlfeatures}	&	\ref{fig:waql}	\\
SV Aqr	&	Unique	&	Asymmetric dust feature	&	\ref{sec:svaqrdisc}	&	\ref{fig:svaqr}	\\
U Del	&	\edits{Resolution-limited}	&	Minimal signal in UD-subtracted map	&	\ref{sec:roundconts}	&	\ref{fig:udel}	\\
$\pi^1$ Gru	&	Unique	&	Companion seen in extended, tail of dust in its wake	&	\ref{sec:pigrudisc}	&	\ref{fig:pi1gru}	\\
U Her	&	Extended	&	Irregular extended distribution	&	\ref{sec:uherdisc}	&	\ref{fig:uher}	\\
R Hya	&	Extended	&	Strong detection	&	\ref{sec:rhyadisc}	&	\ref{fig:rhya}	\\
T Mic	&	\editss{Elongated}?	&	Mostly round and compact	&	\ref{sec:bipolar}	&	\ref{fig:tmic}	\\
S Pav	&	Extended	&	Strong detection	&	\ref{sec:spavdisc}	&	\ref{fig:spav}	\\
IRC+10011	&	\edits{Extended}	&	Likely significant resolved-out flux	&	\ref{sec:wxpscdisc}	&	\ref{fig:irc+10011}	\\
V PsA	&	\edits{Resolution-limited}	&		&	\ref{sec:roundconts}	&	\ref{fig:vpsa}	\\
RW Sco	&	\edits{Resolution-limited}	&		&	\ref{sec:roundconts}	&	\ref{fig:rwsco}	\\
KW Sgr	&	\edits{Resolution-limited}/unique	&	Tentative detection of narrow extended dust feature	&	\ref{sec:roundconts}	&	\ref{fig:kwsgr}	\\
VX Sgr	&	\editss{Elongated}?	&	More flux to the south than north	&	\ref{sec:bipolar}	&	\ref{fig:vxsgr}	\\
AH Sco	&	Unique	&	Possible dust shell or artefacts from resolved-out flux	&	\ref{sec:ahscofeatures}	&	\ref{fig:ahsco}	\\
		\hline
	\end{tabular}
\tablefoot{The ``Sect.'' column gives the subsection in which each star is discussed in more detail.}
\end{table*}
}

\subsection{Resolved-out flux}\label{sec:resolved-out}
\edits{As noted in Sect.~\ref{sec:12mobs}, ALMA} is not guaranteed to recover all the flux associated with a source and may resolve out some smooth large scale flux. 
%The maximum recoverable scale (MRS) is the scale at which smooth regions of flux can be recovered with confidence. In general, the MRS is larger for the more compact (lower angular resolution) configurations of ALMA. The MRS for the main ATOMIUM dataset are $\sim0.4$--$0.6\arcsec$ for the extended array data, $1.5\arcsec$ for the mid array data, and 8--$10\arcsec$ for the compact array data. More precise MRS for each star and configuration are given in Table E.2 of \citet{Gottlieb2022}. 
%The combined MRS is generally as good as for the compact configuration data, except for AH Sco and KW Sgr, which were not observed with the compact configuration, and for which the MRS is as good as for the mid configuration data. For the new ACA data presented here, the MRS range from 64--77\arcsec\ and are given for each star in Table \ref{tab:aca}.
\edits{We are able to check the impact of this effect for}
four ATOMIUM stars which are also in the DEATHSTAR survey \citep{Ramstedt2020}. DEATHSTAR observed a large sample of AGB stars using the ALMA ACA in Bands 6 and 7. For their Band 6 observations, covering similar frequencies to ATOMIUM albeit over a smaller range, the MRS was $25\pm 4 \arcsec$ \citep{Ramstedt2020}. We compared the self-calibrated Band 6 continuum fluxes from DEATHSTAR with those from our ATOMIUM combined and compact continuum images, for the overlapping stars (IRC$-$10529, SV Aqr, T Mic and IRC+10011). We found no evidence of more flux being resolved out for the ATOMIUM data than for the DEATHSTAR data. 

However, this does not mean that all the flux has necessarily been recovered. \textsl{Herschel}/PACS images of dust at 70 and 160 \micron\ are available for some of the stars in our sample. For W Aql, \pigru, R Hya, and T Mic, extended flux is seen on scales of $\sim1\arcmin$ from the position of the star, indicating significant dust at larger scales than can be recovered with ALMA \citep{Cox2012}. This is significant because at longer wavelengths (i.e. 1.2--1.3~mm for ALMA Band 6) we might detect cooler dust, which we would expect to be located further from the star. However, since ALMA cannot recover flux on such large scales --- scales that can be larger than the ALMA field of view for a single pointing --- we would not expect the majority of flux from such a large dust shell to be recovered, even when using the ACA. Comparing our total combined fluxes (Table~\ref{tab:combined}) with single-dish observations from \cite{Altenhoff1994} and \cite{Dehaes2007}, we can estimate the recovered fluxes for R~Aql, R~Hya, and IRC+10011 as around 60\%, 70\%, and 6--12\%, respectively. The large amount of resolved out flux for IRC+10011 (88--94\% lost) fits with the negative artefacts we see in Fig.~\ref{fig:irc+10011}.

To give an indication of \edits{what fraction of flux comes from regions offset from the continuum peak,} we compared the UD flux density (see Table~\ref{tab:ud} and Sect.~\ref{sec:ud}) with the total flux recovered in our combined images (Table~\ref{tab:combined}). 
%In Fig.~\ref{fig:fluxratio} we plot the ratio between the total flux in the combined data and the UD flux density, against distance, excluding SV~Aqr, for which the UD flux density was difficult to fit (see Sect.~\ref{sec:ud}). 
Overall, we found no \edits{significant} correlation with distance \edits{to the stars}. Excluding the RSGs, we found a weak positive correlation between the flux ratio and distance, \edits{which is expected, since a more distant source will appear smaller on the sky and hence be less susceptible to resolved-out flux. Also, for the closest sources, the full extent of the circumstellar envelope may be larger than the primary beam of the ALMA 12-m array.}
Across our sample, there is a large spread in the \edits{ratios of total flux to UD flux}, from 1.07 (U~Del) to 2.24 (IRC$-$10529). U~Del having the lowest flux ratio fits with what we found from the UD-subtracted map (Fig.~\ref{fig:udel}) which shows that the majority of the continuum flux is  associated with a uniform disc of emission at the stellar position \edits{and that there is essentially no extended flux recovered by ALMA}. The stars with the highest ratio (IRC$-$10529 followed by W~Aql) exhibit flux away from the continuum peak. Some of the other stars with lower flux ratios nevertheless have extended features, likely arising from dust, in their continuum maps, or are known to be dusty based on other observations \citep[e.g. \textsl{Herschel}/PACS imaging][]{Cox2012,Maercker2022}. This discrepancy could be caused by resolved-out flux, amplitude errors (especially for sources with high dynamic ranges $>1000$) or a combination of factors. 

We also checked whether there was any correlation between the ratio of total flux and UD flux density with the apparent or physical UD radii and found no trends. \edits{We also found no correlation between the total flux or UD flux with mass-loss rate, distance, or mass-loss rate scaled by distance.}
%\red{(anything else?)}

In summary, it is likely that there is resolved out flux for at least some of our observed stars. The precise fraction of recovered flux is difficult to estimate from the available data, for the majority of our sample. The stars that are further away or which have smaller dust extents are less likely to be affected.
%\blue{add more detail here if more analysis done esp based on CO}

%\subsection{Spatially compact continuum images}
\subsection{Resolution-limited continuum images}
\label{sec:roundconts}

\edits{
In general, we expect spatially resolved emission to appear larger than the restoring beam of the observing array. For emission with spatial extents much larger than the beam, this is trivially apparent (see for example the ATOMIUM CO emission, \citealt{Decin2020}). However, a bright point-like source with a high dynamic range can appear to be larger in the image plane than the reported beam size, because of the Gaussian nature of the beam. This is explained in detail in Appendix \ref{sec:appext}, where we lay out criteria for checking whether bright sources show evidence of resolved emission.
Note that fitting in the visibility plane, as we did for the UD fits (Sect.~\ref{sec:ud}), avoids apparent extension due to convolution with the beam, but is still S/N limited.}

\edits{The most sensitive continuum images, capable of revealing the most extreme small- and large-scale features, were made from the combined data. Most of our sources exhibit some extended emission in the combined maps, though in some cases, such as KW~Sgr (Fig.~\ref{fig:kwsgr}), the extended emission is not apparent until we subtract a UD from the data and lower the resolution to increase sensitivity (Sect.~\ref{sec:ud}). For the particular example of KW~Sgr, this is discussed further in Sect.~\ref{sec:kwsgrdis}. Two stars, U~Del (Fig.~\ref{fig:udel}) and RW~Sco (Fig.~\ref{fig:rwsco}), do not show any extended emission in the combined maps or the UD-subtracted maps. U~Del is so well-described by the UD that there is no significant flux in the UD subtracted map at all. T~Mic (Fig.~\ref{fig:tmic}) and V~PsA (Fig.~\ref{fig:vpsa}) appear unresolved in the combined maps, but the UD subtracted maps reveal some small (comparable to the beam) regions of emission offset from the continuum peaks. The combined continuum maps for the remaining stars all show more significant extended emission and/or asymmetries, which are described in the following subsections.}

%\red{Unfinished edits:}
\edits{When considering the individual array images in the context of our criteria for resolution, we find that \editss{almost} all of our compact images (Fig.~\ref{fig:compact}) are \editss{close to the resolution criterion defined in Appendix \ref{sec:appext}, i.e.~within 10\%} at the $5\sigma$ level, with only \editss{a few stars such as W~Aql and S~Pav showing evidence of extended emission $\geq20\%$ of the resolution criterion}. For observations with the mid configuration, a larger portion of the images show evidence of (partially) resolved extended emission. Some cases of faint extended emission seem to be below the sensitivity limit of the mid data, but above $3\sigma$ in the combined data, making \editss{sources such as S~Pav (Fig.~\ref{fig:spav})} appear unresolved in the mid images. Most images taken with the extended array show evidence of resolved out flux to varying degrees, with only a few stars for which clumpy extended emission is recovered. A prime example of this is \pigru\ (Fig.~\ref{fig:pi1gru}), which we discuss in detail in Sect.~\ref{sec:pigrudisc}.
}

\subsection{Stars with relatively uniform extended flux}\label{sec:extflux}

\edits{Based on the criteria laid out in Appendix~\ref{sec:appext}, the majority of the stars in our sample have some extended continuum emission associated with them. Of these, relatively uniform
}
extended flux is seen in the combined and UD-subtracted continuum images of U~Her (Fig.~\ref{fig:uher}), R~Hya (Fig.~\ref{fig:rhya}), S~Pav (Fig.~\ref{fig:spav}), and IRC+10011 (Fig.~\ref{fig:irc+10011}). In these cases, the flux is relatively symmetric about the stellar positions, though not perfectly so. U~Her and R~Hya also exhibit some extended emission in their mid continuum maps.

\subsubsection{U Her}\label{sec:uherdisc}

The continuum maps of U~Her in Fig.~\ref{fig:uher} show significant extended emission surrounding the continuum peak, most clearly apparent in the UD-subtracted image, \edits{where the regions of extended emission much larger than the beam}. \edits{The mid and extended maps show asymmetric features at the $5\sigma$ level.}
%Even if we disregard the $3\sigma$ contour, there is still significant excess flux.
%The exception is the compact map (Fig.~\ref{fig:compact}), which is featureless aside from a detached region to the west-northwest at around $3\sigma$ and slightly smaller than the size of the synthetic beam, hence not significant. 
%There are several protrusions detected at levels of $>5\sigma$ surrounding the continuum peak in the extended map. These are asymmetric, so unlikely to arise from the data reduction. However, there are two negative regions (down to $-5\sigma$) which may indicate regions of resolved-out flux. 
%\red{...? Anita? is this a likely explanation?} 

%The combined continuum map shows regions of extended emission \edits{much larger than the beam}, more clearly visible in the UD-subtracted image. 
\edits{In the combined map,}
the larger-scale emission at $\geq 3\sigma$ extends to separations of around 0.4\arcsec\ to the south, west and north, but only out to $\sim0.2\arcsec$ to the east-northeast. The emission is not distributed evenly and is clumpier to the north and northwest. %There is also a small gap to the south of the continuum peak, which is repeated at the extended and mid scales.
%When we more closely compare the U~Her emission at different scales, we find that the emission in the combined map is almost all within the $10\sigma$ contour of the mid map. Most of the emission in the extended map is within the $10\sigma$ contour of the combined map, with the exception of the north-northwest protrusion, which extends into the protruding $3\sigma$ contour of the combined map. These results are generally expected since the combined map should represent the individual array maps, albeit at a lower resolution than the extended map and at a more granular scale than the mid map. What is unexpected is that the lower level of emission to the south is repeated on all scales, at the same position and bordered by emission with a very similar shape. 
In the extended, mid and combined maps, there is a small region (at different scales for each map) with less emission to the south of the continuum peak.
%On the extended map, the eastern inner edge of the gap extended from 0.04\arcsec\ from the continuum peak, out to 0.1\arcsec. On the combined map, the apparent gap outlined by a $\cap$-shape in the $3\sigma$ contour is filled with emission at levels of 1 to $2\sigma$. A ray along the inner edge of the $3\sigma$ contour extends from 0.1\arcsec to 0.14\arcsec. In the mid map, the gap in emission is a comparable width (east to west) as the synthetic beam. The gap is less apparent in the UD-subtracted map, but there is also less emission present there in the same direction.
The cause of this region of lower emission is unclear, but the fact that we see it at all scales suggests that either a continuous or repeating process is causing it. %Possibilities will be discussed further in Sect.~\ref{sec:extdust}.

\subsubsection{R Hya}\label{sec:rhyadisc}

The continuum observations of R~Hya (Fig.~\ref{fig:rhya}) have the highest dynamic range of any star in our sample, exceeding 1000 for all individual array configurations, and reaching 6600 for the combined data (see Tables \ref{tab:extended}--\ref{tab:combined}). This may be partly owing to the fact that R~Hya is the closest star in our sample (distance = 126~pc). The roughly symmetric negative regions surrounding the continuum peak in the extended and combined maps shown in Fig.~\ref{fig:rhya} are the result of amplitude errors. However, the UD-subtracted map shows significant extended flux at levels of $\geq 5\sigma$ out to 0.15--0.3\arcsec\ from the stellar position.
% \red{[right Anita?]}.

The mid continuum map has scattered regions of emission around the central continuum peak. These include emission detected with a certainty of $5\sigma$ extended out to $\sim0.75\arcsec$. 
%The compact map is only marginally resolved and relatively featureless (Fig.~\ref{fig:compact}). It is also the only compact map with a dynamic range $> 1000$.
%
%However, although the compact data is featureless (Fig.~\ref{fig:compact}), the mid and combined data show a lot of extended emission (Fig.~\ref{fig:rhya}), most likely from dust surrounding the star.
%
%The extended continuum has some positive and negative artefacts surrounding the central peak. These are generally symmetric and \red{the result of the high dynamic range}.

Finally, we note that a combined continuum map of R~Hya was previously published by \cite{Homan2021} and showed a highly asymmetric structure in the inner regions, positive to the east and negative to the west-northwest. In carefully re-imaging the data, we discovered that these features resulted from a misalignment of the mid array data and did not represent real features. Although some minor artefacts remain in our extended and combined images, they are a more accurate representation of the dust distribution around R~Hya.

%Finally, we note that a combined continuum map of R Hya from the ATOMIUM data was previously published by \cite{Homan2021}. In preparing this work, we discovered that in the earlier work the different array continua were misaligned, leading to the ``void'' reported in the earlier work. The only regions of negative flux we report (notably in the extended data) arise from phase error owing to the large dynamic range of these observations. \red{Small negative regions in the combined map likely arise from ??? also phase errors?} 

\subsubsection{S Pav}\label{sec:spavdisc}

The S~Pav extended and mid continuum images (Fig.~\ref{fig:spav}) are relatively featureless, but the compact map shows some asymmetric protrusions to the northwest and southeast (Fig.~\ref{fig:compact}). The combined map and the UD-subtracted map show significant extended flux, with the $3\sigma$ contour located at around $0.2\arcsec$ from the continuum peak. This corresponds to a physical projected separation of 37~au from the continuum peak. No clear structures are resolved in this extended emission in the UD-subtracted image. Even when we consider the un-subtracted combined image with a smaller synthetic beam, no obvious structures become apparent and all irregular regions are of comparable in size to the beam or smaller, making it difficult to discern their actual sizes.

%\blue{Compare LSR velocity with \url{https://vizier.cds.unistra.fr/viz-bin/VizieR-5?-ref=VIZ675a462728a6d&-out.add=.&-source=III/252/table8&recno=29335} who have $-22\pm3.1$~\kms}

\subsubsection{IRC+10011 (WX Psc)}\label{sec:wxpscdisc}

Examining the continuum maps for IRC+10011 in Fig.~\ref{fig:irc+10011}, we find that the mid map has two small protrusions at $3\sigma$ that are smaller in width than the beam and are approximately rotationally symmetric. There are also protrusions in the extended map to the southeast and northeast, at a level of $5\sigma$ and to the northwest at a level of $3\sigma$, which are not very symmetric. Similar features, at larger scales, can be seen in the combined map. However, the combined map shows more symmetric features and some regions of negative flux surrounding the continuum peak. These are most likely the result of phase and/or amplitude errors.

The UD-subtracted map should not suffer from sidelobe or dynamic range artefacts since the stellar contribution has been subtracted. We find that the flux is roughly evenly distributed around the continuum peak, with extensions to the north and south. The shape of the 3 and $5\sigma$ contours enclosing the UD-subtracted flux corresponds well with the central part of the combined map, excluding possible artefacts to the east and west.

%In the compact continuum map (Fig.~\ref{fig:compact}) we find two small regions of flux detected to the northwest and southeast of the continuum peak. These regions are only detected at a level of $3\sigma$, and with the $3\sigma$ contours smaller than the synthetic beam, making them unreliable detections. Checking a UD-subtracted combined image made with a coarse taper and comparable synthetic beam to the compact data does not recover these features. Hence we must conclude they are not real. This coarse taper combined image does recover an extension to the north, increasing our confidence that this feature is real.

\subsection{Elongated continuum features}\label{sec:bipolar}

Identifying \editss{elongated} structures in the continuum maps can be challenging, as any symmetric features (of positive or negative flux) may be the result of amplitude errors. These are more likely to be present for higher dynamic ranges. Since our sample contains many bright sources, we must be careful when interpreting any symmetric \editss{elongated} features. With this in mind, we have \edits{tentatively} identified a group of stars which exhibit some \editss{elongated} features. These are IRC$-$10529 \edits{(Fig.~\ref{fig:irc-10529})}, T~Mic \edits{(Fig.~\ref{fig:tmic})}, and VX~Sgr \edits{(Fig.~\ref{fig:vxsgr})}.

\subsubsection{IRC$-$10529 (V1300~Aql)}\label{sec:bipolar-irc-10529}

The extended map of IRC$-$10529 (Fig.~\ref{fig:irc-10529}) features some small symmetric protrusions to the northwest and southeast, which may be amplitude errors rather than real features. However, there is a small region of extended flux to the southwest, \edits{in the extended and mid maps,} which is likely real and is not mirrored on the other side. 
%The mid image, suffering from less resolved out flux, also reveals a region of extended flux to the southwest. 
\edits{The combined map shows an elongated emission structure, with some emission extending from the continuum peak}
to the west-southwest and east-northeast.
\edits{This elongated structure, which runs along an axis $\sim70\deg$ east from north, is detected with more confidence in the UD-subtracted map}, which emphasises lower surface brightness features thanks to its lower angular resolution. \edits{This axis is comparable to the axis about which the positive and negative velocities are separated in the SiO moment 1 map shown in \citeapos{Decin2020} Figure S60.} There are also some perpendicular extensions which are not symmetric about the location of the continuum peak, \edits{and which are somewhat reminiscent of} the shape to the Red Rectangle protoplanetary nebula \citep{Cohen2004}, but on a much smaller scale. 
%A line along the same axis ($\sim70\deg$ east from north) passes between the four outer ``points'' of the UD-subtracted map (e.g. assuming a similar shape to the Red Rectangle protoplanetary nebula, \citealt{Cohen2004}, but on a much smaller scale).

\subsubsection{T Mic}\label{sec:bipolar-tmic}

\begin{figure}
	\includegraphics[width=0.45\textwidth]{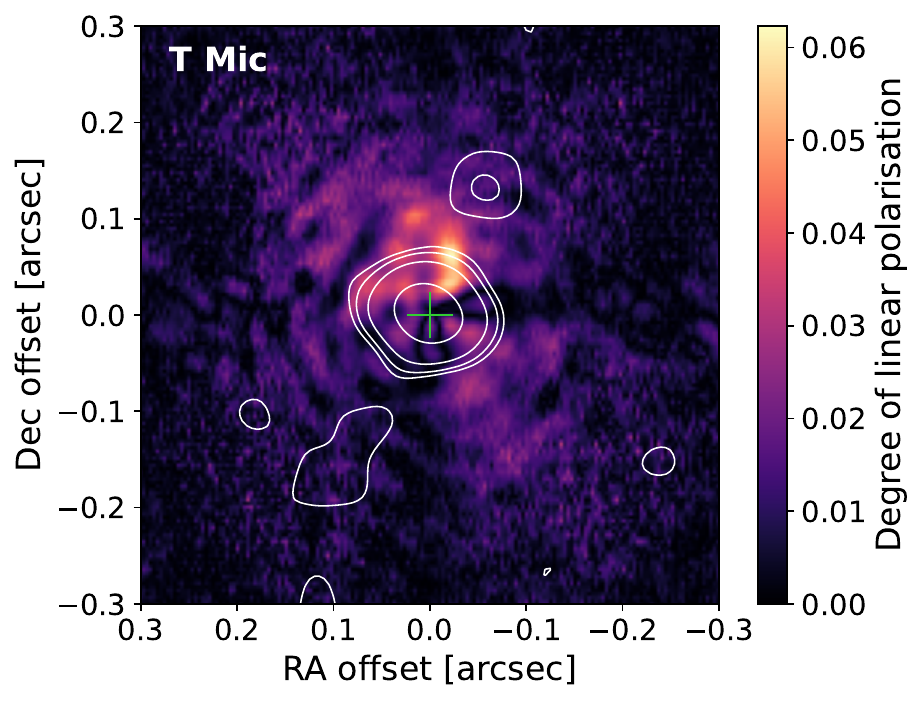}
    \caption{\edits{Degree of linear polarisation around T~Mic observed with SPHERE (colours, from \citealt{Montarges2023}) overplotted with the UD subtracted contours observed with ALMA (see Fig.~\ref{fig:tmic}). The position of the AGB star is indicated by the green cross.}}
    \label{fig:tmicdolp}
\end{figure}

\edits{Although the individual array images of T Mic (Fig.~\ref{fig:tmic}) are unresolved, the
%The mid and combined maps of T Mic (Fig.~\ref{fig:tmic}) are relatively featureless. There are some regions of symmetric low-level flux in the extended map, but this may be the result of amplitude errors rather than real features. 
%A few irregular regions, smaller than the beam size, are present on the compact map (Fig.~\ref{fig:compact}), but their significance is not certain. 
UD subtracted image reveals} a region to the northwest at a significance of 3--$5\sigma$ and a more elongated region, at around $3\sigma$, to the southeast. The coarser beam size of the UD-subtracted image increases the signal to noise of features that are not confidently detected in the combined map with a smaller restoring beam. Drawing a line between the continuum peak and the northern feature in the UD map, we find that it lies approximately $25\deg$ west of north. This is in the same direction as the brightest feature seen in polarised light in the SPHERE map of \cite{Montarges2023}, as we plot in Fig.~\ref{fig:tmicdolp}, raising confidence \edits{that this \editss{elongated} structure is real}. 
\edits{The angle of this feature is approximately perpendicular to the angle separating positive and negative velocities in the $^{30}$SiO ($J=6\to5$) moment 1 map presented in \cite{Decin2020}.}
%Although this line also passes through some of the southern features, we find that the southern features lie within a cone covering ~20--$45\deg$ east of south.
%The best agreement for both northern and southern features is described by a line running $30\deg$ west of north, which is also in reasonable agreement with the (uncertain) northern and southern features on the extended map.

\subsubsection{VX Sgr}\label{sec:bipolar-vxsgr}

Some asymmetric features are seen for all the continuum maps of VX~Sgr (Fig.~\ref{fig:vxsgr}). These are most pronounced for the extended and combined maps, where emission is seen predominantly to the south and north of the continuum peak. Once again, this emission is most pronounced in the UD subtracted map. Considering only the emission closest to the location of the continuum peak (e.g. the emission enclosed by the $5\sigma$ contour), we find that the \editss{elongation} axis is close to running due north-south, with an uncertainty of only a few degrees ($<5\deg$). \edits{We note that this emission could be the result of episodic mass loss, as discussed in Sect.~\ref{sec:rsgdis}.}

\subsection{Highly asymmetric continuum features}

Some of the stars in our sample exhibit features in their continuum maps which are neither uniformly extended nor bipolar. Generally, the resultant maps are unique, so we discuss these stars individually below.

\subsubsection{GY Aql}\label{sec:gyaqldisc}

GY Aql is unusual in that the continuum maps from the extended and compact array configurations \edits{show unresolved emission} (Fig.~\ref{fig:gyaql}, Fig.~\ref{fig:compact}), but a strongly detected feature is seen to the southeast in the mid and combined maps. In the mid map, this feature appears attached to the continuum peak and is detected at a level $>30\sigma$. 
%\edits{It is likely that this feature is not seen in the extended map owning to a combination of resolved-out flux and low surface brightness.}
When constructing the combined image, we found that choosing a restoring beam that was too fine would result in this feature having a very low surface brightness and not being detected above the noise, beyond a small region with the brightest flux (visible at an offset of 0.5,$-0.5$ in Fig.~\ref{fig:gyaql}). Hence, we chose a coarser resolution for the combined map than was used for most other stars (see Table \ref{tab:combined}) to emphasise this feature, which appears detached and which we will refer to as a ``bar'', though it may be part of a spiral, since spiral-like structures are seen in the CO emission \citep{Decin2020}. As can be seen in the UD subtracted map, the continuum flux mainly originates from this bar and from regions close to the star, likely indicating a significant concentration of dust in this bar.

%\blue{Refer to Louise's paper probably. Check with Fabrice about its progress. If not referring here, should refer in the discussion section.}

\subsubsection{R Aql}\label{sec:raqldisc}

\edits{In constructing the combined continuum image, we found  suspiciously regular positive and negative patches of emission, larger than the synthesised beam, at levels of $\gtrsim3\sigma$ around the continuum peak.  This suggests resolved-out emission. Tapering to give more weight to shorter baselines, i.e. coarsening the resolution,
%In constructing the combined continuum image, we found that a significant amount of flux was resolved out, resulting in substantial regions (larger than the beam) at levels of $\gtrsim3\sigma$ around the continuum peak. Using a coarser restoring beam 
alleviated some of this effect and allowed us to recover more flux and structures in the combined map, which is shown in Fig.~\ref{fig:raql}. The extended continuum emission is patchy and irregularly distributed around the continuum peak. Hints of these structures can also be seen in the marginally resolved mid continuum map.}
The extended continuum map of R~Aql has a very high dynamic ranges, in excess of 2000, resulting in increased uncertainty such that features in the map $<~0.1\%$ of the peak should be treated with caution, especially if symmetric. \edits{All the larger scale structures visible in the mid and combined maps are resolved-out in the extended map.}
%The extended and combined continuum images of R~Aql are relatively featureless, but the mid image shows some irregular extended emission around the peak (Fig.~\ref{fig:raql}), possibly indicating dusty regions offset from the stellar position. Some similar extended emission is also seen in the compact continuum image (Fig.~\ref{fig:compact}). The UD subtracted image shows a significant amount of flux centred on the continuum peak, in addition to fainter regions of flux at angular separations of 0.2--0.4\arcsec\ from the star. 

\subsubsection{W Aql}\label{sec:waqlfeatures}

W Aql has a known main sequence (F9) companion at a projected separation of $\sim0.5\arcsec$, corresponding to a physical separation $\approx 200$~au \citep{Danilovich2015,Danilovich2024}. The individual higher resolution 12m array continuum images, plotted in Fig.~\ref{fig:waql}, show a secondary peak, detected with a certainty of $5\sigma$, to the south west of the primary continuum peak. Taking the primary peak as the position of the AGB star, the secondary peak corresponds well to the location of the F9 companion, which was observed with VLT/SPHERE contemporaneously with the extended array data \citep[9 July 2019,][a day after the ALMA observations were completed]{Montarges2023}. We indicate the position of the companion as observed with SPHERE by the yellow cross in Fig.~\ref{fig:waql}. While \edits{some extended emission is resolved around} the primary peak with the extended array, the secondary peak appears as a point source in the extended and combined continuum maps. It is only marginally resolved from the primary peak in the mid continuum and not at all in the compact continuum (Fig.~\ref{fig:compact}). At this wavelength (1.24~mm), the F9 companion is well below the detection threshold, so this significant detection indicates \edits{a concentration of dust close to the F9 star, the origin of} which is discussed in more detail in Sect.~\ref{sec:waqldiscussion}.

There is some additional extended continuum emission surrounding the primary peak, especially to the east and northeast. In the mid and compact (Fig.~\ref{fig:compact}) continuum maps there is some emission extending further than what is visible in the extended and combined maps, suggesting that some larger scale flux may be resolved out at those higher resolutions. The UD subtracted image also shows areas of extended emission towards the south and west, albeit at low surface brightness and only detected at levels of $\sim 3 \sigma$. The peak flux in the UD subtracted image is very slightly offset from the continuum peak but indicates substantial emission close to the AGB star.

\subsubsection{SV Aqr}\label{sec:svaqrdisc}

As can be seen in Fig.~\ref{fig:svaqr}, there are some small regions of extended flux around the continuum peak of SV~Aqr. In the extended map, there are some protrusions to the west and south-southwest of the continuum peak, and some smaller regions scattered at larger distances ($\sim0.2\arcsec$) from the continuum peak. In the mid image, there is a small protrusion to the northwest. The combined map shows an elongation in the west and south-southwest directions, in similar positions to the protrusions seen in the extended map, and some scattered emission to the northwest, aligned with the direction of the protrusion in the mid map. Hence, the combined map represents the emission at both larger and smaller scales well.

The peak flux in the UD subtracted image is to the southwest and not centred on the continuum peak in the unsubtracted image. The UD subtraction reveals an arc of emission close to the stellar position in the south and southeast, which appears to sweep anticlockwise to the northwest, where it is located further from the star (the $3\sigma$ contour extends out to $0.12\arcsec$). This strongly asymmetric feature likely contributed to our difficulty in fitting a uniform disc to SV Aqr (see Sect.~\ref{sec:ud}) and we stress that the features presented in the UD subtracted image are not reliable, aside from where they agree with the individual array images.
% is the likely reason for the non-circular uniform disc fit for SV Aqr (ratio between major and minor axes $r=0.79$, see Sect.~\ref{sec:ud}).

\subsubsection{$\pi^1$~Gru}\label{sec:pigrudisc}

\cite{Homan2020} previously presented the extended continuum map of \pigru, which shows a close secondary peak to the south-west of the primary continuum peak. Here we present the continuum maps obtained from the other ALMA arrays and the combined continuum map for the first time (Figs.~\ref{fig:pi1gru}, \ref{fig:compact}). We also present the ACA Band 3 continuum map (Fig.~\ref{fig:aca}).

The secondary peak is marked by a black cross on the extended and combined continuum plots in Fig.~\ref{fig:pi1gru}. We henceforth refer to this feature as the C companion, following \cite{Montarges2025}.  The AGB star, which we associate with the position of the primary continuum peak, is the A component and the B component is a long-known main sequence G0V star \citep{Feast1953}, which, in the Gaia DR3 epoch (2016.0), is separated from the AGB star by 2.68\arcsec\ \citep{Gaia-EDR3}, corresponding to a projected separation of 440~au. The B component is outside of the plotted field of views of the extended and combined data, but we indicate its approximate relative position (calculated from the Gaia catalogue by assuming that the separation between A and B did not significantly change between the Gaia epoch and our ALMA observations) as a yellow cross on the plot of the mid continuum image (Fig.~\ref{fig:pi1gru}) and the compact continuum image in Fig.~\ref{fig:compact}. 
%We note that the compact continuum emission is relatively featureless. 
%In both the mid and compact images there are small regions of emission in the direction of the B companion but, although they are within $3\sigma$ contours, they are smaller than the synthetic beam, so may not be significant. However, their presence in two separate observations suggests that there may be some gravitational influence from the B companion \citep[similar to what was seen at the larger-scales examined for W~Aql, as discussed in][]{Danilovich2024}. 
We do not detect any emission above the noise near the position of the B companion, in contrast with W Aql, which may be because the companion is more distant and situated in a less dense wind (e.g. compare the mass-loss rates in Table~\ref{tab:massloss}).

The C component is visible as a secondary peak in the extended continuum map, but is not resolved in the combined map. Instead, the combined map shows a central feature elongated in the direction of the C component. There is also some additional extended emission seen around the central peak in all directions except for the southwest. This ``tail'' of extended emission corresponds well with the tail seen in the contemporaneously observed VLT/SPHERE image showing the degree of linear polarisation presented by \cite{Montarges2023} and is discussed further in Sect.~\ref{sec:pigrudis}. There is also some scattered emission present in all directions around the continuum peak in the mid image with, as noted above, a slight bias in the direction of the B component. 

For the UD subtracted image shown in Fig.~\ref{fig:pi1gru}, we chose a comparable synthetic beam to the combined image to better emphasise the tail of emission. Once the flux from the AGB star is subtracted from the continuum map, the peak flux is found at the position of the C component and has the highest relative flux (i.e., the highest dynamic range, see Table \ref{tab:udsub}) of any UD subtracted image in our sample, by a factor of $\sim5$ or higher. For finer resolutions of the UD-subtracted image, we find that the low surface brightness of the tail feature caused it to be lost in the noise. At coarser resolutions, the tail is not necessarily resolved from the dust close to the A and C components. Additionally subtracting a point-like function representing the C component does not reveal other structures.

%\blue{Probably include separate image with the C component also subtracted?}

%\subsection{KW Sgr}

\subsubsection{AH Sco}\label{sec:ahscofeatures}

The extended continuum map of AH~Sco, shown in Fig.~\ref{fig:ahsco}, is relatively featureless, with only a few asymmetric features, smaller than the beam, surrounding the continuum peak. The mid image, in contrast, exhibits significant extended flux around the continuum peak, similar to the combined map of S Pav (Fig.~\ref{fig:spav}). There generally appears to be more extended flux to the northeast and to the southwest. Our understanding of this source is complicated, however, by the reverse-N shaped (or perhaps barred spiral) emission pattern seen in the combined and UD-subtracted maps. It is unclear whether the gaps in emission to the northeast and southwest of the continuum peak are real, for example representing a dust shell, or whether they originate from amplitude errors. %For this reason, we did not group AH~Sco with the AGB stars showing extended emission (Sect.~\ref{sec:extflux}).

The most extended features in the continuum around AH~Sco lie close to the northeast-southwest axis (45--$50\deg$). There is a feature resembling a bar perpendicular to this axis and in which there do not appear to be gaps in the flux. The most extended $3\sigma$ contours lie at 0.3\arcsec\ for the first feature and at 0.25\arcsec\ for the bar-like feature. These values correspond to projected separations of 520 and 430~au.

%Finally, we note that all the continuum maps of AH~Sco have several small regions of flux scattered around the main region of continuum emission. These may represent dispersed clumps of dust in the circumstellar environment, \edits{embedded in a larger, resolved-out dust shell}. 

\section{Analysis}\label{sec:analysis}

\subsection{Spectral indices}

\begin{figure*}
\centering
	\includegraphics[width=0.45\textwidth]{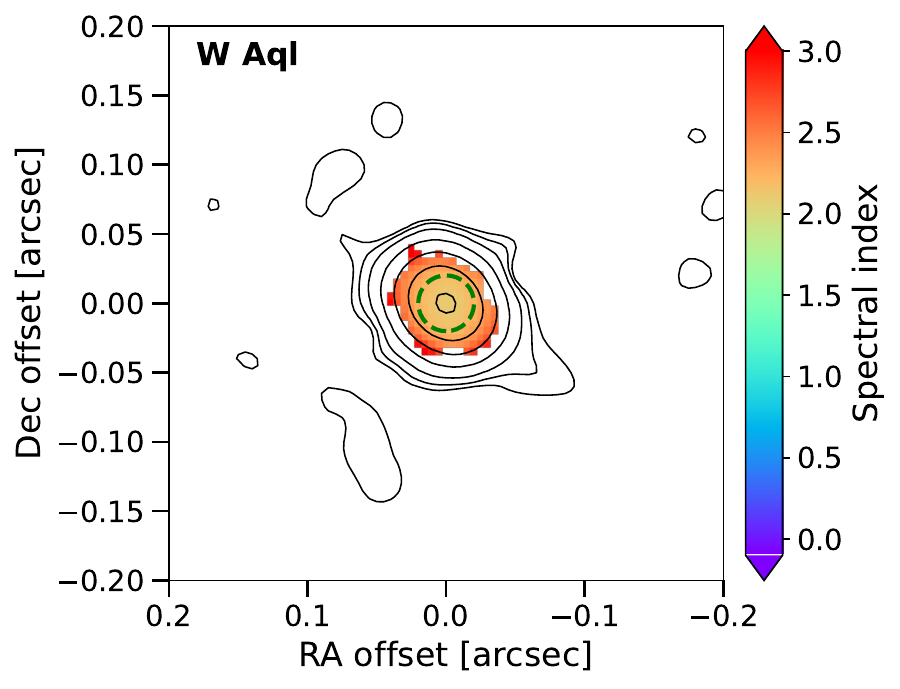}
	\includegraphics[width=0.45\textwidth]{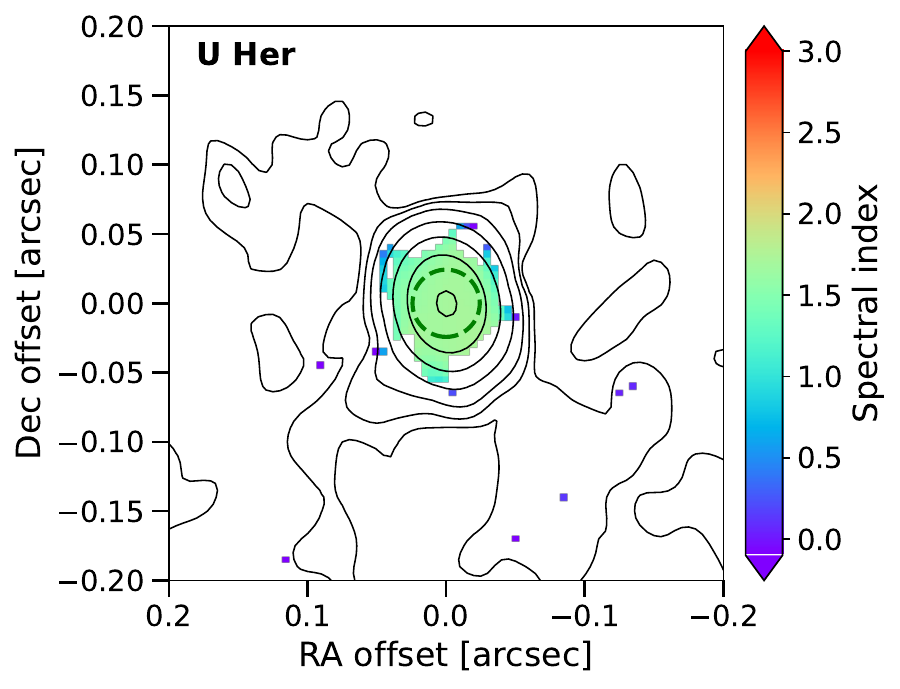}
    \caption{\edits{Examples of spectral index determinations for W Aql (\textit{left}) and U~Her (\textit{right}) computed from the combined Band 6 data. The spectral index (colours) is shown for the pixels where it is well-defined. The dashed green circle indicates the area, centred on the continuum peak, equal to the beam size convolved with the UD radius. Black contours indicate the combined continuum (see Fig.~\ref{fig:waql} and Fig.~\ref{fig:uher}). Similar plots for the remaining ATOMIUM sources are given in Fig.~\ref{fig:spec-all}.}}
    \label{fig:spec-examples}
\end{figure*}

\subsubsection{Intra-Band 6 spectral indices}

\edits{Since our Band 6 observations covered frequencies spanning $\sim56$~GHz, we were able to calculate intra-Band 6 spectral indices for our sample. 
%\red{More details from Anita?} 
In cleaning an image with tclean, as well as identifying the brightest pixels in each cycle, we used the option to fit for the linear dependence of flux with frequency, using the first two terms of a Taylor series, giving total intensity and spectral index (\texttt{alpha}) images.
We use the \texttt{alpha.error} image produced by tclean to exclude unreliable spectral index measurements, but for a well-cleaned image, this is a lower limit to the actual error, since it is based only on the residual noise in the zeroth and first Taylor coefficient images. This does not include systematic errors such as in alignment or calibration (indeed, a well-known diagnostic for tiny spatial misalignments is a systematic gradient in alpha across an image). }

\edits{Without additional calibrators/dedicated flux scale measurements, the accuracy of an individual ALMA observation at Band 6 is 5\% at best.  Each mid and extended configuration data set contains at least 4 observations at 4 sets of frequencies (see \citealt{Gottlieb2022} Fig.~1 for the ranges covered in each tuning).  Our calibration methods force the spectral index to be linear, but if both sets of the lowest and the highest frequencies have 5\% errors in the opposite sense this could tilt the spectral index by  about 0.3, regardless of S/N.  Statistically, this is likely to affect one or two out of our 17 data sets, per configuration. Another factor is the potential presence of residual lines. The combined or extended configuration data possess the most baselines sampling the smallest angular scales, meaning that these can be used for measuring the spectral index of the stellar continuum. The S/N of any other continuum features (e.g. any extended emission) is much too low for a reliable spectral index measurement.}
% for stellar continuum measurements (the S/N of any other continuum is much too low, with the exception of the companion to \pigru).}

\edits{
We find that for most stars, the uncertainty on the spectral index becomes very high further than $\sim1$~beam from the continuum peak. Hence, to compute average spectral indices, $\alpha_\mathrm{B6}$, we only consider pixels (i) which are within the circular region corresponding to the major axis of the beam convolved with the size of the UD fit to the star and (ii) for which the uncertainty on the spectral index is $<0.1$.
Then, for these selected pixels, we construct a histogram of the spectral indices and derive a Gaussian kernel density estimation (KDE). The peak of this KDE is taken to be the spectral index. The uncertainty is then conservatively taken to be the sum in quadrature of the full width at half maximum of the KDE and the maximum uncertainty on the individual pixels. To check the robustness of this estimate, we also considered the spectral index of the pixel closest to the continuum peak and the mean spectral index of all the selected pixels. Both methods agreed well with our final value. }

\edits{
Our calculated $\alpha_\mathrm{B6}$ are give in Table \ref{tab:specind}. 
Example plots of the spectral index for W~Aql and U~Her are given in Fig.~\ref{fig:spec-examples}. The remaining stars are shown in Fig.~\ref{fig:spec-all} and the KDEs of the spectral indices are shown for all stars in Fig.~\ref{fig:spec-kdes}.
We were unable to extract a value of $\alpha_\mathrm{B6}$ for RW Sco from the combined data because the low S/N resulted in very large uncertainties on the spectral index. However, using only the extended array data we were able to estimate a spectral index for RW~Sco.}

\edits{
For \pigru\ we get an anomalously high $\alpha_\mathrm{B6} \gtrsim 3$. This is higher than the Band 6 to Band 7 value, $\alpha_\mathrm{B6-B7}=1.9\pm0.3$, found by \cite{Esseldeurs2025}, and the Band 3 to Band 6 value we find, $\alpha_\mathrm{B3-B6}=1.8\pm0.4$, details in Sect.~\ref{sec:b3specind}. The Band 6 mid data gave a value closer to the multi-band measurements. We suspect that the extended data for this star was affected by all the possible sources of uncertainty mentioned above and thus acts as a caution in using in-band spectral indices. 
}

\begin{table}
	\centering\small
	\caption{\edits{Spectral indices calculated from our observations.}}
	%\red{Should RA, Dec be to more decimal points? OH30 was to 5 for individual arrays, mean to 4}
	\label{tab:specind}
	\begin{tabular}{cccc}
		\hline
Star		& 	$\alpha_\mathrm{B6}$ &	$\alpha_\mathrm{B3-B6}$ & Dust mass [$\msol$]	\\
\hline
GY Aql 	&$	2.4	\pm	0.3	$&$	1.9	\pm	0.6	$&	$3.4\e{-8}$	\\
R Aql 	&$	1.7	\pm	0.1	$&$	2.5	\pm	0.7	$&	$4.3\e{-8}$	\\
IRC$-$10529	&$	2.1	\pm	0.1	$&$	1.7	\pm	0.6	$&	$3.7\e{-7}$	\\
W Aql 	&$	2.1	\pm	0.2	$&$		...		$&	$5.5\e{-8}$	\\
SV Aqr 	&$	2.1	\pm	0.6	$&$		...		$&	$6.2\e{-9}$	\\
U Del 	&$	1.8	\pm	0.1	$&$		...		$&	$1.6\e{-8}$	\\
$\pi^1$ Gru	&$	3.0	\pm	0.1	$&$	1.8	\pm	0.4	$&	$1.4\e{-8}$	\\
U Her 	&$	1.7	\pm	0.1	$&$	1.7	\pm	0.3	$&	$2.9\e{-8}$	\\
R Hya 	&$	2.0	\pm	0.1	$&$		...		$&	$7.3\e{-9}$	\\
T Mic 	&$	2.0	\pm	0.1	$&$	2.0	\pm	0.4	$&	$7.2\e{-9}$	\\
S Pav 	&$	2.1	\pm	0.1	$&$		...		$&	$1.2\e{-8}$	\\
IRC+10011	&$	2.3	\pm	0.1	$&$		...		$&	$1.1\e{-7}$	\\
V PsA 	&$	1.8	\pm	0.1	$&$		...		$&	$9.3\e{-9}$	\\
RW Sco	&$\phantom{^e}	2.3	\pm	0.2	{^e}$&$	\it 1.0	\pm	0.2	$&	$1.7\e{-8}$	\\
KW Sgr 	&$	2.0	\pm	0.1	$&$		...		$&	$1.9\e{-7}$	\\
VX Sgr 	&$	2.1	\pm	0.1	$&$		...		$&	$6.4\e{-7}$	\\
AH Sco 	&$	1.4	\pm	0.3	$&$		...		$&	$1.0\e{-6}$	\\
		\hline
	\end{tabular}
	\tablefoot{\edits{$\alpha_\mathrm{B6}$ is derived from the combined Band 6 continuum with the exception of ($^{e}$) RW Sco, for which $\alpha_\mathrm{B6}$ is derived from extended Band 6 data only. $\alpha_\mathrm{B3-B6}$ is calculated from the ACA Band 3 data and the compact main array Band 6 data. The italic values have a higher uncertainty than the listed formal value. The dust mass estimate is a lower limit in all cases and assumes a dust temperature of 1000~K. See text for details.}}
\end{table}

\subsubsection{Band 3 to Band 6 spectral indices}\label{sec:b3specind}

For the seven stars in our ACA Band 3 sample we also calculate \edits{band-to-band} spectral indices, \edits{$\alpha_\mathrm{B3-B6}$}. Because all the stars are unresolved in the large (10--15\arcsec) synthesised beams of the ACA data, we use the peak fluxes to approximate the total flux. We also fit 2D Gaussians to the observations to extract the integrated flux and found equivalent values, within the uncertainties. 
The peak fluxes are listed in Table \ref{tab:aca}. We adopt 104.96~GHz as the central frequency of the Band 3 ACA continua and \edits{238.45~GHz 
%241.75~GHz 
as the central frequency of the Main Array Band 6 compact configuration continua}\footnote{\edits{Note that this is a slightly different value than for the combined array data because fewer spectral windows were observed with the compact configuration than with the extended and mid configurations; see \cite{Gottlieb2022} for details.}}. The spectral indices are then calculated from these values \edits{and the peak fluxes of the ACA and compact data} (Table \ref{tab:compact}). The resultant spectral indices are listed in \edits{Table \ref{tab:specind}} and fall in the range \edits{1.7--2.5, aside from RW~Sco, for which $\alpha_\mathrm{B3-B6}=1.0\pm0.2$. This is most likely the result of the low S/N measurements for this star.} 

%\edits{Data observed with the compact 12-m array is largely unresolved (Sect.~\ref{sec:roundconts}) and the ACA data is certainly unresolved.
%}
%We note that the 12-m array \edits{compact} Band 6 measurements from the UD fits do not include scattered, extended emission. Since the ACA resolution is greater than the maximum Band 6 MRS, some of the emission detected by the ACA may be either resolved out, or undetected due to low surface brightness at Band 6. On the other hand, the rms noise in the ACA observations is around 0.2 mJy, whilst the 12-m combined rms is mostly $\sim0.01$~mJy, so the ACA flux may be depressed by spatial decorrelation, especially as the S/N was too low for self-calibration.
%To summarise these effects, if the Band 6 data has more missing flux, then the spectral index would be underestimated, an effect that would be systematically worse for the nearby sources. If the Band 3 were decorrelated, the result would be a steeper spectral index, especially for the weaker sources, which are more likely to be more distant. 

\subsubsection{Trends with spectral index}

\edits{Within our uncertainties (aside from RW~Sco and \pigru), the $\alpha_\mathrm{B6}$ and $\alpha_\mathrm{B3-B6}$ spectral indices agree with each other. Aside from the outliers already discussed, we find that the spectral indices for all our stars $\sim 2$. This agrees with the theoretical predictions for unresolved spectral indices calculated by \cite{Bojnordi-Arbab2024}. Since we were only able to reliably derive the $\alpha_\mathrm{B6}$ spectral indices close to the star (essentially only at the continuum peak), we would expect the main contribution to the spectral index to come from the star itself. This is borne out by the predictions of \cite{Bojnordi-Arbab2024}; although they find variations in the unresolved spectral index with pulsation phase, their values are in the range $\sim$1.7--2.05 over frequencies from 100--300~GHz, in good agreement with our results. It is likely that the $\alpha_\mathrm{B3-B6}$ values are also dominated by the stars since, for the same enclosed regions, the flux densities of the standard combined maps (Table~\ref{tab:combined}) are a factor of 2--10 higher than the flux densities of the UD-subtracted maps (Table \ref{tab:udsub}).}
\editss{In general, the value of $\sim2$ of the spectral index is consistent with optically thick blackbody emission, another indication that the stellar contribution dominates.}
%We also find that both the $\alpha_\mathrm{B6}$ and $\alpha_\mathrm{B3-B6}$ values} agree with the unresolved spectral indices calculated by \cite{Bojnordi-Arbab2024} from theoretical models of AGB stars. 

%The spectral index also varies with phase, e.g. \cite{Bojnordi-Arbab2024} show disc-averaged changes of up to $\sim0.4$, with a maximum  $<2.2$ at 200--300~GHz (their Fig.~3). The exact predictions vary with stellar properties and measurement method.

%The stars in our sample with the largest spectral indices ($>1.8$, see Table \ref{tab:specind}) are R~Aql, GY~Aql and T~Mic. We find that R~Aql and T~Mic have strong ACA continuum detections, in excess of $10\sigma$, so are unlikely to suffer from decorrelation. 
%%\red{[Anita to check these statements]} 
%GY~Aql and IRC$-$10529 have comparatively weak ACA continuum detections (5--$7\sigma$) so decorrelation could explain its high spectral index. 

We checked the data for possible correlations between the spectral index and other parameters, \edits{searching for trends or systematics}. We find no significant correlation between spectral index and distance, suggesting that there is no factor uniformly affecting only the closest or most distant stars. For $\alpha_\mathrm{B3-B6}$ we checked for correlations between the spectral index and the individual ACA or 12-m fluxes and found none.
We also found no apparent correlation between spectral index and stellar pulsation period.
\edits{There is a tentative positive correlation between $\alpha_\mathrm{B6}$ and mass-loss rate (see Appendix \ref{app:massloss} for details on the mass-loss rates), with low statistical significance.
%which we plot in Fig.~\ref{fig:spec-mloss}. 
The models of \cite{Bojnordi-Arbab2024} predict a slightly larger spread in unresolved spectral indices for their higher mass-loss rate model ($2\e{-6}\spy$ compared with their canonical $3\e{-7}\spy$ model), but only to 1.55--2.1 in the 100--300~GHz range. We hence suggest that this tentative correlation between spectral index and mass-loss rate requires further observations to confirm.}

\subsection{Dust mass estimates}

\edits{Having calculated the spectral indices for our sample, we can now estimate the mass of the dust visible in our ALMA observations.
Following \cite{Knapp1993} we estimate the dust masses using
\begin{equation}\label{eq:dustmass}
M_d = F_\mathrm{UDsub}\left(\frac{2 a_g \rho_g c^2 d^2}{3 \nu^2 Q_\nu k_B T_d}\right),
\end{equation}
where $F_\mathrm{UDsub}$ is the total flux density of the UD-subtracted continuum image, $a_g$ is the radius of the dust grains, \editss{assumed to be spherical}, which we take as 0.3~\micron, $\rho_g$ is the mass density of the dust grains, which we take as 3.3~g~cm$^{-3}$, $c$ is the speed of light, $d$ is the distance to the star (see Table~\ref{tab:stars}), $\nu$ is the frequency, here 241.75~GHz, $Q_\nu$ is the grain emissivity at $\nu$, $k_B$ is the Boltzmann constant, and $T_d$ is the dust temperature, which we take to be the lower value out of the brightness temperature (Table \ref{tab:ud}) or 1000~K, which is close to the condensation temperature of silicate grains \citep{Suh1999}.
We assume $Q_\nu \propto \nu^\beta$, where $\beta = \alpha -2$ is the dust emissivity spectral index.
%(and where $F_\mathrm{UDsub} \propto \nu^\alpha$). 
We derive $\beta$ for each star from our $\alpha_\mathrm{B6}$ values in Table \ref{tab:specind}, resulting in $\beta \sim 0$ for most stars. For $Q_\nu$, we follow \citeapos{OGorman2015} modification of \citeapos{Knapp1993} estimate and use 
\begin{equation}\label{eq:Qnu}
Q_\nu = 5.65\e{-4} \times \left(\frac{\nu}{274.6~\mathrm{GHz}}\right)^\beta.
\end{equation}}

\edits{Our resultant dust mass estimates are given in Table~\ref{tab:specind} and range from $\sim 6\e{-9}~\msol$ for SV~Aqr to $1\e{-6}~\msol$ for AH~Sco. These are all lower limits on the true dust masses around these stars because, as discussed in Sect.~\ref{sec:resolved-out}, a significant amount of flux has been resolved out for our observations.
This is particularly apparent when comparing our results to dust masses calculated from \textsl{Herschel}/PACS observations by \cite{Cox2012}. For the three stars in both samples, W~Aql, R~Hya, and T~Mic, \cite{Cox2012} find dust masses of 1.6--$1.9\e{-4}~\msol$, $3\e{-5}~\msol$ and 1--$6\e{-5}~\msol$, respectively. This is three to four orders of magnitude larger than our dust masses, in all cases. \cite{Cox2012} assume rather low dust temperatures of 35~K, but even if we use this value in our calculations, we still find about two orders of magnitude less dust.}
\editss{We also investigated the impact of different values of $\beta$ on our dust masses, including the value of $\beta=2$ adopted by \cite{Cox2012} from \cite{Li2001}, and found that varying $\beta$ from $-1$ to 2 resulted in less than a 35\% change in the calculated dust mass.}
\editss{Finally, a more precise estimate of the dust mass would need to take into account non-spherical grains with a range of different sizes, which is beyond the scope of the present work,
%we note that the dust mass as described in equations \ref{eq:dustmass} and \ref{eq:Qnu} may be highly dependent on the choice of dust grain size\footnote{\editss{Although \cite{Hildebrand1983} note that $Q_\nu/a_g$ is roughly independent of $a_g$.}}, which is likely not to be uniform throughout the envelope. Implementing a dust grain distribution is beyond the scope of the present work, 
especially since emission from the larger grains, assumed to be found further from the star, is most likely to be resolved out. }

%\edits{Note that the assumed dust temperature is if the bulk of the grains are significantly cooler than the condensation temperature, the calculated dust mass will 
%}

\subsection{Comparison with NIR stellar radii}\label{sec:rads}

\begin{table*}
	\centering
	\caption{Calculated and measured radii for AGB stars in our sample.}
	\label{tab:radii}
	\begin{tabular}{ccccccccc}
		\hline
Star	&	$T_\mathrm{eff}$	&	$L_\star$	&	Ref. for	&	Calc $R_\star$	& 	Calc diameter	& $R_\mathrm{NIR}$	& NIR diameter & Ref. for	\\
	&	[K]	&	[$\lsol$]	&	$T_\mathrm{eff},L_\star$	&	[cm]	& [mas] & [cm]	& [mas] & NIR	\\
\hline
GY Aql	&	2143	&	6950	&	1, 2	&	4.2e+13	&	13.7	&	...	&	...				&	...	\\
R Aql	&	3008	&	6930	&	1	&	2.1e+13	&	10.7	&	2.2e+13	&	10.9	(	0.30	)	&	9	\\
IRC$-$10529	&	2000	&	11417	&	3, 2	&	6.2e+13	&	8.9	&	1.3e+14	&	18	(	5	)	&	10	\\
W Aql	&	2300	&	7249	&	4, 2	&	3.7e+13	&	13.1	&	3.3e+13	&	11.6	(	1.80	)	&	11	\\
SV Aqr	&	2180	&	3586	&	5	&	2.9e+13	&	8.8	&	...	&	...				&	...	\\
U Del	&	3236	&	7069	&	1	&	1.9e+13	&	7.5	&	2.0e+13	&	7.9	(	0.50	)	&	12	\\
$\pi^1$ Gru	&	3100	&	7200	&	6	&	2.0e+13	&	16.7	&	2.3e+13	&	18.4	(	0.18	)	&	13	\\
U Her	&	2700	&	5800	&	1, 2	&	2.4e+13	&	11.9	&	2.3e+13	&	11.2	(	0.60	)	&	16	\\
R Hya	&	3100	&	10300	&	2	&	2.4e+13	&	26.0	&	2.2e+13	&	23.7	(	1.00	)	&	14	\\
T Mic	&	2856	&	5326	&	1	&	2.1e+13	&	15.8	&	...	&	...				&	...	\\
S Pav	&	2752	&	5564	&	1	&	2.3e+13	&	16.6	&	...	&	...				&	...	\\
IRC+10011	&	1800	&	11082	&	7, 2	&	7.5e+13	&	14.0	&	1.1e+14	&	20	(	5	)	&	10	\\
V PsA	&	2360	&	3176	&	5, 8	&	2.3e+13	&	10.5	&	...	&	...				&	...	\\
RW Sco	&	2500	&	4608	&	15	&	2.5e+13	&	6.0	&	...	&	...				&	...	\\
		\hline
	\end{tabular}
\tablefoot{The ``Calc rad'' column gives the stellar radius in cm as calculated from the stellar effective temperature ($T_\mathrm{eff}$) and luminosity ($L_\star$) using equation \ref{eq:rad}. The conversion between the calculated radii in cm and calculated diameters in milliarcsec are done using the distances given in Table \ref{tab:stars}, and vice versa for the diameters in arcsec and observed radii in cm. The observed diameters are all based on K-band observations except for \pigru, where the observations were carried out at $1.625-1.730~\mu$m. See text for further details.
\textbf{References.} 1:	\cite{McDonald2012}, 2: \cite{Andriantsaralaza2022},
3: \cite{Danilovich2015a}, 4:	\cite{Danilovich2014}, 
5:	\cite{Olofsson2002},
6:	\cite{Mayer2014},
7:	\cite{Ramstedt2014}, 8:	\cite{Guandalini2008}, % comment on how L calcualted if these are kept
9:	\cite{Hofmann2002},
10:	\cite{Blasius2012},
11:	\cite{van-Belle1997},
12:	\cite{Dyck1996},
13:	\cite{Paladini2017},
14:	\cite{Monnier2004},
15: \cite{Groenewegen1999},
16: \cite{van-Belle2002}.
}
\end{table*}

\begin{table}
	\centering
	\caption{Measured radii for RSG stars in our sample.}
	\label{tab:rsgradii}
	\begin{tabular}{cccc}
		\hline
Star	&	  NIR diam & $R_\mathrm{NIR}$	& Ref. for	\\
 & [mas] &	[cm]	 & measurements	\\
\hline
KW Sgr																&	3.9	(	0.25	)	&	6.3e+13	&	1	\\
VX Sgr																&	7.7	(	0.40	)	&	9.0e+13	&	2		\\
AH Sco																&	5.8	(	0.15	)	&	7.6e+13	&	1	\\
		\hline
	\end{tabular}
\tablefoot{References: 1: \cite{Arroyo-Torres2013}; 2: \cite{Chiavassa2010}}
\end{table}

We collected angular diameters measured in the near-infrared (NIR) for our sample of stars. K-band measurements were available for 5 out of the 14 AGB stars and for all 3 RSGs. Additionally, we include the measurement of \pigru\ performed by \cite{Paladini2017} at 1.625, 1.678, and 1.730~\micron, because this is close to the central K-band wavelength of 2.2~\micron. For most of the sources, the sizes were derived by fitting uniform discs and most of these measurements have relatively low uncertainties. For W~Aql, the NIR measurement was done using the technique of lunar limb darkening, as described in \cite{van-Belle1997} and \cite{Richichi2002}. For IRC$-$10529 and IRC+10011 the size was derived using a Gaussian fit to sparse-aperture masking data by \cite{Blasius2012}, who note that both sources are asymmetric and likely surrounded by optically thick dust at 2.2~\micron, introducing additional uncertainties. The well-known variability of AGB stars in luminosity, radius and temperature adds to the uncertainties in both measured and calculated properties.
The collected measurements are given in Table~\ref{tab:radii} for the AGB stars and in Table \ref{tab:rsgradii} for the RSGs.

Since NIR measurements were only available for nine stars in our sample, we additionally calculated radii of all the AGB stars, $R_\star$, from the stellar luminosity, $L_\star$, and effective temperature, $T_\mathrm{eff}$, using the Stefan-Boltzmann relation,
\begin{equation}\label{eq:rad}
R_\star = \frac{1}{T^2_\mathrm{eff}}\sqrt{\frac{L_\star}{4\pi\sigma_\mathrm{sb}}}
\end{equation}
where $\sigma_\mathrm{sb}$ is the Stefan-Boltzmann constant. The temperatures and luminosities were all collected from the literature, preferring results from SED fitting. Where possible, the luminosities were taken from the same paper as the distances \citep[i.e.][]{Andriantsaralaza2022}, otherwise they were scaled by the ratio squared of the literature distance and our adopted distance (Table~\ref{tab:stars}). Where two references are given in the fourth column of Table \ref{tab:radii}, the first refers to the effective temperature and the second to the luminosity. In the case of V~PsA, the luminosity was obtained from the period-luminosity relation given in \cite{Guandalini2008}, though we caution this value is relatively uncertain as the relation was derived for periods 200--600 days. For RW Sco, the temperature is assumed rather than calculated by \cite{Groenewegen1999}.

\begin{figure*}
\centering
	\includegraphics[height=6.6cm]{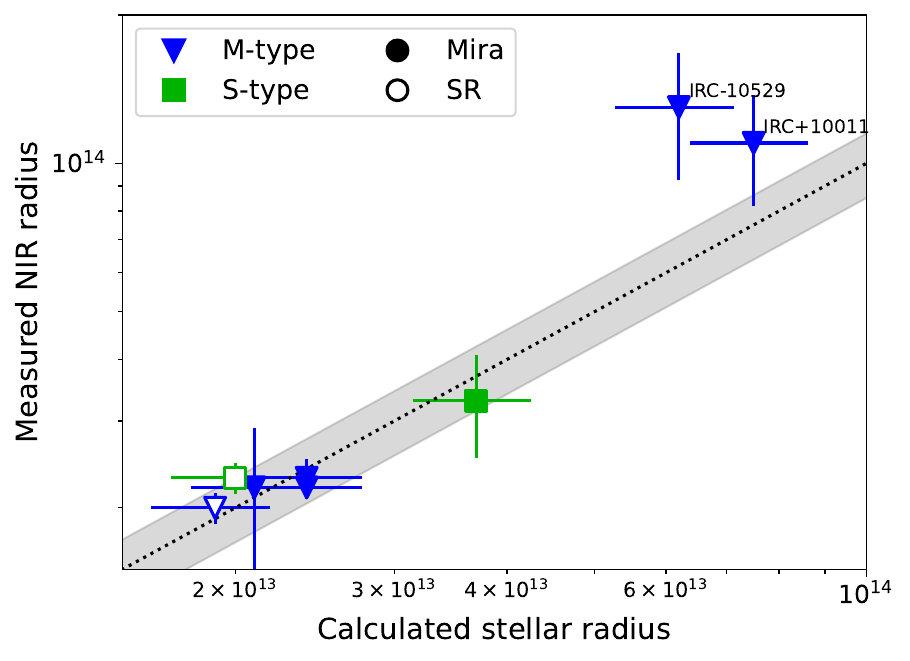}
	\includegraphics[height=6.6cm]{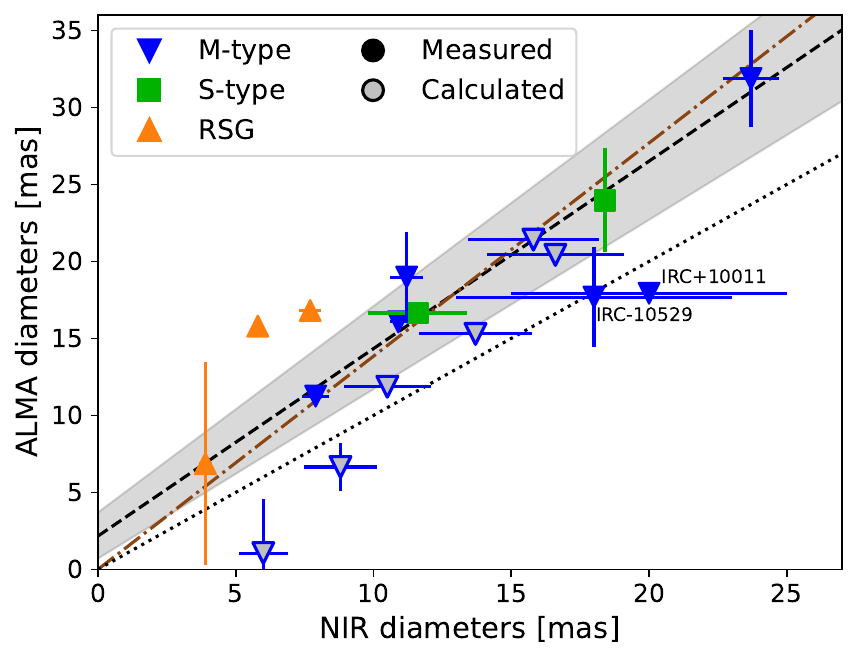}
    \caption{A comparison between measured NIR and calculated stellar radii (left) and between the NIR and ALMA \edits{UD} diameters (right). In each case, the dotted line represents a 1:1 relation. The dashed line in the right panel represents a fit to the measured AGB datapoints using orthogonal distance regression (Eq. \ref{eq:radii}), with the grey region the uncertainty on this fit. \edits{The brown dot-dashed line shows the best linear fit for a relation forced to pass through the origin.} See text for details. The shaded region in the right panel shows the uncertainty on the fit. The shaded region in the left panel shows that the measured NIR radii and the calculated radii for most stars fall within 15\% of a 1:1 relation.}
    %Stars with measured NIR sizes are plotted as filled markers, while stars for which NIR sizes were calculated are plotted as unfilled markers. M-type stars are plotted as blue triangles (point down), S-type as green squares, and red supergiants as orange triangles (point up).}
    \label{fig:radii}
\end{figure*}

Before comparing the calculated and NIR sizes with our new ALMA results, we first checked the available NIR diameters against the corresponding calculated radii. We converted the measured diameters to physical radii using the relation $R_\star = d\theta_\mathrm{NIR}/2$ where the distance, $d$, was taken from Table~\ref{tab:stars} and $\theta_\mathrm{NIR}$ is the NIR diameter. The uncertainties on the NIR radii come from the measurement uncertainties (Table~\ref{tab:radii}) and the distance uncertainties (Table \ref{tab:stars}). The uncertainties on the calculated radii are probably lower limits since uncertainties for stellar effective temperatures and luminosities are not always provided in their literature sources. In the left panel of Fig.~\ref{fig:radii}, we plot the NIR stellar radii against the calculated radii, assuming 15\% uncertainties for the calculated sizes because the majority of stars are scattered around the 1:1 line to within 15\% (grey shaded region). The exceptions are the significantly larger sources IRC+10011 and IRC$-$10529, which have larger measured radii than calculated, by factors of 1.5 and 2 respectively. This is most likely the result of optically thick dust found close to the star \citep{Blasius2012}.

In the right panel of Fig.~\ref{fig:radii} we plot the ALMA UD fit diameters against the measured NIR diameters (uniformly coloured markers), where available, or the calculated stellar diameters (grey filled markers) if there are no NIR measurements. The dotted line in Fig.~\ref{fig:radii} is for a 1:1 correspondence between the two measurements and we see that, in general, the ALMA diameters are larger than the NIR diameters.
% ALMA central wavelength is 1.24~mm

We perform an orthogonal fit to only the AGB stars with measured (not calculated) NIR diameters and find the following relationship:
\begin{equation}\label{eq:radii}
D_{\mathrm{UD}_\mathrm{ALMA}} = a \times D_{\mathrm{NIR}} + b
\end{equation}
where $a=1.22\pm0.12$ and $b=2.2\pm1.5$~mas. This relationship is plotted as the dashed line in the right panel of Fig.~\ref{fig:radii}, with the uncertainties shown by the grey shaded region.
Note that although IRC$-$10529 and IRC+10011 are outliers with larger NIR diameters than ALMA UD diameters, the uncertainties on their NIR radii are so large that they do not significantly contribute to the fit. With the exception of these two stars, there is minimal scatter around the trend line for the stars with measured data. There is more scatter among the stars with calculated optical diameters, but the only true outlier among these is RW~Sco, for which uncertainties come from being unresolved by ALMA (resulting in an uncertain UD diameter, see Table \ref{tab:ud}) and an uncertain effective temperature estimate, as noted above. 

%\red{copied from 5.2}
Of course Eq.~\ref{eq:radii} must only be valid over a limited regime, since a non-zero diameter at one wavelength cannot physically correspond to a diameter of zero in another. However, in the right panel of Fig.~\ref{fig:radii} we plot observed sizes not physical sizes and the real lower limits on both axes are the resolution limits. 
Therefore, we must assume that this relationship only holds over a certain (apparent) size regime, or is not truly linear. To better understand this, we require a larger sample of high quality measurements in both the NIR and at 1.24~mm, ideally observed contemporaneously.

If, \edits{instead of allowing a general linear fit,} we force the line to pass through \edits{the origin}, we find a slope of 1.38, which is still a reasonable fit to the measured data \edits{and is shown as the brown dot-dashed line in Fig.~\ref{fig:radii}}.
Plotting only calculated sizes for all the stars would result in even more scatter in our plot. This can partly be explained by the variations in temperature and radius at different phases of the pulsation period. \cite{Bojnordi-Arbab2024} also showed that the theoretical sizes of AGB stars can vary across pulsation period, which would add to the scatter given that observations of different stars and different configurations were taken at different pulsation phases \citep[e.g. see Table 7 of][]{Baudry2023}.
%Given that the \edits{best-fitting} trend line from Eq.~\ref{eq:radii} does not pass through \edits{the origin}, 

\edits{Finally, in addition to examining the observed sizes of our stars, we tested fitting the physical sizes. This introduces additional uncertainties from the distances but also has the effect of rearranging the data so that less weight is placed on the closest stars which have the largest apparent sizes but similar physical sizes to the majority of our sample. Within the uncertainties, a fit to these parameters results in a 1:1 relation between the ALMA- and NIR-derived physical sizes. (This holds whether or not we include the RSGs and/or the two largest AGB stars, IRC$-$10529 and IRC+10011, in our fit.) The results of this test indicate that the formal uncertainties on our fit (Eq.~\ref{eq:rad}) are smaller than the real uncertainties.}

\subsection{Trends between stellar properties}
% Trends between continuum emission and other stellar properties
% "Sample trends"?

\begin{figure*}
\centering
	\includegraphics[height=7cm]{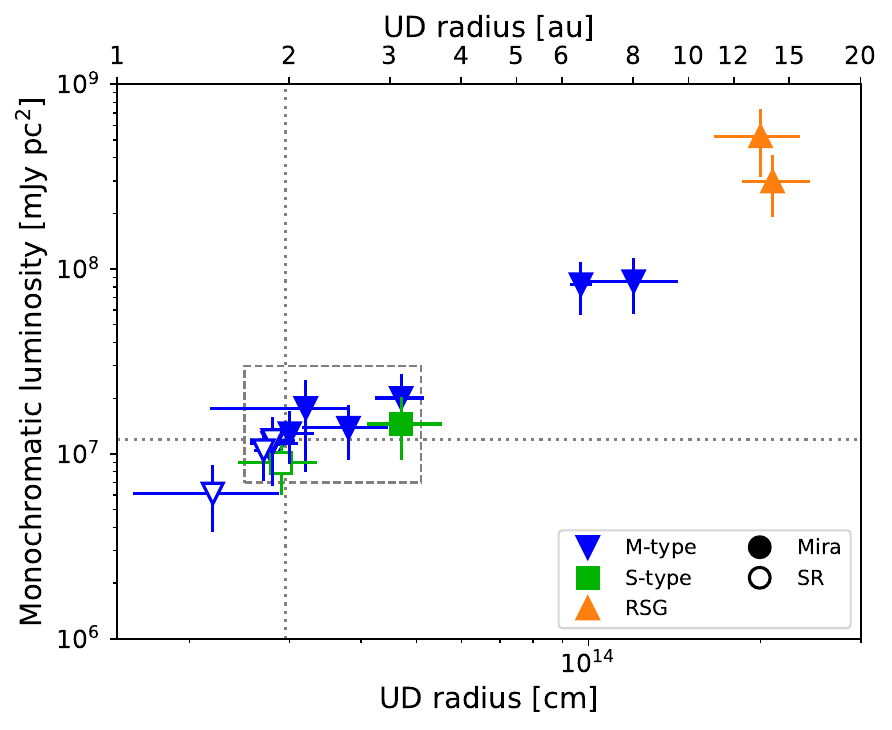}
	\includegraphics[height=7cm]{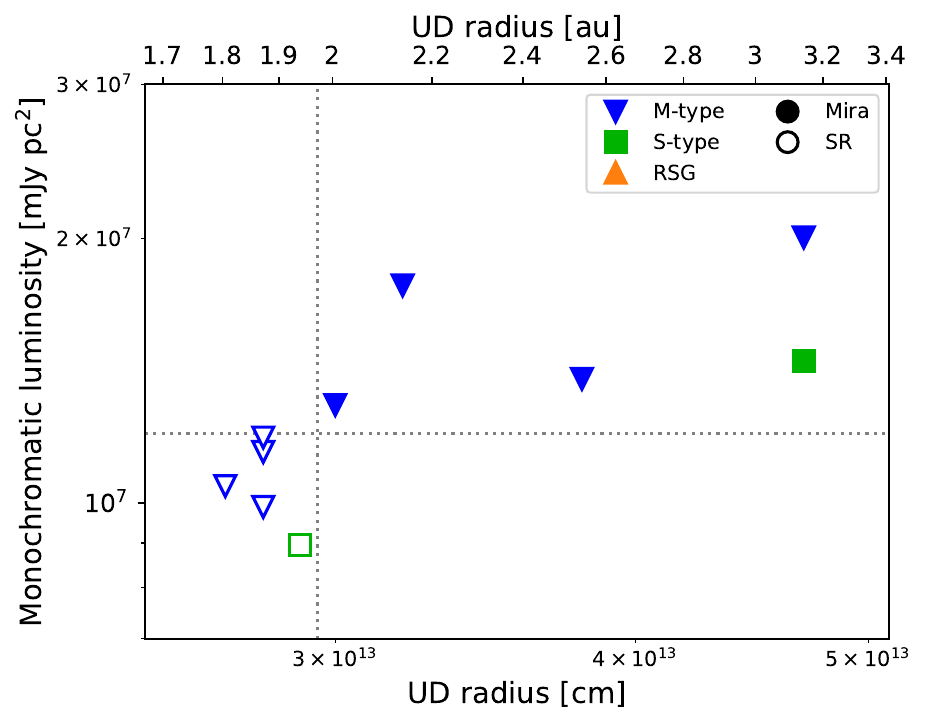}
    \caption{Monochromatic luminosities plotted against physical sizes, as calculated from the UD fits. 
    %Mira variables and red supergiants are plotted as filled markers, while semiregular variables (SRa and SRb) are plotted as unfilled markers. M-type stars are plotted as blue triangles (point down), S-type as green squares, and red supergiants as orange triangles (point up). 
    The left panel shows all the stars and the right panel shows the stars within the dashed rectangle in the left panel. The error bars are omitted in this plot for clarity. The dotted vertical line is plotted at a radius of $2.95\e{13}$~cm and separates the SR AGB stars from the Mira variables, as does the dotted horizontal line plotted at $1.2\e{7}$~mJy pc$^2$.}
    \label{fig:radflux}
\end{figure*}

We checked for any trends between the measured continuum properties, based on the UD fits. In Fig.~\ref{fig:radflux} we plot the monochromatic luminosity, which is essentially a normalised UD flux density, against the measured UD radii for our sample.
The 1.24~mm radio luminosity, which we refer to as the monochromatic luminosity, was calculated using
\begin{equation}\label{eq:fluxdens}
L_\mathrm{mm} = 4\pi F_\mathrm{UD} d^2
\end{equation}
where $F_\mathrm{UD}$ is the UD flux density (Table~\ref{tab:ud}), and $d$ is the distance. The UD radius is calculated from the angular measurements given in Table~\ref{tab:ud} and the distances are listed in Table~\ref{tab:stars}.
We excluded RW~Sco and KW~Sgr from the comparison, since their radii were very uncertain from the UD fits.
In the left panel of Fig.~\ref{fig:radflux} we included a dashed rectangle, which contains the majority of the ATOMIUM stars. We then use the limits of this box in the right panel to more clearly show the clustered data points.
The stars which fall outside of the drawn box are the Mira variables IRC+10011 and IRC$-$10529, the RSGs VX Sgr and AH Sco, and SV~Aqr, which is both fainter and smaller than the other SR variables. The former groups of stars are outliers for their large radii, as shown for IRC+10011 and IRC$-$10529 in Fig.~\ref{fig:radii}, and high fluxes. IRC+10011 and IRC$-$10529 are also known to be very dusty and obscured AGB stars \citep[e.g. both have high dust optical depths,][]{Schoier2013}.

No \edits{robust} trend is seen for the SR AGB stars, although we note that all of the SR stars have radii $<3\e{13}$~cm ($< 2$~au). They also all have monochromatic luminosities $\leq 1.2\e{7}$~mJy~pc$^2$, whereas the Miras all have monochromatic luminosities $> 1.2\e{7}$~mJy~pc$^2$. The dotted vertical and horizontal lines in Fig.~\ref{fig:radflux} illustrate these boundaries between the Miras and SRs. We see a general trend for the Miras of increasing flux with increased radius. The RSGs are both larger and more luminous than the AGB stars, which is to be expected as they are also more massive.

\edits{Considering all the stars plotted in Fig.~\ref{fig:radflux}, there is a clear positive correlation between the UD radius and the monochromatic luminosity. This is to be expected since the UD radius is related to the physical radius of the star (as discussed in Sect.~\ref{sec:rads}) and the monochromatic luminosity has a dependence on the bolometric luminosity. Since the bolometric luminosity is proportional to the square of the physical radius, see Eq.~\ref{eq:rad}, it is then not surprising that a similar correlation would be reflected in the corresponding mm properties. A fit to these properties gives a slope close to 2, as expected for $L_\star \propto R_\star^2$, but with large uncertainties owing to the scatter and uncertainties on our data.}
%$L_\star \propto R_\star^2$

\subsection{Trends with period}

\begin{figure*}
	\includegraphics[height=6.7cm]{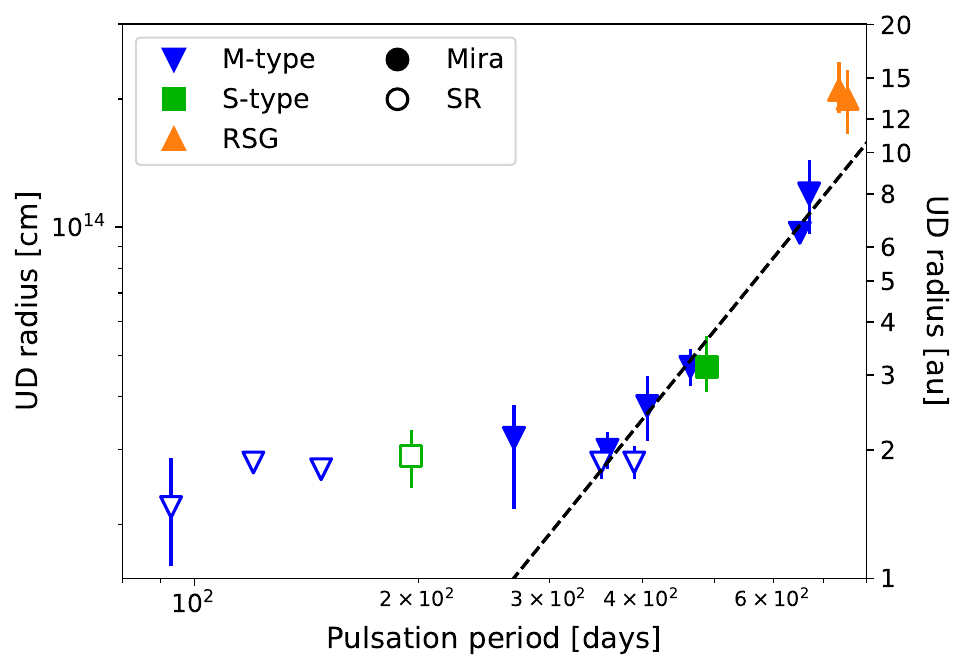}
	\includegraphics[height=6.7cm]{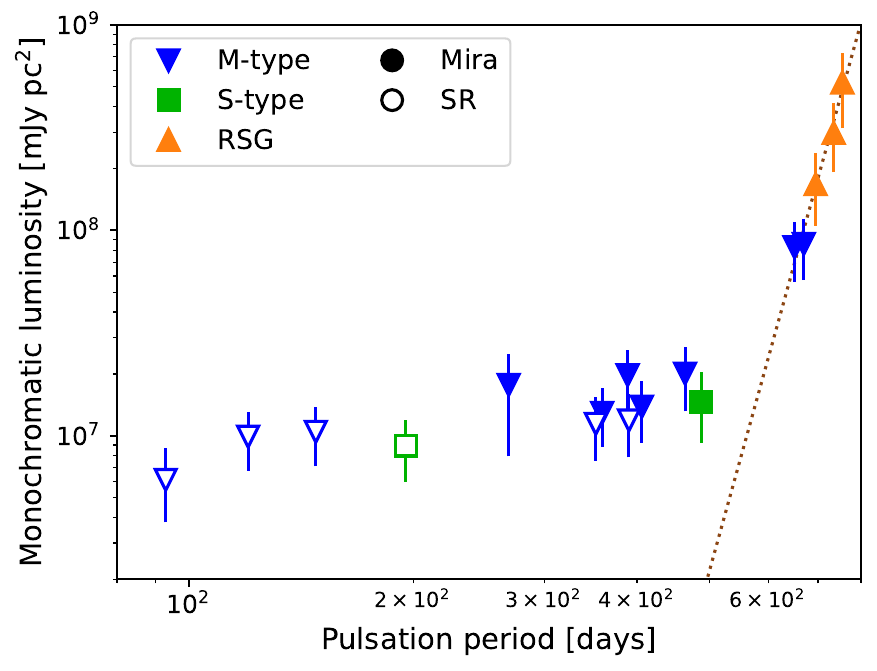}
    \caption{UD radii (left) and monochromatic fluxes (right), plotted against pulsation period. The black dashed line in the left panel is a fit to the Mira variables. The brown dotted line in the right panel is a fit to the RSGs and the two Mira variables IRC$-$10529 and IRC+10011.}
%    Mira variables and red supergiants are plotted as filled markers, while semiregular variables (SRa and SRb) are plotted as unfilled markers. M-type stars are plotted as blue triangles (point down), S-type as green squares, and red supergiants as orange triangles (point up).}
    \label{fig:pulsations}
\end{figure*}

%\red{from Fabrice: \cite{Fernie1964}, \cite{Ahmad2023}}

We checked for any trends between the measured continuum properties and the pulsation period of the stars in our sample. 
%by comparing the period to the UD radii and UD fluxes. 
The left panel in Fig.~\ref{fig:pulsations} shows the physical radii against the primary pulsation periods (Table~\ref{tab:stars}) for our sample. The uncertainties on the radius include the distance uncertainties as well as the uncertainties on our UD fits. There is a positive trend for the Mira variable stars, which the two RSGs also broadly agree with. There is no trend for the SR AGB stars, which all have similar radii (as more clearly seen in Fig.~\ref{fig:radii}) but a range of periods from 89 to 390 days. The only Mira which is an outlier is R~Aql, partly because its distance is relatively uncertain \citep{Andriantsaralaza2022}, adding to the uncertainty in its radius. It is also the only Mira variable with a period $<300$~days. The dashed black line is for a trend to the Mira data, excluding R~Aql, for a relation between the 1.24~mm radius and the period
\begin{equation}\label{eq:P-rad}
\log (R_\mathrm{UD}) = 2.1 \log(P) + 8.1
\end{equation}
with the RSGs found slightly above this trend line.
If we assume that the Mira variables are all fundamental mode pulsators, this is in general agreement with the theoretical models of \cite{Ahmad2023} who found an increasing period with radius based on their CO5BOLD simulations.

In the right panel in Fig.~\ref{fig:pulsations} we plot the monochromatic luminosity (Eq.~\ref{eq:fluxdens}) against the pulsation period. There is a positive relationship between the monochromatic luminosity and the period for the RSGs, with the two most extreme AGB stars (IRC$-$10529 and IRC+10011) agreeing with the trend seen for the RSGs. The dotted brown line plotted is for a relationship
\begin{equation}\label{eq:P-flux}
\log(F_\mathrm{UD}) = 14.3 \log(P) - 32.4
\end{equation}
However, for the majority of the sample, we do not see any clear trends beyond what was already apparent from Fig.~\ref{fig:radflux}: the SR AGB stars have lower fluxes than the Miras. The Miras with periods $\leq490$ days all have comparable monochromatic luminosities with no correlation to the period. 
\edits{There is a weak positive correlation if we consider both Miras with periods $\leq490$ days and SRs, but this arises from the fact that the SRs in our sample generally have shorter periods and all have lower monochromatic luminosities than the Miras.}
%(Note that W Aql has the longest period in this cluster.) 
Fig.~\ref{fig:pulsations} is quite different to period-luminosity diagrams constructed from bolometric luminosities or K-band absolute magnitudes \citep[e.g.][]{Whitelock2008,Trabucchi2017}, which show much clearer trends for both Miras and SR variables. 

Our results are in qualitative agreement with those of \cite{McDonald2016} who used infrared colour as a proxy for dust excess. They find a plateau in their period-colour plot from 120--300 days and an uptick at higher periods. The uptick in our period-monochromatic luminosity plot (right panel of Fig.~\ref{fig:pulsations}) seems to occur at around 500 days rather than 300, but the turning point of a similar uptick in our period-radius plot (left panel of Fig.~\ref{fig:pulsations}) occurs around 300--400 days.

%If we assume that pulsation periods increase with time spent on the AGB, then the (1.24~mm) stellar radius will remain roughly constant until the star reaches a period of around 300 days, when it may transition from being a semi-regular to a Mira variable (Fig.~\ref{fig:pulsations}, left). The flux density, however, remains almost constant until periods $\geq 500$ days (Fig.~\ref{fig:pulsations}, right). Therefore, between -- This was from Iain's email.

\section{Discussion}\label{sec:discussion}

\subsection{Dust features}

With these observations, we present for the first time millimetre stellar and dust continuum emission at high angular resolution for a sizeable sample of AGB and RSG stars. Where our ALMA continuum maps have been discussed phenomenologically in Sect.~\ref{sec:contim}, we now focus on explanations for the observed dust distributions, including some comparisons between the ALMA continuum and observations from the Very Large Telescope's Spectropolarimetric High-contrast Exoplanet Research instrument with the Zurich Imaging Polarimeter \citep[VLT/SPHERE-ZIMPOL,][hereafter referred to as SPHERE]{Beuzit2019,Schmid2018}.
Part of the ATOMIUM campaign involved contemporaneous (within 10 days) observations using the ALMA extended configuration and SPHERE. These have been published in detail by \cite{Montarges2023} and here we pay particular attention to the degree of linear polarisation (DoLP) plots in their Figure 3. \edits{Note that the SPHERE images mainly show scattered light from dust grains located in the plane of the sky \citep[see discussion in][]{Montarges2023}, while our ALMA continuum images recover thermal emission from dust grains, 
in addition to the contribution to the continuum flux from the weak stellar photosphere/chromosphere and the radio photosphere as first described in \cite{Reid1997} --- see Sect.~\ref{sec:litsizes}.}

\subsubsection{\pigru}\label{sec:pigrudis}

\begin{figure}
	\includegraphics[width=0.49\textwidth]{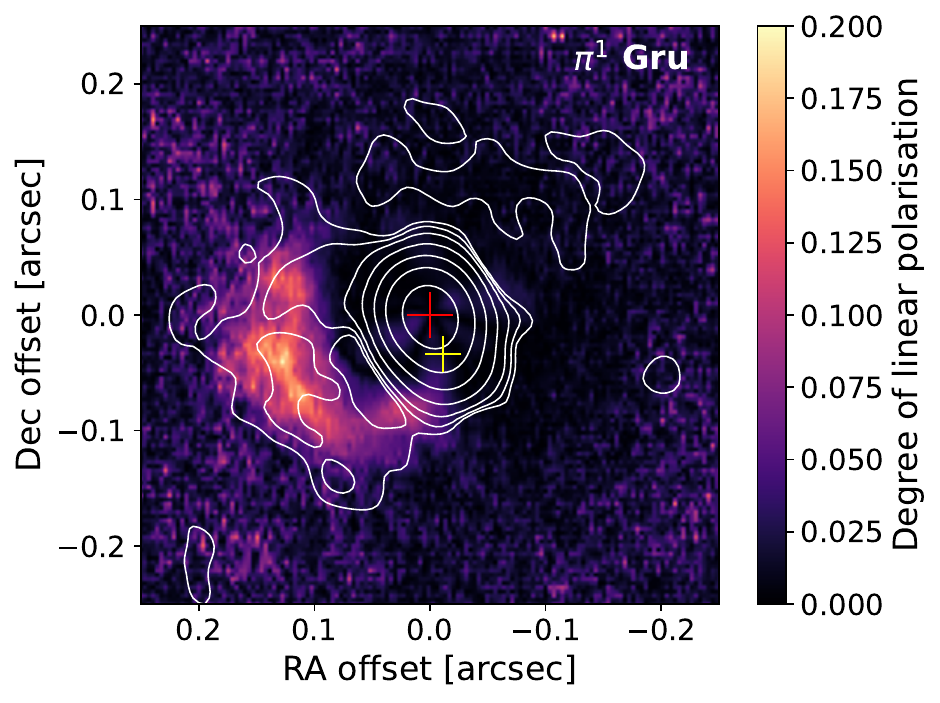}
    \caption{Degree of linear polarisation around \pigru\ observed with SPHERE (colours, from \citealt{Montarges2023}) overplotted with the combined continuum contours observed with ALMA (see Fig.~\ref{fig:pi1gru}), showing good agreement. The position of the AGB star is indicated by the large red cross, and that of the C companion by the smaller yellow cross.}
    \label{fig:pigruDOLP}
\end{figure}

In Fig.~\ref{fig:pigruDOLP} we plot the contours of the combined \pigru\ continuum image over the DoLP image from \cite{Montarges2023}. The tail seen in the combined continuum emission corresponds well to the tail seen in polarised flux. We also note that while the SPHERE image is contemporaneous with the extended ALMA observations, the tail requires the inclusion of the lower resolution configurations which were observed earlier than the extended configuration \citep[extended was observed in June and July 2019, mid in October 2018 and compact in December 2018 and March 2019,][while the SPHERE data were observed in July 2019, \citealt{Montarges2023}]{Gottlieb2022}. Therefore, the combined ALMA image is averaged over a longer period of time than the SPHERE image. Furthermore, the highest degree of polarisation is observed for dust with a $90\deg$ scattering angle, i.e. in the plane of the sky, whereas the continuum emission detected by ALMA does not have that limitation. This could explain why the emission seen to the north and northwest of the AGB star is seen in the ALMA continuum but not in the DoLP map; it is not in the plane of the sky.

This tail of dust lies in the wake of the C companion of the \pigru\ system. By collecting archival SPHERE data, \cite{Montarges2025} have shown that the tail changes position as the companion orbits the AGB star \edits{with a period of 11 years}. Subsequent ALMA observations have confirmed that the continuum feature associated with the C component also changes position in an orbit around the AGB star (Esseldeurs et al, subm.). Using hydrodynamic simulations, \cite{El-Mellah2020} and \cite{Malfait2024} showed that not only do close companions to AGB stars shape the AGB wind into spiral patterns (especially when viewed face on, with the line of sight perpendicular to the orbital plane) but that accretion discs can also form around such companions. \cite{Montarges2025} presented a detailed analysis of the nature of the C companion, and concluded that the feature attributed to the C companion in the extended ALMA continuum likely originates from an accretion disc around the star (which may be a white dwarf of a K dwarf main sequence star). 

\cite{Montarges2025} suggest that the dust tail is seen because dust is concentrated in the wake of the companion or that dust growth is accelerated in that region. Based on the observations presented here of \pigru\ and other stars (see discussion on W~Aql below) we suggest that dust preferentially forms in the wakes of shocks caused by companions moving supersonically through the AGB wind. Since the sound speed is relatively low, even in the inner wind ($\sim3$~\kms), the orbital velocity will almost always be supersonic for a companion within the AGB circumstellar envelope. This is the origin of the shock waves seen in hydrodynamic simulations of companion-wind interactions \citep[e.g.][]{Maes2021,Malfait2021,Malfait2024}. These shocked structures are denser than the surrounding gas, meaning that chemical reactions, including those involved in dust formation, can progress faster. This would also explain the dust distribution around IK~Tau (see discussion in \citealt{Coenegrachts2023}) and is in line with other theoretical modelling results which show that dust forms in the wake of shocks \citep{Freytag2023}. We ran a hydrodynamic model including dust formation to test this point and present some first results in Appendix~\ref{app:hydro}.
A detailed theoretical study of binary-enhanced dust formation over a wider parameter space is forthcoming (Samaratunge et al, in prep).
%\red{INSERT REF TO APPENDIX HERE} Appendix \ref{app:hydro}

\subsubsection{W Aql}\label{sec:waqldiscussion}

W~Aql has a known F9 main sequence companion presently at a separation of around 200~au \citep[see detailed discussion in][]{Danilovich2024}. In Fig.~\ref{fig:waql}, continuum emission is detected close to the position of the companion with a confidence of $5\sigma$. Since we know the nature of the F9 companion, we know that the star itself cannot account for that level of 1.24~mm emission. The hydrodynamic models in \cite{Danilovich2024} do not predict an accretion disc around such a distant companion\footnote{Though we note this could be because those models do not include cooling, see discussion in \cite{Malfait2024}.} \citep[see also][]{El-Mellah2020}, but they do predict a shocked wake close to the companion. The dust observed near the companion could hence be formed in its wake. If this is the case, then we are observing some dust formation $\sim 200$~au from the AGB star. This is much further out than the dust condensation radius, estimated to be around 13~au by \cite{Danilovich2014}, \edits{which is the approximate region within which all dust was previously thought to form}.

The DoLP plot for W~Aql in \cite{Montarges2023} shows large regions of polarised flux detected with high certainty; it has the highest degree of linear polarisation of the ATOMIUM SPHERE sample. The polarised flux is seen on all sides of the AGB star, with the brightest region to the south-southeast. The position of the companion is just outside of the region of polarised flux. It may be contributing to the brighter polarised flux region in the south-southwest by providing additional illumination. The ALMA continuum flux is found in approximately the same general region as the SPHERE polarised flux, but does not follow the regions of highest polarisation.

Overall, we see dust close to the AGB star and further away at the position of the companion. The time scales of the wind expansion and the companion's orbit strongly suggest that dust is forming in the vicinity of the AGB star and in the vicinity of the companion. If dust primarily forms in the wake of shocks, then what we see for W~Aql is dust formation driven by two types of shocks: stellar pulsations close to the AGB star and the wake of the companion moving through the wind supersonically.

%\subsubsection{AGB stars that do not have known companions}

\subsubsection{GY Aql}

The bar seen in the GY~Aql continuum images \edits{(Fig.~\ref{fig:gyaql})} is found at a larger separation from the position of the star \edits{(extending out to $\sim1$\arcsec)} than the DoLP SPHERE image presented in \cite{Montarges2023}, where the majority of the DoLP flux is within $0.2\arcsec$. The SPHERE image shows arcs of polarised flux around the star, especially to the south and northwest. 
Figure 11 of \cite{Montarges2023} shows that the DoLP image may trace out some of the spiral pattern seen in the CO moment 1 map, which represents the velocity field. Since the CO channel maps reveal spiral-like patterns in the wind of GY~Aql \citep{Decin2020}, it is not unreasonable to expect (some of) the dust to also follow the density structures in the wind. Unfortunately, at the separation of the bar structure from the AGB star, the velocity field is messy and it is difficult to draw conclusions through a simple comparison. A more detailed analysis of GY~Aql is forthcoming in a separate paper (Marinho et al, in prep).
%\blue{refer to Louise's paper on GY Aql here?}

\subsubsection{SV Aqr}\label{sec:svaqrdis}

\begin{figure}
	\includegraphics[width=0.5\textwidth]{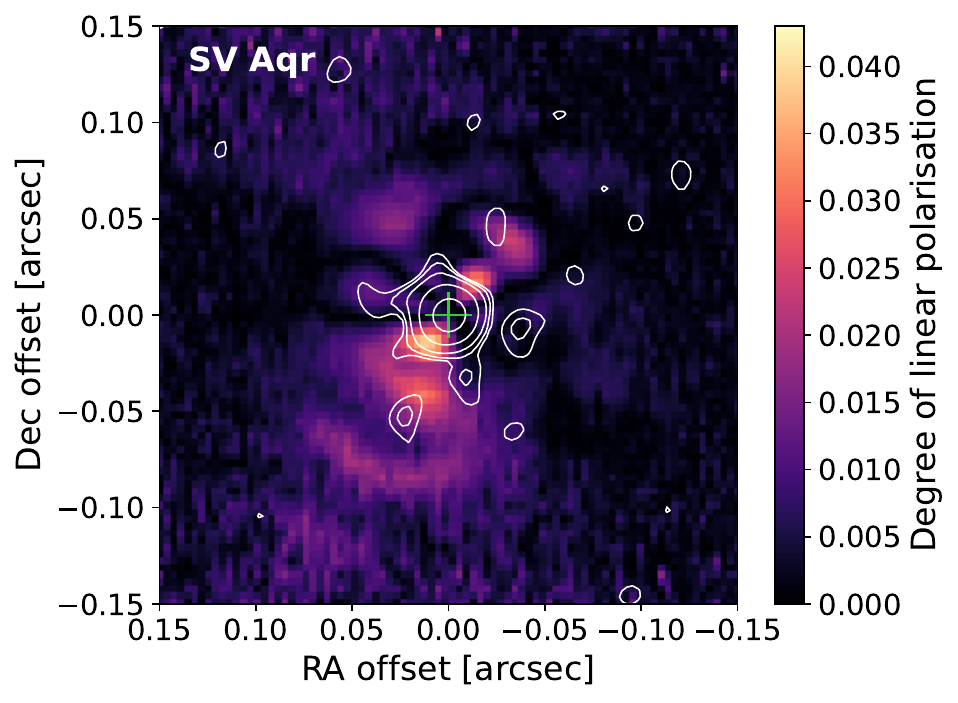}
    \caption{\edits{Degree of linear polarisation around SV~Aqr observed with SPHERE (colours, from \citealt{Montarges2023}) overplotted with the extended continuum contours (white lines) observed with ALMA (see Fig.~\ref{fig:svaqr}). The position of the AGB star is indicated by the green cross.}}
    \label{fig:svaqrdolp}
\end{figure}

The continuum emission observed with ALMA towards SV~Aqr (Fig.~\ref{fig:svaqr}) appears to form an arc. The DoLP image presented in \cite{Montarges2023} has rather low S/N and only a small region of polarised flux above the $3\sigma$ detection threshold, located to the southeast of the star and within $\sim~20$~mas. This is too close to be resolved by the ALMA beam in the extended data, but is close to the asymmetry seen in the $3\sigma$ contour of the extended data. \edits{We reproduce the DoLP image over-plotted with the extended continuum contours in Fig.~\ref{fig:svaqrdolp}, but we caution against over-interpretation of this image, and direct the reader to the discussion of reliability in \cite{Montarges2023} and the $3\sigma$ contours plotted in their Fig.~3.}

Overall, the dust in the vicinity of SV~Aqr follows an irregular pattern. It could be distributed in this way as a result of episodic dust formation, perhaps driven by asymmetric bubbles on the surface, similar to what is seen in the models of \cite{Freytag2023}. \edits{Dust emission forming an arc or loop near the star could} have been formed by a companion shaping the wind and/or enhancing dust in this region. We found SV~Aqr to have an additional period of 231.8 days, with alternating cycles having minima of different depths (see Appendix \ref{app:periods} for details). This could indicate a binary system with a period of 463.6 days. However, the feasibility of this strongly depends on the system mass. If the system mass is only $1~\msol$ (i.e. for a planetary or brown dwarf companion), then the separation for a circular orbit would be only 1.2~au, smaller than the radius we measured with ALMA, even considering our significant uncertainties. A system mass of $2~\msol$ or higher would place a companion (just) outside of our measured radius. \edits{Hence, the available evidence suggests a rather tight binary system. Since it is not presently interacting strongly, this further indicates a system mass $\geq 2\msol$.}

\subsubsection{U Del}\label{sec:udeldis}

The other star in our sample with a long secondary period is U~Del, with a secondary period of 1163 days \citep{Speil2006}. It is the star with the least dust in our sample, especially away from the star, since there was no significant residual flux in our UD subtracted image (Fig.~\ref{fig:udel}). The spectrum of U~Del has the fewest lines from the fewest different molecular carriers of all the stars in the ATOMIUM sample \citep{Wallstrom2024,Baudry2023}. The DoLP image of U~Del in \cite{Montarges2023} shows very little polarised flux above the noise. This is evidence against the long secondary period being caused by a dust-shrouded companion \citep[as described in][]{Soszynski2021}, since in that case we would expect to see more circumstellar dust in general; it is unlikely that a companion would accrete literally all the dust. However, as discussed by \cite{Goldberg2024}, it is possible that a companion might be modulating the AGB dust in a different way. A final possibility is that a companion could be enhancing the formation of the dust on a smaller scale than for \pigru\ --- for example, if the companion has a much lower mass than \pigru~C --- hence not being detected in our ALMA observations and only evident in the long secondary period.
%We finally note that there is likely some dust located close to the AGB star, since the brightness temperature is lower than the stellar effective temperature (Tables \ref{tab:ud} and \ref{tab:radii}), indicating that our UD fit included some contribution from dust.

\subsubsection{KW Sgr}\label{sec:kwsgrdis}

\begin{figure}
	\includegraphics[width=0.5\textwidth]{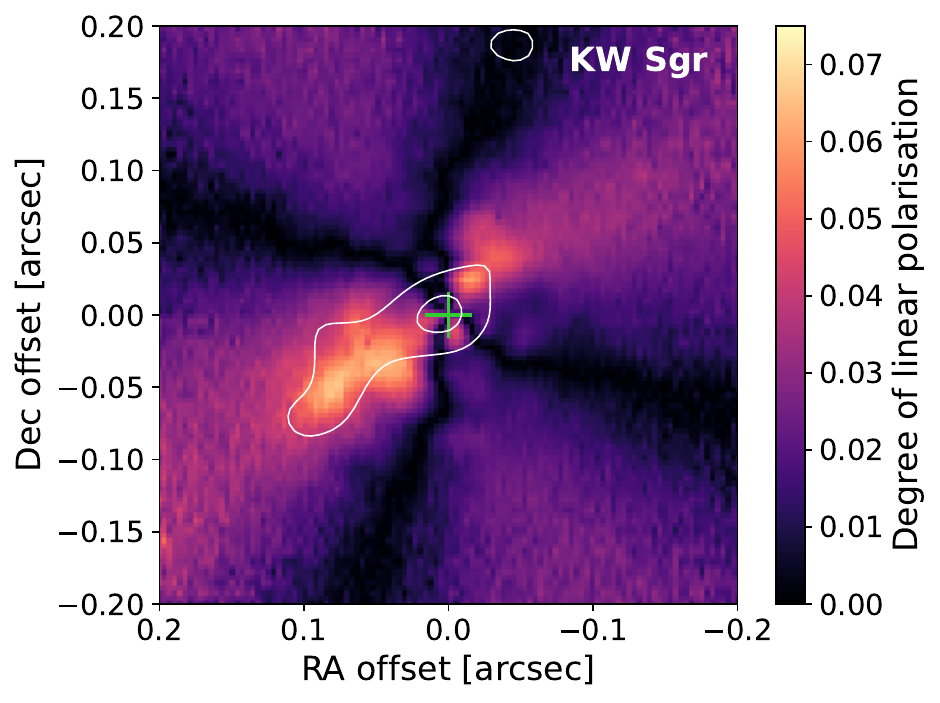}
    \caption{Degree of linear polarisation around KW~Sgr observed with SPHERE (colours, from \citealt{Montarges2023}) overplotted with the UD subtracted contours (at 3 and $5\sigma$) observed with ALMA (see Fig.~\ref{fig:kwsgr}). The emission from both telescopes is in good agreement. The position of the AGB star is indicated by the green cross.}
    \label{fig:kwsgrdolp}
\end{figure}

%The continuum images of KW~Sgr do not have notable features, aside from a protrusion to the east-southeast of the UD subtracted image (Fig.~\ref{fig:kwsgr}). 
KW~Sgr, a RSG \edits{that we did not} observe with the compact configuration of ALMA, has a highly elliptical beam for the mid data and some small asymmetries in the continuum at levels of $\sim3\sigma$ but generally smaller than the area of the beam, and hence not significant. The UD subtracted image (Fig.~\ref{fig:kwsgr}) shows some extended dust to the east south-east of the continuum peak, with a confidence of $3\sigma$, and a little more emission at the position of the continuum peak. Although this protrusion has a low S/N, it agrees well with a similar protrusion seen in the SPHERE DoLP presented by \cite{Montarges2023}, although the DoLP feature also has a low S/N. To emphasise their similarity, we plot the UD-subtracted continuum contours over the SPHERE DoLP image in Fig.~\ref{fig:kwsgrdolp}. The fact that two very different imaging techniques show a similar protrusion, lends some credence to it. Given the high asymmetry of this feature, we suggest that it may be a result of episodic mass loss, which is thought to be characteristic of RSGs (e.g. see \citealt{Levesque2017} and references therein, the phenomenon termed the Great Dimming of Betelgeuse, \citealt{Montarges2021}, and the episodic ejecta of VY~CMa, \citealt{Humphreys2024,Humphreys2025}).

\subsubsection{VX Sgr and AH Sco}\label{sec:rsgdis}

In the SPHERE observations presented in \cite{Montarges2023}, the RSGs VX~Sgr and AH~Sco exhibit moderately extended regions of polarised flux. In both cases, especially AH~Sco, the highlighted regions are partly limited by the S/N of the observations. That is, we see polarised flux mainly to the north and south of AH~Sco with SPHERE but this is partly because the S/N to the east and west is worse, limiting our ability to compare with the significant ALMA flux to the east and west of AH~Sco (\edits{Fig.~\ref{fig:ahsco}}, Sect.~\ref{sec:ahscofeatures}). For VX~Sgr, there appears to be more polarised flux from SPHERE than extended emission from ALMA, but this is likely a result of ALMA resolving out some flux (see Sect.~\ref{sec:bipolar-vxsgr}). The VLT/AMBER imaging of VX~Sgr in the near infrared by \cite{Chiavassa2010} also shows significant complexity in the atmosphere and within a radius of $\sim10$~mas. 
\cite{Chiavassa2010} tentatively show extended structures close to the star ($\sim 10$~mas) to the north and south. Our extended array observations (Fig.~\ref{fig:vxsgr}) show structures to the north and south out to $\sim 100$~mas \edits{($\sim160$~au)}. If these are the same ejecta and we consider the difference in observation times ($\sim$~11 years), then the lower limit on the velocity of these features (ignoring projection effects) would be 60~\kms, comparable to some of the high velocity jets seen for VY~CMa, another extreme RSG \citep{Quintana-Lacaci2023}. If these are not the same structures, then their similarity across physical scales at different times are suggestive of localised episodic mass loss around VX~Sgr.
What we can conclude from the infrared and ALMA observations is that both VX~Sgr and AH~Sco have significant circumstellar dust, as has been shown by earlier observations of, for example, silicate features in infrared spectra \citep{Speck2000b}.

%North-south features in VX Sgr similar to NIR observations of \cite{Chiavassa2010}. Extent in ALMA is much larger (11 years difference).
%
%See also \cite{Chiavassa2022} but this may be less useful.
%
%In general, evidence of episodic mass loss? (see also \cite{Humphreys2022})
%

%\subsubsection{Red supergiants}

\subsection{Angular sizes at millimetre wavelengths}\label{sec:litsizes}

\begin{table*}
	\centering
	\caption{\edits{An overview of UD measurements in the literature.}}
	\label{tab:litUD}
	\begin{tabular}{ccccccccccc}
		\hline
Star	&	Distance	&	Distance	&	Variability	&	Period	&	Period & UD diam. &  Flux &	at freq.& UD  \\
	&	[pc]	&	ref. &	type	&	[days]		& ref.	& [mas] & [mJy] & [GHz] & ref.	\\
\hline
R Dor	&	$	55	\pm	3	$	&	1	&	SRb	&	332, 175	&	5	&	61.8	&	221.4	&	225	&	1	\\
W Hya	&	$	87		^{+11}_{-9}	$	&	2	&	Mira	&	402	&	VSX	&	50	$^{a}$&	210	$^{a}$&	240	&	3	\\
Mira A	&	$	100	\pm	20	$	&	3	&	Mira	&	331	&	VSX	&	43.3	&	99.61	&	229.13	&	3	\\
R Leo	&	$	100	\pm	5	$	&	2	&	Mira	&	312	&	VSX	&	41.88	&	107.83	&	232.53	&	3	\\
R Lep	&	$	471		^{+88}_{-64}	$	&	2	&	Mira	&	437	&	VSX	&	14.2	&	37.3	&	405.0	&	6	\\
CW Leo	&	$	123	\pm	14	$	&	4	&	Mira	&	630	&	VSX	&	53	$^{b}$&	372	&	258	$^{c}$&	7	\\
		\hline
	\end{tabular}
\tablefoot{($^{a}$) Values interpolated from higher and lower frequency observations. ($^{b}$) Parameters given are for a Gaussian fit performed on the data, and hence are not directly comparable to UD fits. ($^{c}$) Approximate estimates.
\textbf{References:} 1: \cite{Vlemmings2024a};
2: \cite{Andriantsaralaza2022};
3: \cite{Vlemmings2019};
4: \cite{Groenewegen2012a};
5: \cite{Bedding1998a};
6: \cite{Asaki2023};
7: \cite{Velilla-Prieto2023};
VSX indicates a period taken directly from the VSX database.
}
\end{table*}

\edits{The apparent angular sizes of AGB stars at radio wavelengths are thought to be dominated by electron-neutral free-free emission out to $\sim2R_\star$ \citep{Reid1997,Matthews2015,Vlemmings2019}. The apparent size of this radio photosphere decreases with increasing frequencies, because of changes in opacity, resulting in apparent sizes $<1.5R_\star$ for $\gtrsim 200$~GHz \citep{Vlemmings2019,Bojnordi-Arbab2024}.}
Prior to the present work, a few studies have examined the angular sizes of AGB stars at sub-mm, mm and longer radio wavelengths. \edits{Pertinent results of such studies are summarised in Table~\ref{tab:litUD}.} To ensure a valid comparison \edits{between our measurements of angular sizes and those of earlier studies}, we focus on the observations closest to the central frequency of our Band 6 continua (241.75~GHz or 1.24~mm). \cite{Vlemmings2019} studied four nearby oxygen-rich AGB stars (W~Hya, $o$~Cet (aka Mira~A), R~Dor, and R~Leo) at high angular resolution across multiple wavelength ranges. Higher resolution observations of W~Hya \citep{Ohnaka2024a} and R~Dor \citep{Vlemmings2024a} are also available. Uniform discs were fit by the authors of these earlier observations, allowing us to directly compare the stellar properties to our sample.

Of these four stars, R~Dor is a SRb type variable, while W~Hya, $o$~Cet and R~Leo are all Mira variables. Situating them in our plot of monochromatic luminosity against UD radius in Fig.~\ref{fig:radflux}, we find that R~Dor sits in the SR quadrant of the plot, with a lower flux and smaller radius than the Mira variables. The three Mira stars lie in the Mira quadrant, upholding the divisions we found. 
We also considered the pulsation periods of these stars and comparing with our plots of UD properties against period, \edits{extending Fig.~\ref{fig:pulsations} in Fig.~\ref{fig:pulsations-lit}}. We find that the Mira variables cluster close to the Miras in our sample (with periods of 402 days, 332 days and 310 days for W~Hya, $o$~Cet, and R~Leo, respectively). R~Dor has two periods of comparable length to T~Mic \citep[332 and 175 days for R~Dor,][and 352 and 178 days for T~Mic, Table~\ref{tab:stars}]{Bedding1998a}, so sits near T Mic and S Pav on the plot of UD radius against period if we consider the longer period as the primary.

\cite{Asaki2023} present high angular resolution continuum images of the carbon-rich Mira variable R Lep, observed with ALMA Bands 8, 9, and 10. These are higher frequencies than our observations, but if we compare their Band 8 results (at 405.0~GHz or 0.74~mm) with our data, we find that R~Lep sits in the Mira quadrant of our Fig.~\ref{fig:radflux}, with a UD radius of $5\e{13}$~cm and monochromatic luminosity of $1\e{8}$~mJy~pc$^{2}$. The parameters of CW~Leo, another carbon-rich Mira, measured by \cite{Velilla-Prieto2023} are also comparable to R~Lep. However, those authors fit a Gaussian rather than a UD to their high angular resolution continuum image (observed with ALMA Band 6), so their results are not directly comparable to ours, especially when it comes to the measured radius.

\edits{Overall, these past studies agree very well with the trends between radius and period that we see in our sample. Aside from CW Leo, which is most likely an outlier because its size was determined from a Gaussian fit, the Mira variables lie near our trend line (Equation~\ref{eq:P-rad}), within the uncertainties. Furthermore, when considering the longer period of R~Dor, described as Mira-like by \cite{Bedding1998a}, it also agrees with the trend line (but we note that its smaller radius makes it more typical of the semiregular variables). When using the shorter period for R~Dor, the star falls among the semiregular variables in the period-radius plot. There is also good agreement between the literature data and the general trends seen between period and monochromatic luminosity for the ATOMIUM sample. In this case, the main outlier is R~Lep, having a larger monochromatic luminosity than other stars with similar periods, but this is probably because the observations were from Band 8 with a higher frequency, making them less directly comparable. We also note that now CW Leo lies close to IRC+10011 and IRC$-$10519, in agreement with our Equation~\ref{eq:P-flux} line, which supports our hypothesis that the stars with the longest periods ($>500$~days) agree with this trend.}

In recent work, \cite{Vlemmings2024} measured the stellar disc of the closest AGB star, R~Dor, with ALMA and compared this diameter with the NIR diameter measured by \cite{Ohnaka2019}. For the ALMA radius at 225~GHz (1.33~mm, close to our 1.24~mm) the measurements of \cite{Vlemmings2024} and \cite{Ohnaka2019} agree well with our Eq.~\ref{eq:radii}. They report for R~Dor $R_\mathrm{UD} = 1.22\pm0.11 R_\mathrm{NIR}$, which is the same gradient as in our Eq.~\ref{eq:radii}, but \edits{passing through the origin}. 
%As mathematically expected, a line with the same slope passing through \edits{the origin} does not agree with our data as well as Eq.~\ref{eq:radii}. 
%\red{this bit has been copied to 4.3, should be removed here:}
When we fit to our data but force the line to pass through \edits{the origin}, we find a slope of 1.38. \edits{Both this result and the relation in Eq.~\ref{eq:radii} are close to being within the uncertainties of the \cite{Vlemmings2024} result for R~Dor.}
%which is a reasonable fit to the measured data. Of course our Eq.~\ref{eq:radii} must only be valid over a limited regime, since a non-zero diameter at one wavelength cannot physically correspond to a diameter of zero in another. However, in the right panel of Fig.~\ref{fig:radii} we plot observed sizes not physical sizes and the real lower limits on both axes are the resolution limits. Fitting to physical sizes is not possible for our dataset because most of the measured stars have very similar radii (see Table \ref{tab:radii} and Fig.~\ref{fig:radflux}).
%If we convert both the ALMA and NIR observations to physical units (e.g. cm or au), it becomes difficult to find a reliable fit since most of the measured stars are clustered together with similar radii (see Table \ref{tab:radii} and Fig.~\ref{fig:radflux}) and a fit to the AGB stars is dominated by IRC+10011 and IRC$-$10529, which have unreliable NIR measurements. Excluding these two sources results in a fit with high uncertainty because of the narrow radius range.

%\edits{In addition to examining the observed sizes of our stars, we tested fitting the physical sizes. This introduces additional uncertainties from the distances }
%
%\red{Fitting to physical sizes is not possible for our dataset because most of the measured stars have very similar radii (see Table \ref{tab:radii} and Fig.~\ref{fig:radflux}).}

Through an analysis of theoretical models, \cite{Bojnordi-Arbab2024} found a variation with pulsation phase of about $\pm20\%$ for the observable radii at 231~GHz of the model AGB star. This would account for the scatter in our comparison between ALMA and NIR data, especially since (i) there is no consistency in the pulsation phase at which measurements were taken and (ii) even if there was, the phase at which the measured radii peak is expected to vary with wavelength, as discussed in detail by \cite{Bojnordi-Arbab2024}. Therefore, to more precisely study the relationship between radio and NIR sizes of AGB stars, high resolution data would need to be taken contemporaneously. This is not possible to achieve with the contemporaneous SPHERE observations of \cite{Montarges2023}, as the stars are mostly smaller than their corresponding point spread functions.

 %\cite{Vlemmings2024} also measured the stellar disc at 338~GHz (ALMA Band 7, 0.89~mm) and found that the disc was 3\% smaller at this wavelength. This value also lies within our uncertainties when paired with the NIR measurement by \cite{Ohnaka2019}.
%($D_{225} /D_{338} = 1.03$)

%Mention comparison done in \cite{Zhao-Geisler2012}

\subsection{Two modes of dust formation: pulsation- and binary-enhanced}\label{sec:dustform}
%\subsubsection{Extended dust distributions}\label{sec:extdust}

We \edits{propose a second} pathway for shock-driven dust formation. \edits{In addition to the well established pathway of} stellar pulsations, \edits{we propose that} the supersonic motion of companion stars through the circumstellar envelope \edits{enhances dust production}. All stars in our sample exhibit some dust in the close stellar environment (even U~Del, \edits{on the basis of its long secondary period} see Sect.~\ref{sec:udeldis}). However, not all stars exhibit regions of continuum flux \edits{offset from the continuum peak}. Excluding RW Sco\footnote{RW Sco is moderately distant (560 pc) and, in general, has a lower signal to noise than the majority of our sample, so it is possible that our observations were not sensitive enough to detect or resolve regions of extended flux.}, all the stars with periods longer than 300 days have some \edits{resolved} regions of continuum flux. For periods shorter than 300 days, only two unusual stars have notable regions of \edits{resolved} flux: \pigru\ and SV~Aqr, discussed above in Sections \ref{sec:pigrudis} and \ref{sec:svaqrdis}, respectively. In both of these cases the dust emission \edits{offset from the continuum peak} can be attributed to a known (\pigru) or suspected (SV~Aqr, see Appendix~\ref{app:periods}) companion.

\subsubsection{Pulsation-enhanced dust formation}

\begin{figure*}
\centering
	\includegraphics[width=0.49\textwidth]{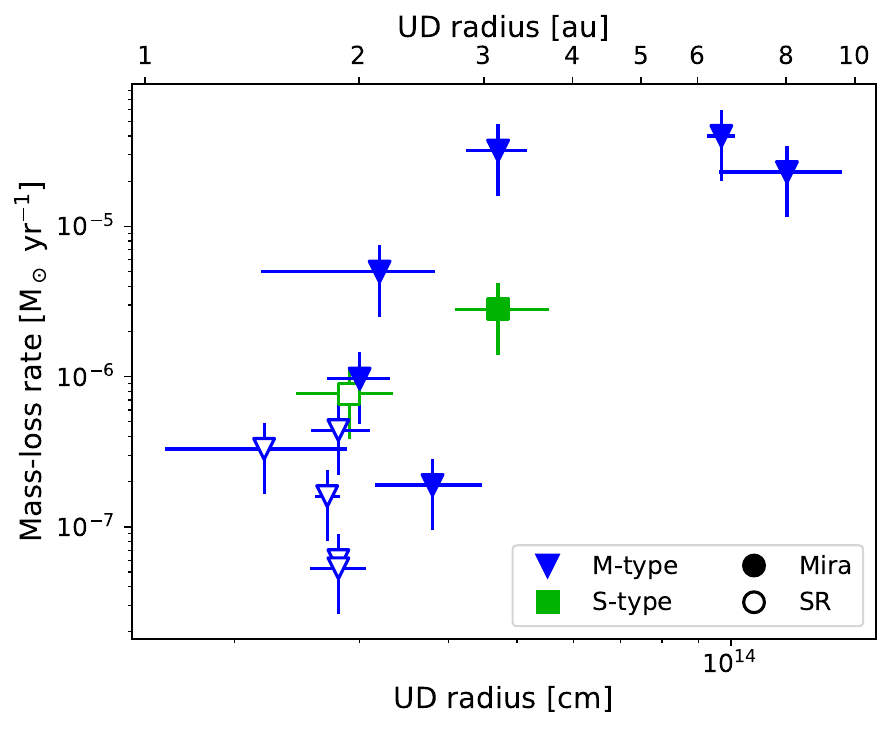}
	\includegraphics[width=0.49\textwidth]{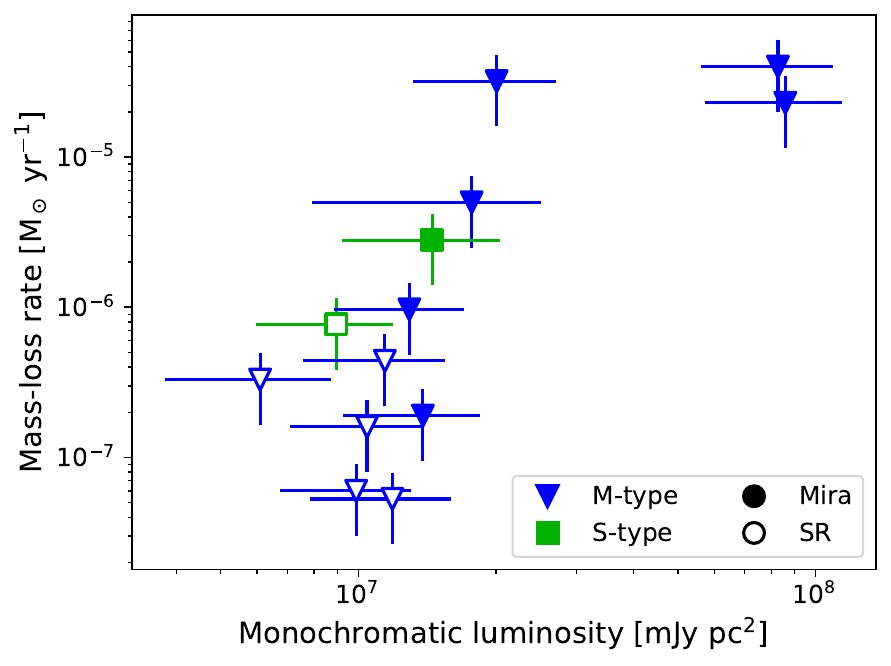}
    \caption{\edits{Mass-loss rates plotted against physical sizes (\textit{left}), as calculated from the UD fits, and monochromatic luminosities (\textit{right}), for the AGB stars. The vertical error bars indicate assumed 50\% uncertainties on the mass-loss rates.}}
    \label{fig:UDvsMLR}
\end{figure*}

Pulsation periods of $\gtrsim 300$ days are thought to be necessary for an AGB star to fully develop a pulsation-enhanced dust-driven wind \citep{Winters2000,McDonald2016}, a hypothesis supported by observations of both oxygen-rich and carbon-rich stars, where a positive correlation between mass-loss rate and period is seen above 300 days \citep{Groenewegen1998a,Olofsson2002,Ramstedt2009,McDonald2016}. 
%In this context, we should expect to see more significant mass-loss for stars with periods above 300 days. 
This is also in agreement with the relation found by \cite{Wood1990} for an exponential increase in mass-loss rate with pulsation period\footnote{The relation found by \cite{Wood1990} is $\log \dot{M} = -11.4 + 0.0125 P$}, mainly holding between 300 and 500 day periods, above which the mass-loss rate no longer increases with period \citep{Vassiliadis1993}.

To check whether such a correlation holds for the ATOMIUM sample, we collected mass-loss rates from the literature, as described in Appendix \ref{app:massloss} and listed in Table~\ref{tab:massloss}. Since the ATOMIUM sample populates a period--mass-loss rate diagram too sparsely to draw conclusions, we supplement the data with the additional 37 stars from the SUCCESS sample, including carbon stars \citep{Danilovich2015a}. We plot the mass-loss rates against pulsation periods for the combined ATOMIUM and SUCCESS samples in Fig.~\ref{fig:period-ml}. We also plot the \cite{Wood1990} relation for periods between 300 and 500 days (brown dashed line). The ATOMIUM stars with periods above 300 days agree with the underlying trend seen in the SUCCESS sample. Below 300 days, there is no trend. Above 500 days there may be a bimodal trend with one mode including most of the carbon stars and the longest-period S-type star, and the other mode including high mass-loss rate oxygen-rich stars and the carbon star II Lup.

\cite{McDonald2016} also noted that stars with pulsation periods above 500 days were the most extreme stars in the sample and included RSGs and known binaries. When considering the period--radius and period--monochromatic luminosity plots in Fig.~\ref{fig:pulsations}, we see an uptick in radius among the Miras for periods above 300 days, possibly suggesting that longer pulsation periods correlate to larger radii as well as increased dust production for this period range. However, the uptick in monochromatic luminosity is only seen above $\sim 500$ day periods, suggesting that it is only these most extreme stars which have higher mm luminosities. This could possibly be the result of the UD fits to the 5 extreme stars (two extreme and distant AGB stars and our RSG subsample) including more circumstellar dust than the closer stars in our sample. This is partly reflected in the rather low brightness temperatures obtained for IRC$-$10529 and AH~Sco. \cite{Kiss2006} also noted that VX Sgr (and to a lesser extent AH Sco\footnote{Note also that the \cite{Kiss2006} sample did not include KW~Sgr.}) has a particularly large pulsation amplitude, which they associated with increased pulsation-enhanced mass loss and dust production. This fits with the finding by \cite{Freytag2023}, based on their 3D theoretical models, that dust forms in the wake of shocks within a few stellar radii of the star; larger-amplitude pulsations would produce stronger shocks.
All three RSGs in ATOMIUM agree well with a trend with pulsation period, even though KW~Sgr appears to be the least dusty of the three \edits{(see estimated dust mass in Table~\ref{tab:specind})}. IRC$-$10529 and IRC+10011 are also known to be extremely dusty \citep{Justtanont2013,Reiter2015}, which fits with the ``extreme'' label noted by \cite{McDonald2016}. From these observations we can hypothesise that for stars with periods above 300 days we see an increase in both mass-loss rate and mm radius, while for stars with periods above 500 days we see an increase in monochromatic luminosity and another increase in mass-loss rate.

\edits{To check this hypothesis, we also plotted mass-loss rates for the AGB stars against each of UD radius and monochromatic luminosity, as shown in Fig.~\ref{fig:UDvsMLR}. In both plots, the two more extreme sources, IRC$-$10529 and IRC+10011, sit apart from the rest of the sample, having the largest UD radius and monochromatic luminosity. For the other stars, there is a trend of increasing mass-loss rate with increasing UD radius and increasing monochromatic luminosity. This suggests that a maximum mass-loss rate has been reached by IRC$-$10529 and IRC+10011 (and perhaps GY~Aql, see Table~\ref{tab:massloss}), and further increases in UD radius, monochromatic luminosity and period (Fig.~\ref{fig:period-ml}) do not contribute to increases in mass-loss rate. Considering the semi-regular variables only, there is no correlation between the mass-loss rate or either of the UD radius or monochromatic luminosity.}

\subsubsection{Binary-enhanced dust formation}

\pigru\ and R~Aql have the two highest mass-loss rates for stars with periods below 300 days (Table~\ref{tab:massloss}). For R Aql this can be explained by its declining period \citep{Zhao-Geisler2012}; in 1915 the period was $\sim320$ days, placing it within the trend of the other longer-period stars. The mass-loss rate calculated from CO lines is an average over several hundred years so will not have altered appreciably in that time. For \pigru, the \edits{gas} mass-loss rate calculated by \cite{Doan2017} takes the complex circumstellar structure into account. The high mass-loss rate, compared with other stars with pulsation periods around 200~days, can be explained by the extra dust produced in the wake of the companion, which orbits with a semi-major axis of $\sim 7$~au (approx. 3--4 stellar radii, \citealt{Montarges2025,Esseldeurs2025}). Once more dust is formed,  radiation pressure from the AGB star will push it outwards with the rest of the dust. If it is within the wind acceleration region, this additional dust can enhance the mass loss as part of the dust-driven wind, since there will be more dust available to collide with and drag the gas outwards. 

The 3D single star models of \cite{Freytag2023} show that some gas still falls back onto the star out to radii of $\sim 9$~au, so additional dust produced within 9~au could contribute to enhanced mass-loss. This phenomenon could explain moderate mass-loss rates for the majority of surveyed SR stars with shorter periods (<300 days). Because the shocks induced by the companion persist further out in the wind (e.g. in spiral patterns, see Appendix \ref{app:hydro} and e.g. \citealt{Kim2017,Kim2019}, \citealt{Malfait2021}), some dust will also form well outside of the wind acceleration region (see the middle panel of Fig.~\ref{fig:hydro}), and will increase the overall dust-to-gas ratio compared with that of an otherwise similar single star. This could also contribute to the over-estimation of mass-loss rates calculated from dust properties and assumed dust-to-gas ratios, and could explain the discrepancy between mass-loss rates calculated from dust and from CO for the dustiest stars \citep[e.g. see comparisons in Table 4 of][where mass-loss rates from the two methods differ by around an order of magnitude]{Justtanont2006}.

Many surveys, including both ATOMIUM and SUCCESS, are biased towards stars that have previously been detected (strongly) in CO, meaning they are likely to have higher mass-loss rates. If single stars with short pulsation periods have low levels of dust production and low mass-loss rates \citep[e.g.][]{McDonald2018} they would be less likely to be included in biased surveys. This explains the apparently high prevalence of binary AGB stars judging by their CO morphologies in high resolution ALMA observations \edits{(see also Sect.~\ref{sec:other-binary})}. An unbiased survey such as the Nearby Evolved Star Survey \citep[NESS,][]{Scicluna2022} is hence likely to include stars with much lower mass-loss rates than stars that have been more frequently studied for their brighter emission lines. Another example is the small fraction of dusty field stars with luminosities below the RGB tip \citep{McDonald2012,McDonald2017}. These have been assumed to be the lowest-mass AGB stars, since they have similar luminosities to the lowest-luminosity dusty AGB stars in globular clusters \citep{Boyer2009,McDonald2009,McDonald2011}. However, these could instead be stars that have dust production triggered by companions. The corollary to this would be that the AGB stars with the lowest mass-loss rates (perhaps $<1\e{-8}\spy$ or lower) are least likely to have close companions.

This principle of two dust production channels can also partly explain the scatter in the period--mass-loss rate diagram and perhaps the bimodal distribution for stars with periods longer than $\sim 500$ days. If dust production and hence mass loss only depended on pulsation period, then we would expect to see minimal scatter beyond that caused by different types of dust forming around the different chemical types (oxygen-rich, carbon-rich, S-type). Companions driving the formation of additional circumstellar dust may increase the dust production by different amounts depending on orbital configuration. For example, the companion to W~Aql seems to have produced a relatively small trail of dust $\sim200$~au from the AGB star, while the closer companion to \pigru, at $\sim 6.5$~au \citep{Montarges2025} has produced a longer dust tail in its wake; compare Figures \ref{fig:waql} and \ref{fig:pi1gru}. Hence we can theorise that although the present contribution of W~Aql B to the total dust budget is low, the companion will have temporarily driven dust formation more vigorously during its periastron passage, in addition to driving chemical diversity \citep{Danilovich2024}. 

%Another interesting example is the dusty but relatively low mass-loss rate AGB star L$_2$~Pup (the lowest mass-loss rate source in Fig.~\ref{fig:period-ml}). \cite{Kervella2016} found a low mass of $\sim12~M_\mathrm{Jup}$ for the companion, which is embedded in a dusty, nearly edge-on disc \citep{Kervella2015,Ohnaka2015}. The dust around \ltpup\ is not distributed spherically, suggesting that the companion has strongly influenced its distribution. However, with a low mass, the gravitational potential of the companion is low, and it will cause weaker shocks as it orbits the AGB star \citep{Maes2021}, contributing to less dust formation than if it were more massive. This explains why the mass-loss rate of \ltpup\ remains low ($2\e{-8}\spy$, \citealt{Danilovich2015a} and Appendix \ref{app:massloss}).

At $1~\msol$, around half of stars are expected to have companions with masses at least 10\% that of the primary \citep{Moe2017}. The proportion of stars with such companions increases to over 100\% for stars of $8~\msol$, meaning that some such stars are expected to be in triple systems. Close companions are more likely for higher-mass stars \citep{Moe2017}. However, the closer the companion, the more intense the shockwaves generated as it travels through the wind and hence the more dust formation triggered, in addition to the pulsation-driven dust formation. \edits{This additional dust can enshroud the companion, making it more difficult to detect with optical telescopes.}
W~Aql B suffers from $\sim2$~mag of extinction $\sim200$~au from the AGB star \citep{Danilovich2015} but can still be detected with Hubble and SPHERE \citep{Ramstedt2011,Montarges2023}. The close companion to \pigru\ has only been detected by the dust associated with it in the ATOMIUM continuum (Fig.~\ref{fig:pi1gru} and \citealt{Homan2020}) and in scattered polarised light \citep{Montarges2023,Montarges2025}. The optical light (whether the companion is a white dwarf or a main sequence star) has been completely attenuated by the dust. 

\pigru\ is the second closest star in the ATOMIUM sample and the companion is only just resolved from the AGB continuum in the extended data. If the system were a little further away, or oriented less favourably, or the projected separation between the two stars a little smaller, then we would not be able to distinguish the companion from the AGB continuum flux. Therefore, we cannot rule out the presence of companions for the majority of the ATOMIUM sample just because we have not (yet) directly observed them. Unseen companions could \edits{partly} explain the extended dust distributions seen for U~Her, R~Hya, and S~Pav, as well as the less symmetric emission seen for GY~Aql, \edits{R~Aql}, IRC$-$10529, SV~Aqr, T~Mic and IRC+10011. 

\subsection{Other indications of binarity}\label{sec:other-binary}

Indirect evidence of AGB binarity has previously been found in resolved CO emission, including from the ATOMIUM data \citep{Decin2020} and earlier studies \edits{of other stars} such as \edits{R~Scl} \citep{Maercker2012}, \edits{CW~Leo} \citep{Cernicharo2015a} and \edits{AFGL~3068} \citep{Kim2017}. Such structures can be reproduced --- or partially reproduced --- through hydrodynamic modelling \citep[e.g.][]{Mastrodemos1998,Mastrodemos1999,Kim2019,El-Mellah2020,Malfait2021}. Molecular signatures can also be used to indirectly infer the presence of UV-emitting binaries \citep{Van-de-Sande2022,Siebert2022,Danilovich2024}. 
Studies surveying AGB stars in the UV or X-ray have detected emission for some stars \citep[e.g.][]{Ramstedt2012a,Sahai2015,Ortiz2021,Schmitt2024} which ought not to come from the AGB star itself. This emission is thought to arise from companion stars, most likely from accretion discs that have formed around the companions \citep{Sahai2015}. None of the stars in the ATOMIUM sample are included in the aforementioned studies as confirmed detections\footnote{\cite{Schmitt2024} include three ATOMIUM stars (\pigru, R~Hya and S~Pav) in their list of surveyed stars, but note optical contamination, which cannot be disentangled from potential X-ray detections, \edits{making their results for these stars ambiguous}.}.

Out of the 14 AGB stars in the ATOMIUM sample, we have suggested that 11 of them might have binary companions, 9 of which have not been directly detected. Given that the ATOMIUM sample was selected to cover a range of mass-loss rates and pulsation behaviours \citep{Gottlieb2022}, there was no particular expectation for the stars to have binary companions. 
%W~Aql, \pigru\ and R~Hya were already know to have wide binary companions (see Sect.~\ref{sec:sample}) and \pigru\ has been found to be a triple system, which was suspected prior to the ATOMIUM observations (e.g. \citealt{Chiu2006}). 
%We can also add R~Hya as a star with a known companion on a wide orbit (Gaia DR3 6195030801634430336, but we will refer to this star as R~Hya~B, confirmed as a companion by \citealt{Kervella2022} and found at a projected separation of 21.1\arcsec\ or 2660~au from the AGB component R~Hya~A).
Population statistics indicate that at least half of stars in the AGB mass range should have companions \citep{Moe2017}, so earlier assumptions of the majority of AGB stars being single stars (i.e. before the advent of resolved imaging of the circumstellar environments) are statistically unlikely.

Recent data and developments from Gaia have allowed us to better constrain distances and infer the presence of companions \citep[including some which can be resolved][]{Kervella2022}. Based on the results of \cite{Kervella2022}, \cite{Montarges2023} used a proper motion analysis combining \textsl{Gaia} Early Data Release 3 and \textsl{Hipparcos} data to estimate the probability of detecting close companions around the AGB stars. Note that the proper motion anomaly is not sensitive to wide binaries.
They determine that GY Aql, R Aql, \pigru, and R~Hya, are likely to have a detectable close companion based on the large S/N of the proper motion anomaly, \edits{in agreement} with our predictions.

\section{Summary and conclusions}\label{sec:concl}

We present the continuum data observed for a sample of 14 AGB and 3 RSG stars as part of the ATOMIUM ALMA Large Programme. For each star, we show the continuum observations from each individual array configuration as well as the combined image. Different features are apparent on different scales, owing to resolution, maximum recoverable scales and noise levels. We identify stars with extended continuum emission \edits{on scales of 10s to 100s of au}, \editss{elongated} structures and a few unique structures. Notably, we see a tail of dust in the wake of the close companion to \pigru\ and some dust forming in the wake of the distant companion to W~Aql.
%Out of 14 AGB stars in our sample, we see some evidence of binarity in the continuum images of 11 of them.
In addition to the main ALMA array data, we present new observations taken at lower frequencies and lower resolution with the ACA, for a subsample of 7 AGB stars. 
%For these stars we calculate spectral indices, which range from 1.5--2.5, in agreement with theoretical expectations.

For each star, we fit uniform discs to the \edits{compact emission at the continuum peaks}, \edits{which is dominated by the stellar flux but, for some stars, is likely to include contributions from dust located close to the stellar surface.} We present residual images after subtracting the uniform discs from the combined continuum maps, \edits{many of which better highlight the emission offset from the continuum peaks.}

We compare the uniform disc sizes and flux densities with stellar properties such as pulsation periods \edits{and mass-loss rates}. Among the AGB stars, we find that all the semiregular variables in our sample have lower monochromatic luminosities and smaller UD radii than the Mira variables. As expected, the RSGs have even larger radii and higher monochromatic luminosities than the Mira variables. \edits{There is an overall correlation between the UD radii and monochromatic luminosities, as is expected from the general relationship between luminosity and radius.}

Comparing our newly derived stellar properties with literature data, we find a linear relationship between ALMA stellar diameters at 1.24~mm and diameters obtained from near infrared observations. We find that there is no trend among the semiregular variables between their pulsation periods and either radius or monochromatic luminosity. For the Mira variables, there is a positive trend between period and radius above 300 days, which matches a previously found uptick in dust production for periods above 300 days. There is no clear trend between monochromatic luminosity and period except for the two AGB stars with the highest fluxes, which agreed with a positive trend seen for the RSGs. This may correspond to extreme dust production being seen for stars with pulsation periods above 500 days. 

%Collecting mass-loss rates from the literature for the ATOMIUM sample, we find that there is no correlation between pulsation period and mass-loss rate for periods below 300 days, and a positive trend above 300 days, as has been previously seen. 
Comparing the ATOMIUM sample to the larger SUCCESS survey, we find that among semiregular stars with low periods, \pigru\ has an especially large mass-loss rate, and conclude this is likely because of the additional dust forming in the wake of the shocks generated by the passage of its close companion. Hence, we propose two channels for dust formation around AGB stars: in the wake of shocks caused by pulsations and in the wake of shocks caused by the supersonic passage of a companion through the circumstellar envelope.
Consequently, we propose that dust emission at mm wavelengths, \edits{offset from the continuum peak and showing coherent structure, e.g.~reminiscent of a wake or shock, could be} an indirect indicator of potential binarity. \edits{Deeper observations, which can detect low surface-brightness features, for more stars with known binary companions are needed to confirm this hypothesis.}

\section*{Data availability}
\edits{The observational data used here are openly available through the
ALMA Science Archive: \url{https://almascience.nrao.edu/aq/}. The images, other data products and scripts for the standard data processing are available from the ATOMIUM Landing Page in the ALMA Science Archive: \url{https://almascience.eso.org/alma-data/lp/atomium}. The fully aligned images and other custom ALMA data products that were produced for this study are available at: \url{doi.org/10.5281/zenodo.17015654}.}

%%%%%%%%%%%%%%%%%%%%%%%%%%%%%%%%%%%%%%%%%%%%%%%%%%%%%%%%%%%%%%
\begin{acknowledgements}

%\blue{Acknowledge Alex Wallace for useful discussions regarding binarity if he's not an author?}
We thank the anonymous referee for their careful reading and thoughtful comments on the manuscript.
TD and NS are supported in part by the Australian Research Council through a Discovery Early Career Researcher Award (DE230100183).
TC acknowledges funding from the Research Foundation - Flanders (FWO), grant 1166724N. 
LD and CL acknowledge support from the KU Leuven C1 excellence grant BRAVE C16/23/009. LD acknowledges support from KU Leuven Methusalem grant SOUL METH/24/012, and the FWO research grants G099720N and G0B3823N.
FDC is a Postdoctoral Research Fellow of the Research Foundation - Flanders (FWO), grant 1253223N.
IEM acknowledges support from ANID/FONDECYT, grant 11240206
ME acknowledges funding from the FWO research grants G099720N and G0B3823N. 
TJM acknowledges support from STFC grant no. ST/T000198/1
HSPM thanks the Deutsche Forschungsgemeinschaft (DFG) for support through the collaborative research center SFB~1601 (project ID 500700252) subprojects A4 and Inf.
JMCP was supported by STFC grant number ST/T000287/1.
DP acknowledges Australian Research Council funding via DP220103767. We also thank the Australia French Association for Research and Innovation (AFRAN) for financially supporting the 5th Phantom users workshop.
RS's contribution to the research described in this publication was carried out at the Jet Propulsion Laboratory, California Institute of Technology, under a contract with NASA. RS thanks NASA for financial support via GALEX GO and ADAP awards.
LS is an FNRS senior researcher.
MVdS acknowledges the Oort Fellowship at Leiden Observatory.
OV acknowledges funding from the Research Foundation - Flanders (FWO), grant 1173025N.
KTW acknowledges support from the European Research Council (ERC) under the European Union's Horizon 2020 Research and Innovation programme (grant agreement number 883867, project EXWINGS).
This work is funded by the French National Research Agency (ANR) project PEPPER (ANR-20-CE31- 0002).
We thank the Australian-French Association for Research and Innovation (AFRAN) for financial support.
The research leading to these results has received funding from the European Research Council (ERC) under the European Union's Horizon 2020 research and innovation program (project UniverScale, grant agreement 951549). 
This paper makes use of the following ALMA data: ADS/JAO.ALMA\#2018.1.00659.L and 2019.1.00187.S. ALMA is a partnership of ESO (representing its member states), NSF (USA) and NINS (Japan), together with NRC (Canada), MOST and ASIAA (Taiwan), and KASI (Republic of Korea), in cooperation with the Republic of Chile. The Joint ALMA Observatory is operated by ESO, AUI/NRAO and NAOJ.
We acknowledge excellent support from the UK ALMA Regional Centre (UK ARC), which is hosted by the Jodrell Bank Centre for Astrophysics (JBCA) at the University of Manchester. The UK ARC Node is supported by STFC Grant ST/P000827/1.
Part of this work was performed on the OzSTAR national facility at Swinburne University of Technology. The OzSTAR program receives funding in part from the Astronomy National Collaborative Research Infrastructure Strategy (NCRIS) allocation provided by the Australian Government, and from the Victorian Higher Education State Investment Fund (VHESIF) provided by the Victorian Government.
This research has made use of the SIMBAD database and the VizieR catalogue access tool, CDS, Strasbourg Astronomical Observatory, France \citep{Wenger2000,Ochsenbein2000}.
This research has made use of the International Variable Star Index (VSX) database, operated at AAVSO, Cambridge, Massachusetts, USA.
This research made use of Astropy \citep{Astropy2013,Astropy2018}, SciPy \citep{Virtanen2020}, Matplotlib \citep{Hunter2007} and NumPy \citep{Harris2020}.

\end{acknowledgements}

% Please note that we have included the references below in
% order to compile the document, but we ask you to:

\bibliographystyle{aa}
\bibliography{master} % if your bibtex file is called example.bib

% %%%%%%%%%%%%%%%%%%%%%%%%%%%%%%%%%%%%%%%%%%%%%%%%%%%%%%%%%%%%%%
% Example below of non-structurated natbib references  
% To use the v8.3 macros with this form of composition of bibliography,
% the option "bibyear" should be added to the command line
% "\documentclass[bibyear]{aa}".
% %%%%%%%%%%%%%%%%%%%%%%%%%%%%%%%%%%%%%%%%%%%%%%%%%%%%%%%%%%%%%%

% \begin{thebibliography}{}

%   \bibitem[1966]{baker} Baker, N. 1966,
%       in Stellar Evolution,
%       ed.\ R. F. Stein,\& A. G. W. Cameron
%       (Plenum, New York) 333

%    \bibitem[1988]{balluch} Balluch, M. 1988,
%       A\&A, 200, 58

%    \bibitem[1980]{cox} Cox, J. P. 1980,
%       Theory of Stellar Pulsation
%       (Princeton University Press, Princeton) 165

%    \bibitem[1969]{cox69} Cox, A. N.,\& Stewart, J. N. 1969,
%       Academia Nauk, Scientific Information 15, 1

%    \bibitem[1980]{mizuno} Mizuno H. 1980,
%       Prog. Theor. Phys., 64, 544
   
%    \bibitem[1987]{tscharnuter} Tscharnuter W. M. 1987,
%       A\&A, 188, 55
  
%    \bibitem[1992]{terlevich} Terlevich, R. 1992, in ASP Conf. Ser. 31,
%       Relationships between Active Galactic Nuclei and Starburst Galaxies,
%       ed. A. V. Filippenko, 13

%    \bibitem[1980a]{yorke80a} Yorke, H. W. 1980a,
%       A\&A, 86, 286

%    \bibitem[1997]{zheng} Zheng, W., Davidsen, A. F., Tytler, D. \& Kriss, G. A.
%       1997, preprint
% \end{thebibliography}

%%%%%%%%%%%%%%%%%%%%%%%%%%%%%%%%%%%%%%%%%%%%%%%%%%%%%%%%%%%%%%%
% Appendices must be placed after   \end{thebibliography}
% They will be placed automatically on a new page.
%%%%%%%%%%%%%%%%%%%%%%%%%%%%%%%%%%%%%%%%%%%%%%%%%%%%%%%%%%%%%%%
\begin{appendix}
%%%%%%%%%%%%%%%%%%%%%%%%%%%%%%%%%%%%%%%%%%%%%%%%%%%%%%%%%%%%%%%
% In the PDF output, floats should be placed
% under their own appendix, not before the title, nor after the
% title of the next appendix.

% In short appendices, onecolumn floats (\figure*
% or \table*) will generate a blank page.
% To prevent this behaviour, a few examples are provided here. 

% In case you have a lot of floating objects for little text and the 
% LaTeX engine moves the floats away from their context, the command
% \FloatBarrier of the “placeins” package will empty the
% float buffer and place all stored floats in the continuity.

% If you still encounter problems with wide floats placement,
% just use the onecolumn environment throughout the appendices.
%%%%%%%%%%%%%%%%%%%%%%%%%%%%%%%%%%%%%%%%%%%%%%%%%%%%%%%%%%%%%%%

\section{Further details of pulsational periods}

\subsection{Overview of pulsational periods}\label{app:periods}

%Some of the stars in our sample have two pulsation periods listed in Table~\ref{tab:stars}.
\edits{Here we clarify some details for stars which have two or more pulsation periods reported in the literature, and stars which are marked with an asterisk (*) in Table~\ref{tab:stars}.}
For the SRb variables \pigru\ and V~PsA, it is possible that previously reported additional periods (of 128 and 105 days, respectively) are in fact aliasing effects, as they are roughly at frequencies of $|f_{peak} \pm n f_{window peak}|$, where the $f_{window peak}$ is the peak frequency in the power spectrum of the survey window (roughly 1 yr). T~Mic has a main period of 352 days, and a secondary period of 178 days. 
Although this period may be from a similar window aliasing effect as mentioned above, the light curve appears to be a superposition of two periods suggesting that two pulsation modes are indeed present.
U~Del is a case of a semiregular variable with a LSP of 1163 d, with its shorter pulsation period being 120 days \citep{Speil2006,Cadmus2024}.

\begin{figure}
    \includegraphics[width=0.49\textwidth]{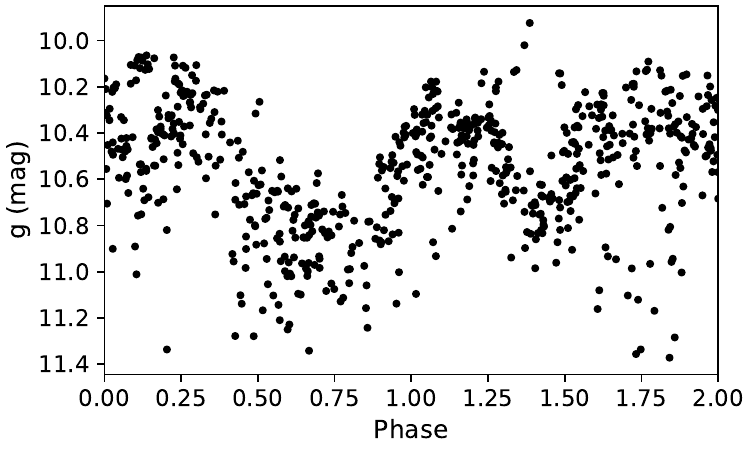}
    \caption{Phase plot of SV~Aqr showing two cycles of the longer period of 231.8 days. Data is from ASAS-SN; see text for details.}\label{fig:svaqrphase}
\end{figure}

SV~Aqr was reported in earlier ATOMIUM papers as a long period variable \citep[e.g.][]{Gottlieb2022}. It is classified as an irregular variable (LB) in VSX, with a listed period of 89 days from ASAS. From inspection of the ASAS and ASAS-SN light curves, we reclassify this star as a SRb variable owing to its relatively smaller amplitude and irregular periodicity. For the present work, we looked further into the pulsations of SV~Aqr and were able to define a primary period of 93 days, using the Lomb-Scargle periodogram \citep{VanderPlas2018}. The VSX entry also includes a GCVS note that a cycle between 200--300 days is possible. We cleaned the ASAS-SN g-band light curve to remove recent saturated photometry using simple magnitude cuts, and found a periodogram peak at 231.8 days. The phase plot in Fig.~\ref{fig:svaqrphase}, including two cycles of this period, shows structure resembling that of an eclipsing variable, with minima of alternating depth.
%Interestingly, the phased light curve of the ASAS-SN g-band light curve, plotted in Fig.~\ref{fig:svaqrphase}, shows structure resembling that of an eclipsing binary, with minima of alternating depth. 

\begin{figure}
    \includegraphics[width=0.49\textwidth]{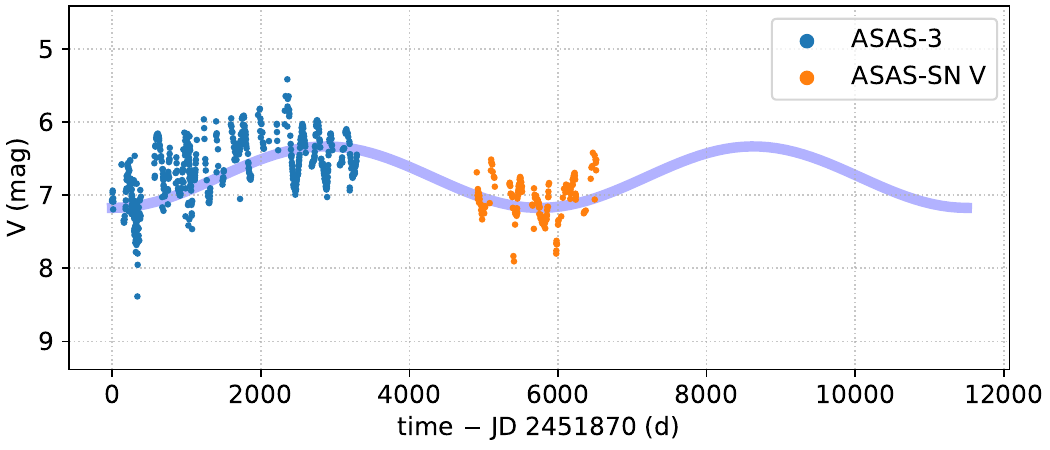}
    \caption{Phase plot of \pigru\ showing two cycles of the longer period of 5,750 days. The V-band magnitudes come from ASAS-3 (blue points) and ASAS-SN (orange points); see text for details.}\label{fig:pigruphase}
\end{figure}

After examining archival photometric data, we find evidence of a long secondary period for \pigru.
In Fig.~\ref{fig:pigruphase} we plot the combined ASAS-3 and ASAS-SN V-band light curve for \pigru, which shows an LSP of 5750 days ($\sim15.7$ years). We measured this period using the Lomb-Scargle periodogram, with a frequency grid between 10 and 10,000 days. The light blue line is a sinusoid with this period, and an amplitude calculated from the unnormalised power of the periodogram peak. We note that LSPs are not sinusoidal in the optical bands \citep[see][]{Soszynski2021}, so the line is included only as a guide. Also, the Lomb-Scargle long period (low frequency) limit was set to be beyond the total light curve observation time, which also adds to the uncertainty of the period. A similar period can be found in the AAVSO data, though it is difficult to see directly due to lower photometric precision. \cite{Montarges2025} and Esseldeurs et al (subm.) found orbital periods of around 11--12 years for \pigru, shorter than the LSP we report. We checked these shorter periods against the ASAS-3 and ASAS-SN data and found that they were not in good agreement with the observations.

The Mira variables R~Hya, R~Aql and W~Aql, and the SRa variable S~Pav all have been reported to have long-term period changes. R~Aql and R~Hya have been reported to have shortening periods \citep{Wood1981,Greaves2000,Joyce2024}, possibly owing to undergoing thermal pulses in the recent past (a few hundred or so years ago). This is discussed further in Sect.~\ref{sec:tps}. Despite being classified as SRa, S~Pav is a case of a ``Mira-like'' semiregular variable, as described in \cite{Kerschbaum1996}. Its visual amplitude is in the Mira regime (> 2.5 mag), and it has periods reported between 380--390 days; a thorough investigation of whether its period is indeed varying may be useful. A careful examination of historical light curve of W~Aql (from AAVSO) reveals that it is probably a wandering Mira \citep{Templeton2005}, albeit with some unexplained reductions in peak magnitude in the 1970s and again since 2006.

 The longest period Mira variables IRC+10011 (WX~Psc) and IRC$-$10529 (V1300~Aql) did not have V-band light curves available, as they are obscured in the visual bands. Their periods are sourced from GCVS, being 660 and 675 days respectively. Both stars were also studied in the IR observations of \cite{Harvey1974}; IRC+10011 is assumed to have a period of 650 days, and IRC$-$10529 670 days. These periods were verified with recent r-band light curves from the Zwicky Transient Facility \citep{Bellm2019,Masci2019}: we found best periods of 651 days for IRC+10011, and 670 days for IRC$-$10529 using the FINK Data Science Portal periodogram tool \citep[][\url{https://fink-portal.org/ZTF22abhcrfy}]{VanderPlas2018,Moller2021}.

Finally, the three RSGs are, by definition, SRc variables, all with variability periods > 600 days. Of these, VX~Sgr and AH~Sco show high peak-to-peak amplitudes of up to 4 mag in the visual bands \citep{Kiss2006}, but often as low as 2 mag, from their historic AAVSO light curves. The period in Table~\ref{tab:stars} of 695 days for KW~Sgr is based on combined ASAS and ASAS-SN data and we note that the amplitudes of the pulsations are close to 1 mag (AAVSO). 

%We examined the V-band light curve of SV~Aqr from the All Sky Automated Survey \citep[ASAS,][]{Pojmanski2002} and found a period of 93 days (similar to the value of 89 reported in VSX). We also examined the g-band light curve from the ASAS-SN Variable Stars Database \citep{Shappee2014,Jayasinghe2019}. After removing the saturated photometry, we found a secondary period of 230 days. The phase plot in Fig.~\ref{fig:svaqrphase}, showing two cycles of this period, somewhat resembles an eclipsing variable with alternating cycles have minima of varying depths. \green{[Data in figure is from ASAS-SN?]} 
%
%\green{Yoshi to check/expand. + add a couple of sentences about the comparison with the OGLE period-luminosity diagram}

\subsection{R~Aql and R~Hya: recent thermal pulses?}\label{sec:tps}

If R~Aql and R~Hya both underwent a relatively recent (past few hundred years) thermal pulse, as suggested by \cite{Wood1981}, \cite{Greaves2000} and \cite{Zijlstra2002}, then we would perhaps expect to see some similarities in their circumstellar envelopes. 
\edits{Their continuum images are indeed qualitatively similar. Although R~Hya (Fig.~\ref{fig:rhya}) has a larger continuous region of extended flux than R~Aql (Fig.~\ref{fig:raql}), both stars have some clumpy regions of continuum flux offset from the continuum peak (e.g. up to $\gtrsim1\arcsec$), most likely originating from dust clumps.}
%This is not true of their continuum images (R~Aql: Fig.~\ref{fig:raql}; R~Hya: Fig.~\ref{fig:rhya}), where R~Aql \red{\bf has a relatively featureless continuum}, while R~Hya has a lot of extended circumstellar dust.  While it is possible that R~Aql suffers from more resolved out flux, it is a little over twice as far away, suggesting a smaller angular size (all else being equal) and hence being less prone to resolved out flux. 
The ALMA continuum observations are in general agreement with the SPHERE observations of \cite{Montarges2023}, which reveal a larger region of polarised flux for R~Hya than for R~Aql. 

The physical radii of both stars, as derived from our UD fits, and their monochromatic luminosities are close to being the same, within our uncertainties. They are also the two Mira stars with the smallest radii when comparing physical units, both from our UD fits (Table~\ref{tab:ud}) and from literature measurements taken in the NIR (Table~\ref{tab:radii}). From their study of dust in the infrared, \cite{Zhao-Geisler2012} estimated that the inner edges of the IR dust shells around these stars are $>220$~au and $>130$~au from the corresponding stars, for R~Aql and R~Hya respectively. Strikingly, R~Aql shows an increase in flux in the far infrared (in the ISO LWS spectrum, see Figure 2 of \citealt{Zhao-Geisler2012}), suggesting a shell of cool dust, as might be expected for a detached (or detaching) dust shell. The same is not seen for R~Hya \citep[see also the earlier results of][]{Decin2008a}. If R~Aql truly has a burgeoning detached shell while R~Hya --- which has a significant amount of dust close to the star --- does not, that would imply that R~Hya had a more recent thermal pulse.  This is counter to the analysis based on luminosity done by \cite{Wood1981}, which predicts that less time has passed since the thermal pulse of R~Aql than R~Hya. However, \cite{Maercker2016,Maercker2024} presented ALMA observations that demonstrate that for some of the surveyed (carbon-rich) AGB stars the shells thought to have been formed through thermal pulses (1000--8000 years ago) were followed by a moderate decrease in mass loss, rather than an abrupt decrease seen in other cases. It is possible that R~Hya has undergone a more moderate decrease in mass-loss post-thermal pulse, while R~Aql has undergone a more drastic change. \edits{The mass-loss rates calculated by \cite{Bergman2020} show a mass-loss rate for R~Aql about five times higher than for R~Hya (see Table~\ref{tab:massloss}), which could be because the historic mass loss probed by CO lines mainly captures the imprint of enhanced mass loss during the thermal pulse for R~Aql, but more of the lower non-thermal pulse generated mass loss for R~Hya.}

On larger scales than our ALMA observations, a dust shell around R~Hya was proposed by \cite{Hashimoto1998a} based on IRAS spectroscopy and photometric imaging. Higher resolution \textsl{Herschel}/PACS imaging reveals a fermata-style bow shock around R~Hya \citep{Cox2012}, which \cite{Maercker2022} modelled as a partial ellipsoid shell approximately 100\arcsec from the AGB star. This more extended dust cannot be captured in our ALMA observations, partly because much of it falls outside of our field of view, and partly because large scale structures tend to be resolved out by ALMA. Similar observations of R~Aql are not available; R~Aql was not imaged with \textsl{Herschel}/PACS.

The CO channel maps of of R~Aql and R~Hya, presented in \cite{Decin2020}, show unique patterns for both stars, dissimilar to each other and to the other stars in the ATOMIUM sample. R~Aql has a rose-like pattern, which \cite{Malfait2024a} was unable to reproduce through a hydrodynamical model of a hierarchical triple star system\footnote{\cite{Malfait2024} found that a qualitatively similar pattern seen in the earlier hydrodynamic models of \cite{El-Mellah2020} and \cite{Malfait2021} was most likely caused by periodic instabilities arising from an absence of cooling functions aside from adiabatic expansion.}. A possibility is that the rose-like structure was caused by a thermal pulse and that the irregular arc-like pattern is an imprint of the convective flows in the stellar atmosphere (compare with the gas densities in the models of \citealt{Freytag2023} and the irregularities in the otherwise thin detached shells seen by \citealt{Maercker2024}). From the angular extent of the shapes in the CO emission, the distance to the star, and the expansion velocities \citep{Decin2020,Wallstrom2024}, this would imply a thermal pulse $\sim700$--1000 years ago. 

R~Hya has at least three pairs of bipolar structures, as described in detail by \cite{Homan2021}. \cite{Decin2008a} estimated a recent decrease in mass loss by a factor of 20 for R~Hya, which they attribute to a thermal pulse period ending 220 years ago, in agreement with the timeline proposed by \cite{Wood1981}. However, it is unclear how the complicated circumstellar structure revealed by the ATOMIUM molecular lines \citep{Homan2021} and the dust in the continuum observations (Fig.~\ref{fig:rhya}) can be reconciled with a recent thermal pulse. A possible explanation could be a companion star in the inner wind, disrupting the more spherical shell that is expected to form during a thermal pulse.

%____________________________________________________________
%       Wide floats at the start of an appendix: first method
%-------------------------------------------------------------
% To prevent a blank page after the start of an appendix:
% - Switch to one \onecolumn first
% - Declare the section title
% - Declare the onecolumn float with the parameter [h!]
% - Revert to \twocolumn at the end of the section

\section{Additional tables and plots}

\begin{figure*}
\centering
	\includegraphics[width=0.32\textwidth]{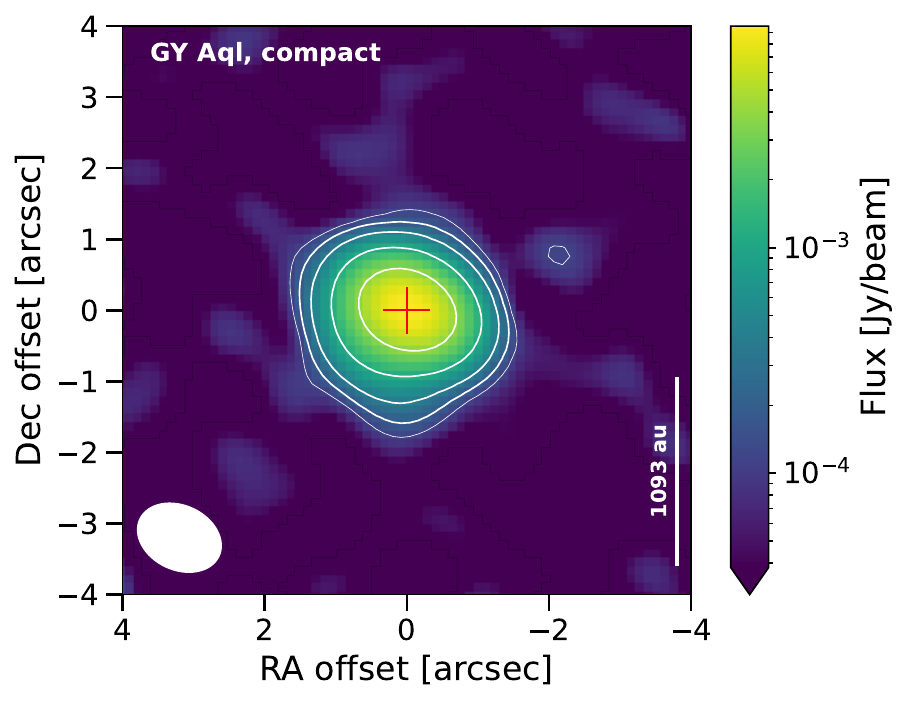}
 	\includegraphics[width=0.32\textwidth]{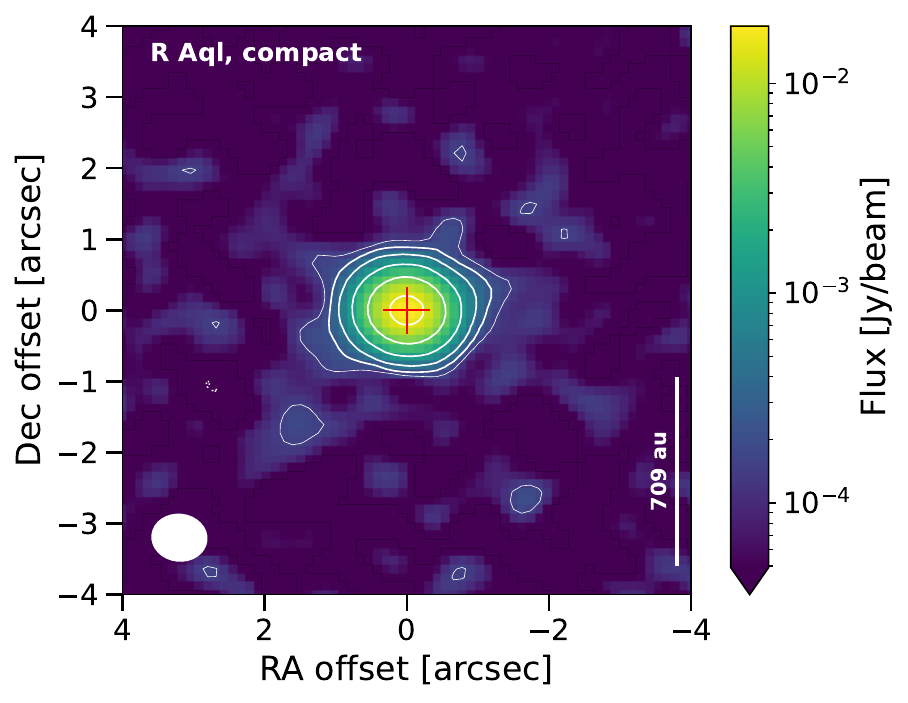}
  	\includegraphics[width=0.32\textwidth]{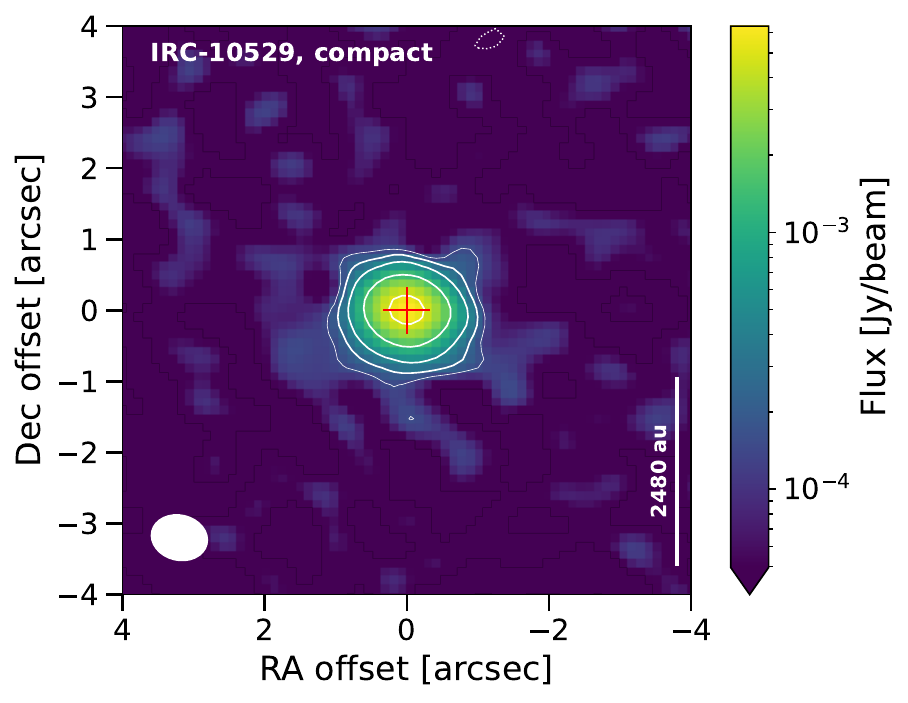}
   	\includegraphics[width=0.32\textwidth]{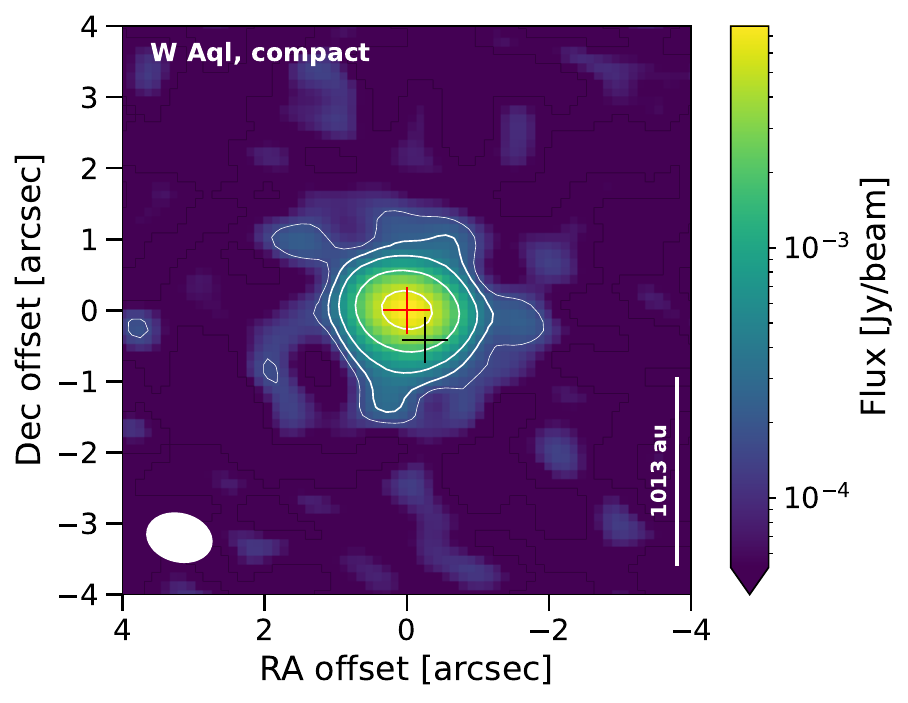}
    \includegraphics[width=0.32\textwidth]{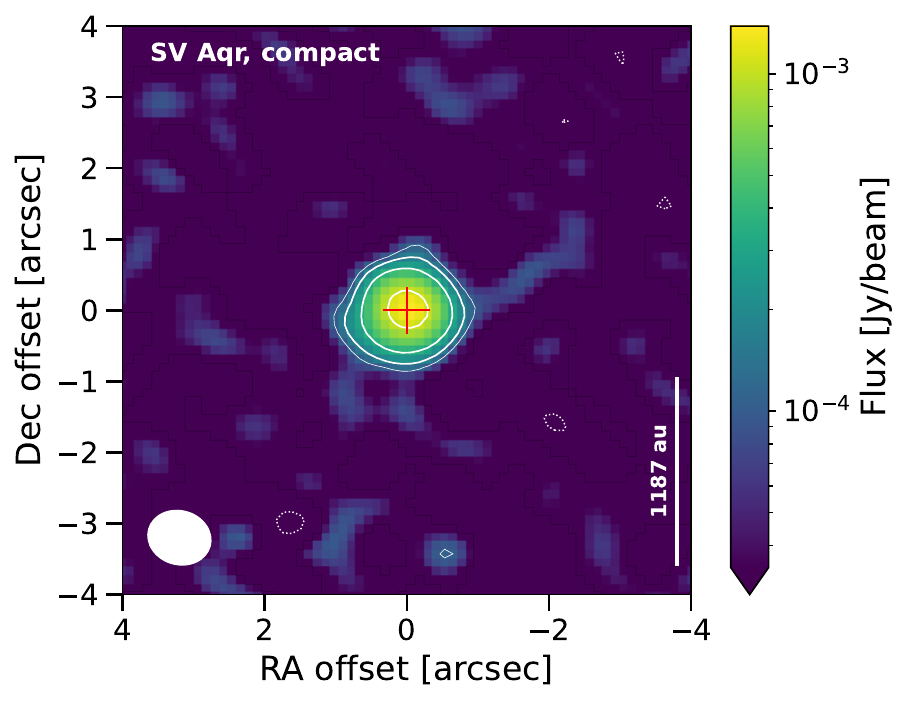}
    \includegraphics[width=0.32\textwidth]{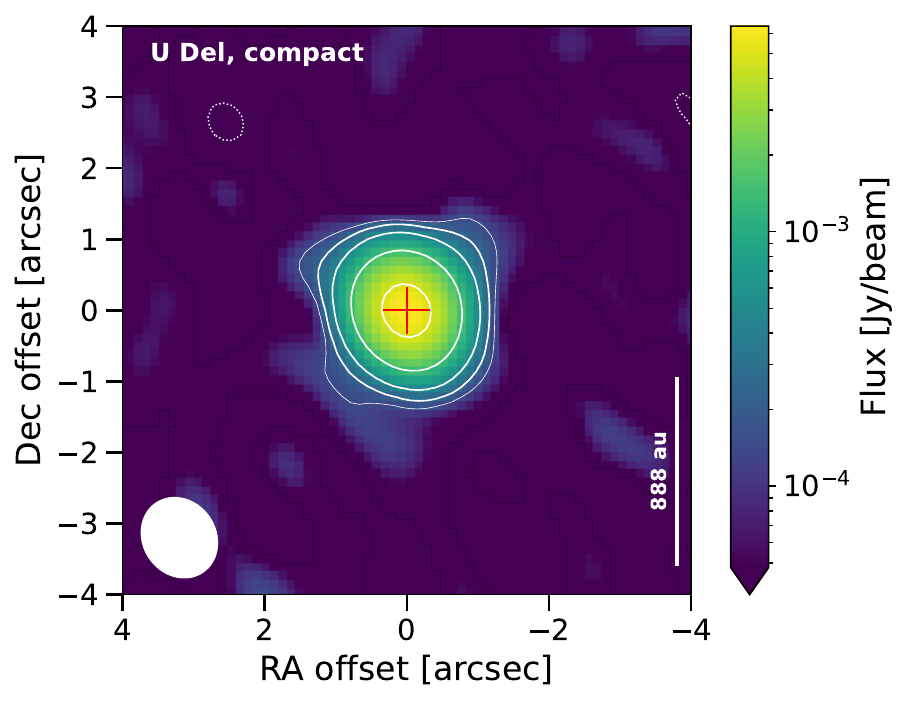}
    \includegraphics[width=0.32\textwidth]{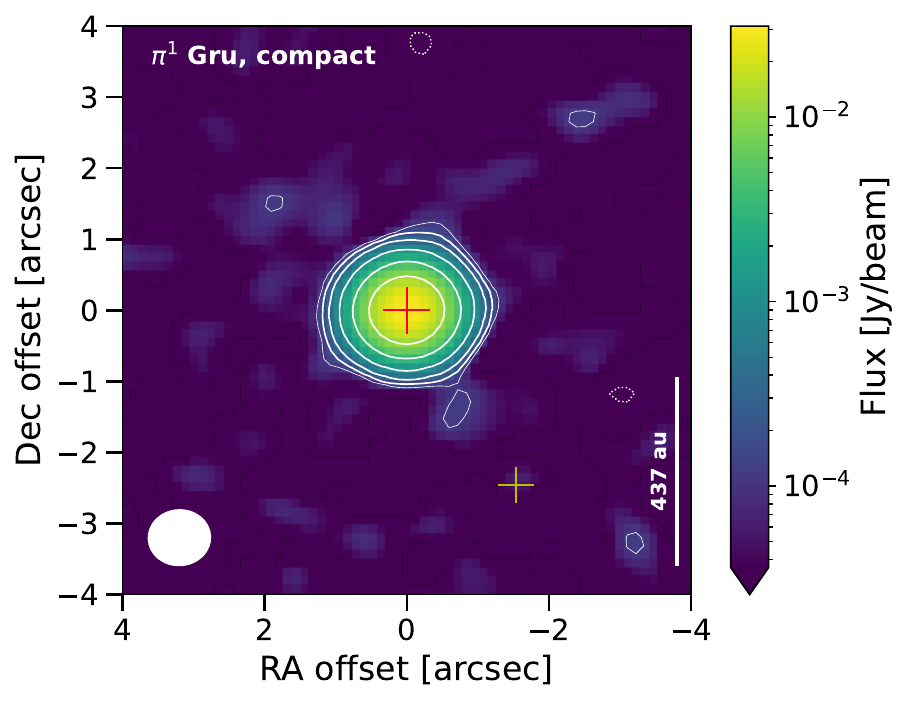}
    \includegraphics[width=0.32\textwidth]{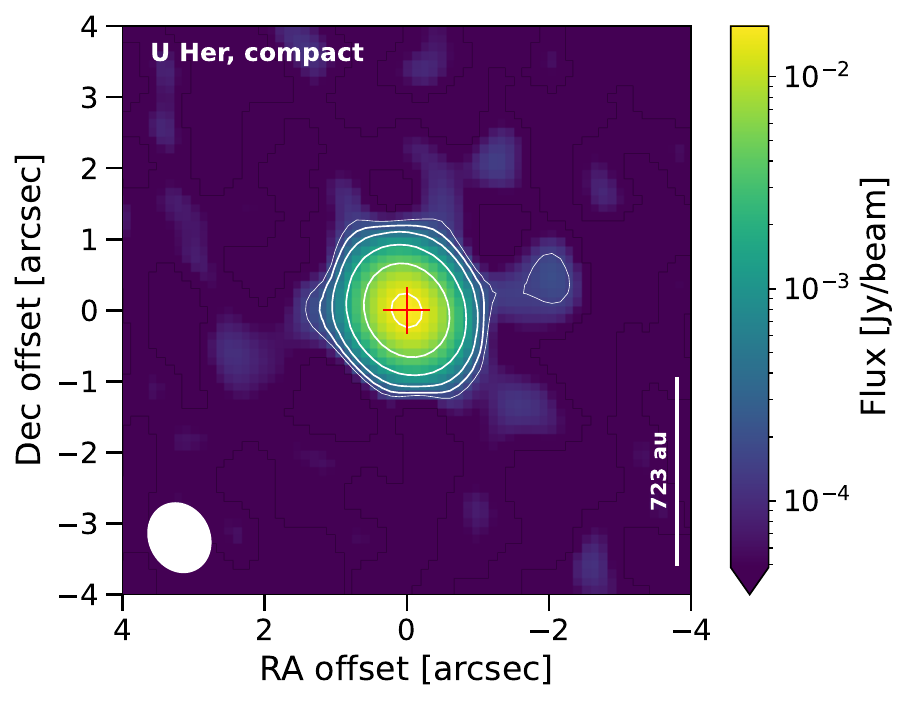}
    \includegraphics[width=0.32\textwidth]{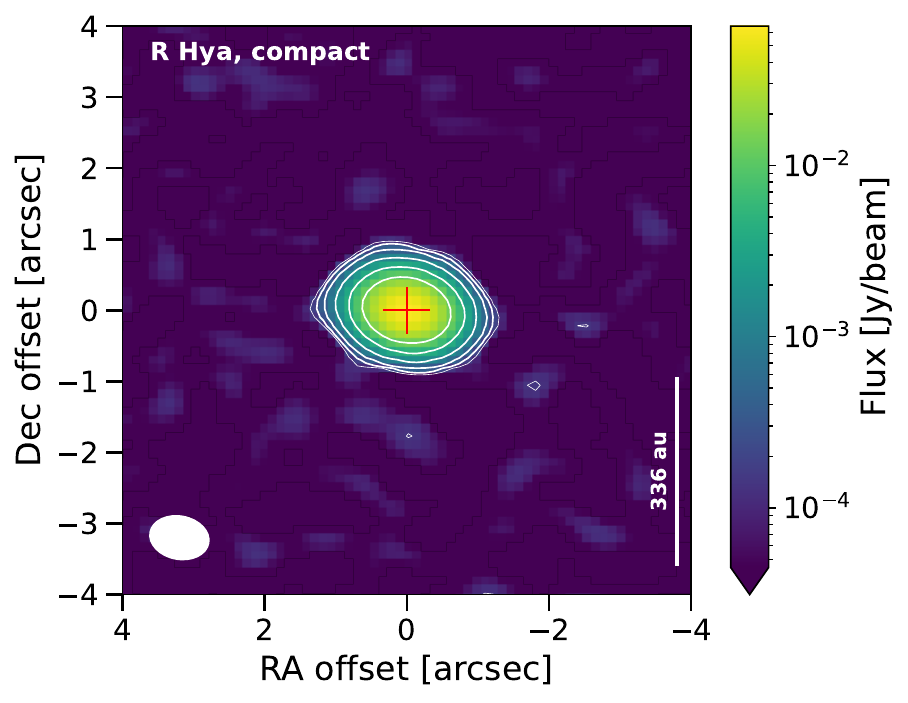}
    \includegraphics[width=0.32\textwidth]{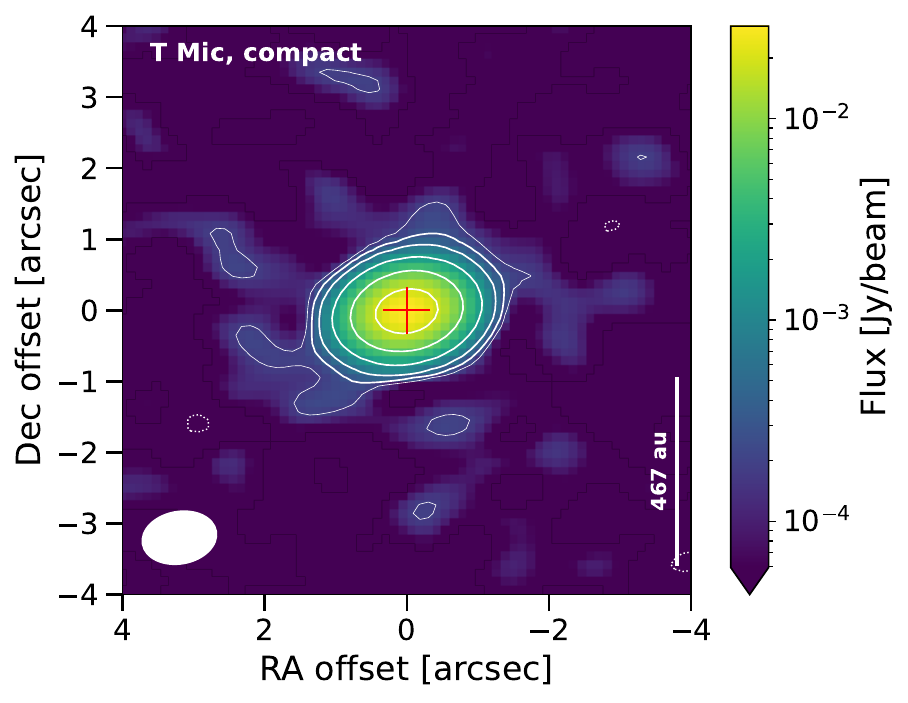}
    \includegraphics[width=0.32\textwidth]{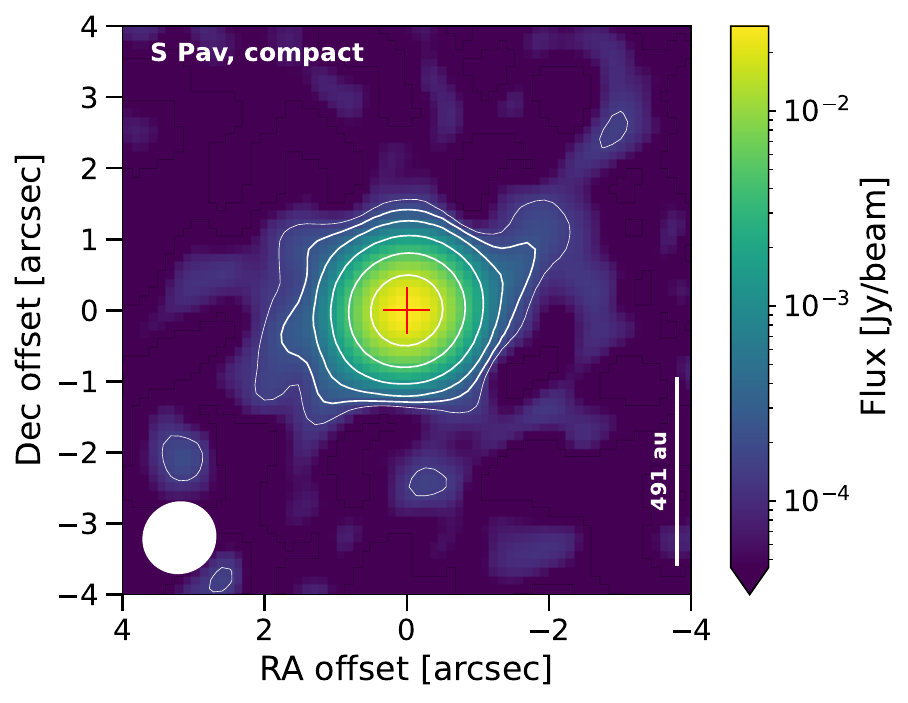}
    \includegraphics[width=0.32\textwidth]{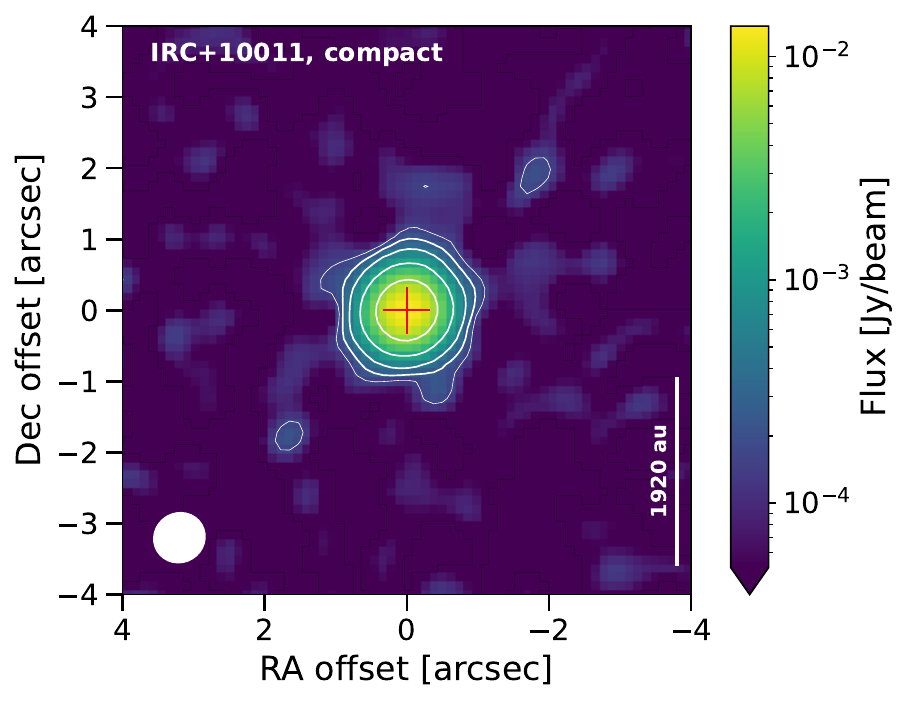}
    \includegraphics[width=0.32\textwidth]{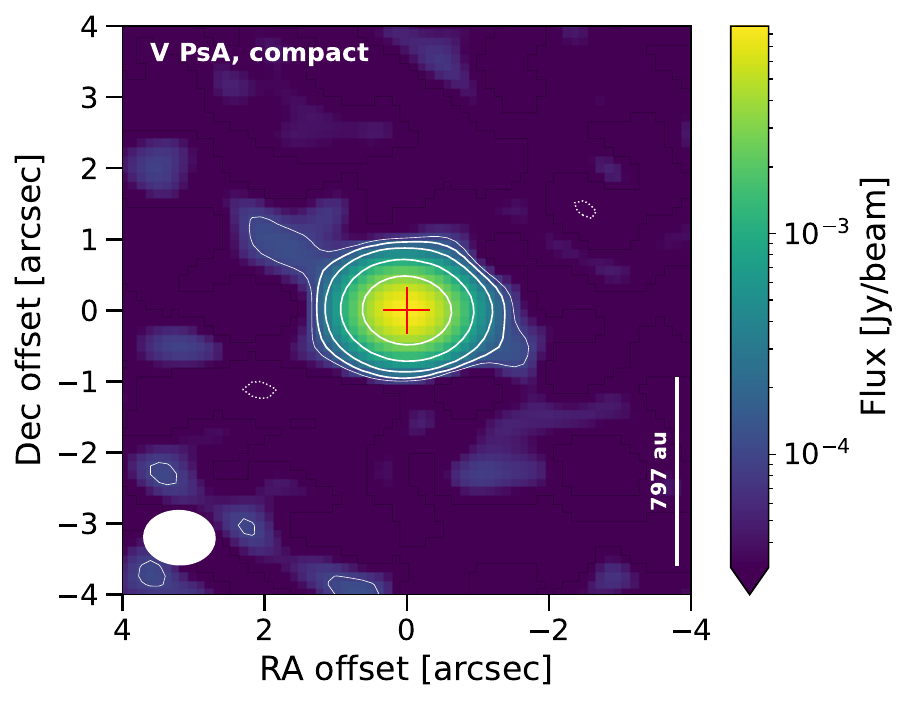}
    \includegraphics[width=0.32\textwidth]{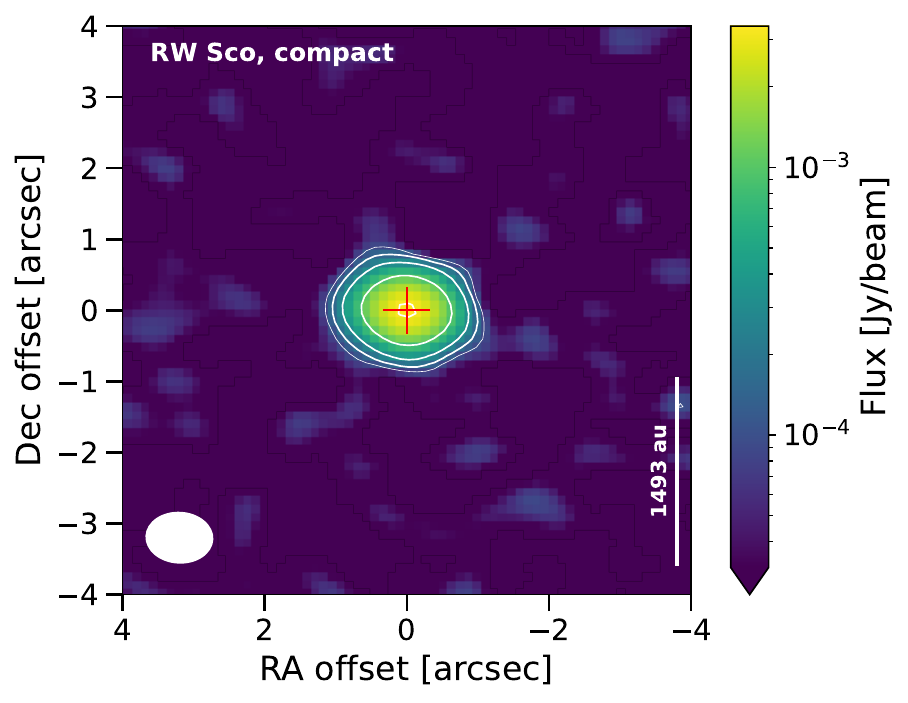}
    \includegraphics[width=0.32\textwidth]{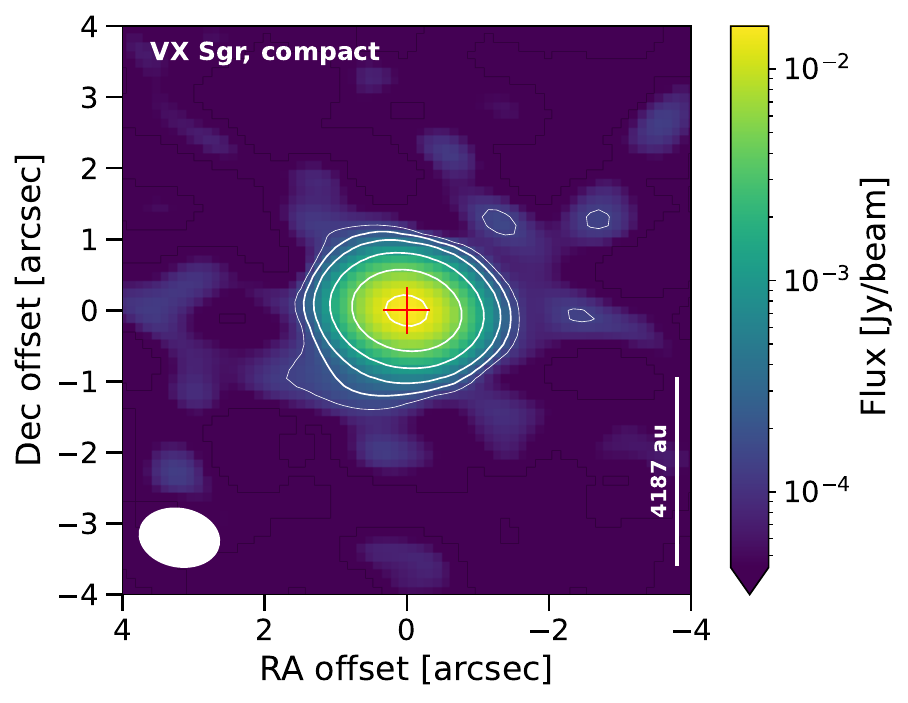}
    \caption{Continuum maps taken with the compact configuration of ALMA. The star is indicated in the top left of each image. The solid contours indicate levels of 3, 5, 10, 30, 100, and 300$\sigma$, and the dotted contours indicate levels of $-3\sigma$. The continuum peak is indicated by the red cross. For W~Aql, the black cross indicates the position of the main sequence (F9) companion. For \pigru, the yellow cross indicates the position of the B companion (G0V). The synthetic beams are indicated by white ellipses in the bottom left corners. \edits{The white bars on the lower right give indicative sizes in physical units for a third of the box length, based on the distance in Table~\ref{tab:stars}.} North is up and east is left.}
    \label{fig:compact}
\end{figure*}

\begin{figure*}
%\centering
	\includegraphics[width=0.32\textwidth]{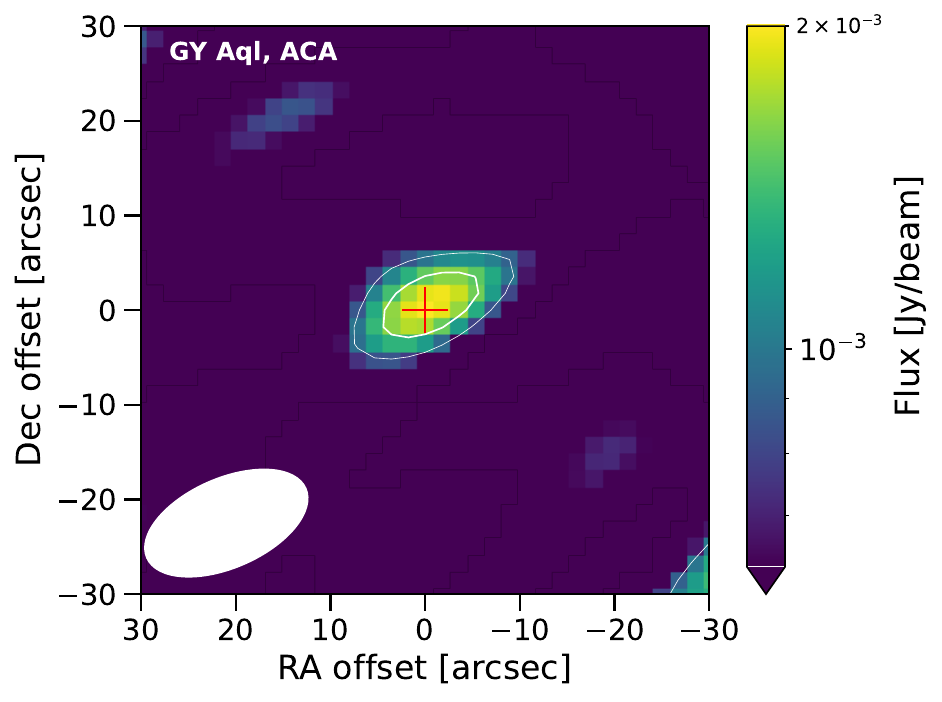}
 	\includegraphics[width=0.32\textwidth]{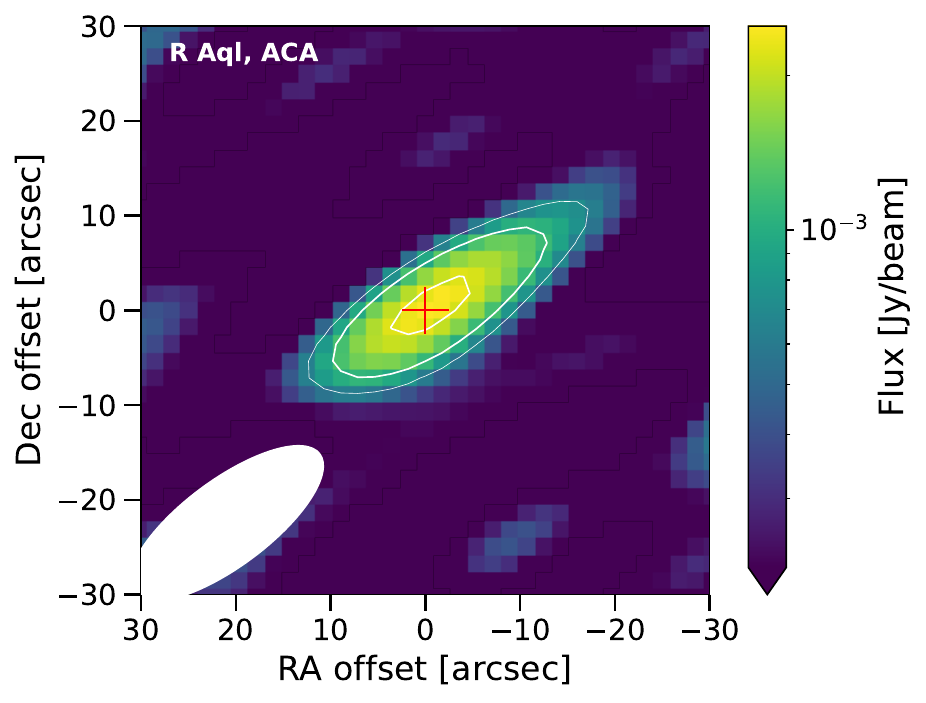}
  	\includegraphics[width=0.32\textwidth]{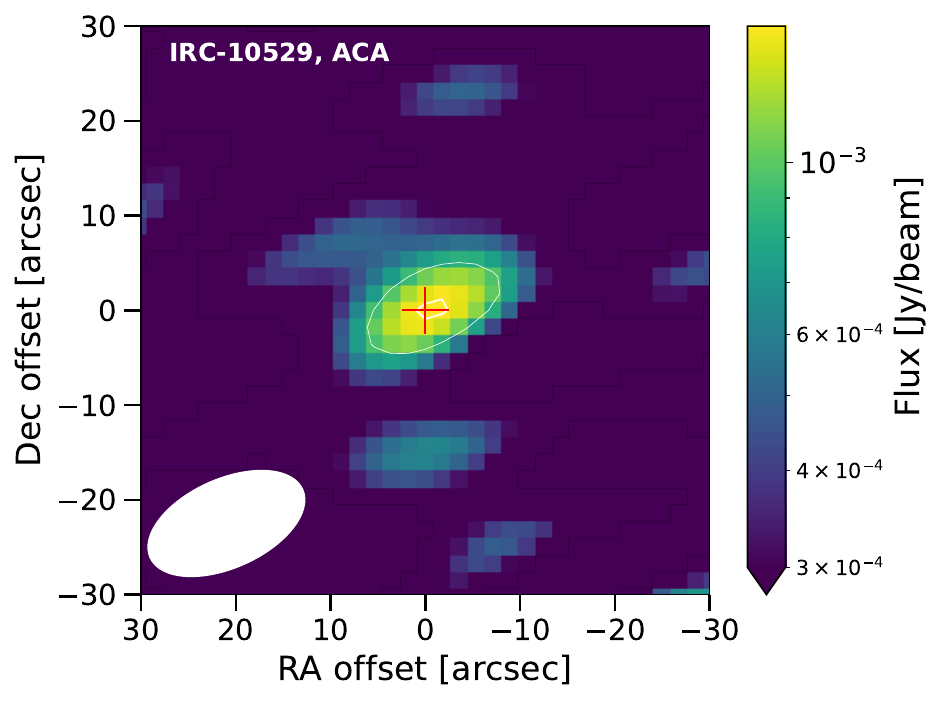}
    \includegraphics[width=0.32\textwidth]{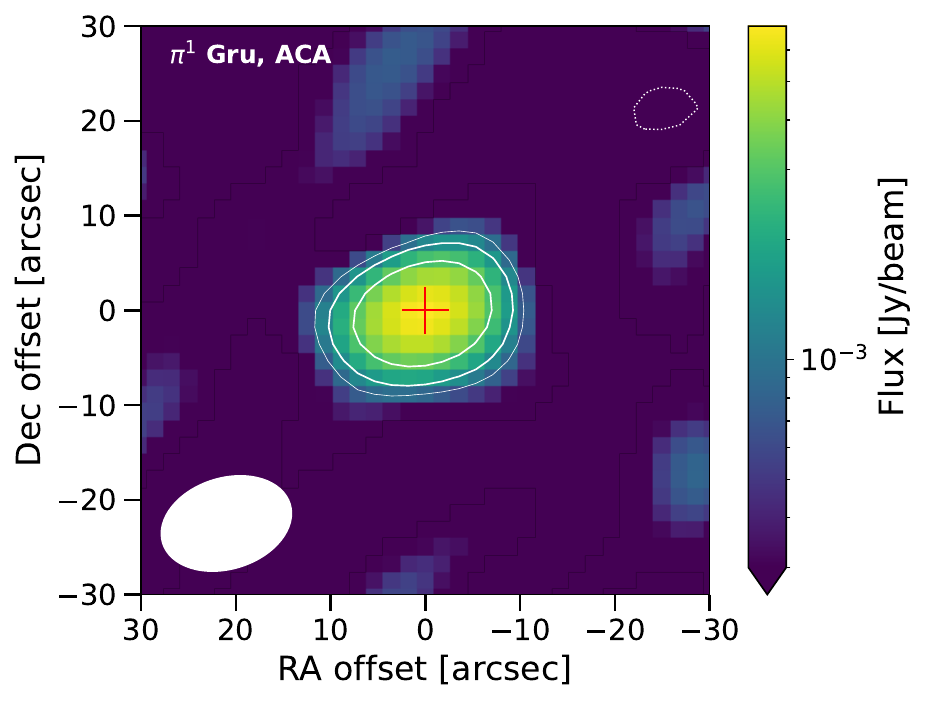}
    \includegraphics[width=0.32\textwidth]{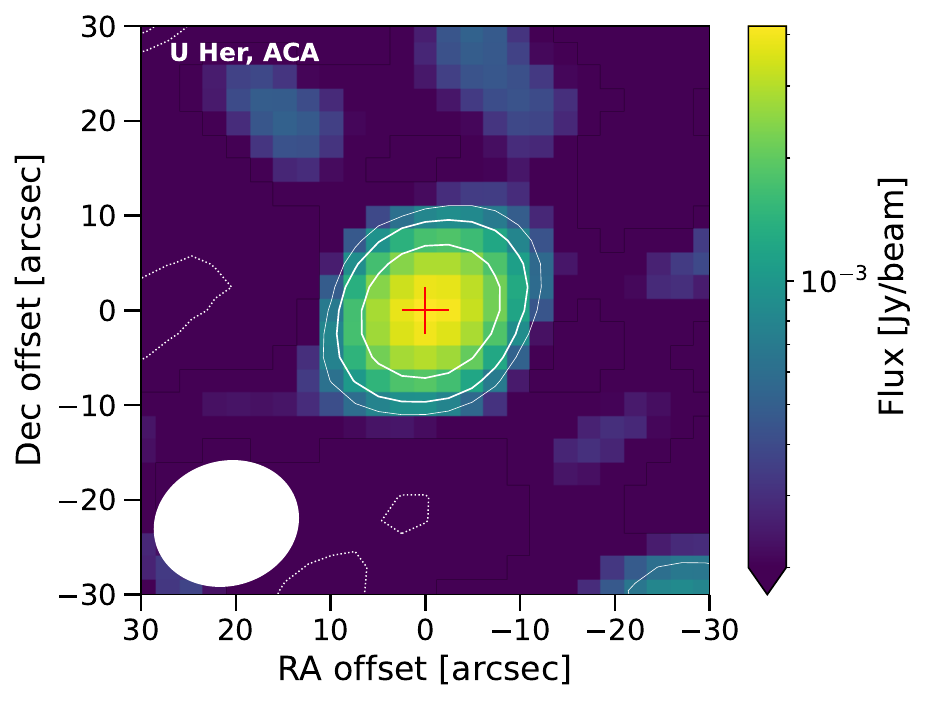}
    \includegraphics[width=0.32\textwidth]{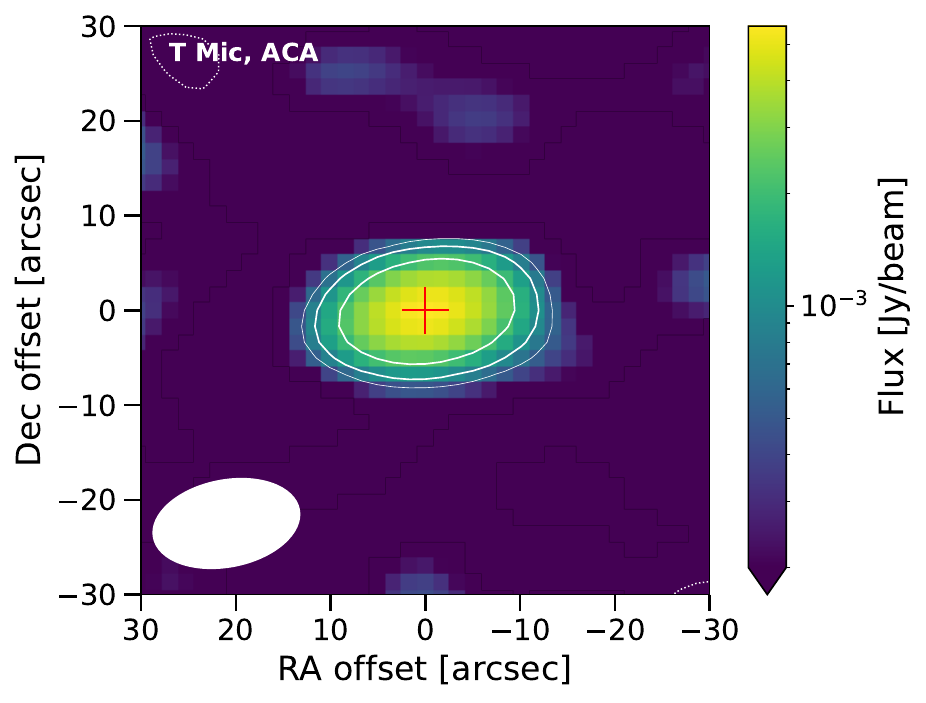}
    \includegraphics[width=0.32\textwidth]{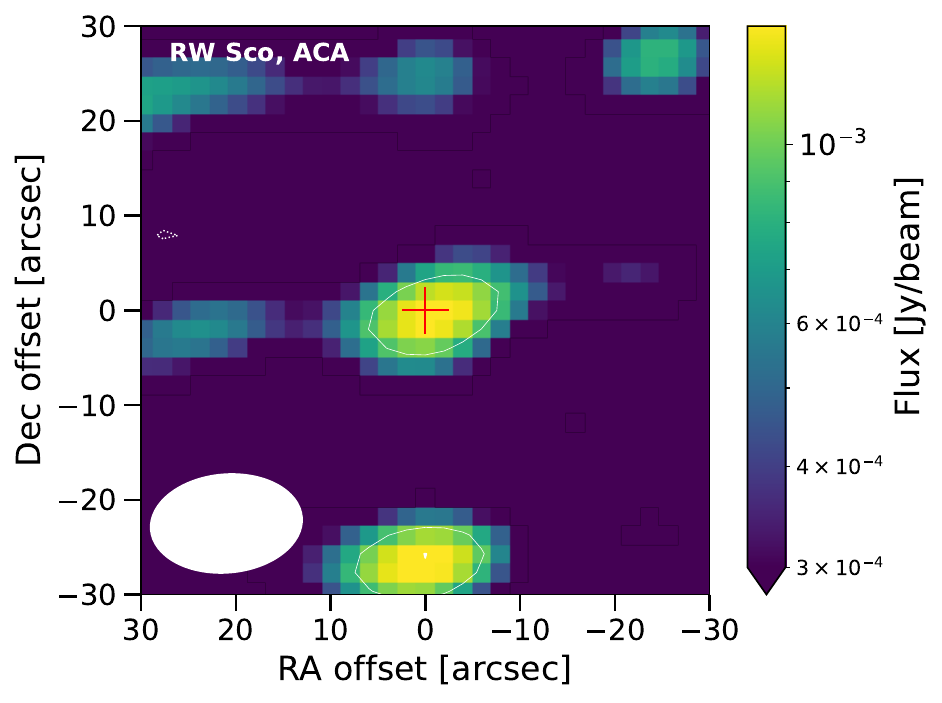}
    \caption{Continuum maps taken with the ACA. The star is indicated in the top left of each image. The contours indicate levels of 3, 5, and 10$\sigma$, and the dotted contours indicate levels of $-3\sigma$. The continuum peak is indicated by the red cross. The synthetic beams are indicated by white ellipses in the bottom left corners. North is up and east is left.}
    \label{fig:aca}
\end{figure*}

In Tables \ref{tab:extended}, \ref{tab:mid}, \ref{tab:compact}, and \ref{tab:combined}, we give overviews of the continua observed with the ALMA 12m extended, mid, compact, and combined configurations, and with the ACA in Table \ref{tab:aca}. The compact continuum images are shown in Fig.~\ref{fig:compact} and the ACA continuum images are shown in Fig.~\ref{fig:aca}. In Table~\ref{tab:udsub} we give details of the imaging properties of the UD-subtracted images. The UD fits themselves are shown with their corresponding visibility amplitudes in Fig.~\ref{fig:udfits}.

\edits{In Fig.~\ref{fig:pulsations-lit} we reproduce Fig.~\ref{fig:pulsations} with the addition of the literature data given in Table~\ref{tab:litUD}. See Sect.~\ref{sec:litsizes} for details.}

\begin{table*}
	\centering
	\caption{An overview of the continua observed with the extended array configuration \edits{(representative frequency 241.75~GHz)}.}
	\label{tab:extended}
	\begin{tabular}{ccrccc}
		\hline
Star	&	Beam size	&	Beam PA	&	rms	&	peak flux	&	Dynamic range	\\
	&	[\arcsec]	&	\multicolumn{1}{c}{[$\deg$]}	&	[Jy/beam]	&	[Jy/beam]		&		\\
\hline
GY Aql	&$	0.0254	\times	0.0223	$&$	-56.18	$&	$1.65\e{-5}$	&	$9.01\e{-3}$	&	545	\\
R Aql	&$	0.0236	\times	0.0215	$&$	-13.59	$&	$7.45\e{-6}$	&	$1.65\e{-2}$	&	2219	\\
IRC$-$10529	&$	0.0265	\times	0.0226	$&$	-55.67	$&	$1.99\e{-5}$	&	$6.63\e{-3}$	&	332	\\
W Aql	&$	0.0237	\times	0.0207	$&$	-47.05	$&	$5.09\e{-6}$	&	$6.63\e{-3}$	&	1301	\\
SV Aqr	&$	0.0217	\times	0.0211	$&$	43.85	$&	$8.71\e{-6}$	&	$1.46\e{-3}$	&	167	\\
U Del	&$	0.0296	\times	0.0206	$&$	-25.79	$&	$9.22\e{-6}$	&	$6.51\e{-3}$	&	706	\\
$\pi^1$ Gru	&$	0.0195	\times	0.0185	$&$	60.16	$&	$1.32\e{-5}$	&	$1.79\e{-2}$	&	1355	\\
U Her	&$	0.0245	\times	0.0182	$&$	8.27	$&	$1.16\e{-5}$	&	$1.17\e{-2}$	&	1005	\\
R Hya	&$	0.0342	\times	0.0247	$&$	67.29	$&	$1.18\e{-5}$	&	$4.63\e{-2}$	&	3937	\\
T Mic	&$	0.0241	\times	0.0207	$&$	-73.99	$&	$1.10\e{-5}$	&	$2.09\e{-2}$	&	1899	\\
S Pav	&$	0.0250	\times	0.0196	$&$	-13.31	$&	$8.98\e{-6}$	&	$2.17\e{-2}$	&	2416	\\
IRC+10011	&$	0.0275	\times	0.0195	$&$	31.42	$&	$1.87\e{-5}$	&	$1.20\e{-2}$	&	643	\\
V PsA	&$	0.0228	\times	0.0205	$&$	-77.91	$&	$8.56\e{-6}$	&	$8.36\e{-3}$	&	976	\\
RW Sco	&$	0.0243	\times	0.0198	$&$	-70.73	$&	$1.71\e{-5}$	&	$6.21\e{-3}$	&	363	\\
KW Sgr	&$	0.0219	\times	0.0202	$&$	-66.01	$&	$7.46\e{-6}$	&	$2.75\e{-3}$	&	368	\\
VX Sgr	&$	0.0276	\times	0.0200	$&$	89.34	$&	$1.42\e{-5}$	&	$1.46\e{-2}$	&	1031	\\
AH Sco	&$	0.0233	\times	0.0230	$&$	70.19	$&	$1.06\e{-5}$	&	$7.26\e{-3}$	&	684	\\
		\hline
	\end{tabular}
\end{table*}

\begin{table*}
	\centering
	\caption{An overview of the continua observed with the mid array configuration \edits{(representative frequency 241.75~GHz)}.}
	\label{tab:mid}
	\begin{tabular}{ccrccc}
		\hline
Star	&	Beam size	&	Beam PA	&	rms	&	Peak flux	&	Dynamic range	\\
	&	[\arcsec]	&	\multicolumn{1}{c}{[$\deg$]}	&	[Jy/beam]	&	[Jy/beam]	&		\\
\hline
GY Aql	&$	0.324	\times	0.247	$&$	-70.91	$&	$2.97\e{-5}$	&	$1.01\e{-2}$	&	339	\\
R Aql	&$	0.306	\times	0.238	$&$	-54.59	$&	$3.04\e{-5}$	&	$1.99\e{-2}$	&	656	\\
IRC$-$10529	&$	0.146	\times	0.113	$&$	-63.08	$&	$2.31\e{-5}$	&	$8.10\e{-3}$	&	351	\\
W Aql	&$	0.351	\times	0.223	$&$	-68.13	$&	$2.76\e{-5}$	&	$5.79\e{-3}$	&	210	\\
SV Aqr	&$	0.124	\times	0.104	$&$	-75.72	$&	$2.23\e{-5}$	&	$2.01\e{-3}$	&	90	\\
U Del	&$	0.316	\times	0.235	$&$	-33.78	$&	$2.63\e{-5}$	&	$7.24\e{-3}$	&	276	\\
$\pi^1$ Gru	&$	0.248	\times	0.235	$&$	30.08	$&	$3.18\e{-5}$	&	$3.19\e{-2}$	&	1003	\\
U Her	&$	0.267	\times	0.195	$&$	-33.48	$&	$4.25\e{-5}$	&	$1.50\e{-2}$	&	354	\\
R Hya	&$	0.256	\times	0.223	$&$	70.93	$&	$2.45\e{-5}$	&	$5.37\e{-2}$	&	2191	\\
T Mic	&$	0.268	\times	0.225	$&$	-89.56	$&	$2.23\e{-5}$	&	$2.99\e{-2}$	&	1341	\\
S Pav	&$	0.304	\times	0.234	$&$	56.16	$&	$2.18\e{-5}$	&	$3.06\e{-2}$	&	1405	\\
IRC+10011	&$	0.112	\times	0.100	$&$	38.52	$&	$2.54\e{-5}$	&	$9.44\e{-3}$	&	371	\\
V PsA	&$	0.283	\times	0.229	$&$	85.25	$&	$1.88\e{-5}$	&	$9.10\e{-3}$	&	485	\\
RW Sco	&$	0.147	\times	0.120	$&$	-86.29	$&	$4.20\e{-5}$	&	$5.29\e{-3}$	&	126	\\
KW Sgr	&$	0.155	\times	0.102	$&$	-77.20	$&	$1.75\e{-5}$	&	$2.78\e{-3}$	&	159	\\
VX Sgr	&$	0.162	\times	0.095	$&$	-75.77	$&	$3.02\e{-5}$	&	$1.62\e{-2}$	&	537	\\
AH Sco	&$	0.159	\times	0.100	$&$	-79.55	$&	$1.86\e{-5}$	&	$7.55\e{-3}$	&	405	\\
		\hline
	\end{tabular}
\end{table*}

\begin{table*}
	\centering
	\caption{An overview of the continua observed with the compact configuration of the ALMA main array \edits{(representative frequency 238.45~GHz)}.}
	\label{tab:compact}
	\begin{tabular}{ccrccc}
		\hline
Star	&	Beam size	&	Beam PA	&	rms	&	Peak flux	&	Dynamic range	\\
	&	[\arcsec]	&	\multicolumn{1}{c}{[$\deg$]}	&	[Jy/beam]	&	[Jy/beam]	&		\\
\hline
GY Aql	&$	1.22	\times	0.90	$&$	64.99	$&	$3.82\e{-5}$	&	$9.61\e{-3}$	&	252	\\
R Aql	&$	0.76	\times	0.65	$&$	83.87	$&	$4.92\e{-5}$	&	$1.88\e{-2}$	&	382	\\
IRC$-$10529	&$	0.79	\times	0.63	$&$	76.78	$&	$4.96\e{-5}$	&	$6.36\e{-3}$	&	128	\\
W Aql	&$	0.92	\times	0.67	$&$	76.04	$&	$5.28\e{-5}$	&	$7.68\e{-3}$	&	145	\\
SV Aqr	&$	0.89	\times	0.75	$&$	74.49	$&	$3.44\e{-5}$	&	$1.38\e{-3}$	&	40	\\
U Del	&$	1.16	\times	1.01	$&$	33.98	$&	$4.80\e{-5}$	&	$6.41\e{-3}$	&	134	\\
$\pi^1$ Gru	&$	0.87	\times	0.77	$&$	-86.96	$&	$3.62\e{-5}$	&	$3.12\e{-2}$	&	861	\\
U Her	&$	1.00	\times	0.84	$&$	26.50	$&	$4.85\e{-5}$	&	$1.73\e{-2}$	&	357	\\
R Hya	&$	0.83	\times	0.60	$&$	79.35	$&	$4.47\e{-5}$	&	$6.52\e{-2}$	&	1458	\\
T Mic	&$	1.05	\times	0.73	$&$	-79.98	$&	$5.91\e{-5}$	&	$2.88\e{-2}$	&	488	\\
S Pav	&$	1.03	\times	0.98	$&$	-56.29	$&	$4.56\e{-5}$	&	$2.73\e{-2}$	&	599	\\
IRC+10011	&$	0.72	\times	0.69	$&$	-59.03	$&	$5.15\e{-5}$	&	$1.37\e{-2}$	&	265	\\
V PsA	&$	0.99	\times	0.75	$&$	87.87	$&	$3.08\e{-5}$	&	$8.68\e{-3}$	&	282	\\
RW Sco	&$	0.93	\times	0.70	$&$	86.91	$&	$3.20\e{-5}$	&	$3.37\e{-3}$	&	105	\\
VX Sgr	&$	1.13	\times	0.81	$&$	79.83	$&	$4.38\e{-5}$	&	$1.60\e{-2}$	&	365	\\
		\hline
	\end{tabular}
\end{table*}

\begin{table*}
	\centering
	\caption{An overview of the continua from the combined 12m arrays \edits{(representative frequency 241.75~GHz)}.}
	\label{tab:combined}
	\begin{tabular}{ccrccccc}
		\hline
Star	&	Beam size	&	Beam PA	&	rms	&	Peak flux	&	Dynamic	& Total flux$^{a}$ & Total flux$^{b}$\\
	&	[\arcsec]	&	\multicolumn{1}{c}{[$\deg$]}	&	[Jy/beam]	&	[Jy/beam]	&	range	& [Jy] & region [\arcsec]\\
\hline
GY Aql	&$	0.160	\times	0.143	$&$	-53.03	$&	$1.97\e{-5}$	&	$1.03\e{-2}$	&	524	&	$1.3\e{-2}$	&	2.7	\\
R Aql	&\edits{$	0.165	\times	0.142$}	&\edits{$	-30.32	$}&	\edits{$1.15\e{-5}$}	&	\edits{$2.03\e{-2}$}	&	\edits{1765}	&	\edits{$2.9\e{-2}$}	&	\edits{5.0}	\\
IRC$-$10529	&$	0.044	\times	0.040	$&$	72.87	$&	$1.83\e{-5}$	&	$7.63\e{-3}$	&	416	&	$1.8\e{-2}$	&	4.4	\\
W Aql	&$	0.040	\times	0.033	$&$	45.91	$&	$6.83\e{-6}$	&	$7.59\e{-3}$	&	1110	&	$1.6\e{-2}$	&	3.6	\\
SV Aqr	&$	0.036	\times	0.028	$&$	45.34	$&	$8.12\e{-6}$	&	$1.57\e{-3}$	&	194	&	$2.4\e{-3}$	&	0.3	\\
U Del	&$	0.049	\times	0.041	$&$	-1.09	$&	$9.67\e{-6}$	&	$6.90\e{-3}$	&	714	&	$8.9\e{-3}$	&	1.7	\\
$\pi^1$ Gru	&$	0.047	\times	0.041	$&$	25.14	$&	$1.18\e{-5}$	&	$2.60\e{-2}$	&	2214	&	$3.4\e{-2}$	&	0.3	\\
U Her	&$	0.050	\times	0.040	$&$	12.22	$&	$1.26\e{-5}$	&	$1.38\e{-2}$	&	1099	&	$2.3\e{-2}$	&	2.8	\\
R Hya	&$	0.054	\times	0.044	$&$	63.15	$&	$8.35\e{-6}$	&	$5.51\e{-2}$	&	6597	&	$7.2\e{-2}$	&	4.4	\\
T Mic	&$	0.049	\times	0.042	$&$	65.23	$&	$1.07\e{-5}$	&	$2.80\e{-2}$	&	2625	&	$3.5\e{-2}$	&	2.8	\\
S Pav	&$	0.048	\times	0.039	$&$	5.02	$&	$8.89\e{-6}$	&	$2.63\e{-2}$	&	2960	&	$3.7\e{-2}$	&	4.4	\\
IRC+10011	&$	0.107	\times	0.100	$&$	27.93	$&	$2.18\e{-5}$	&	$1.21\e{-2}$	&	554	&	$1.9\e{-2}$	&	4.7	\\
V PsA	&$	0.047	\times	0.043	$&$	57.17	$&	$8.36\e{-6}$	&	$9.31\e{-3}$	&	1114	&	$1.2\e{-2}$	&	4.4	\\
RW Sco	&$	0.039	\times	0.037	$&$	-16.51	$&	$1.61\e{-5}$	&	$6.14\e{-3}$	&	395	&	$6.2\e{-3}$	&	0.05	\\
KW Sgr	&$	0.032	\times	0.029	$&$	38.03	$&	$7.34\e{-6}$	&	$2.80\e{-3}$	&	382	&	$3.7\e{-3}$	&	0.6	\\
VX Sgr	&$	0.045	\times	0.035	$&$	76.21	$&	$1.59\e{-5}$	&	$1.73\e{-2}$	&	1089	&	$2.6\e{-2}$	&	4.0	\\
AH Sco	&$	0.085	\times	0.082	$&$	-52.50	$&	$1.13\e{-5}$	&	$8.36\e{-3}$	&	739	&	$1.4\e{-2}$	&	1.9	\\
		\hline
	\end{tabular}
\tablefoot{($^{a}$) The total flux is the flux density in a large circular region, of radius given in the rightmost column. ($^{b}$) The radius was chosen to enclose all the flux surrounding the star without significant negative regions or the less sensitive part of the primary beam.
%a lowered beam response reducing the enclosed flux value.
}
\end{table*}

\begin{table*}
	\centering
	\caption{An overview of the continua observed with the ACA \edits{(representative frequency 104.96~GHz)}.}
	\label{tab:aca}
	\begin{tabular}{ccrccccc}
		\hline
Star	&	Beam size	&	Beam PA	& Observing date &	MRS$^{a}$ & rms	&	Peak flux	&	Dynamic \\
	&	[\arcsec]	&	\multicolumn{1}{c}{[$\deg$]}	& [yyyy-mm-dd] &	[\arcsec]	&	[Jy/beam]	&	[Jy/beam]	&	range \\
\hline
GY Aql	&$	18.26	\times	9.39	$&$	-66.41	$&	2019-11-07	&	69	&	$3.02\e{-4}$	&	$2.01\e{-3}$	&	7	\\
R Aql	&$	24.53	\times	9.07	$&$	-53.34	$&	2019-11-19	&	77	&	$1.83\e{-4}$	&	$2.51\e{-3}$	&	14	\\
IRC$-$10529	&$	17.55	\times	9.30	$&$	-65.90	$&	2019-11-02	&	66	&	$2.97\e{-4}$	&	$1.57\e{-3}$	&	5	\\
$\pi^1$ Gru	&$	14.00	\times	9.55	$&$	-72.98	$&	2019-10-16	&	64	&	$3.30\e{-4}$	&	$6.89\e{-3}$	&	21	\\
U Her	&$	15.29	\times	12.97	$&$	-73.56	$&	2019-12-12	&	73	&	$2.30\e{-4}$	&	$4.26\e{-3}$	&	19	\\
T Mic	&$	15.59	\times	9.10	$&$	-79.05	$&	2019-11-11	&	66	&	$1.13\e{-4}$	&	$5.64\e{-3}$	&	50	\\
RW Sco	&$	15.97	\times	10.35	$&$	-85.19	$&	2019-12-12	&	65	&	$3.29\e{-4}$	&	$1.47\e{-3}$	&	4	\\
		\hline
	\end{tabular}
\begin{flushleft}
\textbf{Notes.} ($^{a}$) Maximum recoverable scale.
\end{flushleft}
\end{table*}

\begin{table*}
	\centering
	\caption{Imaging properties of UD-subtracted images \edits{of the combined data}.}
	\label{tab:udsub}
	\begin{tabular}{ccrcccccc}
		\hline
Star	&	Beam size	&	Beam PA	& rms	&	Peak flux & \edits{Total flux}	&Dynamic\\
	&	[mas]	&	\multicolumn{1}{c}{[$\deg$]}	&	[$\mu$Jy/beam]	&	[$\mu$Jy/beam]	& \edits{[mJy]}&	range \\
\hline
GY Aql	&$	170.3	\times	141.8	$&$	-52.2	$&	19.6	&	730	&	2.90	&	37	\\
R Aql	&$	\edits{165.3	\times	141.8}	$&$ \edits{	-30.3}	$&	\edits{11.6}	&	\edits{437}	&	\edits{9.60}	&	\edits{38}	\\
IRC$-$10529	&$	81.1	\times	75.1	$&$	-84.5	$&	20.1	&	433	&	4.93	&	22	\\
W Aql	&$	60.0	\times	53.9	$&$	52.2	$&	7.2	&	314	&	4.96	&	44	\\
SV Aqr	&$	67.8	\times	62.2	$&$	63.2	$&	10.3	&	112	&	0.47	&	11	\\
U Del	&$	69.6	\times	61.5	$&$	-0.7	$&	11.2	&	49	&	2.31	&	4	\\
$\pi^1$ Gru	&$	42.7	\times	36.0	$&$	28.6	$&	11.7	&	3288	&	7.06	&	280	\\
U Her	&$	84.7	\times	69.4	$&$	-6.8	$&	15.5	&	434	&	6.18	&	28	\\
R Hya	&$	82.1	\times	71.1	$&$	84.1	$&	10.7	&	512	&	6.93	&	48	\\
T Mic	&$	74.2	\times	67.8	$&$	72.8	$&	12.2	&	601	&	3.53	&	49	\\
S Pav	&$	72.2	\times	60.6	$&$	0.6	$&	10.9	&	673	&	5.34	&	62	\\
IRC+10011	&$	84.0	\times	76.2	$&$	29.9	$&	20.3	&	517	&	3.07	&	25	\\
V PsA	&$	71.2	\times	68.7	$&$	79.7	$&	9.7	&	195	&	1.62	&	20	\\
RW Sco	&$	78.3	\times	74.6	$&$	-33.8	$&	21.1	&	1243	&	0.77	&	59	\\
\hline															
KW Sgr	&$	64.9	\times	62.8	$&$	38.5	$&	8.5	&	48	&	0.61	&	67	\\
VX Sgr	&$	81.1	\times	70.5	$&$	84.4	$&	14.9	&	569	&	3.87	&	34	\\
AH Sco	&$	94.6	\times	89.4	$&$	-56.3	$&	11.3	&	505	&	5.39	&	45	\\
		\hline
	\end{tabular}
	\tablefoot{\edits{The total flux column gives the flux density measured in circular regions, centred on the continuum peaks, defined by the radii given in Table~\ref{tab:combined}. }}
\end{table*}

\begin{figure*}
	\includegraphics[width=0.49\textwidth]{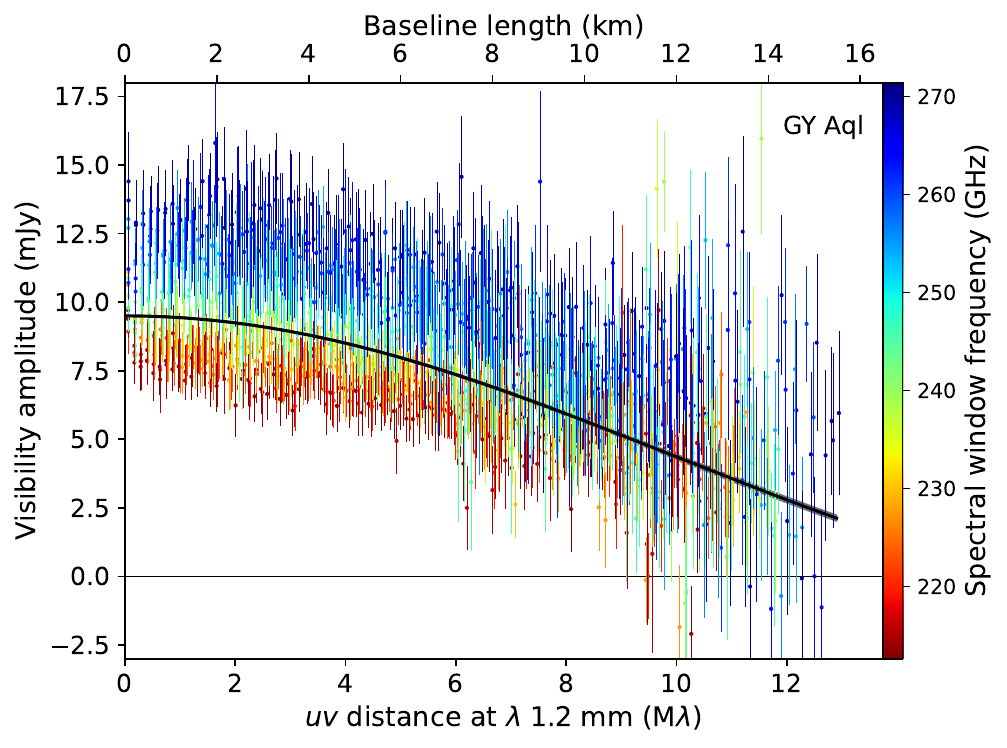}
 	\includegraphics[width=0.49\textwidth]{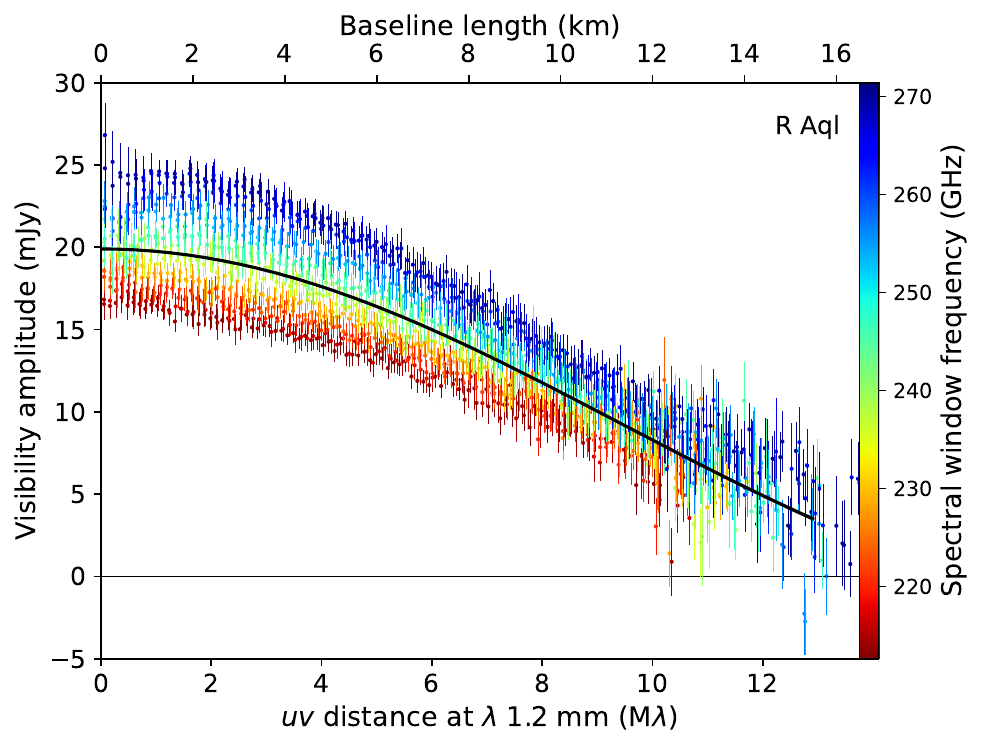}
  	\includegraphics[width=0.49\textwidth]{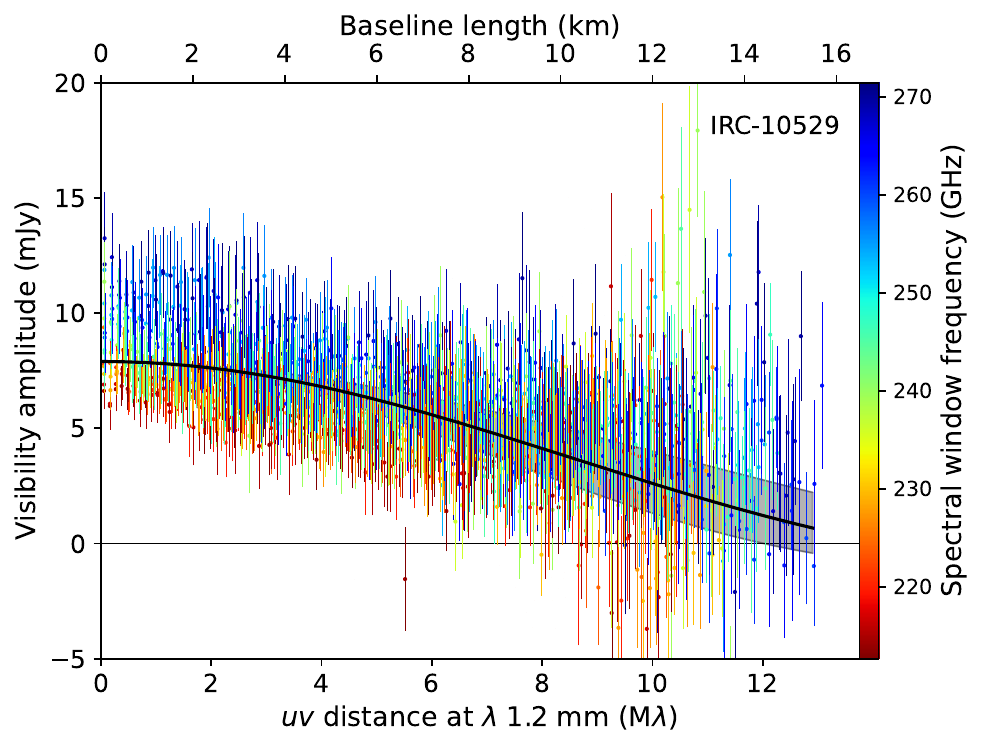}
   	\includegraphics[width=0.49\textwidth]{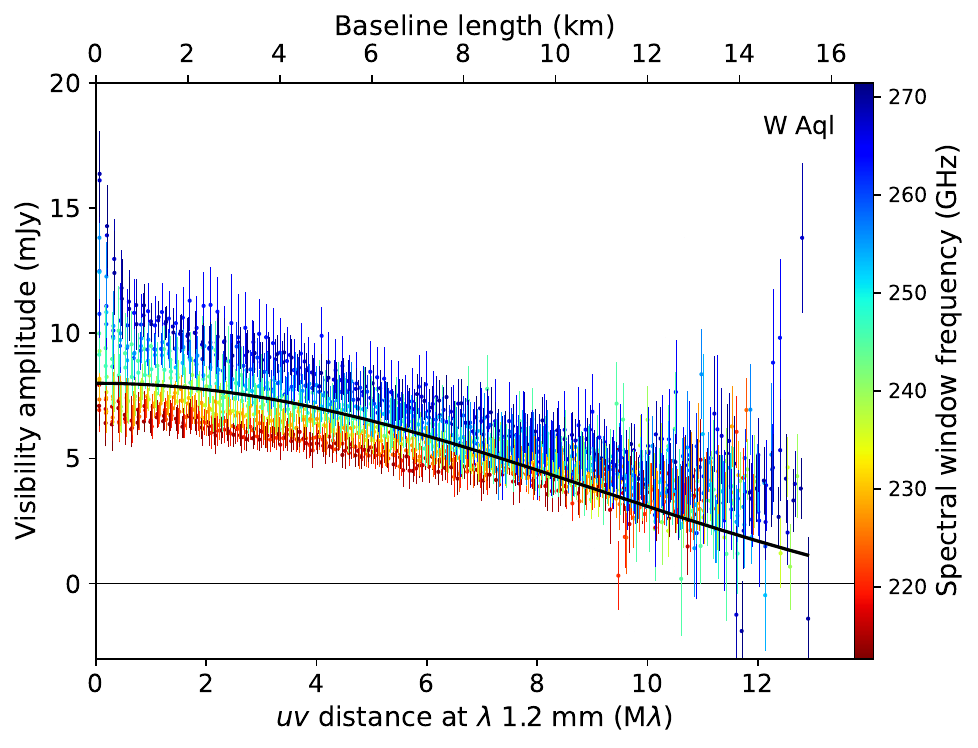}
    \includegraphics[width=0.49\textwidth]{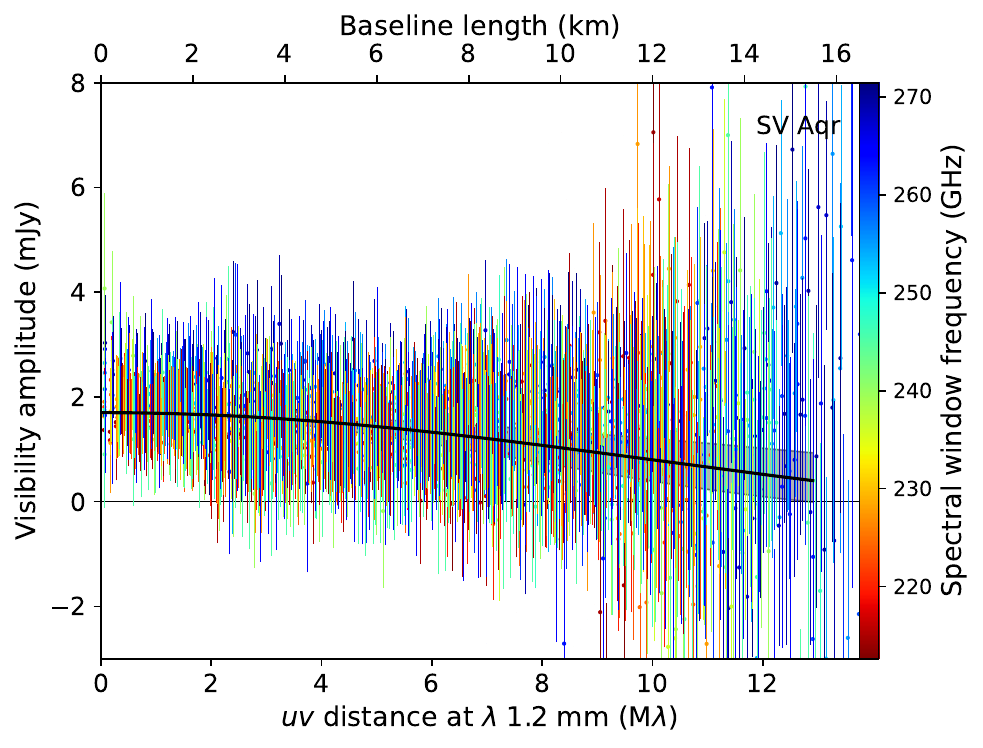}
    \includegraphics[width=0.49\textwidth]{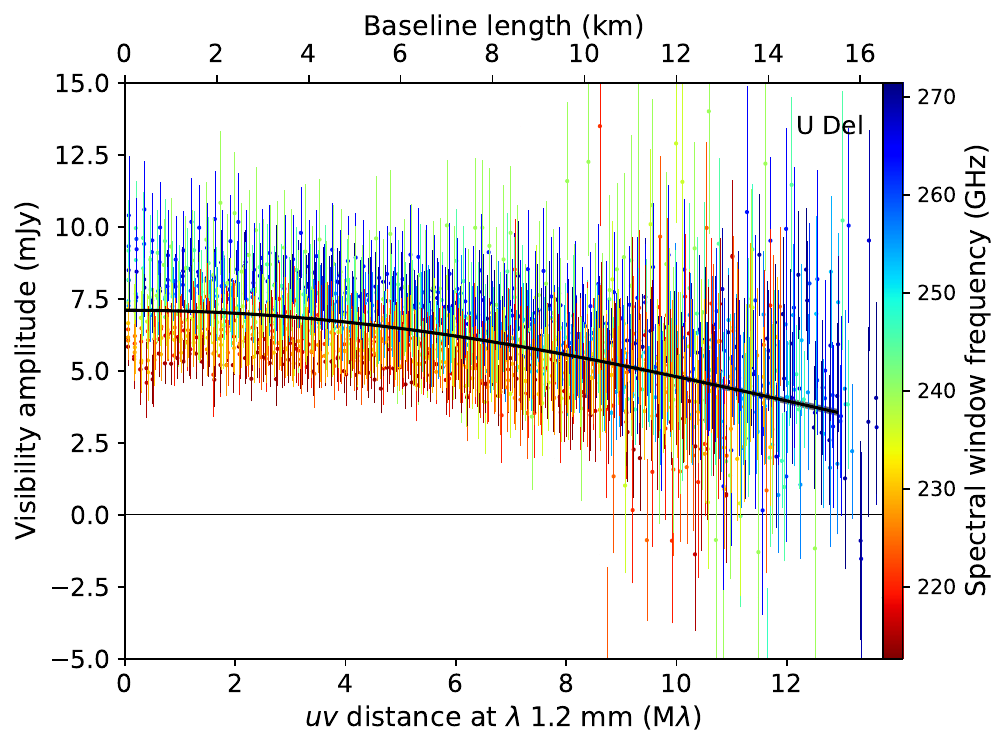}
    \caption{Observed visibility amplitudes of the continuum data as a function of the uv distance (bottom axes) and baseline length (top axes). Each panel corresponds to the combined data of the star noted in the top right corner. Points are colour-coded by their spectral windows, with the lower frequencies in red and the higher frequencies in blue. The black lines are fits of uniform disc models, as described in the text, and the shaded regions represent the fitting errors.}
    \label{fig:udfits}
\end{figure*}
\addtocounter{figure}{-1}
\begin{figure*}
    \includegraphics[width=0.49\textwidth]{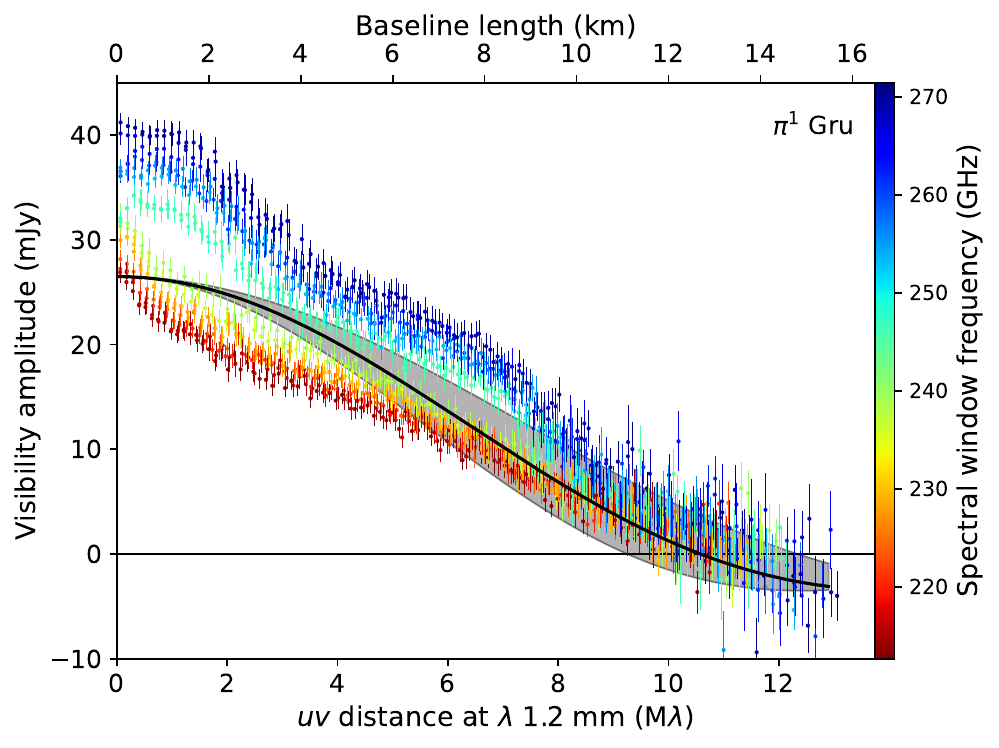}
    \includegraphics[width=0.49\textwidth]{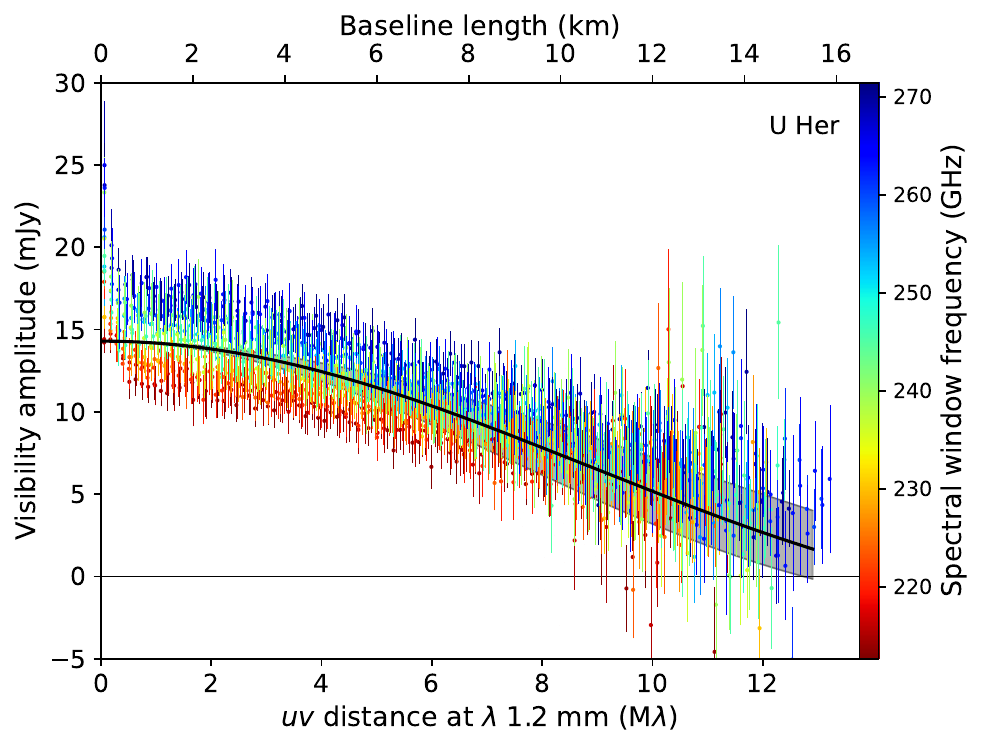}
    \includegraphics[width=0.49\textwidth]{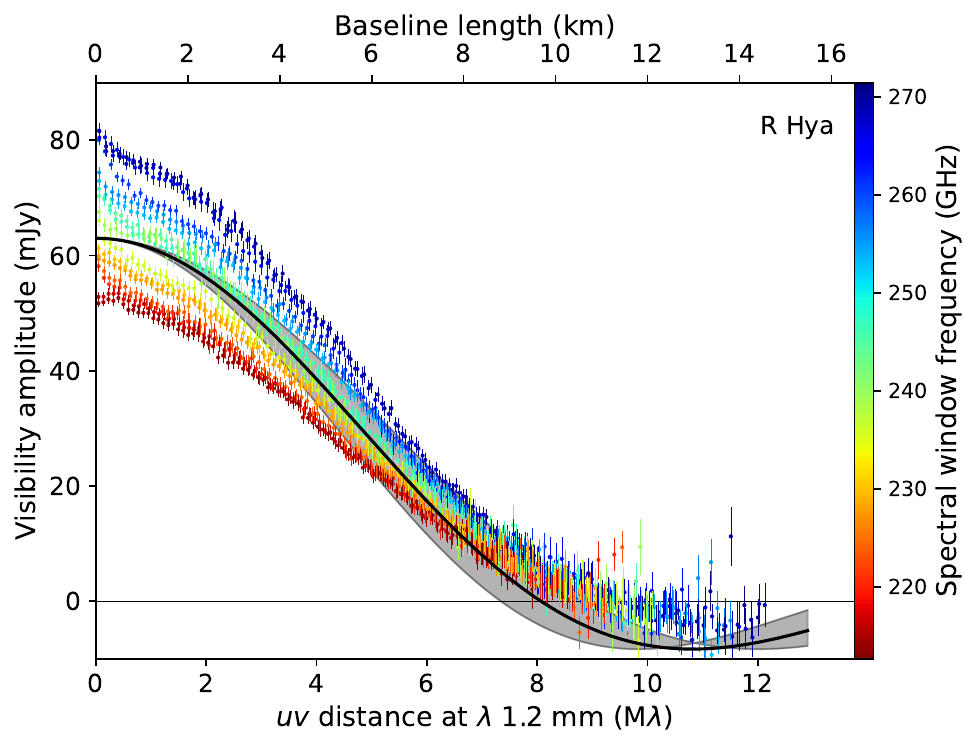}
    \includegraphics[width=0.49\textwidth]{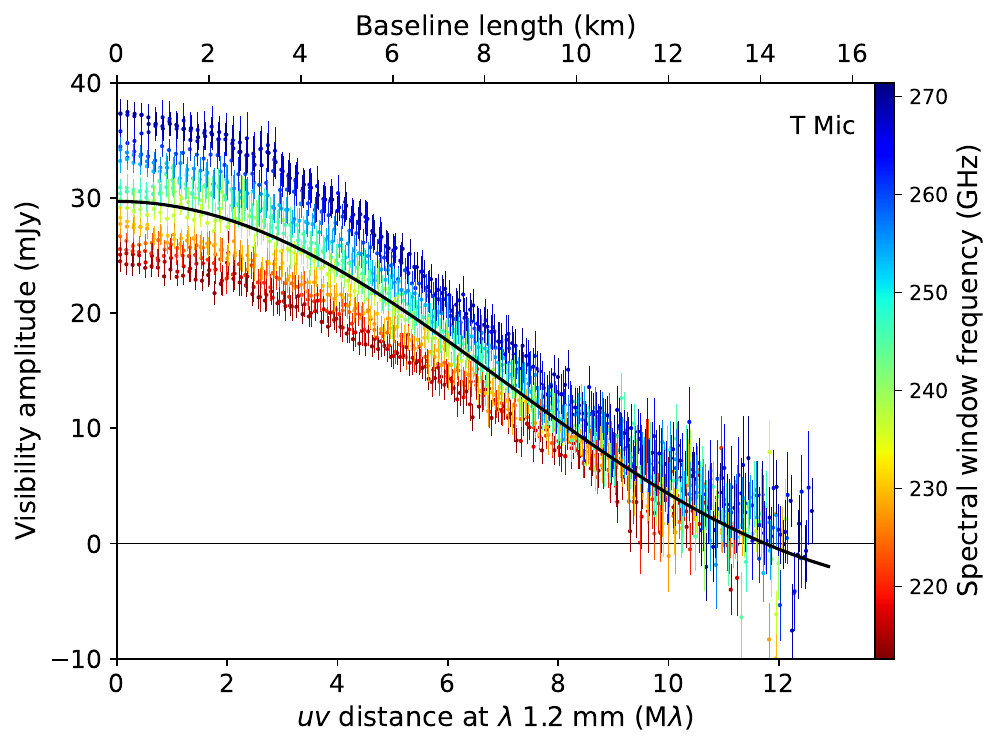}
    \includegraphics[width=0.49\textwidth]{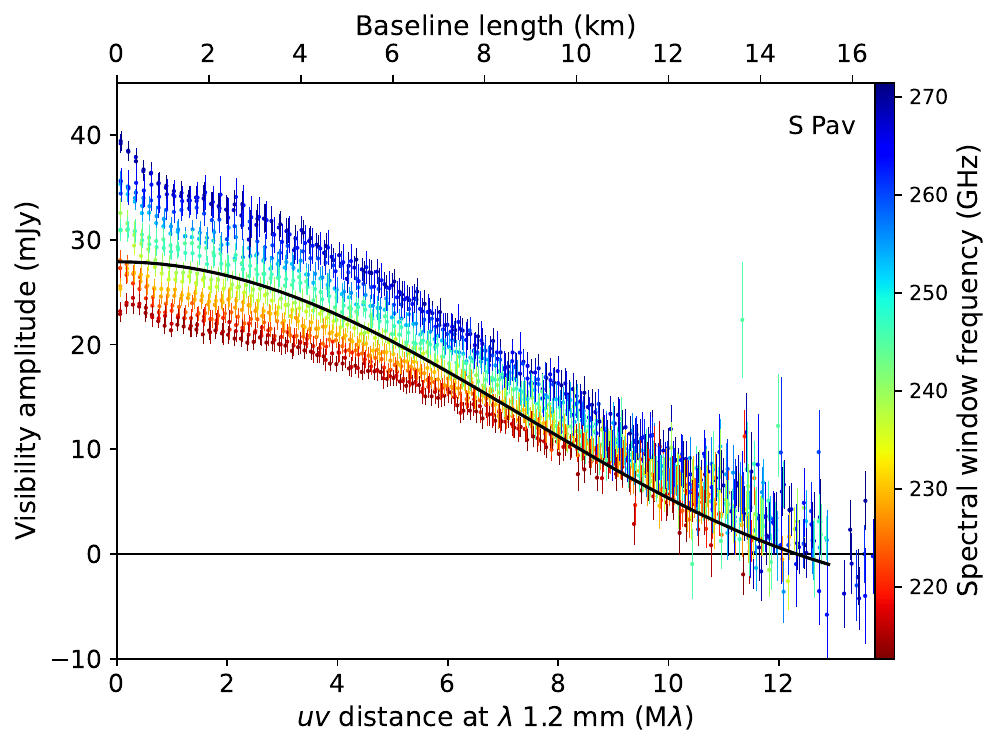}
    \includegraphics[width=0.49\textwidth]{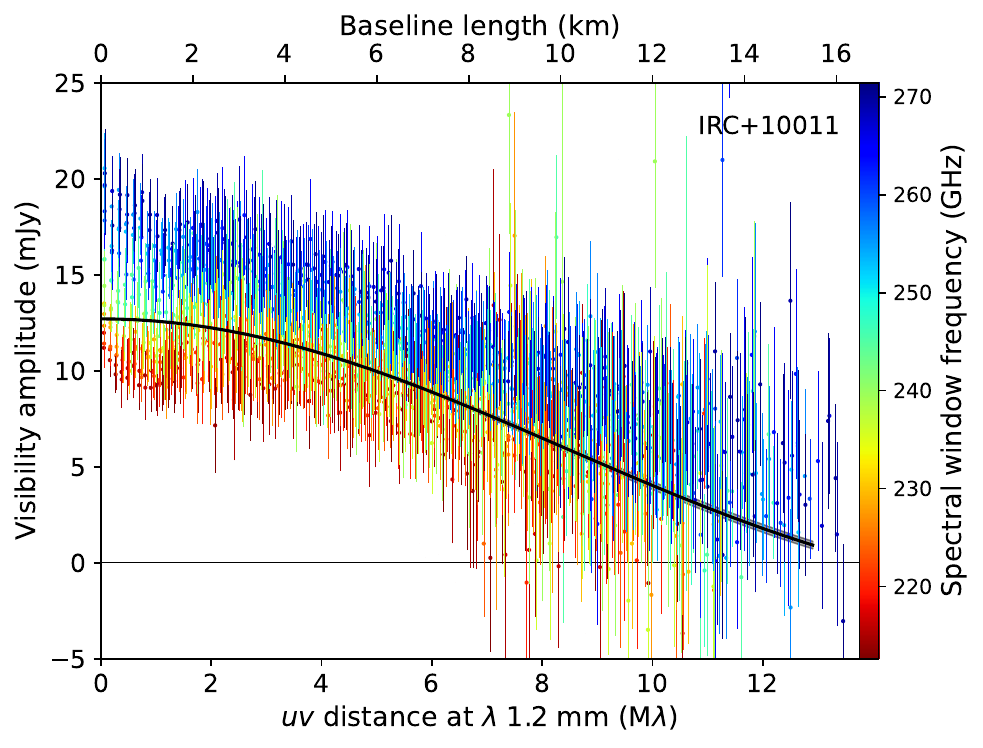}
    \caption{Continued.}
\end{figure*}
\addtocounter{figure}{-1}
\begin{figure*}
    \includegraphics[width=0.49\textwidth]{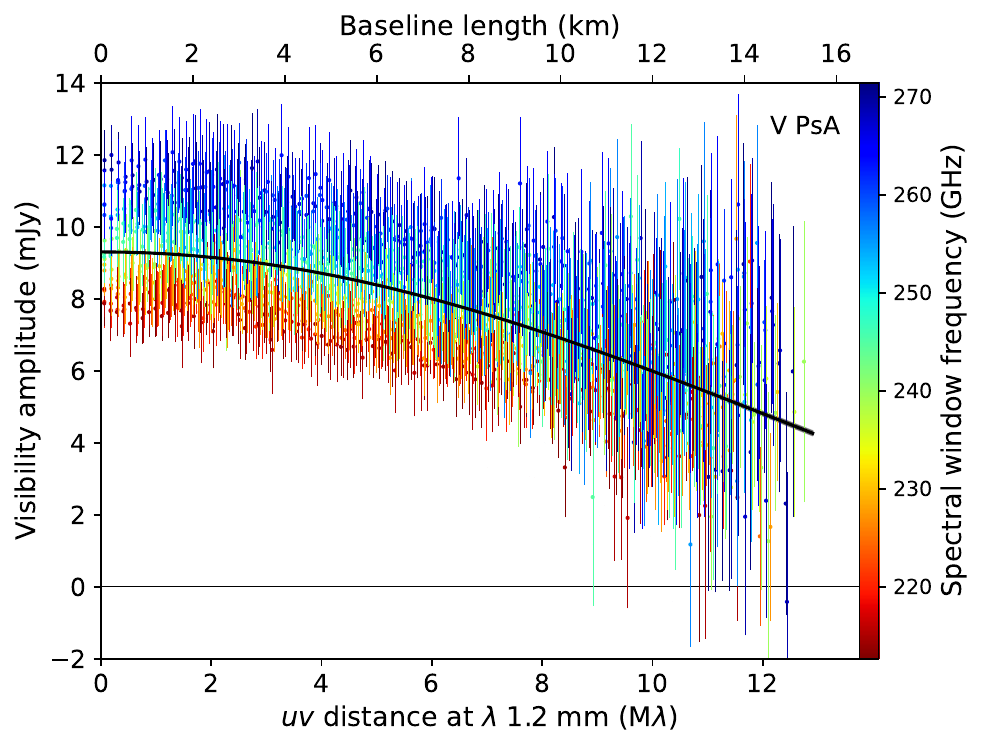}
    \includegraphics[width=0.49\textwidth]{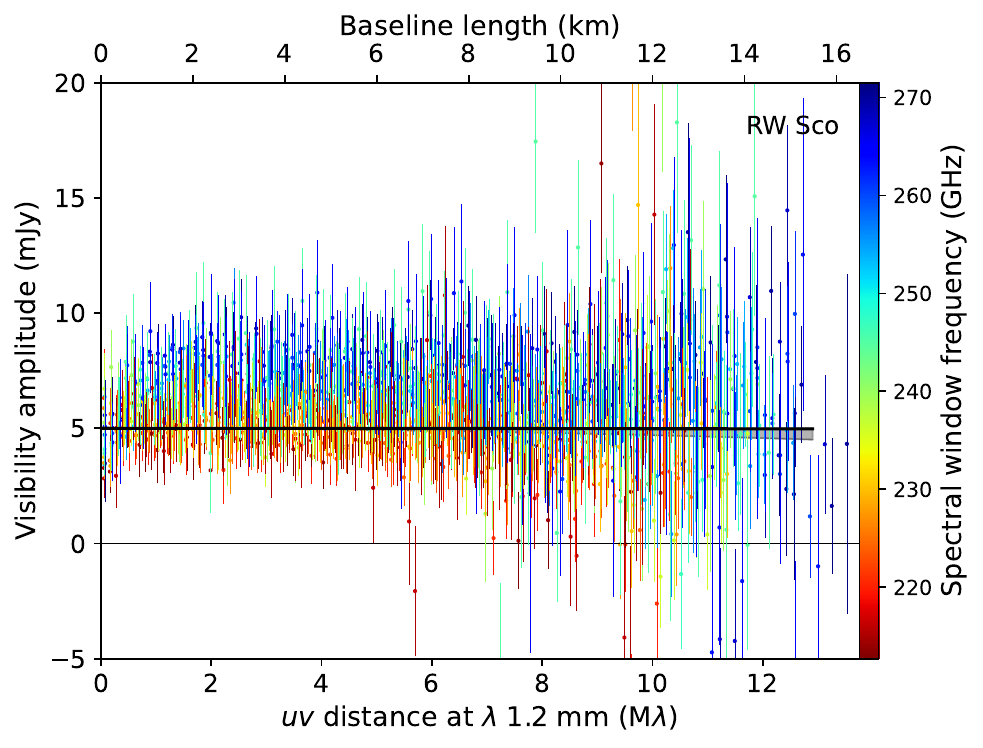}
    \includegraphics[width=0.49\textwidth]{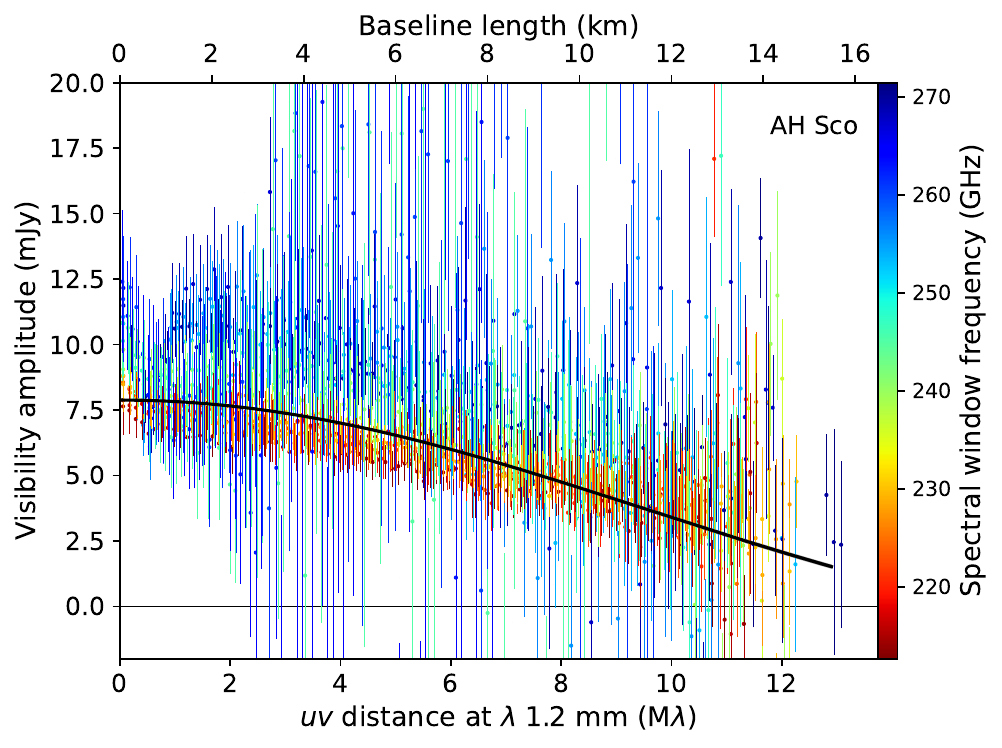}
    \includegraphics[width=0.49\textwidth]{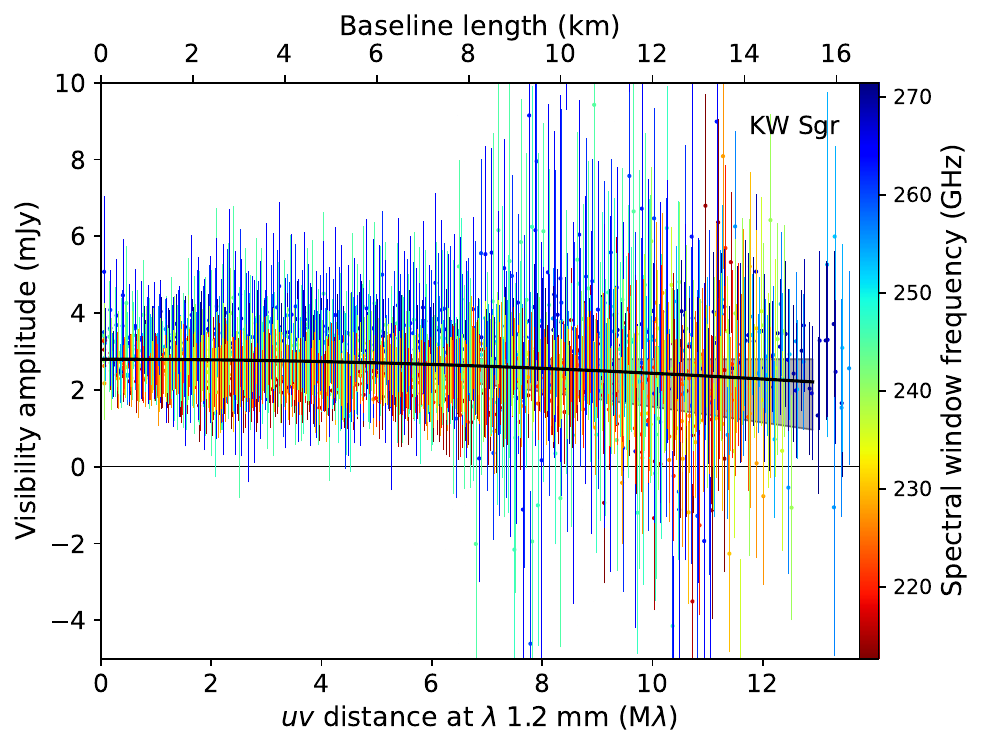}
    \includegraphics[width=0.49\textwidth]{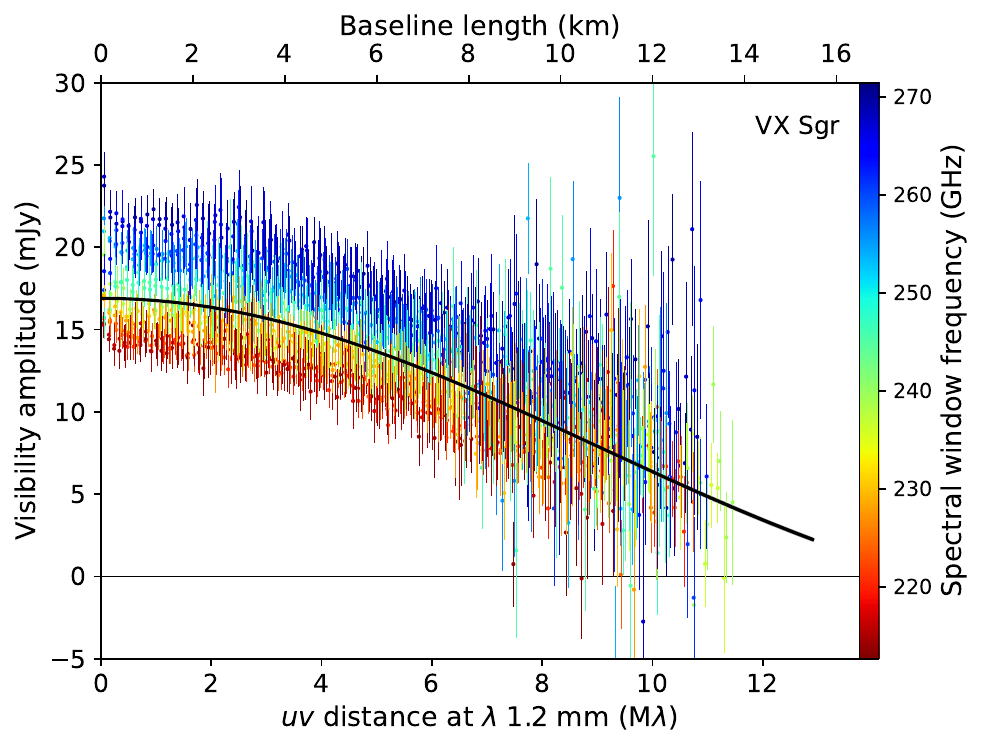}
    \caption{Continued.}
\end{figure*}

\begin{figure*}
	\includegraphics[height=6.7cm]{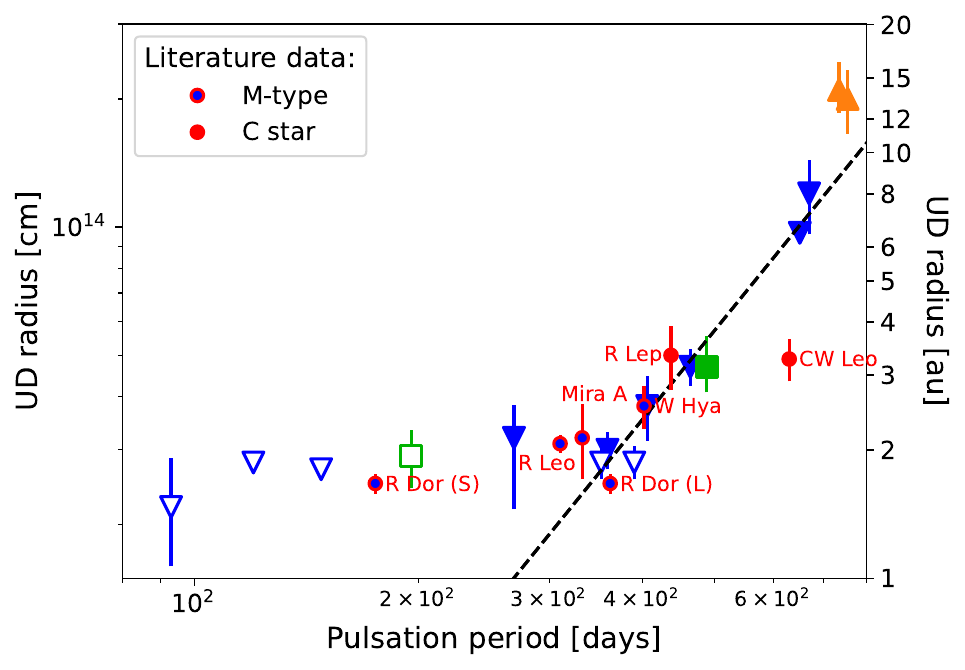}
	\includegraphics[height=6.7cm]{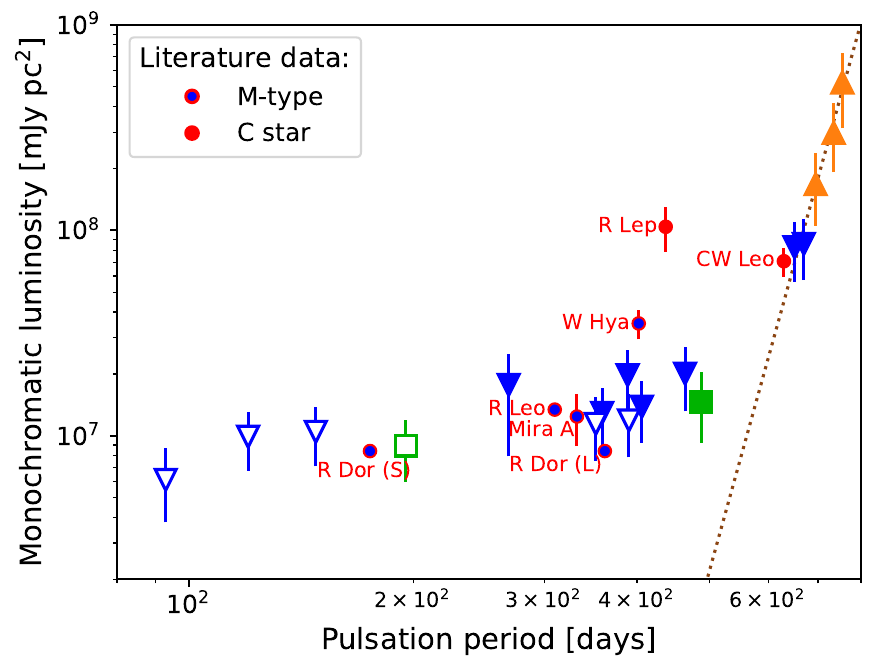}
    \caption{\edits{UD radii (left) and monochromatic fluxes (right), plotted against pulsation period of the ATOMIUM sample (large markers, see Fig.~\ref{fig:pulsations}) and some literature data (smaller red circular markers, see Table \ref{tab:litUD}). For R~Dor, `S' and `L' indicate the shorter and longer periods found for this star \citep{Bedding1998a}.}}
    \label{fig:pulsations-lit}
\end{figure*}

\begin{figure*}
\centering
	\includegraphics[width=0.32\textwidth]{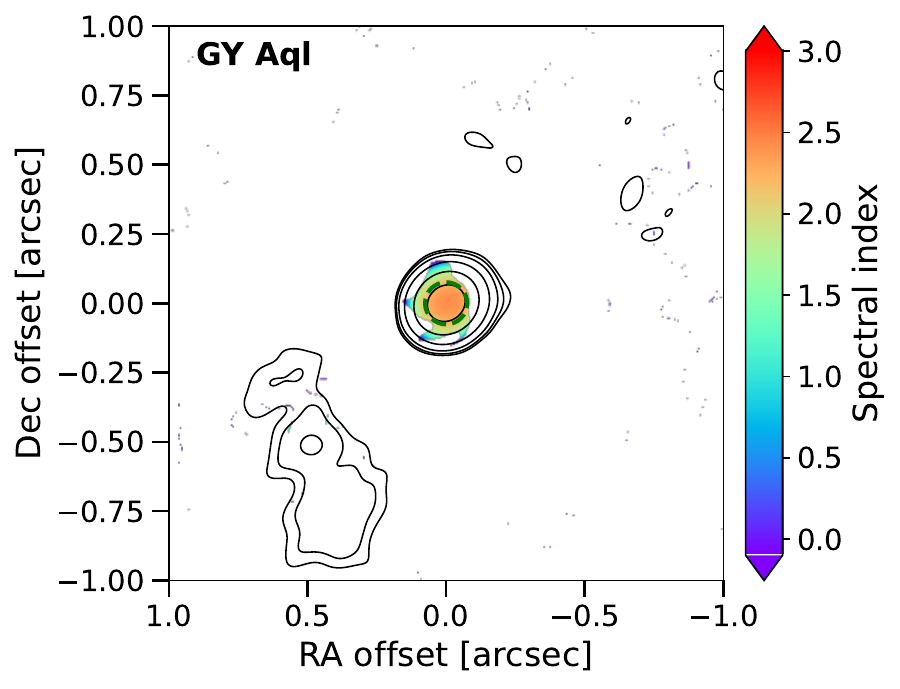}
 	\includegraphics[width=0.32\textwidth]{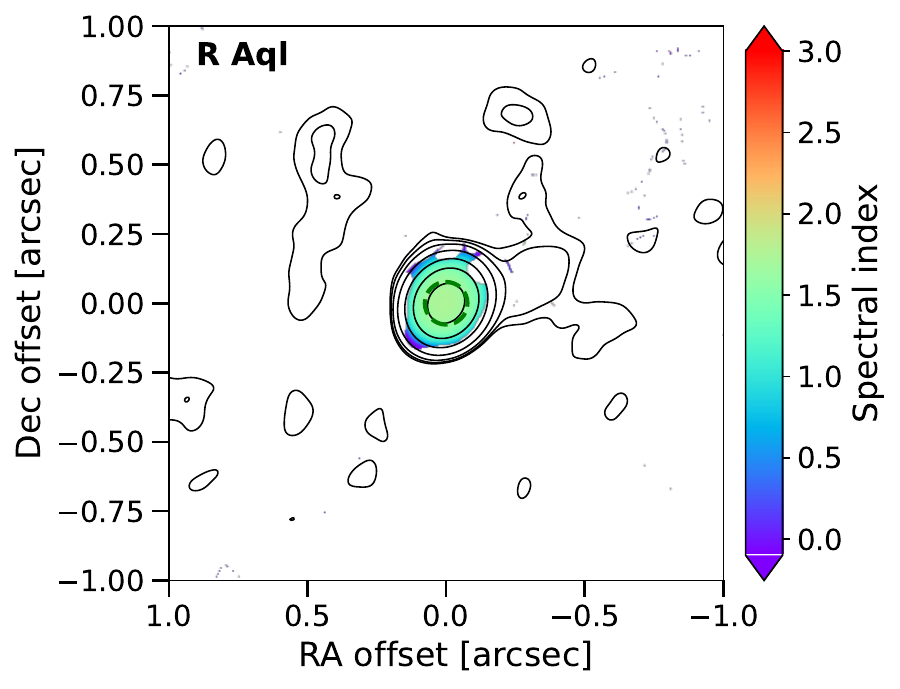}
  	\includegraphics[width=0.32\textwidth]{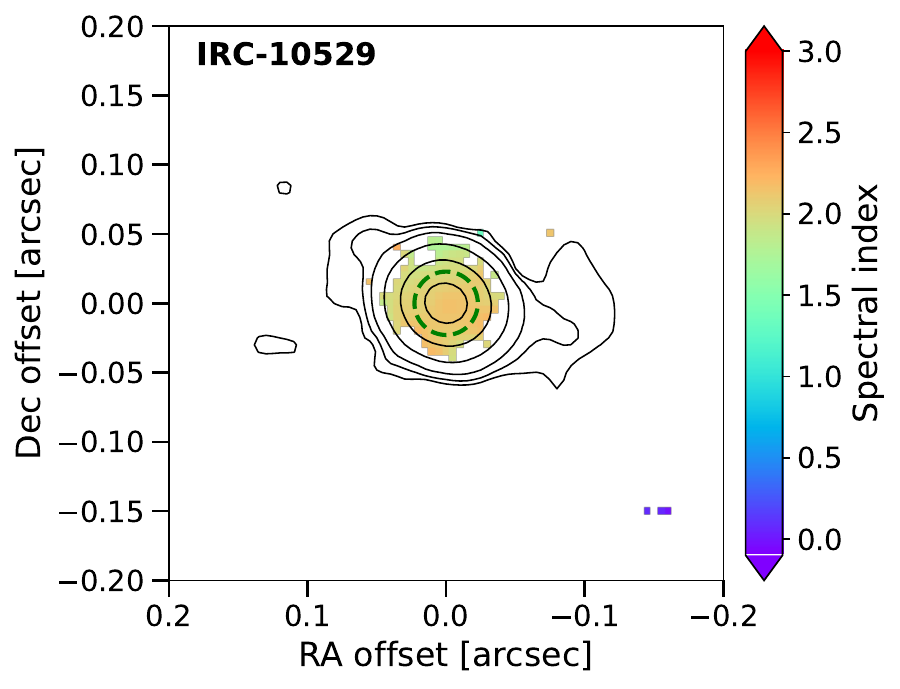}
    \includegraphics[width=0.32\textwidth]{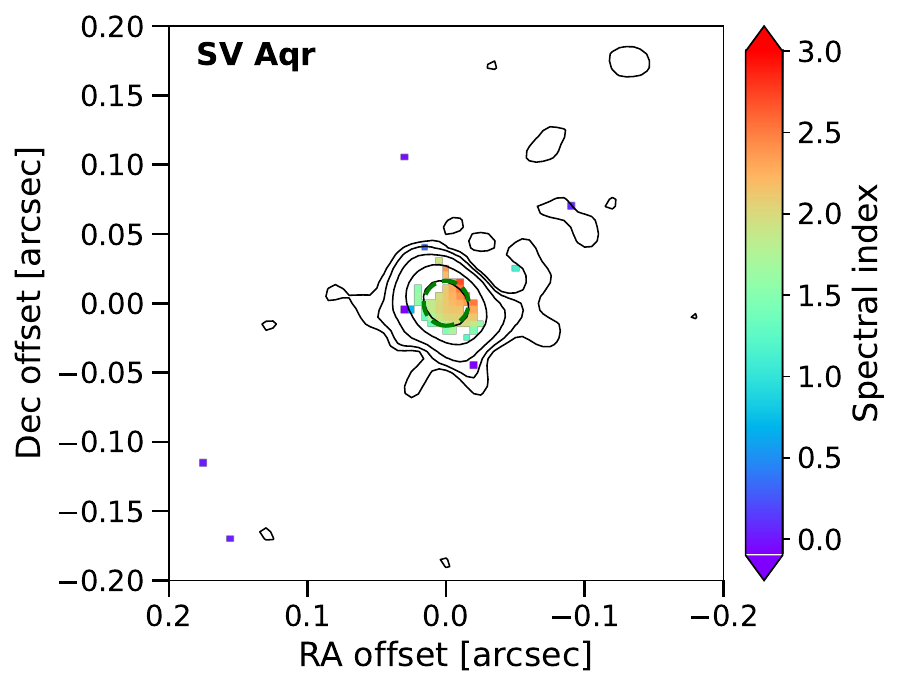}
    \includegraphics[width=0.32\textwidth]{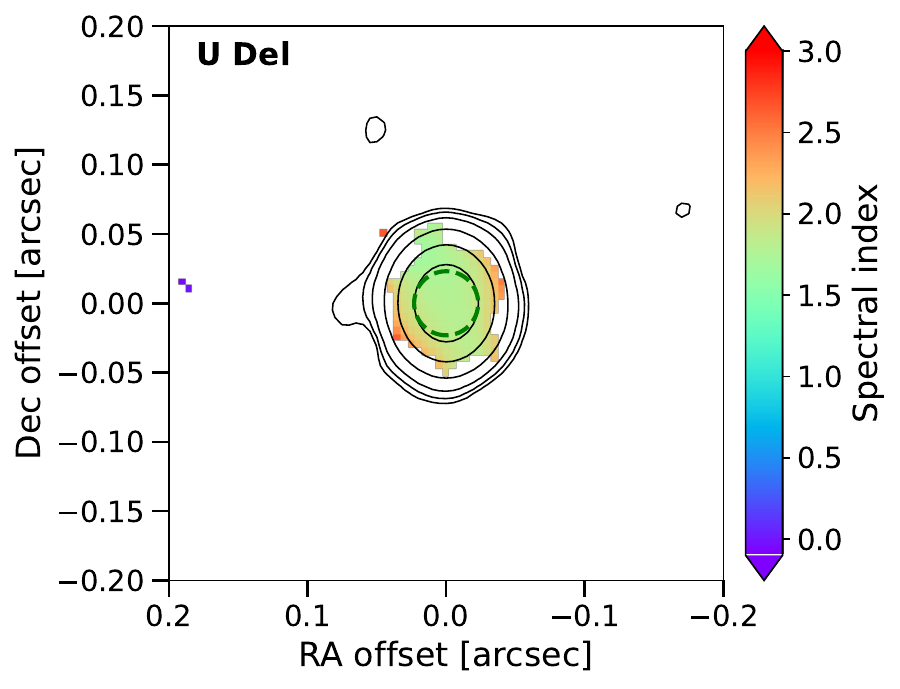}
    \includegraphics[width=0.32\textwidth]{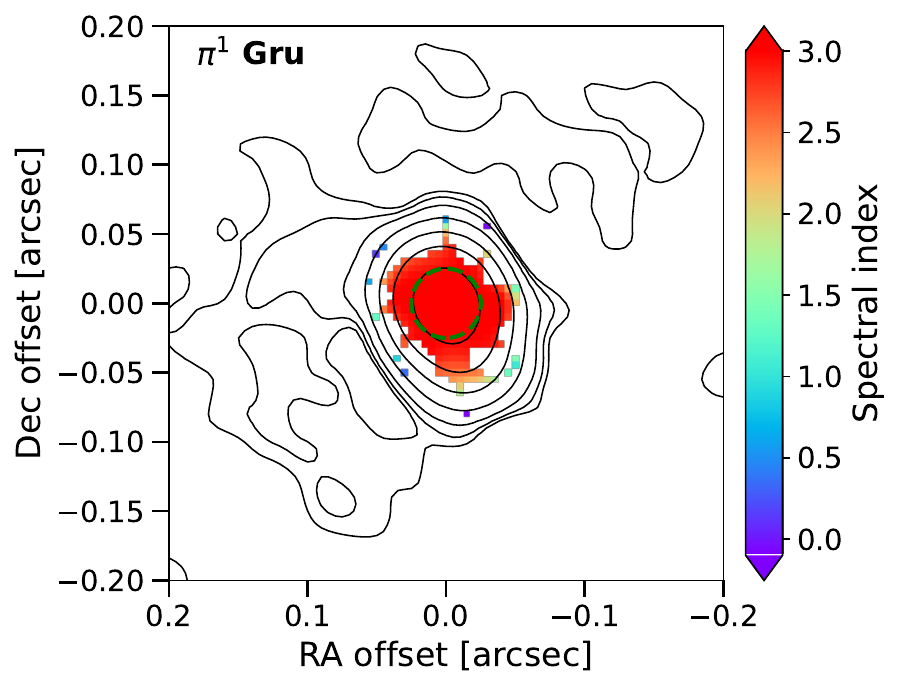}
    \includegraphics[width=0.32\textwidth]{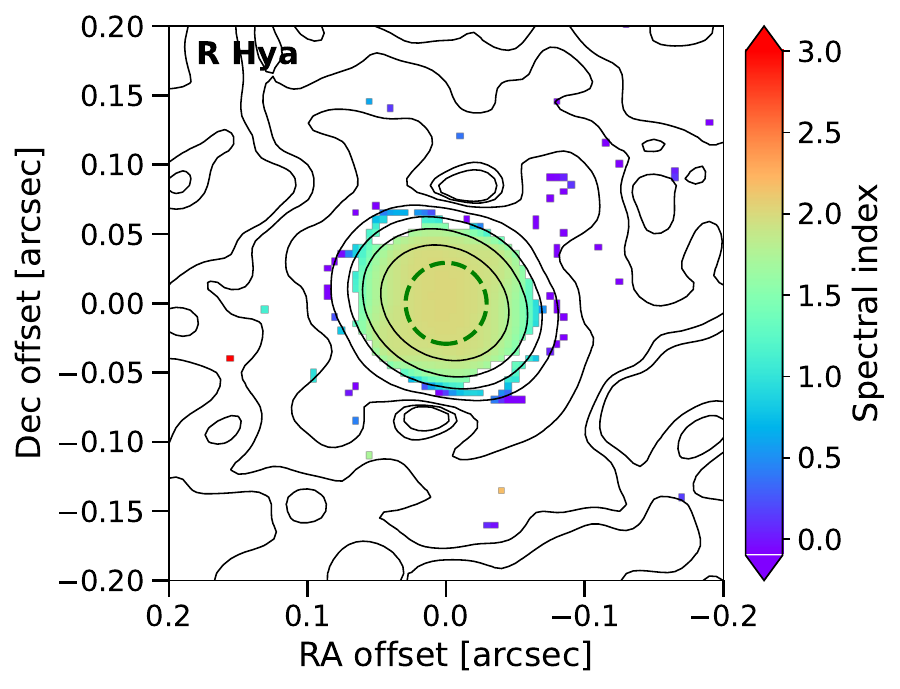}
    \includegraphics[width=0.32\textwidth]{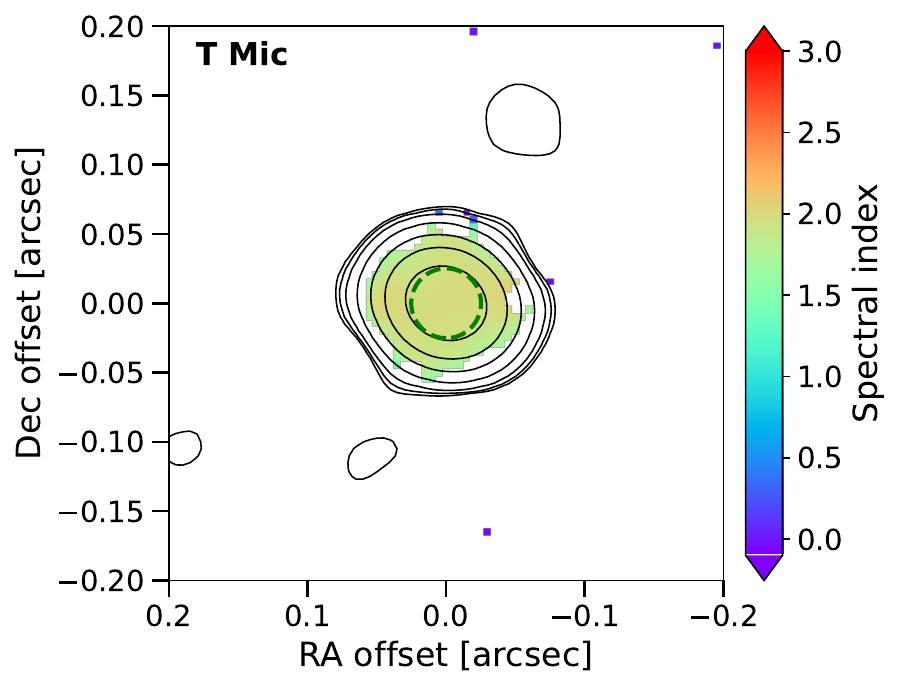}
    \includegraphics[width=0.32\textwidth]{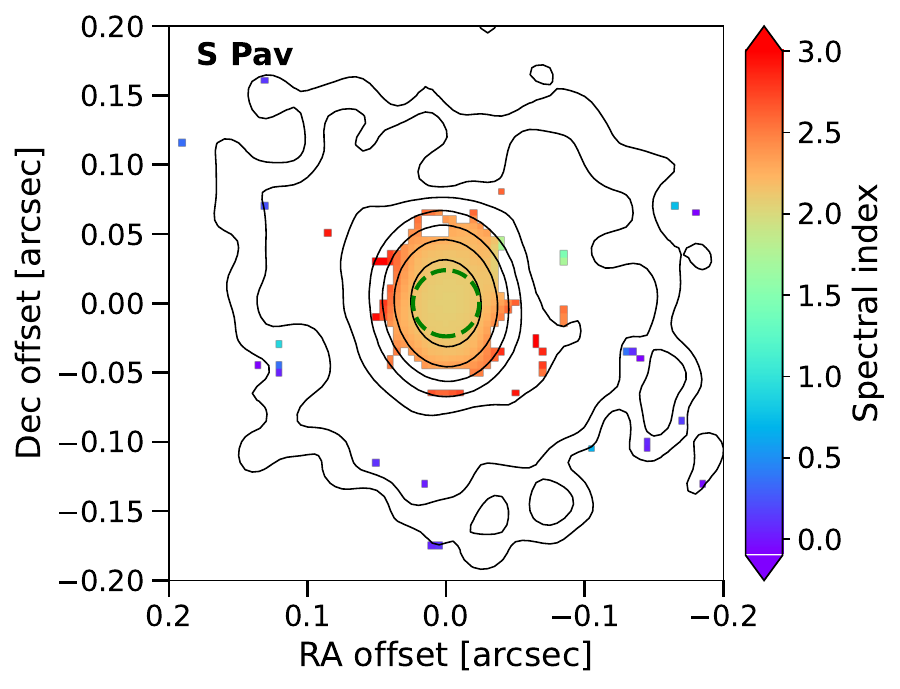}
    \includegraphics[width=0.32\textwidth]{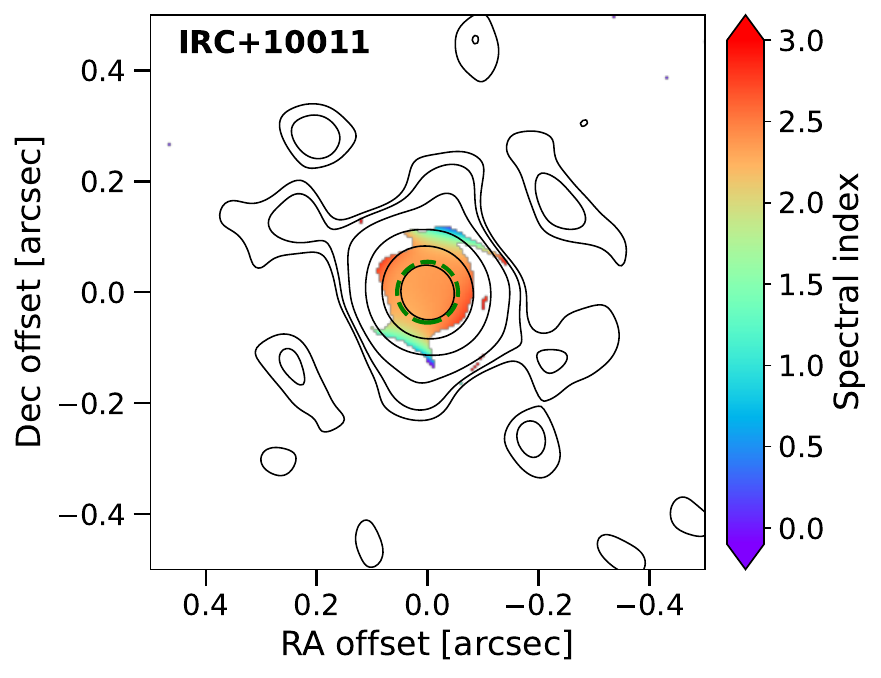}
    \includegraphics[width=0.32\textwidth]{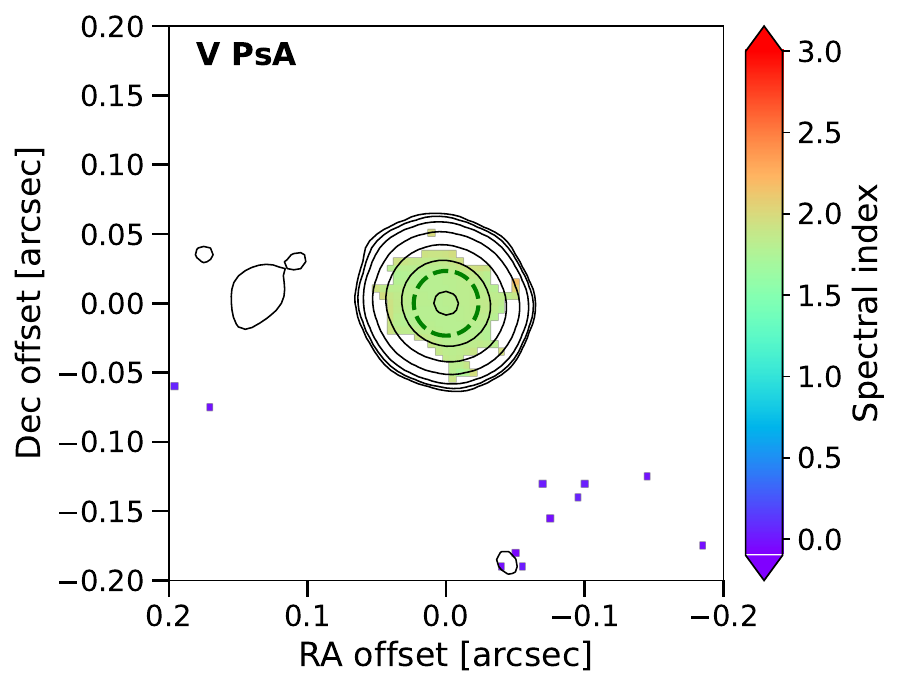}
    \includegraphics[width=0.32\textwidth]{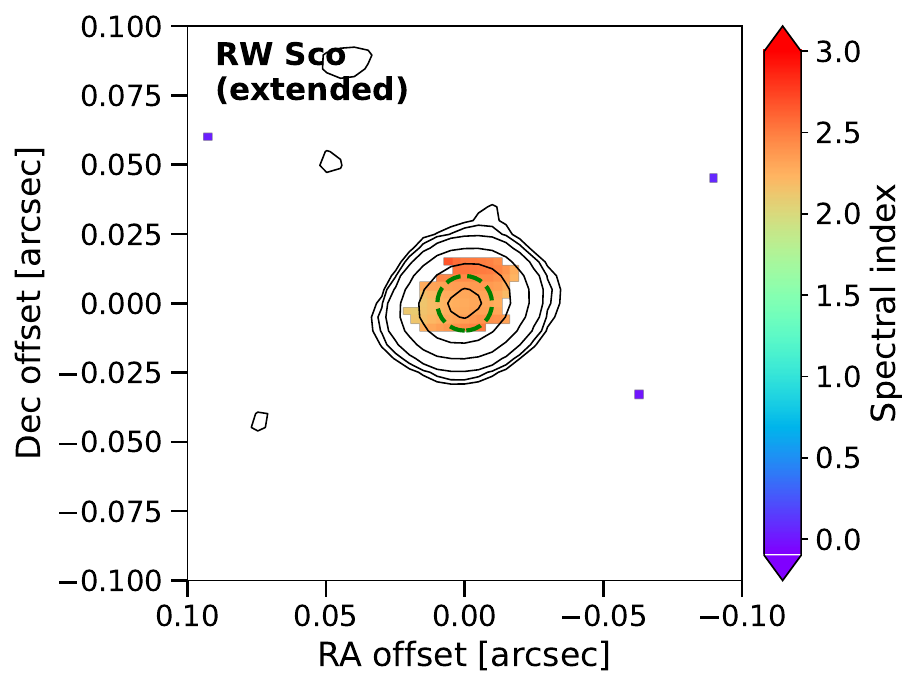}
    \includegraphics[width=0.32\textwidth]{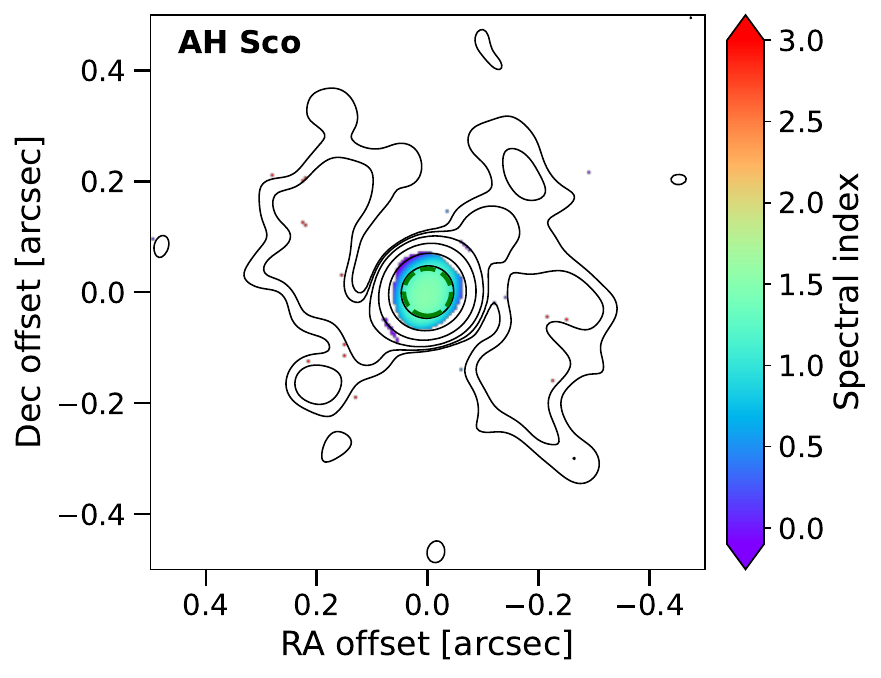}
    \includegraphics[width=0.32\textwidth]{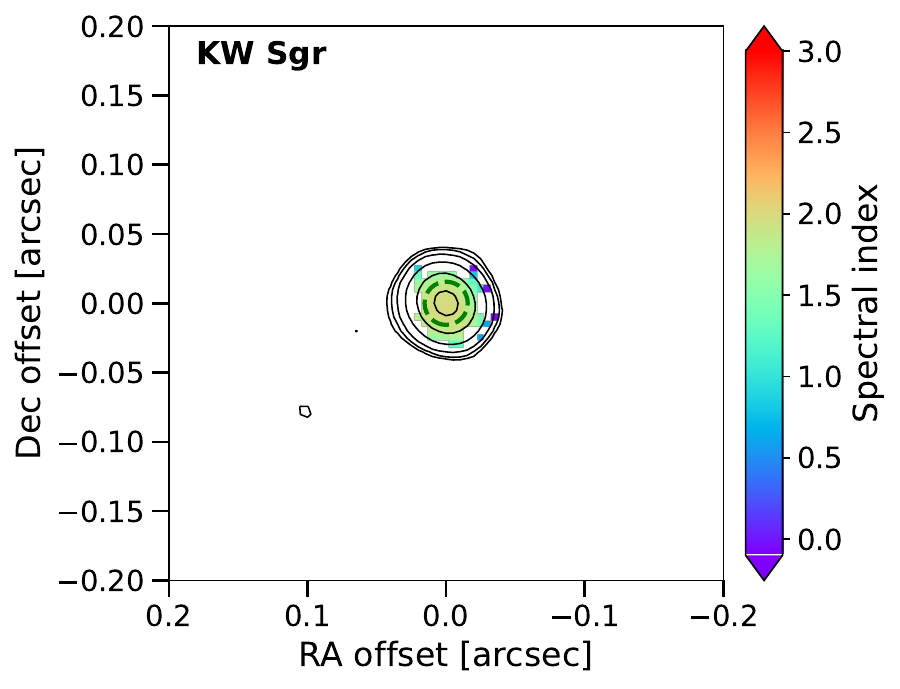}
    \includegraphics[width=0.32\textwidth]{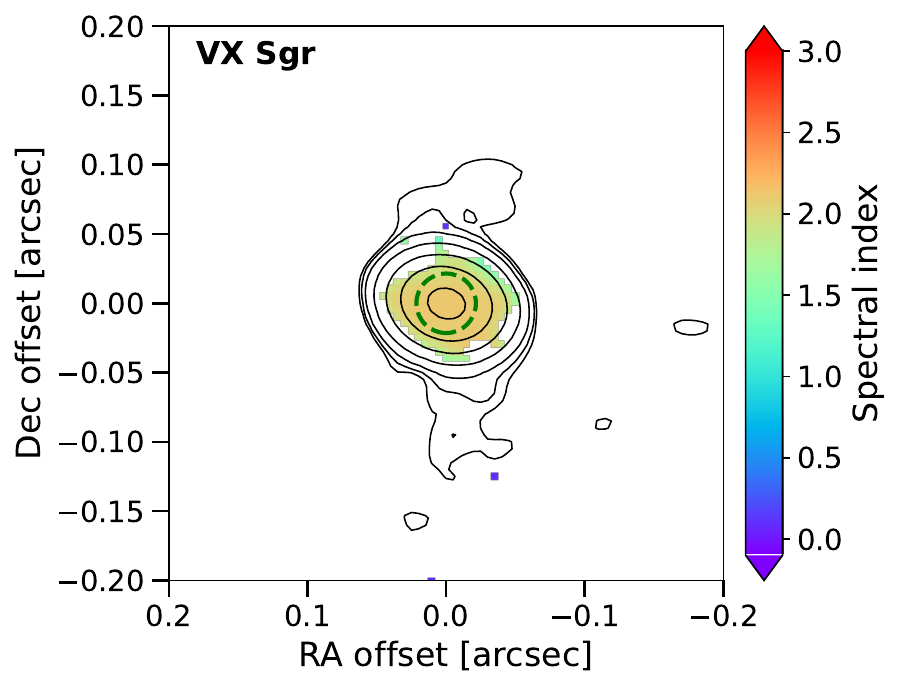}
    \caption{\edits{Spectral indices computed from the combined Band 6 data (or extended data in the case of RW Sco). The spectral index (colours) is shown for the pixels where it is well-defined. The dashed green circle indicates the area, centred on the continuum peak, equal to the beam size convolved with the UD radius. Solid black contours indicate the combined continuum. Similar plots for W~Aql and U~Her are given in Fig.~\ref{fig:spec-examples}.}}
    \label{fig:spec-all}
\end{figure*}

\begin{figure*}
\centering
	\includegraphics[width=0.245\textwidth]{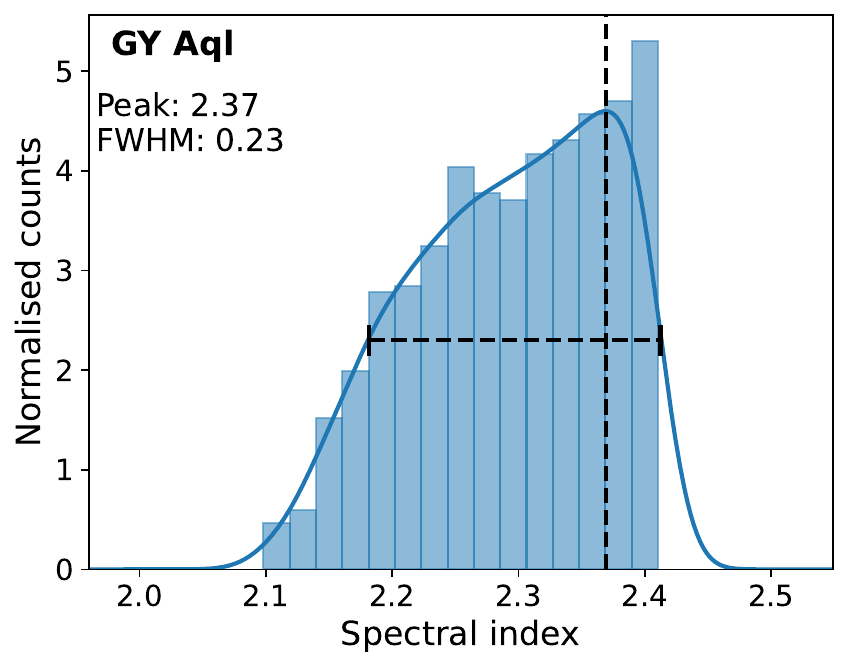}
 	\includegraphics[width=0.245\textwidth]{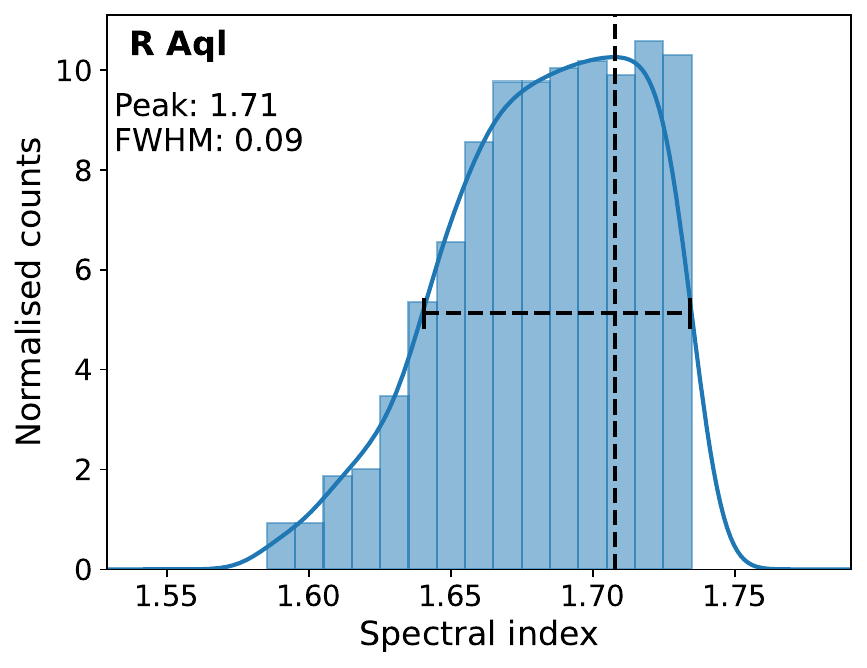}
  	\includegraphics[width=0.245\textwidth]{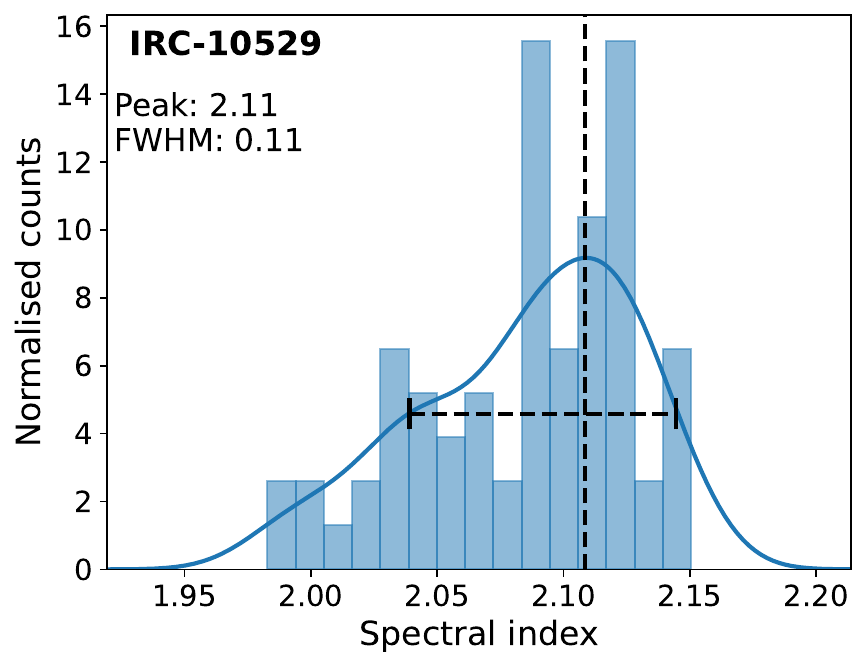}
	 \includegraphics[width=0.245\textwidth]{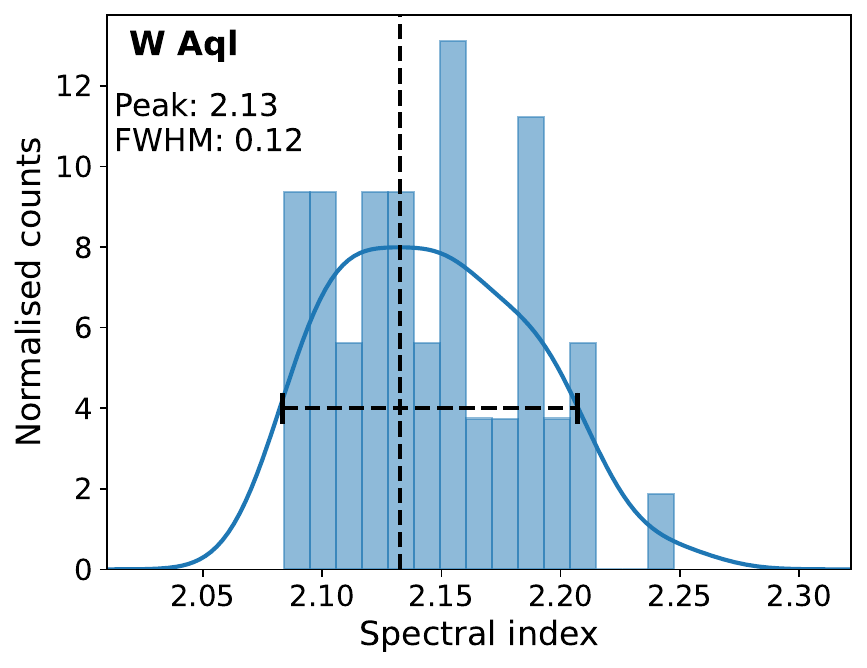}
    \includegraphics[width=0.245\textwidth]{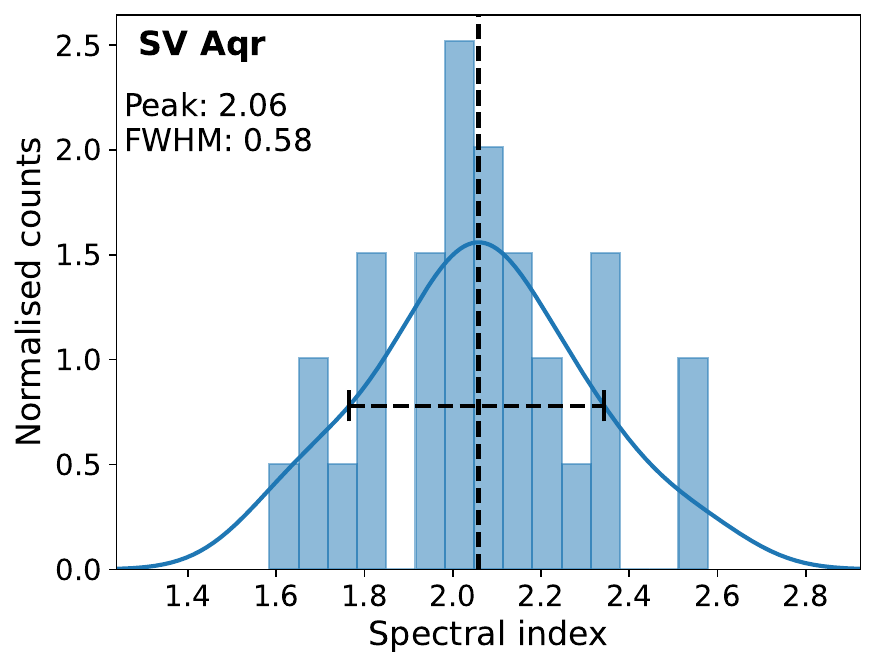}
    \includegraphics[width=0.245\textwidth]{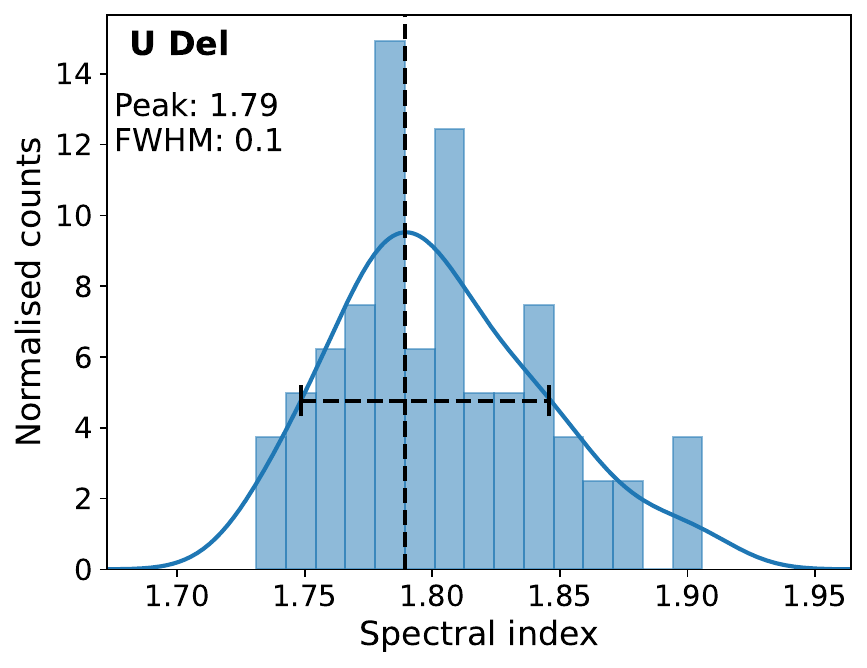}
    \includegraphics[width=0.245\textwidth]{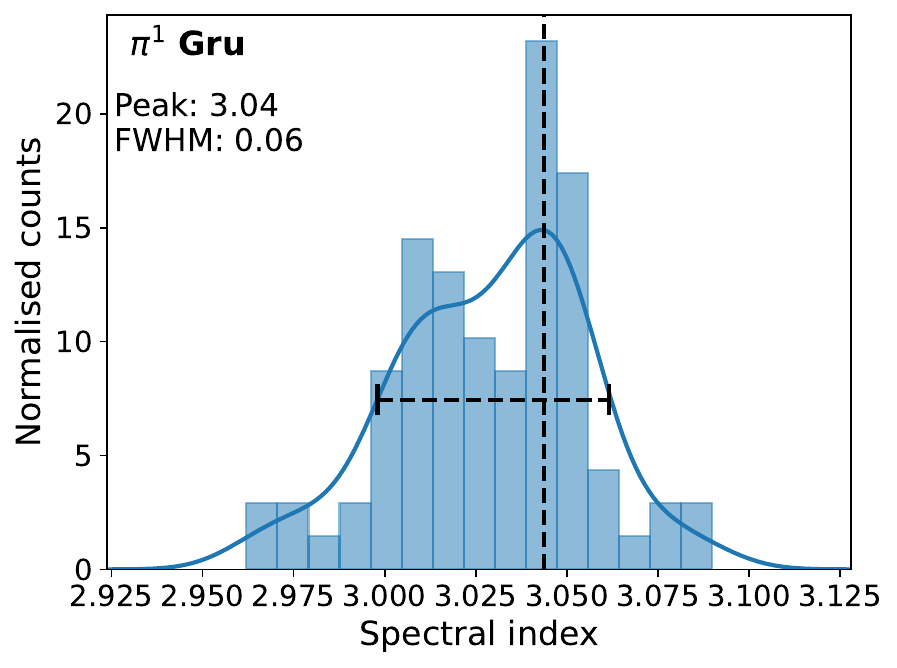}
    \includegraphics[width=0.245\textwidth]{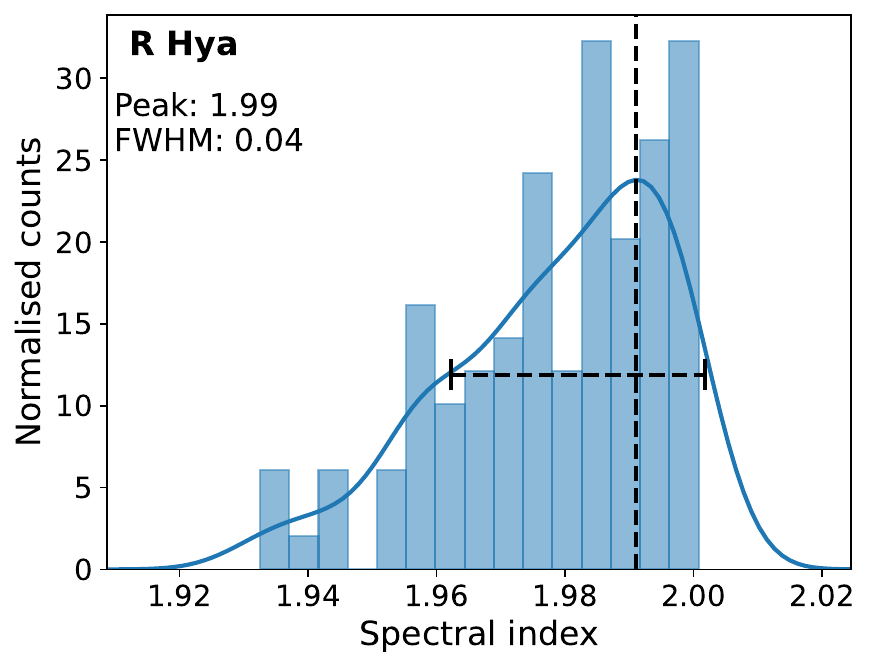}
    \includegraphics[width=0.245\textwidth]{R_Hya_B6_combined_spec_index_histogram_centre-conv-beam.pdf}
    \includegraphics[width=0.245\textwidth]{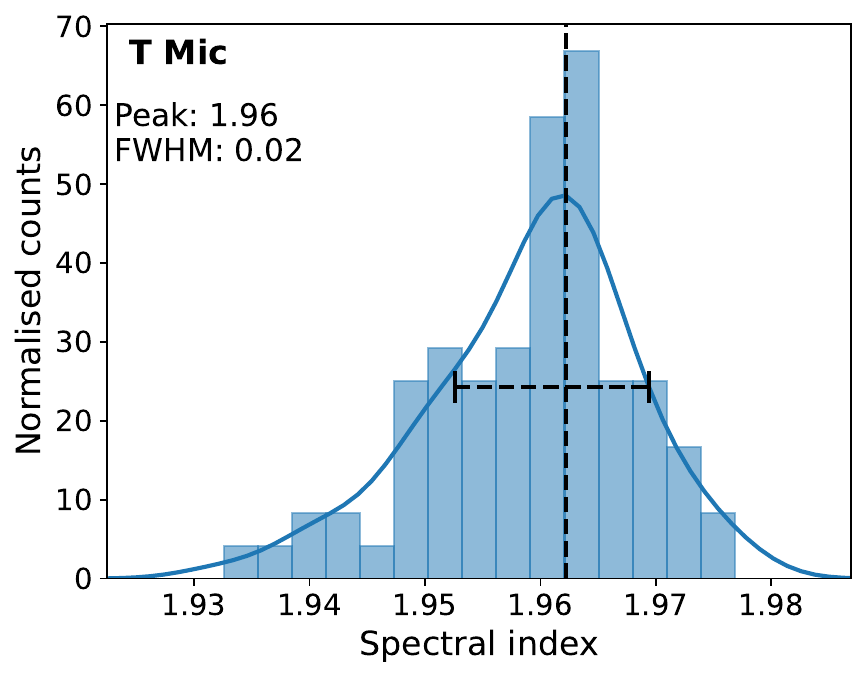}
    \includegraphics[width=0.245\textwidth]{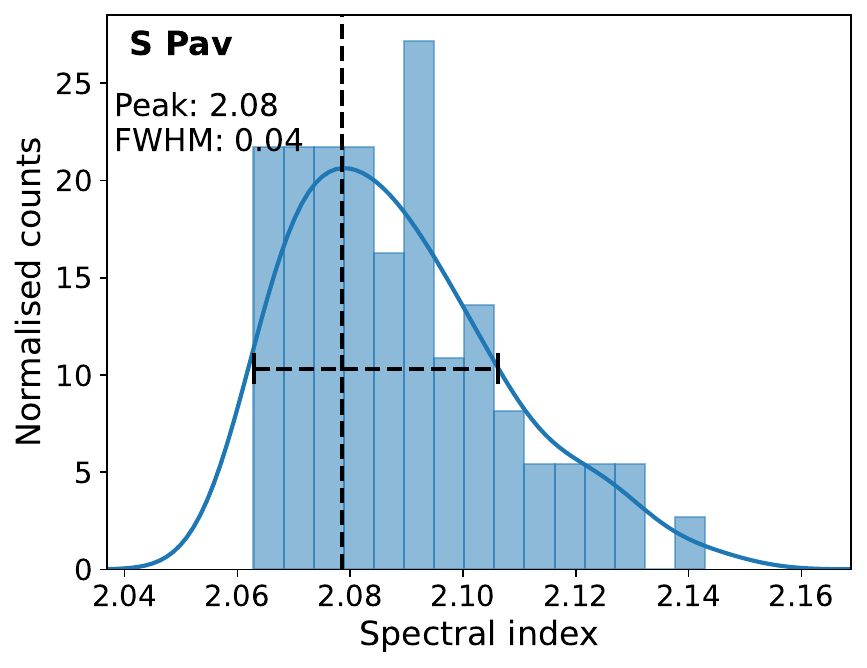}
    \includegraphics[width=0.245\textwidth]{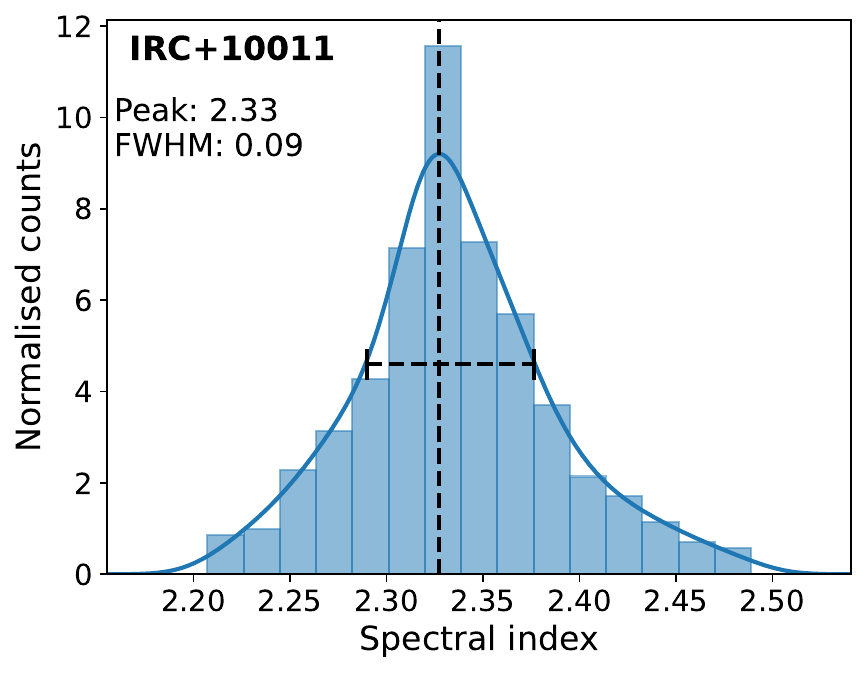}
    \includegraphics[width=0.245\textwidth]{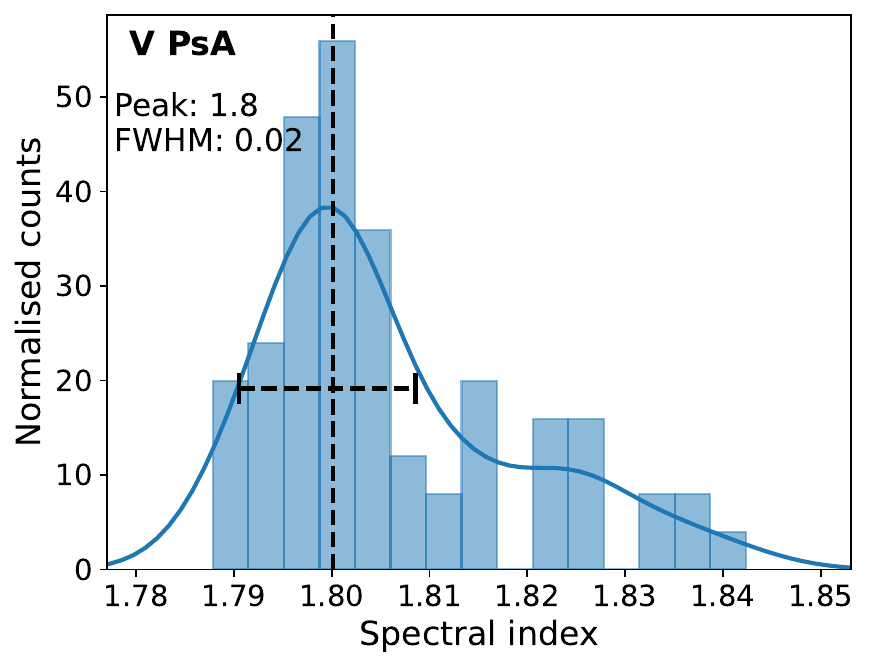}
    \includegraphics[width=0.245\textwidth]{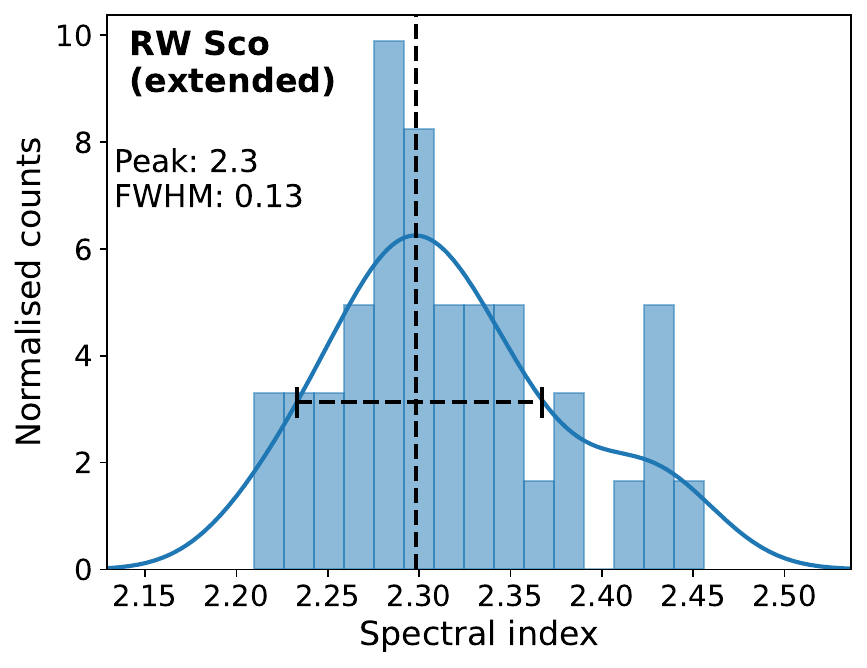}\\
    \includegraphics[width=0.24\textwidth]{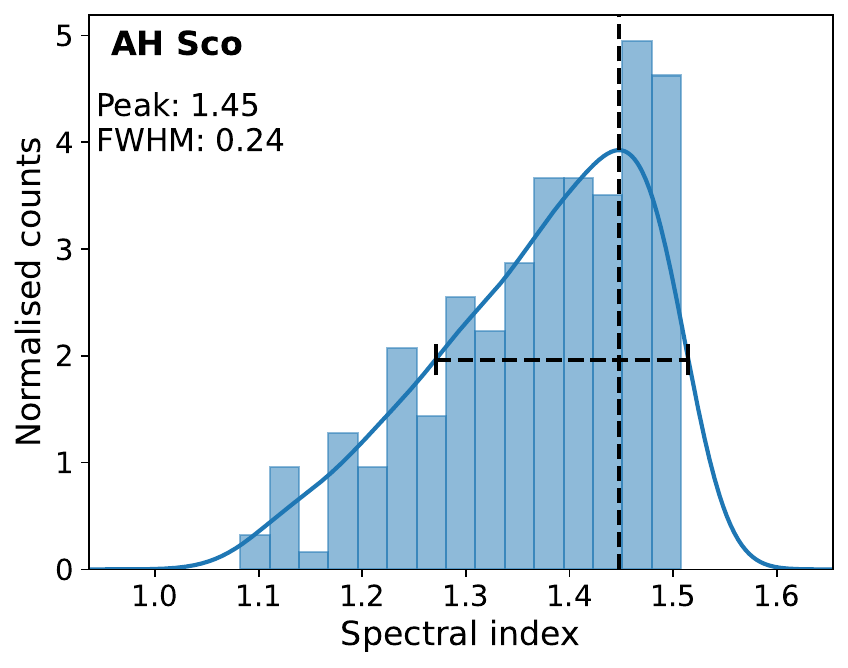}
    \includegraphics[width=0.25\textwidth]{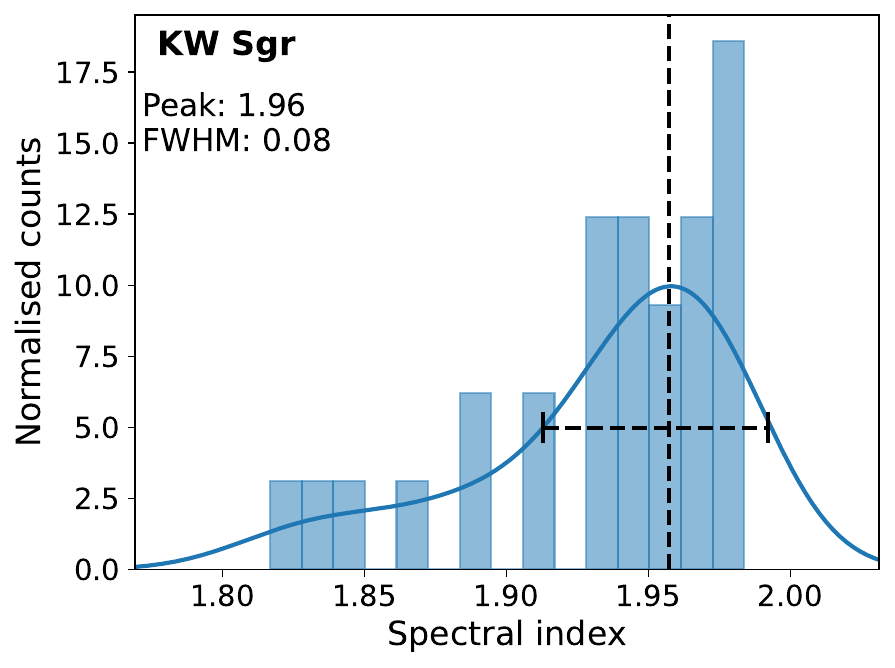}
    \includegraphics[width=0.245\textwidth]{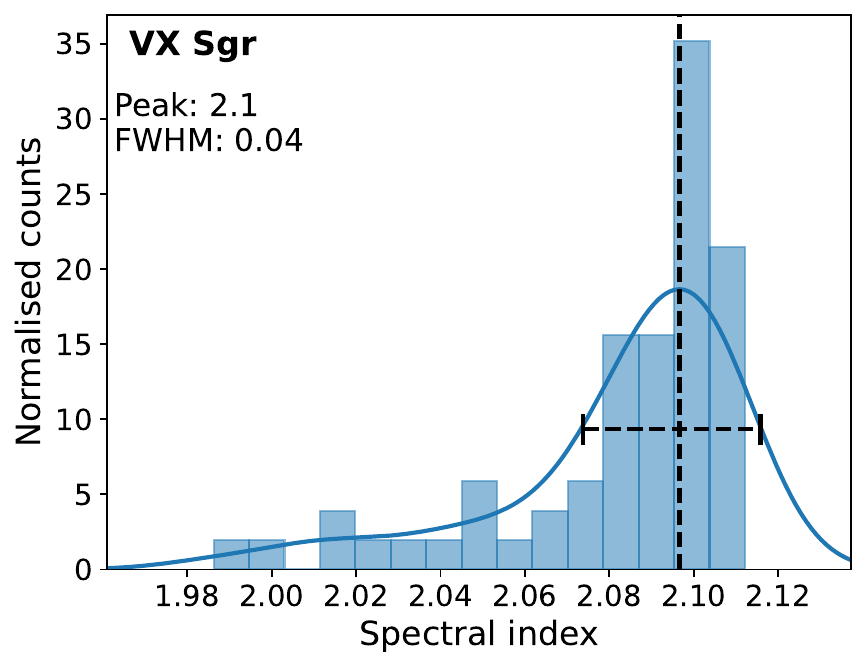}
    \caption{\edits{Kernel density estimates (KDEs) for the spectral indices computed from the combined Band 6 data (or extended data in the case of RW Sco). The vertical dashed line marks the peak of the KDE and the horizontal dashed line indicates the full width at half maximum (FWHM). Note the differing x-axes. The bottom row contains the three RSG stars.}}
    \label{fig:spec-kdes}
\end{figure*}

%\begin{figure*}
%\centering
%    \includegraphics[width=0.75\textwidth]{Massloss_vs_spec_ind.pdf}
%%    \hspace{2cm}
%    \caption{\edits{The intra-Band 6 spectral indices plotted against mass-loss rates. The dashed black line traces a weak positive correlation following $\alpha_\mathrm{B6} = 0.16 \log \left(\dot{M}\right)+3.0$. Star names are given along the bottom of the plot. \red{don't keep this plot?}}}
%    \label{fig:spec-mloss}
%\end{figure*}

\FloatBarrier
\clearpage
%\afterpage{\clearpage}

\section{RW Sco ACA continuum}\label{res:rwsco}

All the continuum maps for RW Sco are relatively featureless, with no significant extended emission above $5\sigma$ (Fig.~\ref{fig:rwsco}). The exception is the ACA Band 3 map, shown in Fig.~\ref{fig:aca}, where a second peak is seen 27\arcsec\ south of the AGB continuum peak (outside of the 12m array field of view). To mitigate the impact of this feature we produced a continuum map that was not primary beam corrected, unlike the majority of the other maps plotted in this study. This is what is shown in Fig.~\ref{fig:aca} and was necessary to avoid the secondary peak dominating the continuum. 

From a simple 2D Gaussian fit, we determine the position of the secondary peak to be RA, Dec = 17:14:51.76, $-$33:26:21.61 (ICRS).
We cross-checked the position of the secondary peak with SIMBAD \citep{Wenger2000} and with Gaia DR3 \citep{Gaia-DR3-2023}. There were no candidates with entries in SIMBAD, but we did find two possibilities in Gaia: DR3 5979057409127540096, which is 1.02\arcsec\ from the ALMA peak and DR3 5979057409141332608, which is 1.79\arcsec\ from the ALMA peak and appears to be associated with a faint source visible in the 2MASS Ks image, but with no assigned 2MASS designation. This second source has a five-parameter solution in Gaia and a calculated distance of $\sim5$~kpc \citep{Bailer-Jones2021}. Based on this and the Gaia colours, we judge that it is likely to be an early red giant branch star, and hence unlikely to be of comparable brightness to an AGB at 105 GHz.

Our other candidate has a slightly better agreement with the ALMA position, so we tentatively assign Gaia DR3 5979057409127540096 at RA, Dec = 17:14:51.68, $-$33:26:21.18 to the secondary ACA peak. The Gaia G-band mean magnitude of this source is 20.9 mag, very close to its detection threshold \citep[G = 20.7 mag,][]{Gaia-2016}. %\blue{indicating that it may be a sub-mm galaxy? - Laura Driessen will hopefully check low freq radio catalogue}
For all calculations based on the ACA data, we ignore this secondary peak and focus only on the AGB star.

%We concluded \red{because of Anita - check notes or ask her} that this secondary peak is associated with a background sub-mm galaxy and is not connected with the AGB star, except for through a coincidental line of sight alignment.
%
%\blue{Anita: RW Sco I have no recollection of a background galaxy.... did we check the position in Simbad? what is the position? Or if you give me a flux density I can estimate a probability.}

\section{ACA Band 3 molecular lines}\label{sec:acalines}

\begin{figure*}
	\includegraphics[width=\textwidth]{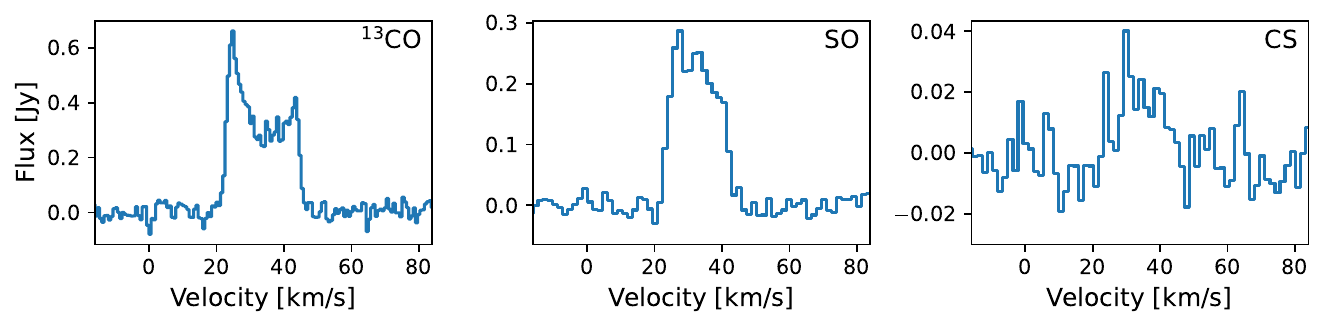}
    \caption{Spectral lines detected with the ACA towards GY Aql. Spectra were extracted for a circular aperture centred on the star, with a radius of 20\arcsec\ for $^{13}$CO and SO, and 10\arcsec\ for CS.}
    \label{fig:gyaqlmol}
\end{figure*}

\begin{figure*}
	\includegraphics[width=\textwidth]{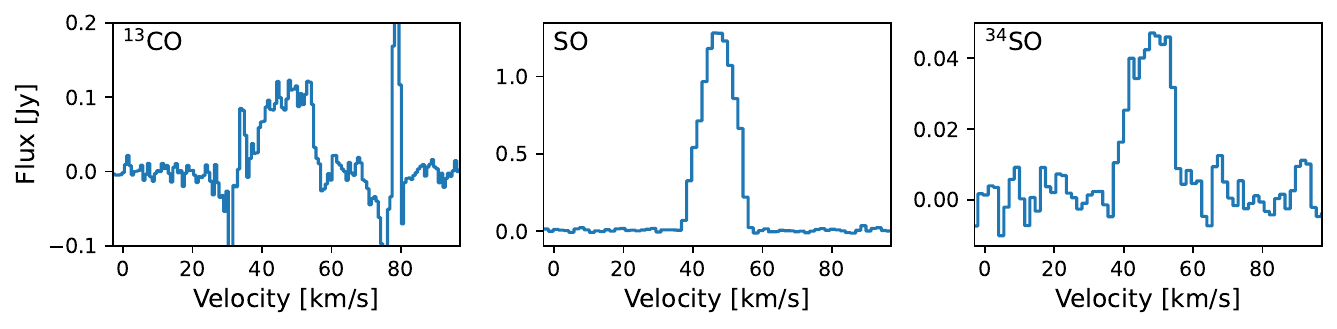}
    \caption{Spectral lines detected with the ACA towards R Aql. Spectra were extracted for a circular aperture centred on the star, with a radius of 20\arcsec\ for SO, and 10\arcsec\ for $^{13}$CO and CS. The narrow features in the $^{13}$CO spectrum are ISM contamination.}
    \label{fig:raqlmol}
\end{figure*}

In addition to the continuum images, we also obtained some line observations using the ACA.
We detected lines of $^{13}$CO, C$^{17}$O, $^{12}$CS, $^{32}$SO, $^{34}$SO, and, tentatively, C$^{18}$O towards at least one source each. For all sources at least one line was detected, although for \pigru\ we only tentatively detect one $^{13}$CO line. The molecular line detections, including transition details and integrated fluxes, are given in Table \ref{tab:acalines}. The line spectra are plotted in Figs.~\ref{fig:gyaqlmol}--\ref{fig:rwscomol}. Table \ref{tab:acalines} lists the lines detected in our ACA observations and their integrated fluxes. 

The $^{13}$CO line towards R~Aql suffers from ISM contamination, which we also see around the C$^{18}$O line frequency. However, aside from one narrow peak that we attribute to a contaminating foreground or background source, we do not detect circumstellar C$^{18}$O above the noise for R Aql. An analysis of the oxygen isotopic ratios for IRC$-$10529 is given in Sect.~\ref{sec:orats}, although we cannot constrain the initial mass to better than $\lesssim 4~\msol$.
%where we find an initial mass for this star of \red{$2.25~\msol$ -- check final value}.

%The observed spectra for GY~Aql, R~Aql, IRC$-$10529, U~Her, T~Mic and RW~Sco are plotted in Figs.~\ref{fig:gyaqlmol}--\ref{fig:rwscomol}.
%\ref{fig:raqlmol}, \ref{fig:irc-10529mol}, \ref{fig:uhermol}, \ref{fig:tmicmol}, and

\subsection{Oxygen isotopic ratios}\label{sec:orats}

\begin{figure*}
	\includegraphics[width=\textwidth]{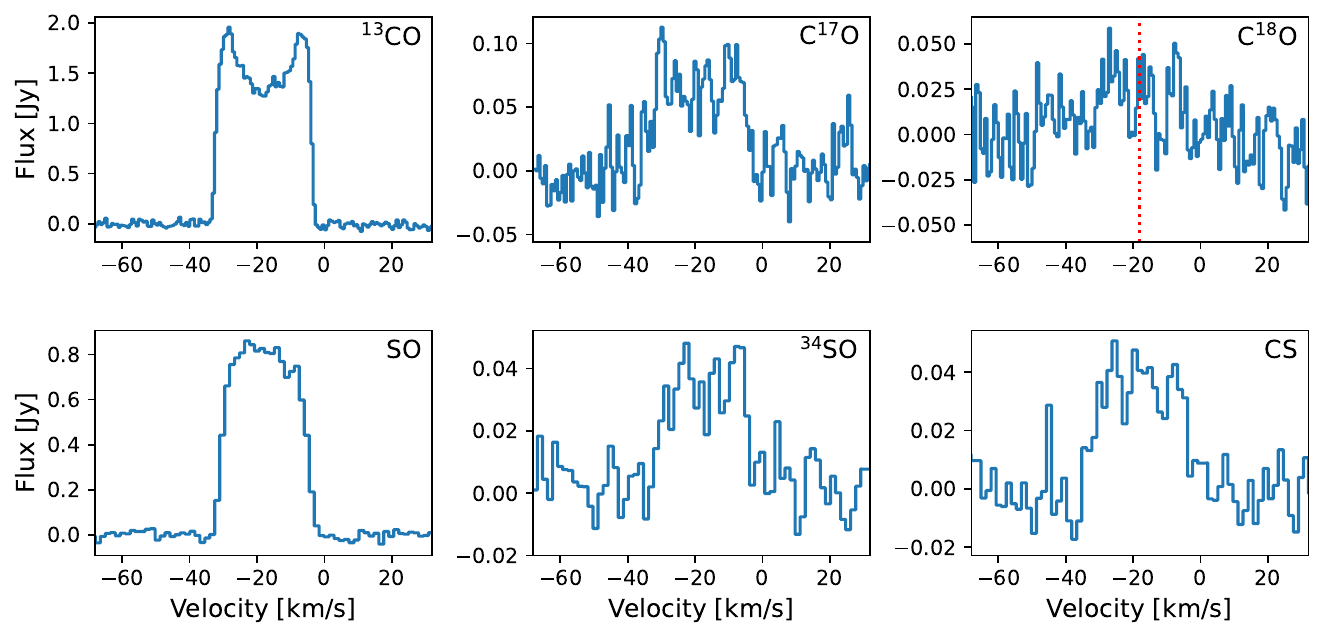}
    \caption{Spectral lines detected with the ACA towards IRC $-$10529. Spectra were extracted for a circular aperture centred on the star, with a radius of 20\arcsec\ for $^{13}$CO and SO, and 10\arcsec\ for the other lines. For the tentative C$^{18}$O line, we also plot a dashed red vertical line to indicate the LSR velocity of the star.}
    \label{fig:irc-10529mol}
\end{figure*}

\begin{figure*}
	\includegraphics[width=\textwidth]{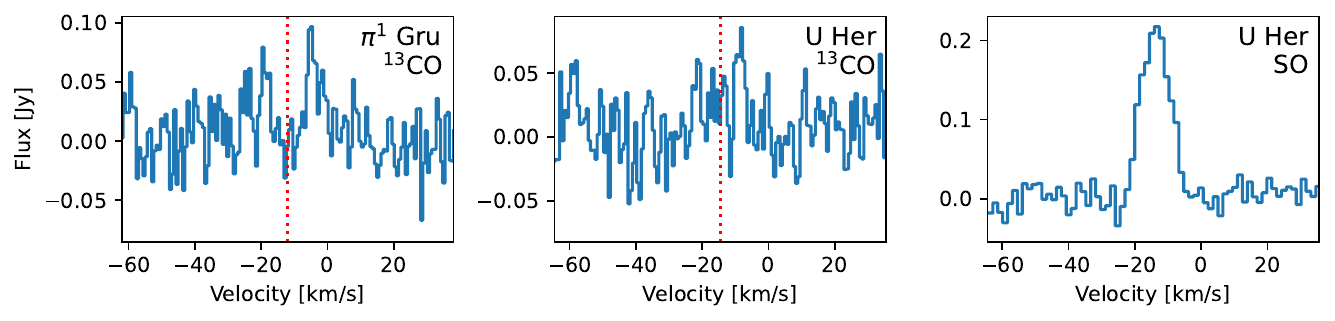}
    \caption{Tentative detections of $^{13}$CO with the ACA towards \pigru\ and U Her, and a certain detection of SO towards U~Her. Spectra were extracted for a circular aperture centred on the star, with a radius of 15\arcsec. For the tentative $^{13}$CO we also plot a dashed red vertical line to indicate the LSR velocity of the stars.}
    \label{fig:uhermol}
\end{figure*}

\begin{figure*}
\centering
	\includegraphics[width=0.66\textwidth]{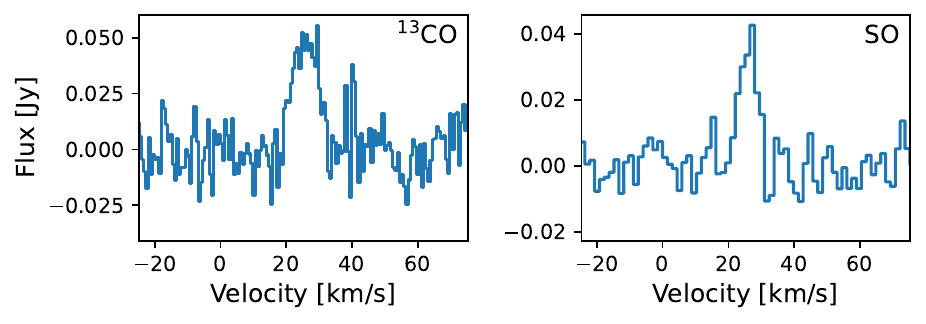}
    \caption{Spectral lines detected with the ACA towards T Mic. Spectra were extracted for a circular aperture centred on the star, with a radius of 10\arcsec.}
    \label{fig:tmicmol}
\end{figure*}

\begin{figure*}
\centering
	\includegraphics[width=0.66\textwidth]{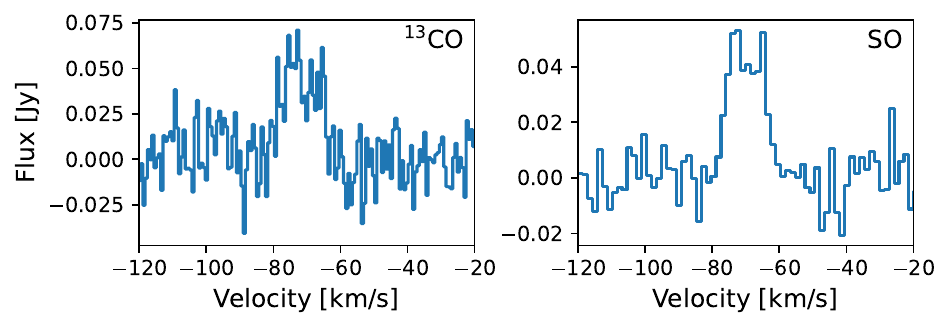}
    \caption{Spectral lines detected with the ACA towards RW Sco. Spectra were extracted for a circular aperture centred on the star, with a radius of 10\arcsec.}
    \label{fig:rwscomol}
\end{figure*}

The original goal of our ACA observations was to estimate initial stellar masses based on the oxygen isotopic ratios $^{17}$O/$^{18}$O \citep{Karakas2016,De-Nutte2017,Danilovich2017} from CO observations. However, as noted in Table~\ref{tab:acalines}, we only detected C$^{17}$O towards one source, IRC$-$10529, and only tentatively detected C$^{18}$O towards the same source (Fig.~\ref{fig:irc-10529mol}). Hence, we can only attempt to estimate the initial stellar mass of this source.

We estimate the isotopic abundance ratio based on the CO isotopologue ratio C$^{17}$O/C$^{18}$O. This is valid because we do not expect chemical fractionation between the different isotopologues and because the lines of these low-abundance species are optically thin \citep[as suggested by the clearly double-peaked $^{13}$CO line profile,][which is expected to have a higher abundance than C$^{17}$O and C$^{18}$O]{Olofsson2003}. To obtain the isotopic abundance ratio, we also need to account for the difference in line strengths between the two isotopologues, using \citep{Danilovich2020}
\begin{equation}
^{17}\mathrm{O}/^{18}\mathrm{O} = \frac{F_{17}}{F_{18}}\left( \frac{\nu_{18}}{\nu_{17}}\right)^2
\end{equation}
where $F$ is the integrated flux of the C$^{17}$O or C$^{18}$O line and $\nu$ is the corresponding line frequency (given in Table \ref{tab:acalines}).

Although the $^{13}$CO line has a high S/N (65), the S/N are much lower for C$^{17}$O (5) and C$^{18}$O (1.5), when considering spectra extracted for a circular aperture of 10\arcsec. This makes the line flux determinations less reliable for the rarer isotopologues, especially C$^{18}$O, even when we choose the integration limits based on the well-defined limits of the $^{13}$CO line ($-34$ to $-2$~\kms). We find $F_{17}=1.96\pm0.57$~Jy~\kms\ and $F_{18}=0.59\pm0.54$~Jy~\kms\ with uncertainties based on the rmses of the spectra. This gives a ratio of $3.2\pm3.1$ owing to the large uncertainty on the C$^{18}$O line. This does not allow us to put tight constraints on the initial mass of IRC$-$10529, and we find $M_i = 2.25_{-1.25}^{+1.75}~\msol$, assuming solar metallicity.
The fact that we do weakly detect the C$^{18}$O indicates that the star has not undergone hot bottom burning, which destroys $^{18}$O, and hence puts an upper limit on the initial mass of $\lesssim 4~\msol$ \citep{Karakas2016}.

Not all the ATOMIUM stars were included in our ACA observations, in some cases because initial mass estimates based on oxygen isotopic ratios had already been made. This includes W~Aql and IRC+10011, which \cite{De-Nutte2017} estimated to have initial mass of $1.6~\msol$ for W~Aql and $\sim1~\msol$ for IRC+10011. However, based on the higher S/N observations of \h2$^{18}$O and \h2$^{17}$O with \textsl{Herschel}/HIFI \citep{Justtanont2012}, the initial mass of IRC+10011 may be closer to $1.5~\msol$.

%IRC $-$10529 mass from O isotopic ratios: $2.25$? 17/18O ratio is 3.13, which is close to the peak in the stellar models \cite{De-Nutte2017}. May suggest a lower metallicity \cite{Karakas2016} and Fig.~\ref{fig:oratios}.

\begin{table*}
	\centering
	\caption{An overview of the molecular lines observed with the ACA.}
	\label{tab:acalines}
	\begin{tabular}{lcrcccccccc}
		\hline
Molecule	&	Line	&	Freq [GHz] & Ref	&	GY Aql	&	R Aql	&	IRC $-$10529	&	$\pi^1$ Gru	&	U Her	&	T Mic	&	RW Sco	\\	
\hline
$^{13}$CO	&	$J=1\to0$	&	110.201	&	1	&	8.5	&	1.8 (ISM)	&	44.8	&	0.8 T	&	0.6 T	&	0.4	&	0.6	\\
C$^{17}$O	&	$J=1\to0$	&	112.359	&	2	&	x	&	x	&	2.1	&	x	&	x	&	x	&	x	\\
C$^{18}$O	&	$J=1\to0$	&	109.782	&	3	&	x	&	ISM	&	0.6	T &	x	&	x	&	x	&	x	\\
CS	&	$J=2\to1$	&	97.981	&	4	&	0.3 T	&	x	&	1.0	&	x	&	x	&	x	&	x	\\
SO	&	$N_J=2_3 \to 1_2$	&	99.300	&	5	&	4.2	&	15.4	&	20.0	&	x	&	2.3	&	0.2	&	0.6	\\
$^{34}$SO	&	$N_J=2_3 \to 1_2$	&	97.715	&	6	&	0.3	T &	0.9	&	1.0	&	x	&	x	&	x	&	x	\\
		\hline
	\end{tabular}
\tablefoot{Non-detections are marked with `x' and `T' indicates tentative detections. Numbers in the star columns are the integrated flux in Jy \kms. ISM refers to contamination in the spectrum, in the form of narrow features thought to arise from foreground or background sources. The `Ref' column indicates references for the line frequencies. The numbers correspond to: (1) \cite{Cazzoli2004};
(2) \cite{Cazzoli2002};
(3) \cite{Cazzoli2003};
(4) \cite{Gottlieb2003};
(5) \cite{Tiemann1974};
(6) \cite{Tiemann1982}.}
\end{table*}

%\begin{figure}
%    \includegraphics[width=0.5\textwidth]{Oratios_Karakas.pdf}
%    \caption{INSERT HERE. Data from \citet{Karakas2016}}\label{fig:oratios}
%\end{figure}

\FloatBarrier
\clearpage
%\afterpage{\clearpage}

\section{Calculating brightness temperatures}\label{sec:Tbright}

The brightness temperature, in SI units, is given by \citep{Condon2016}:
\begin{equation}
T_{\mathrm{B}} = \frac{S_{\mathrm{W}} \lambda^2} {2 k_{\mathrm{B}} \Omega}
\label{eq:Tbasic}
\end{equation}
where Boltzmann's constant $k_{\mathrm{B}} = 1.38\times10^{-23}$ W K$^{-1}$ Hz$^{-1}$, and the flux density, $S_\mathrm{W}$ in Watts, is measured at wavelength, $\lambda$ in m, from an area, $\Omega$ in steradian.

The area of a uniform disc is $\pi \times R_\mathrm{UD}^2$ where the radius, $R_\mathrm{UD}$, is in radians.  For a flux density of $S_\mathrm{mJy}$ in mJy, at frequency $\nu$ GHz, and for a UD with diameter $D_\mathrm{mas}$ in mas,   Eq~\ref{eq:Tbasic} becomes
\begin{eqnarray}
T_{\mathrm{B}} &=& \frac{\frac{S_\mathrm{mJy}}{10^{29}} \left(\frac{c}{10^9\nu}\right)^2}
{2 k_{\mathrm{B}}\,\, \pi \left(\frac{0.5D_\mathrm{mas}}{3600000} \frac{\pi}{180}\right) ^2}\\
         & = & 1.764\times10^9 \,\, \frac{S_\mathrm{mJy}}{\nu^2 D_\mathrm{mas}^2}
\label{eq:Tbfull}
\end{eqnarray}
where in the second line all the constant terms have been gathered. 
The uncertainty is then given by the uncertainties on the flux and diameter: $T_{\mathrm{B}} \times \sqrt{(\sigma_{\mathrm{S}}/S)^2 + 2 (\sigma_\mathrm{D}/D)^2}$.

%This was done using the formula \citep{Vlemmings2017}
%\begin{equation}
%T_b = \frac{1.96\e{4} S_\mathrm{mJy} c^2}{\nu^2 \theta^2}
%\end{equation}
%where $S_\mathrm{mJy}$ is the UD flux density (in mJy), $\theta$ is the UD diameter (in arcsec), $\nu=241.75$~GHz is the central frequency of our combined data, and $c$ is the speed of light. 

\section{Apparent extent of bright unresolved sources}\label{sec:appext}

\edits{
The apparent size of a source observed with ALMA (or another radio interferometer) depends on the true size of the source, the dimensions of the synthetic beam, and the intensity of the source. We can think of the latter quantity in terms of the dynamic range or the S/N ratio. When giving the size of the synthetic beam (as in Tables \ref{tab:extended}--\ref{tab:aca}) we give the FWHM along the major and minor axes. For very bright sources, the apparent size of the emission when considering the $3\sigma$ or $5\sigma$ contours, where $\sigma$ is the rms noise, can be much larger than the beam, even if the source itself is smaller than the beam. If we consider a Gaussian distribution of flux along one of the beam axes:
\begin{equation}\label{eq:gauss}
F(r) = F_\mathrm{peak} \exp\left(-\editss{b}\frac{r^2}{\theta^2}\right)
\end{equation}
where $F_\mathrm{peak}$ is the peak flux, $r$ is the offset from the centre of the beam, $\theta$ is the FWHM of the beam along the major (or minor) axis, \editss{and $b=4\log_e(2)$, chosen to account for the fact that $\theta$ is related to the Gaussian standard deviation, $\sigma_G$, by $\theta = 2\sqrt{2 \log_e} \sigma_G$}. We can then calculate that, for a point source, the $N\sigma$ contour, where $\sigma$ is the rms noise, will lie at an offset $r$ that depends on $N$ and the dynamic range of the observation, $R_\mathrm{SNR}$. In this case, $F_\mathrm{peak} = \sigma R_\mathrm{SNR}$ and we are interested in $F(r) = N\sigma$. Then equation \ref{eq:gauss} can be rearranged to find
\editss{\begin{equation}
%N \sigma &=& \sigma R_\mathrm{SNR} \exp\left(-\frac{r^2}{2\theta^2}\right)\\
r =  {\theta} \sqrt{\frac{\log_e \left(\frac{R_\mathrm{SNR}}{N}\right)}{b}} = 0.60\,{\theta} \sqrt{{\log_e \left(\frac{R_\mathrm{SNR}}{N}\right)}}
\end{equation}}
}

\edits{
From this we can see that for a very bright point source (e.g. large $R_\mathrm{SNR} = 1000$), the $5\sigma$ contour would be located at an offset of \editss{$r=1.3\theta$ from the centre of the emission. Such a source would appear to have a diameter of $2.6\theta$.} Thus, we cannot conclude that such a source is resolved if the $5\sigma$ contour \editss{ has a diameter of $\sim2.6\theta$. This effect can be significant even for more modest dynamic ranges; for example, for $R_\mathrm{SNR}=15$, the $3\sigma$ contour would have a diameter of around $1.5\theta$.}
}

\section{Mass-loss rates}\label{app:massloss}

There has not yet been a single consistent study of mass-loss rates covering the entire ATOMIUM sample. To allow us to make some comparisons between mass-loss rate and other star properties, we have collected mass-loss rates from the literature for the ATOMIUM AGB stars. To ensure some uniformity, we have exclusively used mass-loss rates derived through radiative transfer modelling of CO line emission. We have also scaled the literature values to account for our adopted distance, i.e.
\begin{equation}
\dot{M} = \dot{M}_\mathrm{lit} \left( \frac{D}{D_\mathrm{lit}} \right)^2 
\end{equation}
where $\dot{M}$ is the mass-loss rate and $D$ is our adopted distance, given in Table~\ref{tab:stars}. Note that this scaling is most valid for sources with optically thin CO lines and that such scaling was not necessary for \pigru\ or IRC+10011, for which the literature distances match those used here. The mass-loss rate values and their literatures sources are given in Table \ref{tab:massloss}.

\begin{figure}
	\includegraphics[width=0.5\textwidth]{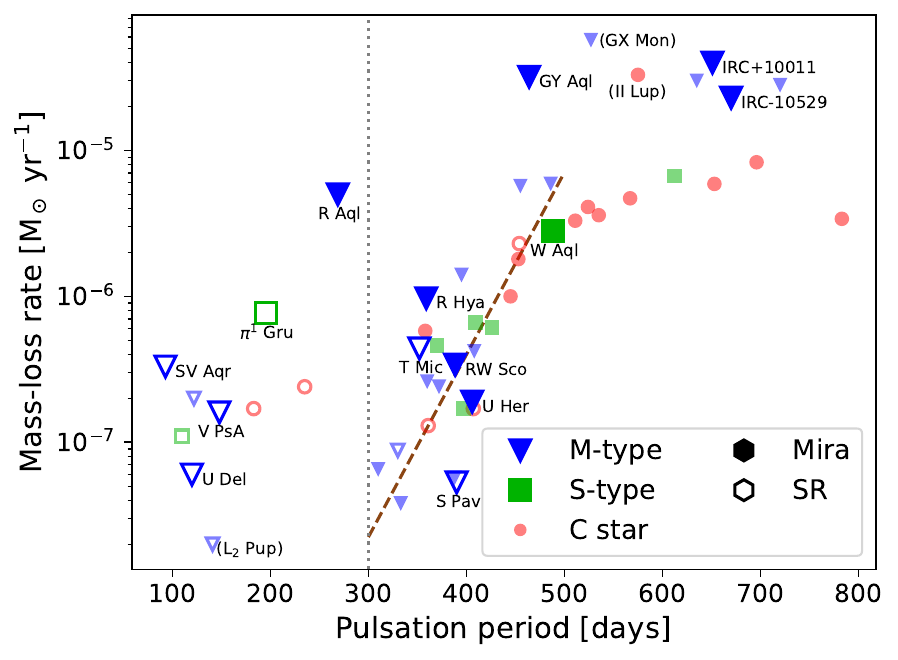}
    \caption{The mass-loss rates plotted against periods for the AGB stars in the ATOMIUM sample (large markers, individually labelled) and the SUCCESS sample (fainter, small markers). 
    %Mira variables are plotted as filled markers, while semiregular variables (SRa and SRb) are plotted as unfilled markers. M-type stars are plotted as blue triangles, S-type as green squares, and carbon-rich stars as red circles. 
    The dotted vertical line is plotted at a period of 300 days. The brown dashed line shows the period--mass-loss rate relationship found by \cite{Wood1990}, for periods between 300 and 500 days.}
    \label{fig:period-ml}
\end{figure}

We used the same method to scale the mass-loss rates from the SUCCESS sample plotted in Fig.~\ref{fig:period-ml}. In that case, the distances we used came from \cite{Andriantsaralaza2022}, if available, and \cite{Bailer-Jones2021} if possible. Where the star didn't appear in either of these studies (V701 Cas) we retained the distance and mass-loss rate given in \cite{Danilovich2015a}.

%For R~Hya, 

\begin{table}
	\centering
	\caption{Mass-loss rates from the literature for the ATOMIUM sample}
	\label{tab:massloss}
	\begin{tabular}{lcc}
		\hline
Star	&	Mass-loss rate	&	Reference 	\\
& [$\spy$] &\\
\hline
GY Aql	&	$3.2\e{-5}$	&	\cite{Wannier1986}	\\
R Aql	&	\edits{$5.0\e{-6}$}	&	\cite{Bergman2020}	\\
IRC$-$10529	&	$2.3\e{-5}$	&	\cite{Danilovich2015a}	\\
W Aql	&	$2.8\e{-6}$	&	\cite{Ramstedt2017}	\\
SV Aqr	&	$3.3\e{-7}$	&	\cite{Olofsson2002}	\\
U Del	&	$6.0\e{-8}$	&	\cite{Olofsson2002}	\\
$\pi^1$ Gru	&	$7.7\e{-7}$	&	\cite{Doan2017}	\\
U Her	&	\edits{$1.9\e{-7}$}	&	\cite{Bergman2020}	\\
R Hya	&	\edits{$9.7\e{-7}$}	&	\cite{Bergman2020}	\\
T Mic	&	$4.4\e{-7}$	&	\cite{Olofsson2002}	\\
S Pav	&	$5.3\e{-8}$	&	\cite{Olofsson2002}	\\
IRC+10011	&	$4.0\e{-5}$	&	\cite{Danilovich2017a}	\\
V PsA	&	$1.6\e{-7}$	&	\cite{Olofsson2002}	\\
RW Sco	&	$3.4\e{-7}$	&	\cite{Groenewegen1999}	\\
		\hline
	\end{tabular}
\tablefoot{Mass-loss rates have been scaled to account for the differences between the source paper and the distances used here, with details given in Appendix \ref{app:massloss}.}
\end{table}

\subsection{Notes on a few individual SUCCESS stars in light of binary-enhanced dust formation}

An interesting example is the dusty but relatively low mass-loss rate AGB star L$_2$~Pup (the lowest mass-loss rate source in Fig.~\ref{fig:period-ml}). \cite{Kervella2016} found a low mass of $\sim12~M_\mathrm{Jup}$ for the companion, which is embedded in a dusty, nearly edge-on disc \citep{Kervella2015,Ohnaka2015}. The dust around \ltpup\ is not distributed spherically, suggesting that the companion has strongly influenced its distribution. However, with a low mass, the gravitational potential of the companion is low, and it will cause weaker shocks as it orbits the AGB star \citep{Maes2021}, contributing to less dust formation than if it were more massive. This explains why the mass-loss rate of \ltpup\ remains low ($2\e{-8}\spy$, \citealt{Danilovich2015a} and Appendix \ref{app:massloss}).

We also note that the carbon star with the highest mass-loss rate in the SUCCESS sample, comparable to the aforementioned oxygen-rich stars, is II Lup, around which \cite{Unnikrishnan2024} recently tentatively identified SiN, which suggests the presence of a binary companion \citep{Van-de-Sande2022}. The oxygen-rich star with the highest mass-loss rate in the SUCCESS sample is GX~Mon, around which earlier observations have shown a circumstellar spiral, also indicative of a binary companion \citep{Randall2020}.

\section{Hydrodynamical model}\label{app:hydro}

\begin{figure*}
	\includegraphics[width=0.33\textwidth]{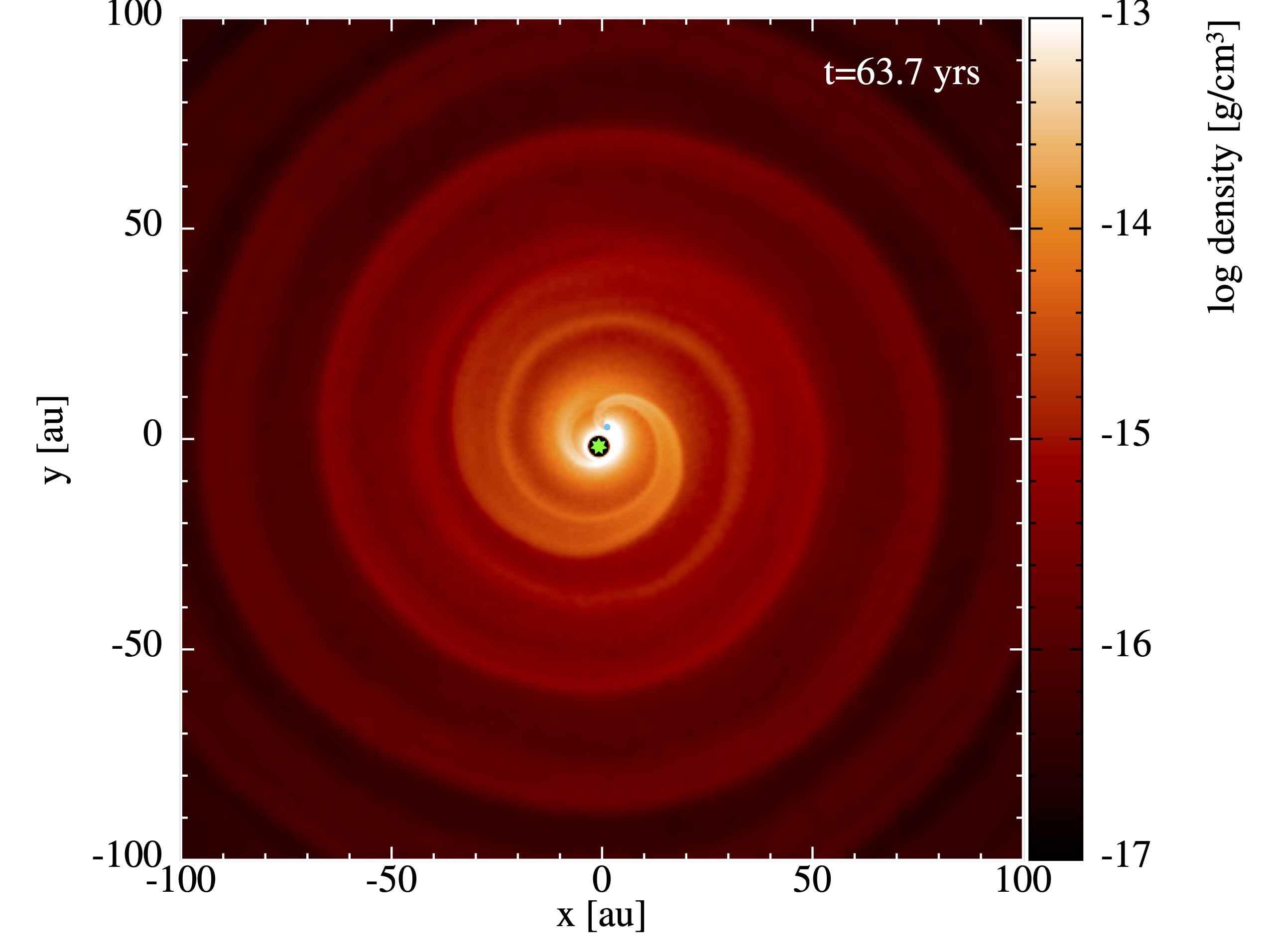}
	\includegraphics[width=0.33\textwidth]{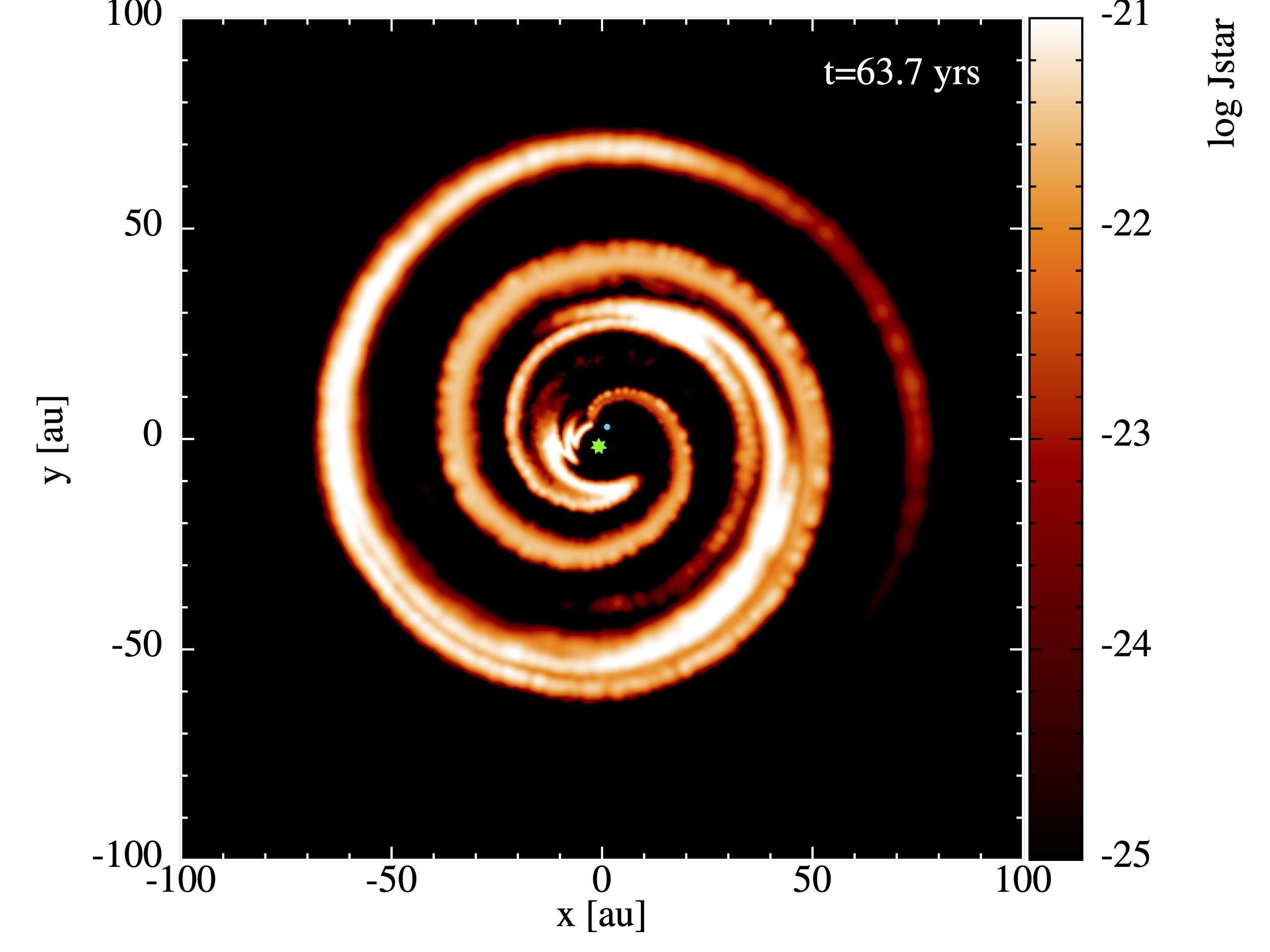}
	\includegraphics[width=0.33\textwidth]{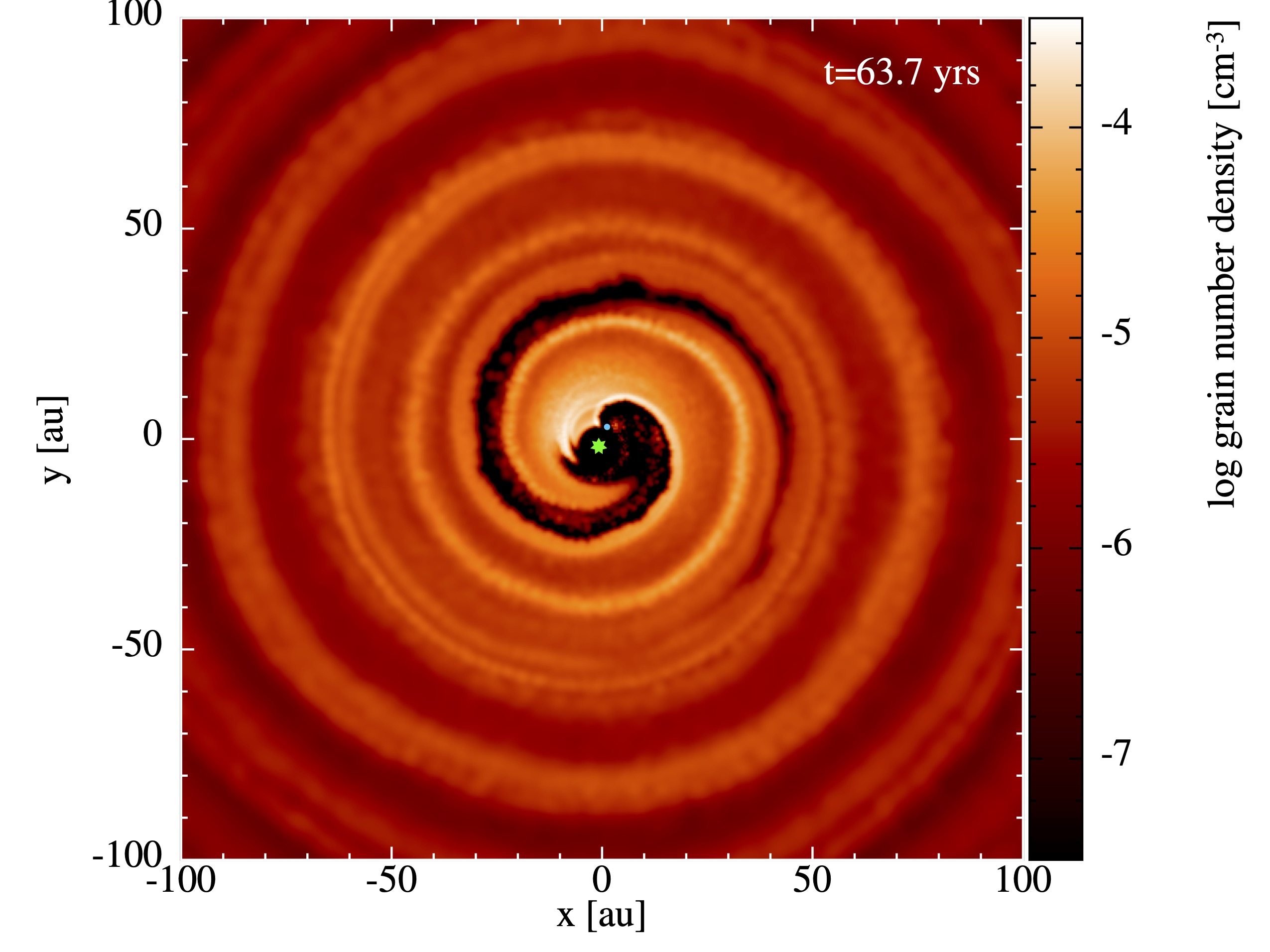}
    \caption{A snapshot of our hydrodynamical simulation. Each panel plots quantities for a slice through the orbital plane ($z=0$) with a thickness of 2~au in the $z$ direction. The position of the AGB star is indicated by the central green star the companion by the blue dot above and to the right of it. The companion is moving in an anti-clockwise direction relative to the AGB star. \textit{Left:} Total density distribution. \textit{Centre:} The normalised rate of dust nucleation (see details in \citealt{Siess2022} and \citealt{Bermudez-Bustamante2024}). \textit{Right:} The number density of dust grains with masses $\geq 2\e{-20}$~g formed during the simulation.}
    \label{fig:hydro}
\end{figure*}

\begin{table}
	\centering
	\caption{SPH model parameters}
	\label{tab:sph}
	\begin{tabular}{lc}
		\hline
Parameter	&	Value 	\\
\hline
AGB mass & 1.6~$\msol$	\\
Companion mass & 1~$\msol$	\\
Semi-major axis & 5~au\\
Eccentricity & 0\\
Mass-loss rate & $3\e{-5}\spy$\\
Wind injection radius & 2.7~au\\
Wind injection velocity & 5~\kms \\
AGB C/O ratio & 1.4\\
SPH particle mass (resolution) & $2.4\e{-9}~\msol$\\
Adiabatic index & 1.2\\
		\hline
	\end{tabular}
\end{table}

To check how a companion star impacts dust formation around an AGB star, we performed three-dimensional hydrodynamic simulations using Phantom \citep{Price2018}, a smoothed particle hydrodynamics code. Dust formation is implemented into Phantom as described by \cite{Siess2022}, and was tested by \cite{Bermudez-Bustamante2024} in the context of common envelope evolution. Here we test how wind-binary interaction impacts dust formation for a system where the two stars do not directly interact. We show a single test model because a full parameter study is beyond the scope of the present work and will be presented in a forthcoming paper (Samaratunge et al, in prep).

Our model system parameters are given in Table~\ref{tab:sph}. In summary, we simulated a $1.6~\msol$ AGB star with a $1~\msol$ Sunlike companion in a circular orbit with a semimajor axis of 5~au. The AGB mass-loss rate is $3\e{-5}\spy$, comparable to our three highest mass-loss rate sources (GY Aql, IRC$-$10529 and IRC+10011, see Appendix~\ref{app:massloss} and Table~\ref{tab:massloss}), which all show spiral-like structures in their winds \citep{Decin2020}. The wind is injected at a radius of 2.7~au, comparable to the radii of some of our stars (e.g. see Fig.~\ref{fig:radii}).  The wind was injected with a velocity of 5~\kms\ and accelerated to 15--20~\kms\ using the dust opacity to accelerate the wind on top of the free wind approximation \citep{Siess2022}. The dust formation module we used currently only supports carbonaceous dust through a chemical reaction network described in \cite{Siess2022}. In principle, the dust begins to form when the conditions for nucleation are met \citep[as described in][]{Siess2022,Bermudez-Bustamante2024}, following the chemical nucleation network. We would expect broadly similar behaviour from oxygen-rich silicate dust, and its formation will be implemented in a future release of Phantom. To allow carbonaceous dust to form in our test model, we set a C/O ratio for the AGB star of 1.4, based on that of CW Leo found by \cite{Winters1994}. No pulsations are included in the model, meaning companion-induced dust formation can be studied in isolation. We consider adiabatic cooling, but no other cooling terms \citep[e.g. H\textsc{i} cooling, as described in][was not included in this model]{Malfait2024}.

The model was run for almost 64 years, allowing $\sim 9$ orbits of the binary system.
In Fig.~\ref{fig:hydro} we plot a slice through the orbital plane of the system showing the gas density, a measure of the dust nucleation rate and a measure of the dust mass formed. Details of the latter two quantities are described in \cite{Siess2022} and \cite{Bermudez-Bustamante2024}. The density plot shows a spiral structure, as commonly seen for binaries in AGB winds \citep[e.g.][]{Kim2019,Maes2021,Malfait2021}. The normalised dust nucleation rate, $\hat{J_*}$, which is a measure of the number of seed particles or critical clusters formed, follows the pattern of the shocks induced by the orbital motion of the stars. The dust does not form immediately in the vicinity of the stars, but does start forming from a distance of 5~au from the AGB star. The shocked regions are denser and hence have higher rates of dust formation, since chemical reactions tend to scale with the square of the number density. This is illustrated in the right-most plot in Fig.~\ref{fig:hydro}, where we plot the number density of fully-fledged dust grains, showing that the dust distribution also follows the spiral pattern seen in the density plot. A dust grain is defined as a cluster made up of at least 1000 monomers (carbon atoms, in this case) and corresponds to a grain mass of $\geq 2\e{-20}$~g. We find a companion-induced dust production rate on the order of $2\e{-9}\spy$, which is significant in comparison to total dust production rate estimates \citep[e.g.][]{Scicluna2022}.

In summary, in a hydrodynamic model where dust was allowed to form when the correct conditions were met, we found that dust preferentially formed in the wake of shocks caused by the orbital motion of the stars. This result is in general agreement with our observations, particularly of \pigru\ (see Sect.~\ref{sec:pigrudis}) and W~Aql (Sect.~\ref{sec:waqldiscussion}). For the other stars, we would not necessarily expect to resolve a companion in our observations. Considering the distances in Table~\ref{tab:stars} and our extended observations, for the example of a circular face-on orbit with a 5~au separation, we would only expect to resolve a companion for \pigru, and marginally for T~Mic, R~Hya and S~Pav. For less optimal orbital parameters (e.g. an inclined orbit causing the projected separation of the companion to be closer to the AGB star) we would not expect to resolve such a companion. Our combined images are more sensitive than extended, but generally have lower resolutions, making potential close companions more difficult to detect directly. Hence the absence of such detections does not imply that companions are not present for some of our stars and may be contributing to dust formation, as shown in our model.

%\red{concluding remarks}

% wind injection velocity is 5 km/s, accelerates up to 20 (but some particles remain slower)

\end{appendix}
\end{document}